\documentclass[sigconf,balance=false]{acmart}
\usepackage{popets}
\usepackage{makecell}
\usepackage{booktabs}
\usepackage{soul}

\setcopyright{popets}
\copyrightyear{YYYY}

\acmYear{YYYY}
\acmVolume{YYYY}
\acmNumber{X}
\acmDOI{XXXXXXX.XXXXXXX}
\acmISBN{}
\acmConference{Proceedings on Privacy Enhancing Technologies}
\usepackage[utf8]{inputenc} %
\usepackage[T1]{fontenc}    %
\PassOptionsToPackage{hyphens}{url}\usepackage{hyperref}
\usepackage{booktabs}       %
\usepackage{amsfonts}       %
\usepackage{nicefrac}       %
\usepackage{color}
\usepackage{amsmath}
\usepackage{amsthm}
\usepackage{bm}
\usepackage{graphicx}
\usepackage{textcomp}
\usepackage{xcolor}
\usepackage{float}
\usepackage{threeparttable, tablefootnote}
\usepackage{epstopdf}
\usepackage{multicol}
\usepackage{multirow}
\usepackage{subfigure}
\usepackage{bbm}
\usepackage[bottom,hang]{footmisc}
\usepackage{adjustbox}
\usepackage{wrapfig}
\usepackage[labelfont=bf, justification=justified]{caption}
\usepackage[normalem]{ulem}
\usepackage{arydshln}
\usepackage[linesnumbered,lined,vlined,ruled]{algorithm2e}
\usepackage{xpatch}
\usepackage{enumitem}

\def\eqref#1{equation~\ref{#1}}

\def\1{\bm{1}}

\def\vmu{{\bm{\mu}}}

\def\vm{{\bm{m}}}

\def\vp{{\bm{p}}}
\def\vq{{\bm{q}}}

\def\vv{{\bm{v}}}

\def\vx{{\bm{x}}}

\def\vz{{\bm{z}}}

\def\mI{{\bm{I}}}

\def\mQ{{\bm{Q}}}

\def\mW{{\bm{W}}}
\def\mX{{\bm{X}}}

\def\mSigma{{\bm{\Sigma}}}

\DeclareMathAlphabet{\mathsfit}{\encodingdefault}{\sfdefault}{m}{sl}
\SetMathAlphabet{\mathsfit}{bold}{\encodingdefault}{\sfdefault}{bx}{n}

\def\gA{{\mathcal{A}}}

\def\gC{{\mathcal{C}}}
\def\gD{{\mathcal{D}}}

\def\gM{{\mathcal{M}}}
\def\gN{{\mathcal{N}}}

\def\gQ{{\mathcal{Q}}}

\def\gS{{\mathcal{S}}}

\newcommand{\Ls}{\mathcal{L}}

\DeclareMathOperator*{\argmin}{arg\,min}

\theoremstyle{definition}
\newtheorem{theorem}{Theorem}[section]
\newtheorem{definition}{Definition}[section]

\xpretocmd{\proof}{\setlength{\parindent}{0pt}}{}{}

\newcommand{\myparagraph}[1]{
\vspace{0.1cm}\noindent
\textbf{#1.}
}

\newcommand{\rongauss}[1][ ]{\texttt{RON-Gauss}#1}
\newcommand{\privsyn}[1][ ]{\texttt{PrivSyn}#1}
\newcommand{\pgm}[1][ ]{\texttt{Private-PGM}#1}
\newcommand{\gan}[1][ ]{\texttt{GAN}#1}
\newcommand{\vae}[1][ ]{\texttt{VAE}#1}

\def\epfive{$\varepsilon=5$}

\def\epfifty{$\varepsilon=50$}
\def\ephundred{$\varepsilon=100$}

\renewcommand{\sectionautorefname}{Section}
\renewcommand{\figureautorefname}{Fig.}

\setitemize{topsep=0pt,itemsep=0pt,leftmargin=*}
\setenumerate{topsep=0pt,itemsep=0pt,leftmargin=*}

\begin{document}
\title[]{Towards Biologically Plausible and Private\\Gene Expression Data Generation}


\author{Dingfan Chen}
\orcid{0000-0001-7279-6624}
\affiliation{%
  \institution{CISPA Helmholtz Center for Information Security}
  \city{Saarbrücken}
  \country{Germany}}
\email{dingfan.chen@cispa.de}
\authornote{Equal contribution}

\author{Marie Oestreich}
\orcid{0000-0002-4754-1301}
\affiliation{%
  \institution{German Center for Neurodegenerative Diseases (DZNE)}
  \city{Bonn}
  \country{Germany}}
\email{marie.oestreich@dzne.de}
\authornotemark[1]

\author{Tejumade Afonja}
\orcid{0000-0003-0639-9668}
\affiliation{%
  \institution{CISPA Helmholtz Center for Information Security}
  \city{Saarbrücken}
  \country{Germany}}
\email{tejumade.afonja@cispa.de}
\authornotemark[1]

\author{Raouf Kerkouche}
\orcid{0000-0002-1458-7805}
\affiliation{%
  \institution{CISPA Helmholtz Center for Information Security}
  \city{Saarbrücken}
  \country{Germany}}
\email{raouf.kerkouche@cispa.de}

\author{Matthias Becker}
\orcid{0000-0002-7120-4508}
\affiliation{%
  \institution{German Center for Neurodegenerative Diseases (DZNE)}
  \city{Bonn}
  \country{Germany}}
\email{matthias.becker@dzne.de}
\authornote{Joint last authorship}

\author{Mario Fritz}
\orcid{0000-0001-8949-9896}
\affiliation{%
  \institution{CISPA Helmholtz Center for Information Security}
  \city{Saarbrücken}
  \country{Germany}}
\email{fritz@cispa.de}
\authornotemark[2]


\renewcommand{\shortauthors}{chen, oestreich, afonja et al.}

\begin{abstract}
Generative models trained with Differential Privacy (DP) are becoming increasingly prominent in the creation of synthetic data for downstream applications. Existing literature, however, primarily focuses on basic benchmarking datasets and tends to report promising results only for elementary metrics and relatively simple data distributions.
In this paper, we initiate a systematic analysis of how DP generative models perform in their natural application scenarios, specifically focusing on real-world gene expression data. We conduct a comprehensive analysis of five representative DP generation methods, examining them from various angles, such as downstream utility, statistical properties, and biological plausibility.

Our extensive evaluation illuminates the unique characteristics of each DP generation method, offering critical insights into the strengths and weaknesses of each approach, and uncovering intriguing possibilities for future developments. Perhaps surprisingly, our analysis reveals that most methods are capable of achieving seemingly reasonable downstream utility, according to the standard evaluation metrics considered in existing literature.
Nevertheless, we find that none of the DP methods are able to accurately capture the biological 
characteristics of the real dataset. This observation suggests a potential over-optimistic assessment of current methodologies in this field and underscores a pressing need for future enhancements in model design.
\end{abstract}


\maketitle
     
\section{Introduction}

Genomic data is considered a goldmine for medical researchers, enabling them to tackle a wide array of challenges. These challenges range from identifying patients at risk of specific diseases, to developing tailored drugs to enhance treatment reliability and reduce care duration. Gene expression data stands as one of the most extensively utilized forms of genomic data. More specifically, in a cell, the instructions on how to build the cell's proteins are encoded in the DNA as genes. In order to produce proteins, copies of the genes are made from the DNA in the form of messenger RNAs (mRNAs) which are then translated into proteins. The more mRNA copies are made of a gene, the more of the corresponding protein can be produced. Conditions such as environmental stimuli or diseases can alter the kind and quantity of proteins that are being produced. Thus, the cell's response to such conditions is reflected in the transcription of genes, i.e., the strength of their expression.  Measuring gene expression has therefore become an essential biomedical tool in order to understand how a cell, tissue or organism responds to the conditions it is exposed to ~\cite{Wang_Gerstein_Snyder_2009, Conesa}.

Nevertheless, the use of gene expression data is not without danger, as it can threaten patient privacy ~\cite{Oestreich_Chen_Schultze_Fritz_Becker_2021}. The precise nature of the information it contains could attract the interest of malicious entities, capable of exploiting it for multiple purposes. For example, an insurance company could choose to raise the coverage cost for a patient predisposed to a serious illness. Additionally, publishing information about a person's genetic predispositions for stigmatized diseases can severely impact their social life and societal acceptance.

In light of these concerns, there arose a need to protect individual privacy and avoid such problems, leading to exploration of methods that are able to generate synthetic data backed by rigorous privacy guarantees. Such approaches involve creating synthetic datasets that reflect the characteristics of real gene expression data while providing strong theoretical differential privacy (DP) guarantees. Nonetheless, employing DP entails introducing randomness during the training process, which inevitably compromises the quality of the produced synthetic data. Furthermore, as we strive for stronger privacy guarantees, the randomness required for privacy increases proportionally, further affecting the quality of the synthetic data to a larger extent. This underscores the well-known trade-off between privacy and utility.

Despite significant advances in DP data generation methods that report both good generation quality and privacy guarantees, the majority of quality assessment have unfortunately focused solely on downstream utility. A notable gap persists in evaluations that overlook the preservation of essential statistical and biological characteristics. These characteristics are, however, crucial for ensuring the fidelity and applicability of the generated data. In real-world scenarios, the challenge becomes even more pronounced due to the vast feature space inherent in gene expression data, which stands in stark contrast to the often limited number of available samples.  Consequently, the effectiveness of existing methods, previously tested primarily 
on basic benchmark datasets with relatively simple distributions, remains unclear when applied to real-world gene expression data.

In this paper, we fill this gap by presenting the first systematic quality assessment of synthetic gene expression data produced by five benchmark DP generation models with diverse characteristics. Our assessment encompasses five metrics, spanning various aspects from downstream utility, statistical fidelity, to biological plausibility. Our extensive experimental results reveal intriguing findings: (1) significant privacy risks do exist if the generative models are trained non-privately, while DP training (even with a high privacy budget of $\varepsilon=100$) greatly mitigates such risks; 
(2) almost all methods manage to achieve seemingly near-perfect performance in terms of standard utility metrics while providing a reasonably strong privacy guarantee (e.g., $\varepsilon\leq 10$), yet none of the DP models succeed in  producing biologically plausible data.

In summary, the key contributions of our study are outlined below:
\begin{itemize}
    \item Our work presents the first comprehensive and systematic analysis of DP generation methods applied to real-world gene expression data. Our extensive investigation encompasses five diverse generation models, five metrics targeting three principal aspects, providing the first comprehensive view for the current state of real-world applicability of DP generation methods.
    
    \item  Our analysis reveals crucial insights, highlighting  the limitations of existing evaluations that predominantly focus on a single aspect, namely, downstream utility. In contrast, our thorough assessment establishes a reliable evaluation framework that effectively addresses the misconceptions arising from these one-dimensional evaluations.
    
    \item Our compelling findings, complemented by an in-depth discussion, offer fresh perspectives for the future development in the related field. With our systematic assessment, we aim to steer DP generation methods towards improved practicality in real-world applications involving sensitive data.
\end{itemize}

\section{Related Work}
\subsection{Models for Synthetic Gene Expression Data}

Various types of generative models have been employed for generating synthetic gene expression data. Variational autoencoders and deep Boltzmann machines have been used to generate data that aids in designing studies and planning analysis for large experiments~\cite{Treppner_Salas-Bastos_Hess_Lenz_Vogel_Binder_2021}. Generative adversarial networks have been exploited for generating gene expression data to combat the challenges of low sample sizes via data augmentation, which is specifically motivated by the unfavorable ratio of samples to features in these datasets~\cite{Lall_Ray_Bandyopadhyay_2022, Marouf_Machart_Bansal_Kilian_Magruder_Krebs_Bonn_2020}. Additionally,  synthetic gene expression data has also been used to train imputation methods for handling missing data~\cite{Pandey_Onkara_2023}. However, none of these methods ensure privacy during the whole data generation process. Given that genome-related data, including the gene expression data, is highly privacy-sensitive~\cite{Oestreich_Chen_Schultze_Fritz_Becker_2021}, applying existing works in real-world scenarios becomes challenging due to privacy regulations. 


To the best of our knowledge, there is a lack of research delving into the differentially private generation of synthetic gene expression data. While some studies, like \cite{torfi2022differentially}, have investigated the private generation of synthetic data within the realm of medical data at large, a dedicated focus on gene expression data remains notably absent.


\subsection{Measuring Quality of Synthetic Gene Expression Data}
A variety of methods have been applied in the past to assess the quality of synthetic gene expression data from a biological standpoint. These methods have been used both in the context of bulk as well as single-cell RNA-seq data. Bulk RNA-sequencing refers to the process of sequencing the mRNA transcripts from a sample containing a collection of many cells~\cite{Li_Wang_2021}. The resulting data thus reports the average expression strength of each gene across these cells. Single-cell RNA-sequencing on the other hand, first separates the cells present in the sample before sequencing each individually, generating an expression profile at cell resolution rather than sample resolution~\cite{Li_Wang_2021,Tang_Barbacioru_Wang_Nordman_Lee_Xu_Wang_Bodeau_Tuch_Siddiqui_et_al}. The methods used for evaluating this data comprise the comparison of expression data distributions~\cite{Assefa_Vandesompele_Thas_2020, Treppner_Salas-Bastos_Hess_Lenz_Vogel_Binder_2021, Zappia_Phipson_Oshlack_2017} by looking at mean and median expressions, proportion of zero counts (in single-cell cases) and coefficients of variation. Also, metrics related to functional biology have been applied~\cite{Viñas_Andrés-Terré_Liò_Bryson_2022, Treppner_Salas-Bastos_Hess_Lenz_Vogel_Binder_2021, Lall_Ray_Bandyopadhyay_2022, Pandey_Onkara_2023, Marouf_Machart_Bansal_Kilian_Magruder_Krebs_Bonn_2020}, including preservation of gene-gene correlations, gene ontology terms, differentially expressed genes and clusters in reduced dimensional space, using for example t-SNE, PCA, UMAP or after feature selection.

\section{Preliminaries}
\subsection{Threat Model}
\label{sec:threat_model}

The objective of an adversary is to infer private information about individuals in the training datasets by launching various privacy attacks, such as membership inference attack (MIA), which aims to ascertain if a particular data point was used in training the dataset.

We consider two common scenarios for synthetic data generation from an attack standpoint:
\begin{itemize}
\item A trained generator generates the synthetic data (e.g., \sectionautorefname~\ref{sec:rongauss}-\ref{sec:pgm}). In this case, the adversary can have either black-box access or white-box access to the generator. Black-box access means the adversary can only access the synthetic data generated by querying the model through an API. White-box access allows the adversary to access the generator's internal state, including its parameters.

\item The synthetic data is directly generated without using any generator (e.g., \sectionautorefname~\ref{sec:privsyn}). In this scenario, the adversary only has access to the synthetic data.
\end{itemize}

While our privacy model protects against the most powerful adversaries, as discussed below in Section \ref{subsec:privacy_model}, our experiments consider the scenario with the most knowledgeable adversary who has white-box access to the trained generator, as well as the practical scenario where only the synthetic data is accessible.  

\subsection{Privacy Model}
\label{subsec:privacy_model}
We aim to develop a solution that protects against potential attacks as delineated in our threat model in \sectionautorefname~\ref{sec:threat_model}. Specifically, we adopt differential privacy (DP), which ensures the difficulty to infer the presence of any record in the training dataset, even when the adversary has white-/black-box access to the trained generator and/or to the synthetic data. As a result, any potential negative impact on an individual's privacy cannot be attributed to their involvement in the training phase (up to $\varepsilon$ and $\delta$). For instance, if an insurance company accesses the generator or the synthetic data (from DP generation methods) and decides to increase an individual's insurance premium, such a decision cannot be attributed to the individual's data presented in the training dataset.

\begin{definition}[$(\varepsilon,\delta)$-DP~\cite{dwork2014algorithmic}] 
\label{def:DP} 
A randomized mechanism $\mathcal{M}$ with range $\mathcal{R}$ is $(\varepsilon,\delta)$-DP, if 
\begin{equation*}
    \Pr[\mathcal{M}(\gD)\in \mathcal{O}] \leq e^\varepsilon \cdot \Pr[\mathcal{M}(\gD')\in \mathcal{O}]+\delta 
\end{equation*}
holds for any subset of outputs $\mathcal{O}\subseteq \mathcal{R}$ and for any adjacent datasets $\gD$ and $\gD'$, where
$\gD$ and $\gD'$ differ from each other by adding or removing one training example, i.e., $\gD' = \gD \cup \{x\}$ or $\gD = \gD' \cup \{x\}$ for a data sample $x$.
The privacy parameter $\varepsilon$ is the upper bound of privacy loss, and $\delta$ is the probability of breaching DP constraints. Smaller values of both $\varepsilon$ and $\delta$ translate to stronger DP guarantees and better privacy protection.
\end{definition}

\begin{definition}[Gaussian Mechanism~\cite{dwork2014algorithmic}]
Let $f: X \rightarrow \mathbb{R}^d$ be an arbitrary $d$-dimensional function with $L_2$-(global) sensitivity $\Delta^2 _f$:
\begin{equation}
    \Delta^2_f= \max_{\gD,\gD'} \Vert f(\gD)-f(\gD')\Vert_2
\end{equation}
The Gaussian Mechanism  $\mathcal{M}_\sigma$, parameterized by $\sigma$, adds noise into the output, i.e.,
    \begin{equation}
  \mathcal{M}_\sigma(\vx) = f(\vx) + \mathcal{N}(0,\sigma^2 
  \mI).
\end{equation}
$\gM_\sigma$ is $(\varepsilon,\delta)$-DP for $\sigma\geq \sqrt{2\ln{(1.25/\delta)}}\Delta^2_ f/\varepsilon$.

\label{def:gaussian_mechanism}
\end{definition}

\begin{theorem}[Post-processing Theorem~\cite{dwork2014algorithmic}]
\label{theorem:post-processing}
 If $\mathcal{M}$  satisfies $(\varepsilon,\delta)$-DP, $F\circ \mathcal{M}$ will satisfy $(\varepsilon,\delta)$-DP for any data-independent function $F$ with $\circ$ denoting the composition operator.\end{theorem}
The post-processing theorem guarantees that 
if a DP generation model is $(\varepsilon,\delta)$-DP, releasing the trained generator and the synthetic dataset will also be privacy-preserving, with the privacy cost bounded by $\varepsilon$ and $\delta$.

\subsection{Biological Criteria}
\myparagraph{Differential Expression}
When diseases and other pathological conditions affect the body, they can alter gene activation within cells, contributing to the manifestation of symptoms. The specific set of genes whose expression levels vary from one disease to another are commonly referred to as \textit{differentially expressed (DE) genes}. Identifying DE genes that distinguish between two conditions is a fundamental step in gene expression analysis~\cite{Anders_Huber_2010, Costa-Silva, Ritchie_2015, Robinson_McCarthy_Smyth_2010}. Differential expression can occur as either \textit{up-regulation} or \textit{down-regulation}, meaning that the expression of genes is significantly \textit{increased} or \textit{decreased} in one condition compared to another, respectively (see \sectionautorefname~\ref{sec:biological_metrics_description} for the formal definition).

\myparagraph{Gene Co-Expression}
Genes that are involved in the same biological pathways often form a functional group or \emph{module}, meaning they collectively respond to a condition by similar changes in the expression strength. 
For example, all genes involved in fighting off a bacterial infection will be activated together when such a pathogen enters the body. Such genes are referred to as \emph{co-expressed}. In order to identify activated or inactivated biological pathways, detecting such modules of co-expressed genes is a common step in the analysis of gene-expression data~\cite{Oestreich_Holsten_Agrawal_Dahm_Koch_Jin_Becker_Ulas_2022, Langfelder_Horvath_2008}. Specifically, co-expression between a pair of genes with indices $j$ and $k$ is quantified using their \textit{Pearson correlation coefficient} $r_{jk}$ with
\begin{align}
r_{jk} = \frac{ \sum_{i=1}^{n}(x^{(i)}_j-\bar{x}_j)(x^{(i)}_k-\bar{x}_k) }{%
        \sqrt{\sum_{i=1}^{n}(x^{(i)}_j-\bar{x}_j)^2}\sqrt{\sum_{i=1}^{n}(x^{(i)}_k-\bar{x}_k)^2}},
\label{coexpression_equation}
\end{align}
where $x^{(i)}_j$ and $x^{(i)}_k$ are the expression values of genes $j$ and $k$ in sample $i$, respectively, while $\bar{x}_j$ and $\bar{x}_k$ are the mean expression values of the two genes across $n$ biological samples.
Groups of genes with high Pearson correlation coefficients are considered $modules$ of co-expressed genes, with $r_{jk}>0.7$ are typically considered as biologically significant co-expressions.

\section{Models}
Given the real dataset $\gD=\{(\vx^{(i)},y^{(i)})\}_{i=1}^n$ consisting of $n$ samples $(\vx^{(i)},y^{(i)})$ with $\vx^{(i)} \in \mathbb{R}^d$ and $y^{(i)} \in \{1, ...,C\}$ denoting the features and class labels respectively, the objective of the generation methods is to capture the real underlying distribution $p(\vx,y)$ and generate synthetic data samples $(\tilde{\vx},\tilde{y})$ that mimic the statistical characteristics of  the real samples from $\gD$. In our case, the feature vector $\vx^{(i)}$ represents the gene expression level and the class label $y^{(i)}$ corresponds to the disease type, with $d$ and $C$ denoting the feature dimension and number of label classes, respectively.

In this work, we explore the most prominent categories of (DP) generation methods found in the literature: (1) \textit{density estimation (probability distribution fitting)}, (2) \textit{graphical models-based} methods, (3) \textit{marginal-based} methods, and (4) \textit{deep generative models}. A summary of these methods and their diverse characteristics can be found in \tableautorefname~\ref{tab:model_summary}.

\begin{table}[!htbp]
\aboverulesep=0ex
\belowrulesep=0ex
    \centering
    \resizebox{\columnwidth}{!}{
    \begin{tabular}{l|ccc}
    \toprule
    Method & Category & Attribute type & DP sanitization \\
    \midrule
    \rongauss & Density estimation & continuous only & one-shot\\
    \vae & Deep generative model & continuous & iterative \\
    \gan & Deep generative model &  continuous & iterative\\
    \pgm & Graphical model & discrete only & one-shot \\
    \privsyn & Marginal & discrete only & one-shot\\
    \bottomrule
    \end{tabular}
    }
    \vspace{.5cm}
    \caption{Summary of Models.}
    \label{tab:model_summary}
    \end{table}

\vspace{-24pt}
\subsection{RON-Gauss}
\label{sec:rongauss}
\rongauss~\cite{chanyaswad2019ron} generates synthetic data by drawing samples from a multivariate Gaussian distribution fitted in a  projected space of the real data. Specifically, it operates by executing the following steps: Firstly, the data is pre-processed to ensure it possesses bounded sensitivity and adheres to the regularity conditions for the Diaconis-Freedman-Meckes effect (which guarantees the data will exhibit Gaussian-like distribution after projection with high probability). Next, a random orthonormal (RON) projection is applied on the pre-processed data, i.e., $\overline{\mX}= \mW^T \mX$ with $\mW\in \mathbb{R}^{d\times p}$ signifying the RON projection matrix and $\mX$ representing the pre-processed data matrix. Subsequently, a multivariate Gaussian model is fitted onto the projected data. During the inference stage, new samples are drawn from the fitted Gaussian distribution and are inversely projected into the original data space to form synthetic data samples. To maintain privacy, DP noise is added into both the mean and covariance of the fitted Gaussian distribution. Moreover, the Gaussian model is independently applied to each label class to facilitate label-conditional generation, which aligns with the concept of a Gaussian mixture model (GMM), where each label class forms a mode of the GMM. The detailed algorithm is presented in Algorithm~\ref{alg:ron_gauss}.
\begin{algorithm}[!htbp]
\LinesNumberedHidden
\caption{\rongauss}
\label{alg:ron_gauss}
\SetAlgoLined
\SetKwInput{KwInput}{Input}
\SetKwInput{KwResult}{Output}
 \KwInput{Dataset $\gD=\{(\vx^{(i)},y^{(i)})\}_{i=1}^n$, projection dimension $p$, noise scale $\sigma$} 
 \KwResult{Synthetic dataset $\gS$}
\For{c \emph{\textbf{in}} $\{1,...,C\}$ }  
{(1) {Extract samples with label class $c$ to form data matrix $\mX_c\in \mathbb{R}^{d\times n_c}$\;}
 (2) Pre-process data and compute the mean:\\
 \begin{itemize}
     \item Pre-normalize:$\vx^{(i)} := \vx^{(i)}/\Vert \vx^{(i)}\Vert_2 \quad \forall \vx^{(i)} \in \mX_c$
     \item Compute the DP mean: $\vmu_c = \frac{1}{n_c}\sum_{i=1}^{n_c} \vx^{(i)} + \gN(0,\sigma^2\mI)$
     \item Center the data: $\vx^{(i)}  := \vx^{(i)}  - \vmu_c \quad \forall \vx^{(i)}  \in \mX_c$
     \item Re-normalize: $\vx^{(i)}  := \vx^{(i)} /\Vert \vx^{(i)} \Vert_2 \quad \forall \vx^{(i)} \in \mX_c$
 \end{itemize}
 (3) Apply RON projection: $\overline{\mX}_c := \mW^T \mX_c \in \mathbb{R}^{p\times n_c}$\;
 $\text{(4) Derive the DP covariance: $\mSigma_c=\frac{1}{n_c}\overline{\mX}_c\overline{\mX}_c^T+\gN(0,\sigma^2\mI)$\;}$
 (5) Synthesize data for class $c$ by drawing samples from the Gaussian distribution $\widetilde{\vx}^{(i)}  \sim \gN(\mW^T\vmu_c,\mSigma_c)$\;
 (6) Inversely project and recenter: $\widetilde{\vx}^{(i)}  :=\mW \widetilde{\vx}^{(i)} + \vmu_c$ and construct the synthetic set $\gS_c=\{(\widetilde{\vx}^{(i)} ,c)\}_{i=1}^{n_c}$\;
}
\KwRet Synthetic dataset $\gS=\gS_1 \cup \cdots \cup \gS_C$
\end{algorithm}
\vspace{-20pt}

\subsection{VAE} 
\label{sec:vae}
The Variational Autoencoder (\vae[])~\cite{Kingma2014} is a type of  deep generative model that consists of both an encoder and a decoder. During training, these two components are cascaded and optimized to reconstruct data under pre-defined similarity metrics such as $L_1$/$L_2$ loss. The encoder (denoted as $q_\phi$) maps input data $\vx$ into a latent space, while the decoder (denoted as $p_\theta$) maps the encoded latent representation back into the data space. Meanwhile, \vae regularizes the encoder by imposing a prior $P_z$ over the latent code distribution. This regularization encourages the latent code to form a simple distribution that is amenable to sampling. During inference, new latent codes $\vz$ are sampled from the prior distribution $P_z$ and then fed into the decoder to generated synthetic samples. The formal VAE objective is composed of a reconstruction term and a prior regularization term:
\begin{align}
\label{eq:vae_loss}
\min_{\theta,\phi}\Ls_{VAE} = -\mathbb{E}_{q_\phi(\vz|\vx)}[p_\theta(\vx|\vz)] + KL(q_\phi(\vz|\vx) \Vert P_z)
\end{align}
where $KL(\cdot \Vert \cdot)$ denotes the KL divergence, $\vz$ and $\vx$ stand for the latent code and the real data, respectively. $q_\phi(\vz|\vx)$ represents the probabilistic encoder parameterized by $\phi$, and $p_\theta(\vx|\vz)$ represents the probabilistic decoder parameterized by $\theta$. 
In practice, the prior $P_z$ is always chosen to be a unimodal Gaussian distribution and $\vz$ is sampled using the reparameterization trick, facilitating  a closed-form derivation of the second term. 

We employ the class conditional (CVAE)~\cite{sohn2015learning} for label-conditional generation. In this framework, both the encoder and the decoder receive additional (one-hot) label information $y$. Formally, the training objective can be expressed as:
\begin{align}
\label{eq:cvae_loss}
\min_{\theta,\phi}\Ls_{CVAE} = -\mathbb{E}_{q_\phi(\vz|\vx,y)}[p_\theta(\vx|\vz,y)] + KL(q_\phi(\vz|\vx,y) \Vert P_z)
\end{align}
During the generation process, labels are generated based on their occurrence rates in the real dataset.
Privacy constraints is incorporated in the training stage by replacing the regular stochastic gradient descent (SGD) update with DP-SGD~\cite{abadi2016deep}, which involves clipping the per-example gradients and adding calibrated random noise to the mini-batch gradients.

\subsection{GAN}
\label{sec:gan}
The Generative Adversarial Network (\gan[])~\cite{goodfellow2014generative} is another widely used type of deep generative model. It comprises two neural network components, a generator $G_\theta$ and a discriminator $D_\phi$, which are trained simultaneously in an adversarial manner. The generator takes random noise $\vz$ (latent code) as input and generates samples that approximate the distribution of the training data. Conversely, the discriminator evaluates both generator-generated samples and real training data samples, aiming to distinguish between the two sources. Throughout training, these two modules engage in a competitive process, each adapting to the other: the generator seeks to generate progressively more realistic samples to deceive the discriminator, while the discriminator learns to distinguish the two sources more accurately. The standard \gan training objective can be formulated as
\begin{equation}
\label{eq:gan_loss}
  \min_{\theta} \max_{\phi} \mathbb{E}_{\vx\sim P_\text{data}}[\log(D_{\phi}(\vx))] + \mathbb{E}_{\vz\sim P_z}[\log(1-D_{\phi}(G_{\theta}(\vz)))]
\end{equation}
where $\theta, \phi$ denote the parameters of the generator and the discriminator respectively. $P_\text{data}$ stands for the real data distribution, and the $P_z$ is the prior distribution of the latent code. The first term in the objective prompts the discriminator to output high scores for real data samples. In contrast, the second term encourages the discriminator to assign lower scores to generated samples, while the generator is optimized to maximize the discriminator's output score.  During inference, the generator will receive new latent code samples $z$ drawn from the known prior distribution $P_z$, often standard  Gaussian, and produce synthetic data samples.

For private training, we adopt the DP Wasserstein \gan (DP-WGAN)~\cite{uclanesl_dp_wgan} implementation and its conditional variant to integrate label information during generation. Specifically, the Wasserstein distance~\cite{arjovsky2017wasserstein} is used as the training objective with the label information acting as auxiliary input for both the generator and discriminator:
\setlength{\abovedisplayskip}{2pt}
\setlength{\belowdisplayskip}{2pt}
\begin{equation}
    \min_\theta \max_\phi \mathbb{E}_{\vx\sim P_\text{data}}[D_\phi(\vx,y)] - \mathbb{E}_{\vz\sim P_z}[D_\phi(G_{\theta}(\vz,y),y)]
\end{equation}
The DP guarantee is ensured by employing DP-SGD for discriminator updates, which in turn guarantee the privacy of the whole \gan model and the synthetic data due to the post-processing theorem (Theorem \ref{theorem:post-processing}).

\subsection{Private-PGM}
\label{sec:pgm}
The Private Probabilistic Graphical Models (\pgm[]) framework~\cite{mckenna2019graphical} is designed to construct undirected graphical models from DP noisy measurements over low-dimensional marginals, which facilitates the generation of new synthetic samples via sampling from the learned graphical model. Specifically, \pgm operates on records consisting of discrete attributes. Formally, a record is denoted as $\vx=(x_1,...,x_d,x_{d+1})$ where each feature attribute $x_i$ for all $i$$\in$$\{1,...,d\}$ and the label $y=x_{d+1}$ fall within a discrete finite domain. Let $\gC$ represent a collection of \textit{measurement sets}, where each $C$$\in$$\gC$ is a subset of $\{1,...,d+1\}$ (i.e., the combinations of attributes), and let $\vv_C$ define the marginal probability vector on $C$. \pgm first obtains DP noisy measurements $\vm_C=\mQ_C\vv_C+\gN(0,\sigma_C^2 \mI)$ with $\mQ_C$ denoting the linear marginal query set over measurement set $C$ and $\gN(0,\sigma_C^2 \mI)$ representing the noise introduced by the Gaussian mechanism (with $\sigma_C$ the noise scale determined by the desired privacy level $\varepsilon_C$ and $\delta$. Refer to Definition \ref{def:gaussian_mechanism}). Subsequently, it estimates the marginal $\hat{\vv}$ that best explain all the noisy measurement $\hat{\vv}= \argmin_\vv \Vert \gQ\vv -\vm \Vert$ where $\gQ$ is a block-diagonal matrix with diagonal blocks $\{\gQ_C\}_{C\in \gC}$ (i.e., combining all the query set $\gQ_C$) and $\vm=(\vm_C)_{C\in\gC}$ the combined vector of measurements. Meanwhile, it estimates the parameter of the graphical model using existing graph inference and learning algorithms such as belief propagation on a junction tree.

In general, $\gQ_C$ can represent a complex set of linear queries expressed over $C$, and its selection can be adaptively tailored to downstream objectives. In our work, we adhere to the default implementation
where $\gQ_C$ is set to be an identity matrix. This configuration renders the measurement $\vm_C$ equivalent to the corresponding noisy marginal $\vv_C+\gN(0,\sigma_C^2\mI)$. Moreover, for computational feasibility, we adopt the basic configuration offered by the official implementation that sets $\gC=\big\{\{1\},...,\{d+1\}\big\}\cup \big\{\{1,d+1\},...,\{d,d+1\}\big\}$, which encompasses all one-way marginals as well as the 2-way marginals associated with the label attribute. The privacy budget is allocated uniformly across each measurement, i.e., $\varepsilon_C=\varepsilon/|\gC|$ with $\varepsilon$ the total privacy cost due to sequential composition.

\subsection{PrivSyn}
\label{sec:privsyn}
Similar to \pgm[], \privsyn~\cite{zhang2021privsyn} operates on data with discrete attributes to obtain measurable (noisy) marginals. However, while \pgm explicitly constructs factorized sparse graphical models, \privsyn directly generates data from the noisy marginal measurements. This approach inherently allows the use of an implicitly dense graphical model, enhancing its expressiveness capacity.

\privsyn is structured to execute the following steps sequentially:
\begin{itemize}
    \item \textit{Marginal selection}: This step selects the most informative marginals from the candidate set to optimize the privacy-utility trade-off.
    \item  \textit{Noise addition}: DP noise is added to the selected marginal measurements, ensuring privacy guarantee.
    \item \textit{Post-processing}: This phase ensures consistency from the noisy measurements. It addresses issues such as negative marginal measurements, cases where probabilities do not sum up to 1, and aligning different marginals that share common attributes.
    \item \textit{Data Synthesis}: Starting with a randomly initialized synthetic dataset, this step iteratively updates it to ensure alignment with the marginal measurements.
\end{itemize}

In our experimental evaluation, we omit the more involved 2-way marginal selection step for our dataset, as this step is prohibited by the significant computation and privacy costs, which scale quadratically with the feature dimensions. Instead, we utilize all 2-way marginals linked with the label attribute, aligning with the approach taken in \pgm to ensure a fair comparison. Apart from this, we adhere to the default configuration of the official implementation, which allocates the privacy budget at a ratio of $1:8$ between publishing the 1-way and 2-way marginals.

\section{Multi-Dimensional Evaluation of Synthetic Gene Expression Data}
\label{sec:metric_description}
Our study delved into a comprehensive assessment of various models. This evaluation was executed through a meticulous analysis of model performance across three main aspects: \textbf{utility} (\sectionautorefname~\ref{sec:utility}), \textbf{statistical} (\sectionautorefname~\ref{sec:statistical_evaluation}), and \textbf{biological} (\sectionautorefname~\ref{sec:biological_metrics_description}) evaluation. Each aspect encompasses distinct metrics: \textit{machine learning efficacy} for \textbf{utility} evaluation, \textit{marginal} (\textit{histogram intersection}) and \textit{joint} (\textit{distance to closest record}) closeness for \textbf{statistical} evaluation, as well as \textit{differential expression} and \textit{gene co-expression} for \textbf{biological} evaluation.

\subsection{Utility Evaluation}
\label{sec:utility}
\subsubsection{Machine Learning Efficacy}
\label{sec:machine_learning_efficacy}
Evaluating the quality of synthetic data typically involves a standard procedure of assessing its performance within a downstream task. This evaluation determines whether the synthetic data, when used as a replacement for the real data, can accomplish the desired task with comparable effectiveness. This is executed by training machine learning models on real (train) data and evaluating their performance on held-out (test) data. Subsequently, a parallel model is trained on synthetic data and evaluated using the same held-out data. The choice of evaluation metrics is determined by the specific nature of the task at hand. 
In our work, we adopt the standard \textit{accuracy} score  for evaluating the disease classification task. 

\subsection{Statistical Evaluation}
\label{sec:statistical_evaluation}
Utility-based metrics, however, often offer an incomplete perspective due to their narrow evaluation lens, presenting a single facet of the model's performance, which can occasionally lead to misleading impressions. In order to address this potential bias, it becomes crucial to incorporate additional statistical metrics that emphasize the fidelity of the generation process. This entails assessing how effectively the model captures both the marginal distribution and the underlying joint distribution of the data, providing a more comprehensive understanding of its performance.

\subsubsection{Histogram Intersection} 
\label{sec:histogram_intersection}
The \textit{histogram intersection} serves as a prevalent qualitative tool for visualizing one-dimensional data (i.e., single columns/attributes), enabling a comprehensive exploration of the data's distribution characteristics. Understanding such single-dimensional distributions can be pivotal for subsequent pre-processing and analysis steps. Prior studies have harnessed this metric to compare the distributions of synthetic and real data by selecting specific attributes from the real dataset and overlaying the histograms of the corresponding real data onto the synthetic ones. This technique, referred to as \textit{distribution matching plots}, provides a qualitative assessment of how closely the two distributions align.

However, relying solely on qualitative measures has its limitations, particularly when confronted with large feature sets like gene expression data. Manually visualizing each column becomes impractical. This necessitates a quantitative approach that maintains a similar essence but can be aggregated to yield a single score. The normalized \textit{histogram intersection metric} proposed in~\cite{afonja2023margctgan} is applicable in such scenario. It is computed as the sum of the minimum probability values between the real data column and the synthetic data column. This sum is subsequently averaged across the various columns in the dataset (see ~\autoref{eq:histogram_intersection}). 
In contrast to other analogous techniques like the \textit{Wasserstein distance}~\cite{kantorovich1960mathematical} or \textit{Jensen-Shannon divergence score}~\cite{Lin1991DivergenceMB}, the \textit{histogram intersection} score demonstrates superior performance and exhibits a strong correlation with other metrics~\cite{afonja2023margctgan} (see Appendix \autoref{fig:marginal_corr})\footnote{
The histogram intersection metric defined here also corresponds to 1 - total variation distance, a popular metric that quantifies the similarity between two probability distributions.}. This quantitative approach strikes a balance between comprehensiveness and practicality, making it an effective tool for evaluating the quality of the generated data.
\setlength{\abovedisplayskip}{2pt}
\setlength{\belowdisplayskip}{2pt}
\begin{align}
&p_c = \frac{s_c}{|\gD|\Delta_i} \quad q_c = \frac{t_c}{|\gS|\Delta_i} \\
&\text{HI}(\vp_i, \vq_i) = \sum_c{\min(p_{c}, q_{c})} \label{eq:histogram_intersection} \\
&\text{Overlap Score} = \frac{1}{d}\sum_i\text{HI}(\vp_i, \vq_i) \label{eq:overlap_score}
\end{align}
where $\mathbf{p}_i$ and $\mathbf{q}_i$ denote the histogram representations of the probability distributions for the real ($\gD$) and synthetic ($\gS$) datasets within feature $i$, respectively. The terms $p_c$ and $q_c$ represent the proportions of category $c$ for feature $i$, with $s_c$/$t_c$ denoting the counts of real/synthetic samples in category $c$. The factor $\Delta_i$ is introduced as a normalization term, specifying the bin size for numerical features.
The term $\text{HI}(\mathbf{p}_i, \mathbf{q}_i)$ represents the histogram intersection score for feature $i$. The dimensionality of the feature space is denoted by $d$. The \textit{Overlap Score} is computed by averaging the histogram intersection scores across all features.

\subsubsection{Distance to Closest Record}
\label{sec:distance_to_closest_record}
The \textit{distance to closest record} metric aims to measure the similarity between the \textit{joint} distribution of real and synthetic data. Obtaining an exact measurement of the joint distribution is inherently challenging and always infeasible, as the underlying probability distribution of the real data is unknown and generally intractable. To circumvent this, we approximate the alignment of joint distributions using k-nearest neighbors (KNN). This involves computing the Euclidean distance between each synthetic data sample and its $k$ nearest neighbors in either the held-out or training set. The objective is to evaluate the plausibility of each synthetic sample being real. The final KNN Distance score is the average across all synthetic dataset samples and various $k$ values, as defined in ~\autoref{eq:knn_distance}.
\setlength{\abovedisplayskip}{2pt}
\setlength{\belowdisplayskip}{2pt}
\begin{align}
&d_k(\tilde{\vx}) =\mathrm{first}_k\Big(\mathrm{sort}\Big(\big\{\Vert \tilde{\vx} - \vx \Vert_2 \,|\, \forall \vx \in \gD_\text{train/test}\big\}\Big)\Big)\\
&\text{KNN Distance Score} = \frac{1}{|\gS| \cdot k }\sum_{\tilde{\vx}\in \gS}\sum_{i=1}^k d_{k,i}(\tilde{\vx}) \label{eq:knn_distance}
\end{align}
where  $\gS$ denotes the synthetic set, $\gD$ the real dataset, $d_k(\tilde{\vx})$ is a sequence contain the \(k\) smallest values of distances (where $\textrm{sort}(\cdot)$ represents the sorting operation in ascending order and $\mathrm{first}_k(\cdot)$ denotes the operation for retrieving the first $k$ elements from the sorted sequence), with $d_{k,i}(\tilde{\vx})$ denoting the i-th element of $d_k(\tilde{\vx})$.

\subsection{Biological Evaluation}
\label{sec:biological_metrics_description}

\subsubsection{Differential Expression}

There are several methods to measure differential expression, but many of them make strong assumptions on the distribution underlying gene expression data~\cite{Anders_Huber_2010, Robinson_McCarthy_Smyth_2010, Love_Huber_Anders_2014}. However, the question of which distribution gene expression data follows has been subject to debate for many years~\cite{de_Torrenté_Zimmerman_Suzuki_Christopeit_Greally_Mar_2020}. To avoid making any (potentially false) assumptions regarding the distribution, we chose a non-parametric test for the identification of differentially expressed genes, namely the \textit{Wilcoxon signed rank test}~\cite{Wilcoxon_1945}. For each pair of conditions in the data, the test was conducted on the expression values of each gene measured across the samples of the respective condition. We ran the test using the pairwise Wilcoxon function from the R-package \textit{scran} (version 1.26.2) and using the alternative hypothesis for each side to differentiate between \textit{up}- and \textit{down-regulation}. We considered a gene as differentially expressed between two conditions if the p-value was at most 0.05. The reconstruction of DE-genes by different generative models $M$ at varying privacy levels $\varepsilon$ is quantified via the mean true positive rate ($TPR$) defined as follows. 
\setlength{\abovedisplayskip}{2pt}
\setlength{\belowdisplayskip}{2pt}
\begin{align}
\text{TPR}_{\mathcal{M}, \varepsilon} & =  \frac{\underset{\{a_i, a_j\}\subset \mathcal{A}, a_i \neq a_j}{\sum}\left(\text{TPR}^\text{up}_{a_i, a_j, \mathcal{M},\varepsilon} + \text{TPR}^\text{down}_{a_i, a_j, \mathcal{M},\varepsilon}\right)}{2 \cdot {\vert \mathcal{A} \vert \choose 2}}
\label{DE_plot_equation}
\end{align}
where $\mathcal{A}$ denotes the set of condition pairs (each pair representing different disease types distinguished by unique label classes in our case). $\text{TPR}^\text{up (down)}_{a_i, a_j, \mathcal{M},\varepsilon}$ signifies the true positive rate for identifying \textit{up-regulated} (or\textit{ down-regulated}) DE-genes within synthetic data generated by the model $\mathcal{M}$ under a given privacy budget $\varepsilon$, in comparison to the actual DE-genes observed in the real dataset for conditions $a_i$ and $a_j$.
${\vert \mathcal{\gA} \vert \choose 2}$ represents the count of all possible unordered condition pairs.

\subsubsection{Gene Co-Expression}

To assess if groups of co-expressed genes that are present in the real data were preserved in the synthetic data, we applied \textit{hCoCena}~\cite{Oestreich_Holsten_Agrawal_Dahm_Koch_Jin_Becker_Ulas_2022}, an R-package that enables the integration of different gene expression datasets, i.e., the real and the synthetic data in our case, and their subsequent joint co-expression analysis. The tool creates a gene co-expression network for each set, 
which is a weighted graph $G$ = ($V$, $E$), where the nodes $V$ represent genes, edges $E$ represent co-expressions and the edges are weighted with the co-expression strength. The weight $w$ is computed as the \textit{Pearson correlation coefficient} $r$ (see \equationautorefname~\ref{coexpression_equation}) between their expression values across samples, such that $w(e_{j,k})$=$r_{jk}$. Afterwards, genes that are not significantly strongly co-expressed according to a user-defined correlation cut-off with any other gene are discarded to only include strong co-expressions that are potentially biologically meaningful. A gene co-expression network is created for each dataset. We then used these co-expression networks to identify the number of co-expressions (i.e., graph edges) that were correctly reconstructed in the synthetic data and the number of spurious co-expressions introduced in the synthetic data that did not exist in the real data. Additionally, modules of strongly co-expressed genes were identified in the network of the real dataset using the \textit{Leiden community detection algorithm}. We investigated the \textbf{mean group fold-changes (GFCs)} for the detected modules across conditions in the real and the synthetic data. GFCs are a metric for the average expression of a module in a group of samples, i.e. all samples of a particular experimental condition, essentially representing the activation or deactivation of the module under the given condition.

\vspace{-5pt}
\section{Evaluation}
\label{sec:evaluation}
\begin{figure*}[!htbp]
\centering
\newcommand{\figwidth}{0.32\textwidth}
\subfigure[]{
\begin{minipage}[b]{\figwidth} \includegraphics[width=1.0\textwidth]{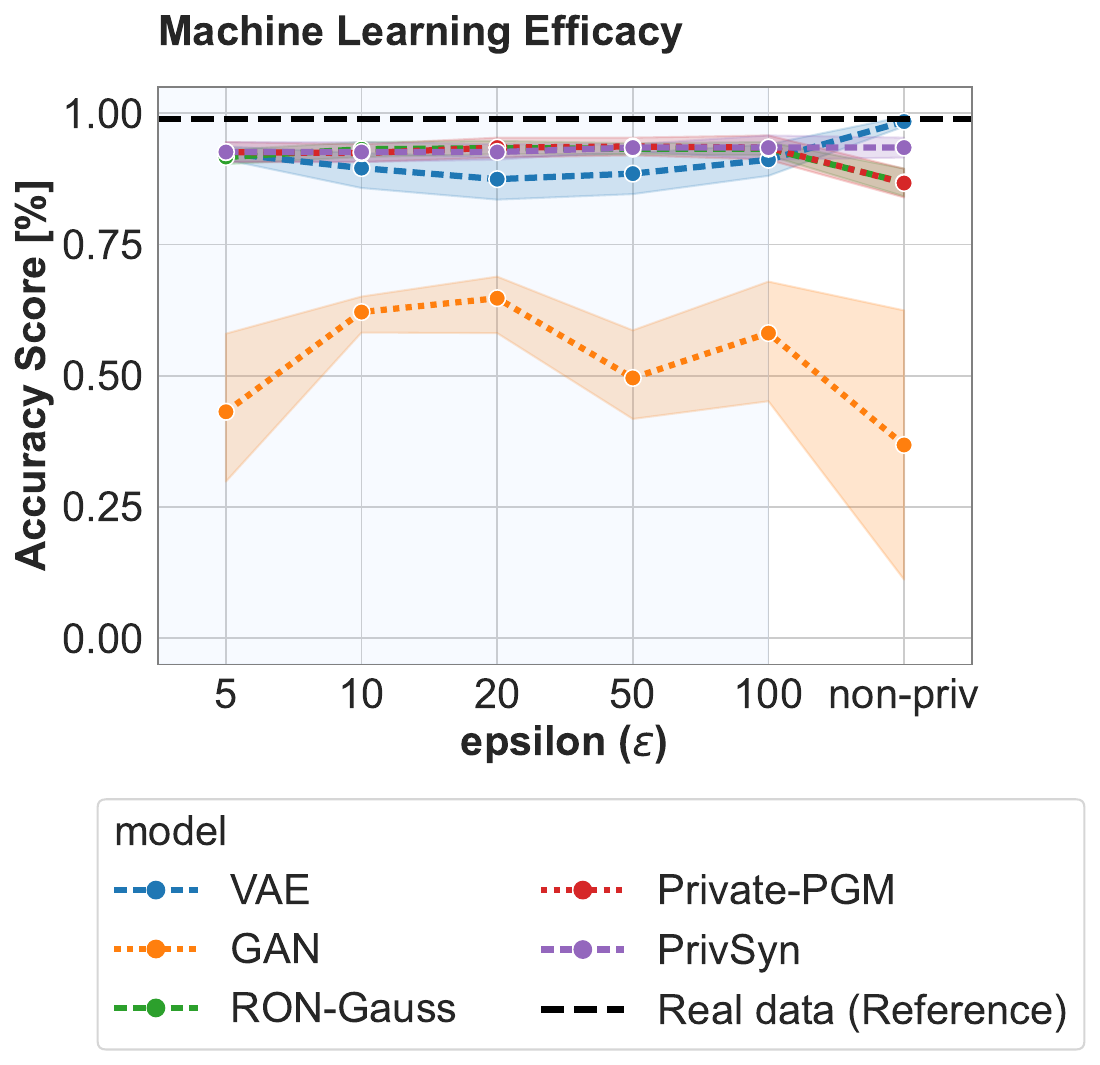}
\label{fig:ml_efficacy_combined}
\end{minipage}
}
\subfigure[]{
\begin{minipage}[b]{\figwidth} \includegraphics[width=1.0\textwidth]{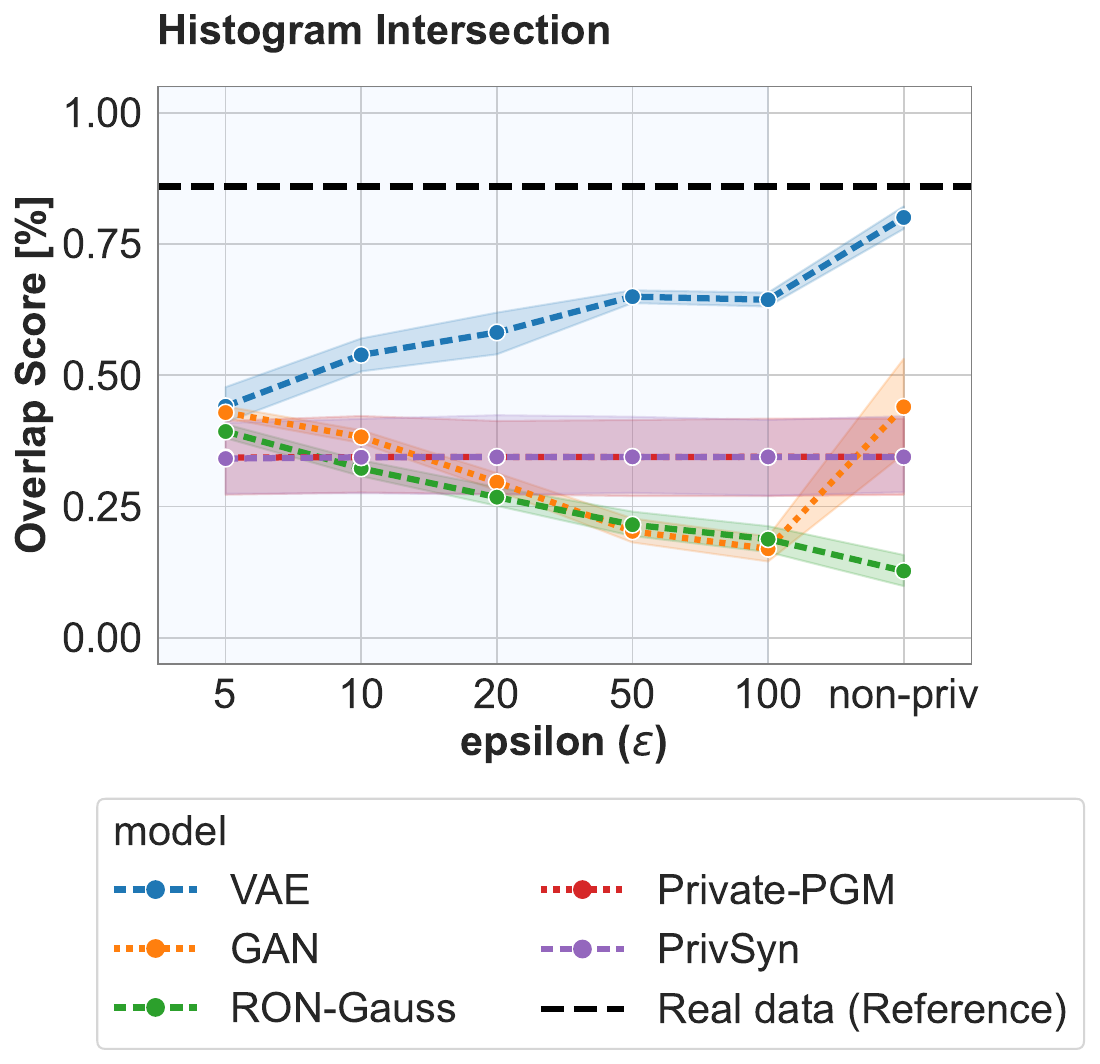}
\label{fig:histogram_intersection_combined}
\end{minipage}
}
\subfigure[]{
\begin{minipage}[b]{\figwidth} \includegraphics[width=1.0\textwidth]{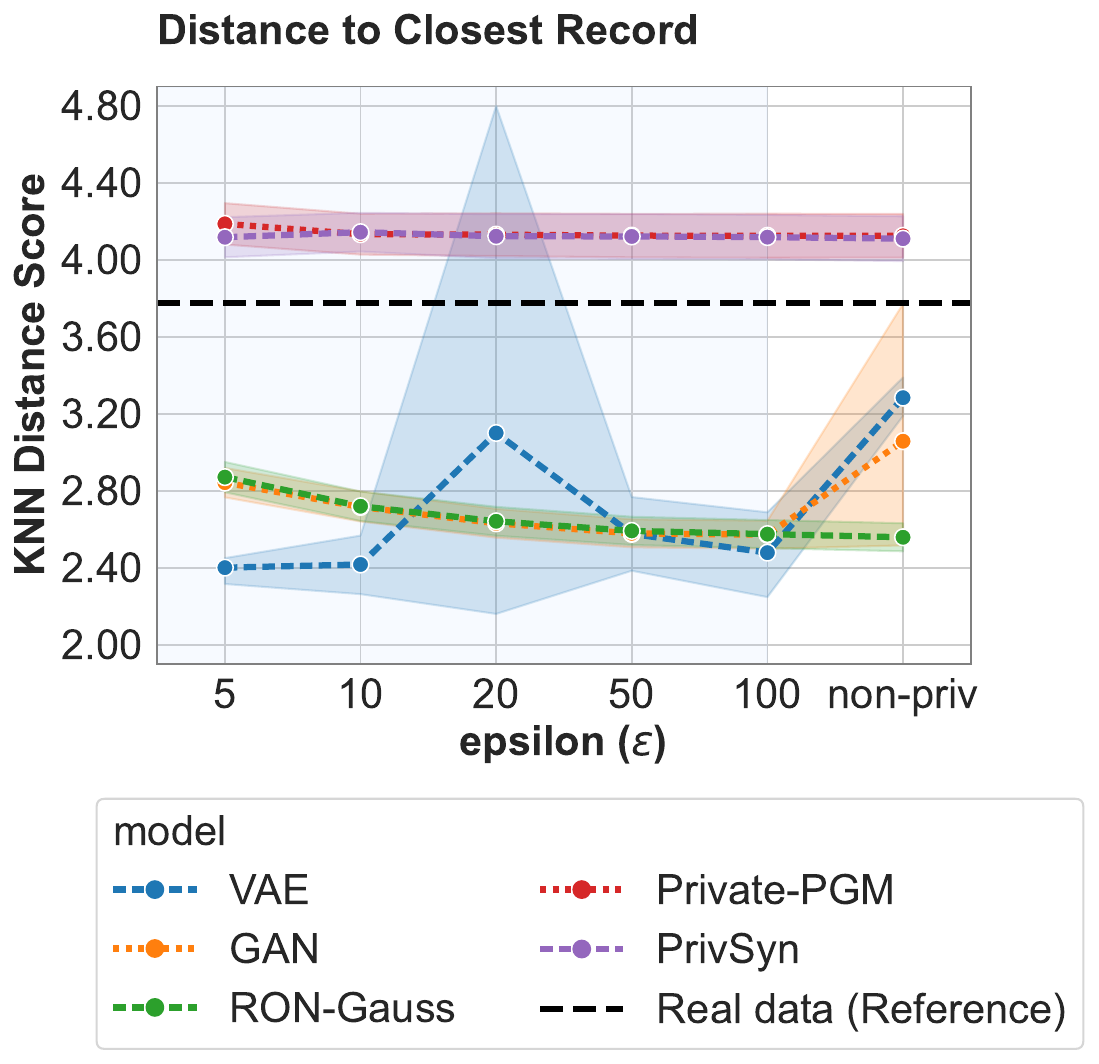}
\label{fig:closeness_approximation_combined}
\end{minipage}
}
\caption{Utility Evaluation by Machine Learning Efficacy, and Statistical Evaluation by Histogram Intersection and Distance to Closest Record. \textmd{Shown in (a) are the Accuracy Scores for the \textit{Machine Learning Efficacy} metric across 5 various models for the DP-case (blue shading) with varying $\varepsilon$ values, alongside the non-private case. Similarly, (b) and (c) display the Overlap Score and K-Nearest Neighbors Distance Score for the \textit{Histogram Intersection} metric and \textit{Distance to Closest Record} metric, respectively. Evaluations encompassed two seeds for training split creation and two synthetic dataset randomizations. The presented values represent means across these randomization seeds. The black dashed line represents the reference score on actual train-test data, signifying the best attainable score.}} 
\vspace{10pt}
\label{fig:utility_and_statistical_combined}
\end{figure*}

\subsection{Dataset}
\begin{table}[!tbp]
\begin{center}
\begin{tabular}{ ccccccc } 
 \toprule
 Class & AML & ALL & CML & CLL & Other & \textbf{Total}\\ 
 \midrule
 \# Samples & 508 & 12 & 14 & 13 & 634 & \textbf{1181}\\ 
 \bottomrule
\end{tabular}
\end{center}
\vspace{.5cm}
\caption{Dataset summary. \textmd{Listed are the different sample classes present in the dataset and the number of samples in each class.}}
\label{table:1}
\vspace{-20pt}
\end{table}
The generative models were trained on a bulk RNA-seq dataset compiled by Warnat-Herresthal \emph{et al.}~\cite{Warnat-Herresthal}, which comprises a reasonable number of independent samples, rendering it suitable for DP training (see \sectionautorefname~\ref{sec:discussion} for detailed discussion). The dataset is structured as a matrix, with rows corresponding to \textit{samples} and columns to \textit{features}. Each row represents a biological specimen obtained from a patient, while each column indicates the expression level of a particular gene. The expression levels are quantified by RNA-seq counts, with higher integer values indicating greater gene activity. 
It comprises samples from 5 disease classes, 4 classes of which are types of leukemia and the fifth class is the category ``\emph{Other}'', which is made up of samples from various other diseases as well as healthy controls. The 4 leukemia types are acute myeloid leukemia (``\emph{AML}''), acute lymphocytic leukemia (``\emph{ALL}''), chronic myeloid leukemia (``\emph{CML}'') and chronic lymphocytic leukemia (``\emph{CLL}''). Sample counts per class are listed in \tableautorefname~\ref{table:1}. 
As per the original publication, the data were normalized with DeSeq2~\cite{Love_Huber_Anders_2014}  
to account for varying sequencing depths and RNA composition, which is necessary to compare expression levels of different samples and conduct a DE-gene analysis. Given the high dimensionality of the features (more than 12k genes) and the comparatively low sample size (1181), we reduced the feature space to 958 genes. Notably, even this reduced feature dimension remains significantly high, especially when compared with standard benchmarking datasets which typically comprise merely dozens of features. 

These 958 genes were not selected randomly but based on their characterization as \emph{landmark genes} in the LINCS L1000 project ~\cite{Subramanian}. The landmark genes were identified as representative genes that, when measured, allow the inference of around 20k other genes.

\myparagraph{Pre-processing and Post-processing} 
In accordance with standard practices, we pre-processed our data prior to model training. For 
\rongauss[], \gan[], and \vae[], which operate on continuous data expected to be well-centered, we standardized each feature by subtracting its mean and dividing by its standard deviation. Conversely, for \pgm and \privsyn which rely on discrete representations for computing marginals, we discretized each feature into four bins based on its quantiles: <25\%, 25\%-50\%, 50\%-75\%, and >75\%. This approach was chosen to accurately represent \textit{up}- and \textit{down-regulation}, while also maintaining a condensed format (resulting in a limited number of bins after discretization) for an optimized privacy-utility trade-off.

After training and generation, we implemented the following post-processing measures:
\begin{itemize}[leftmargin=*]
    \item For \rongauss[], \gan[], and \vae[]: We reverted the standardization by multiplying the generated data features by the standard deviation and adding back the mean (both the standard deviation and the mean were pre-computed on the real dataset).
    \item For \pgm and \privsyn[]: We mapped the generated discrete data back to the original continuous mean value associated with each bin.
\end{itemize}
We verify the efficacy of our approach via our preliminary experiments: the continuous pre- and post-precessing was proved to be lossless, while the discrete one did \textit{not} affect the biological and utility evaluation. 

In line with the common evaluation protocol adopted in DP literature, we do \textit{not} incorporate DP into the pre- and post-processing process, and the label class occurrence ratio is treated as public information and used during generation. This approach aids in producing meaningful evaluation results and offers a more accurate indication of performance, particularly given our challenging setup. However, it is crucial to note that in real-world applications, all such processes including the hyperparameter selection~\cite{papernot2021hyperparameter} would require DP sanitization to ensure stringent privacy protection. Although implementing such sanitization is generally technically straightforward (e.g., either computed on public data or using DP techniques such as Algorithm 2 in \cite{tramer2020differentially} and \cite{gillenwater21a} for DP sanitization techniques applicable to continuous and discrete processing, respectively), it can lead to considerable utility loss in bio-data, mainly due to limited sample sizes, which warrants further discussion and investigation.  

\vspace{-5pt}
\subsection{Setup}
We follow the official  implementation for methods that offer open-source code:~\rongauss[]\footnote{\url{https://github.com/inspire-group/RON-Gauss/tree/master}}, 
\gan[]\footnote{\url{https://github.com/nesl/nist_differential_privacy_synthetic_data_challenge/}},
\pgm[]\footnote{\url{https://github.com/ryan112358/private-pgm}},
\privsyn[]\footnote{\url{https://github.com/usnistgov/PrivacyEngCollabSpace/tree/master/tools/de-identification/Differential-Privacy-Synthetic-Data-Challenge-Algorithms/DPSyn}} and adopt the default hyperparameter setting tuned for general tabular datasets. We adhere to such setting as further attempts at fine-tuning did not give rise to notably better results in our preliminary experiments. We use the RDP accountant implementation from TensorFlow privacy\footnote{\url{https://github.com/tensorflow/privacy/blob/master/research/hyperparameters_2022/rdp_accountant.py}} for \vae[] and \gan[]. For the \vae[] model, which lacks an official DP implementation, we tuned the key hyperparameters (including the weight of the reconstruction loss term, the number of training iterations, the gradient clipping bound, the batch size, and the latent dimension) via grid-search.  We repeat the experiments over different random seeds and report the mean and standard deviation over these seeds by default. For biological evaluation where results from different seed cannot be aggregated, we detail the outcomes for each individual seed separately. The $\delta$ is set to be $10^{-5}$ by default across our experiments.

\section{Experiments}
\begin{figure*}[!htbp]
\includegraphics[width=0.47\textwidth]{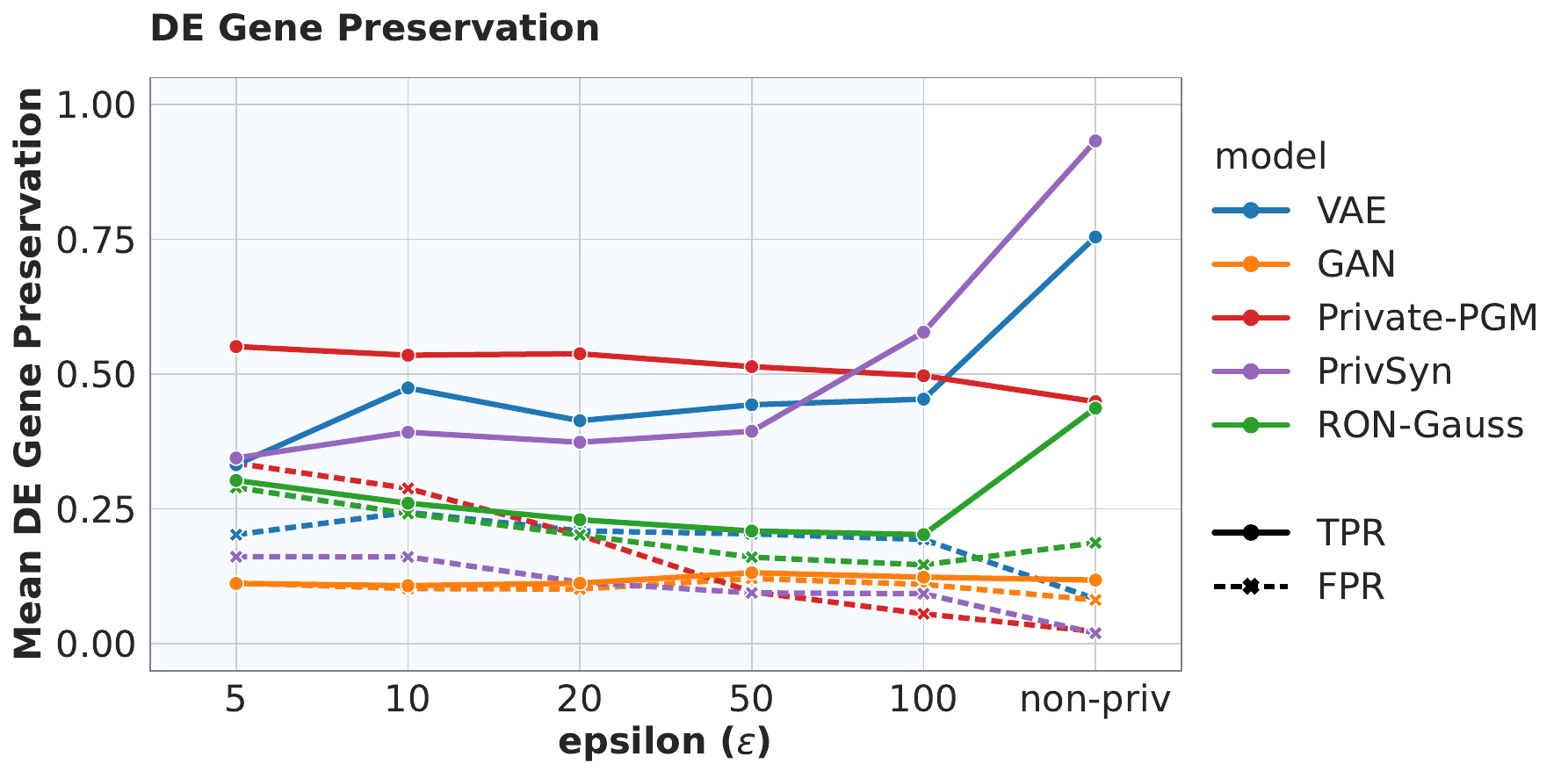}
\includegraphics[width=0.47\textwidth]{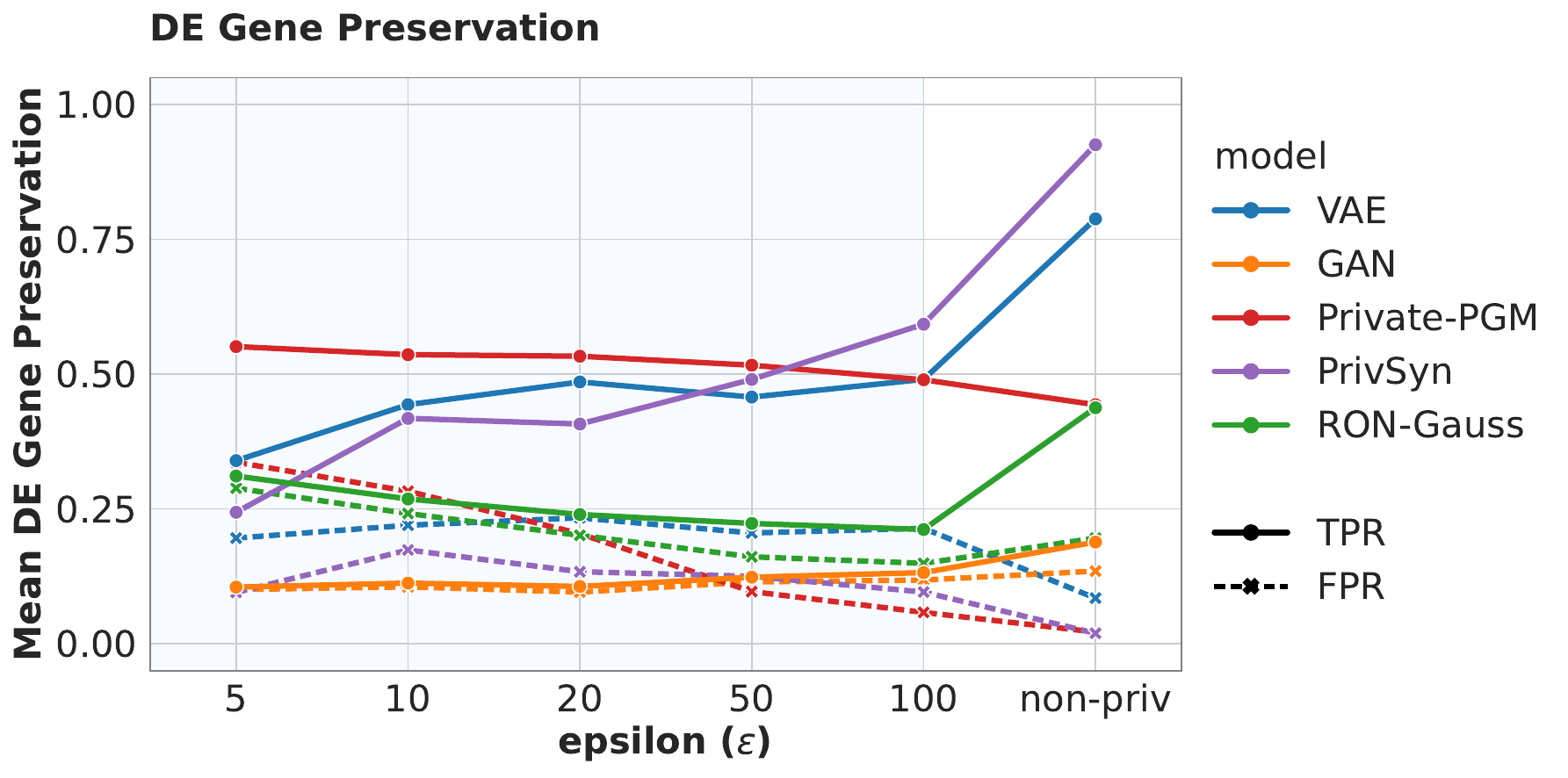}
\caption[Biological Evaluation by DE-Gene Preservation]{Biological Evaluation by DE-Gene Preservation. \textmd{Shown is the preservation of DE-genes 
(true positive rate (TPR): solid lines; false positive rate (FPR): dashed lines) across the tested models for the DP-case (indicated by blue shading) with different values of $\varepsilon$ and the non-private case. The evaluation was performed for two different seeds used for creating the training split (left and right plot). The presented values are means across 
two different seeds set for generating the data (except for \pgm and \privsyn[], where seeding is not possible). }}
\vspace{15pt}
\label{figure:DE-Gene Preservation}
\end{figure*}

\begin{figure*}[!htbp]
\centering 
\includegraphics[width=1.0\textwidth]{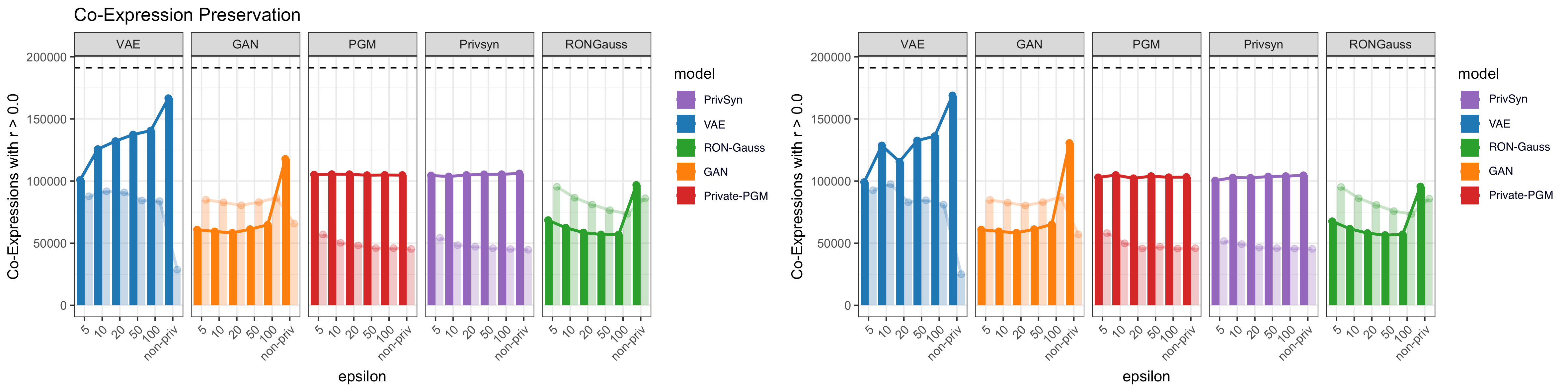}
\caption{Biological Evaluation by Co-Expression Preservation for $r$ > 0. \textmd{Shown is the co-expression preservation across the tested models for different values of $\varepsilon$ as well as the non-private case for two different seeds used for creating the training split (left and right plot). Specifically, non-transparent bars give the number of correctly reconstructed co-expressions with Pearson Correlation Coefficient $r$ > 0 and an associated p-value < 0.05, while semi-transparent bars give the number of co-expressions introduced by the model that did not exist in the real data. The dashed black line indicates the number of co-expressions in the real data. All values shown are means across two different seeds set for generating the data (except for \pgm and \privsyn[], where seeding is not possible). }}

\vspace{15pt} 

\label{figure:Co-Expression Preservation}
\end{figure*}

We study five different generative models: \vae[], \gan[], \rongauss[], \pgm[], and \privsyn[], which encompass diverse categories, attribute types, and DP sanitation approaches, as summarized in \tableautorefname~\ref{tab:model_summary}. Our assessment was conducted under two scenarios: initially, without the imposition of DP constraints, and subsequently, with DP integration using values of epsilon ($\varepsilon$) ranging from 5 to 100, signifying a spectrum from high to low privacy levels. We did not reduce the privacy budget to values smaller than 5, as the models fail to achieve reasonable results at this threshold. For models such as \vae and \gan that were not originally designed with DP protections, we incorporate DP to the gradients following the DP-SGD framework~\cite{abadi2016deep} to create their respective private variants.
Conversely, for inherently privacy-centric models like \rongauss[], \pgm[], and \privsyn[], we set the noise scale to be zero to simulate their non-private counterparts. These diverse models were then evaluated using the metrics detailed in \sectionautorefname~\ref{sec:metric_description}. We set the real data as \textit{Reference} (See the dashed black lines in \figureautorefname~\ref{fig:utility_and_statistical_combined}), which represents the score of each metric when applied to the real training data and then evaluated on the real held-out (i.e., test) data.

\subsection{Utility Evaluation}
For the Machine Learning Efficacy metric, given the classification nature of the task (predicting diverse disease types using gene expression data), we employ a widely-used and straightforward machine learning approach known as logistic regression. This model undergoes training as outlined in \sectionautorefname{~\ref{sec:machine_learning_efficacy}}. The chosen evaluation metric is the \textit{accuracy} score.

\myparagraph{Results and Findings}
The outcomes, depicted in \figureautorefname~\ref{fig:ml_efficacy_combined}, portray the machine learning utility scores for various generative models across differing privacy levels, ranging from high ($\varepsilon=5$) to low ($\varepsilon=100$). In the non-private context (termed as non-priv in \figureautorefname~\ref{fig:ml_efficacy_combined}), we observe that all five models—with the exception of \gan —exhibit a substantial utility score ranging from 86\% to 98\%. This shows a moderate decrease of 0.5\% to 12\% relative to the reference point set by real data (black dashed line). Within the private realm ($\varepsilon=5,10,20,50,100$), models such as \pgm[], \privsyn[], and \rongauss display consistent high utility, encountering a reduction of less than 7.4\% in very high privacy conditions ($\varepsilon=5$) to a 5.8\% drop in situations with lower privacy ($\varepsilon=100$). Notably, these models demonstrate a higher utility as $\varepsilon$ increases. Remarkably, the utility metric easily saturates, even with a simple probabilistic model (i.e., the unimodal Gaussian as in \rongauss[]), while the \vae exhibits slight advantages in the non-private case. The \gan model generally performs worse in terms of the utility metric and exhibits relatively high variance, potentially due to the unstable nature of its adversarial training process, which is exacerbated in our dataset with limited samples.

\subsection{Statistical Evaluation}
\subsubsection{Histogram Intersection}
We initiate by subjecting the numerical column to min-max pre-processing, a technique that rescales values to fit within the range of 0 to 1. Following this normalization, a discretization binning process is employed, utilizing 25, 50, and 100 bin size, which provides an approximated representation of the numerical column's distribution, and thus ensures tractability. No additional pre-processing steps are required for the discrete and categorical columns. Our computation of the \textit{Overlap Score} adheres to the definition in~\autoref{eq:overlap_score}.

\myparagraph{Results and Findings}
\figureautorefname{\autoref{fig:histogram_intersection_combined}} illustrates the overlap score, which serves as the mean of the histogram intersection scores between the columns of real and synthetic data, as detailed in ~\sectionautorefname{~\ref{sec:histogram_intersection}}. In general, across both private and non-private cases, most models exhibit subpar performance on this metric. An exception stands out: the \vae model (depicted by the blue line). Remarkably, for the non-private case, it showcases an impressive overlap score of \textasciitilde{}80\%, experiencing only a 6.8\% relative drop compared to the reference set by the real data (black dashed line). This performance trend consistently improves from the high privacy (\epfive) to the low privacy case (\ephundred), indicating that the synthetic data's marginal distribution increasingly resembles that of the real data, with the relaxation of privacy constraints. However, this does not uniformly apply to all models. For instance, the \rongauss model (represented by the green line) shows an unexpected behavior—its overlap score is higher in the very high privacy case (\epfive) compared to the non-private case, exhibiting an 85\% drop in performance relative to the real data reference. This outcome is surprising given that this model involves continuous attribute types, which should typically lead to a moderately increasing overlap score as $\varepsilon$ increases. Similarly, the \gan model follows a similar trend to \rongauss[], but it demonstrates a higher overlap score in the non-private case, with a reduced relative drop of 48.8\%. We conjecture that such seemingly abnormal behavior may be partially explained by the fact that both \gan and \rongauss did not capture the marginal distributions faithfully even in the non-private case, which making the results more influenced by randomness than by the learnability. The inferior performance of the \pgm and \privsyn models in this metric can potentially be attributed to the loss in precision resulting from the reverse transformation inherent in the discretization process, which may dominate the additional information loss incurred by privacy constraints. 
Interestingly, despite the modest performance in this metric, the same models excel in the private case for the machine learning efficacy metric. This underlines the necessity of evaluating synthetic data from various generative models across an array of metrics to gain a comprehensive understanding of their behavior relative to real data.

\subsubsection{Distance to Closest Record}
This metric aims to approximate the likelihood that a synthetic data sample originates from the distribution of real data samples. This measurement relies on the K-Nearest Neighbors (KNN) approximation technique. In our experimental setup, we specifically set the value of k to 10, which dictates the computation of the KNN Distance Score according to ~\autoref{eq:knn_distance}. We use the scikit-learn KNN estimator\footnote{\url{https://scikit-learn.org/stable/modules/generated/sklearn.neighbors.NearestNeighbors.html}} to compute the nearest neighbors distance of each synthetic sample to the real test data. \figureautorefname{~\ref{fig:closeness_approximation_combined}} shows the averaged 10-NN distance score for different epsilon values (x-axis) and diverse generative models. A higher proximity of this score to the reference established by the real data implies a greater likelihood that the \textit{joint} distribution of real and synthetic data aligns closely. Scores falling below the reference point set by real data imply that the synthetic data samples are closely aligned with the distribution of the real test data. However, it is essential to exercise caution while interpreting these results due to the relatively small size of the test set. Making assertive conclusions based solely on these findings might be premature.

\myparagraph{Results and Findings}
Intuitively, we anticipate that the score for this metric should be lower, indicating closer alignment to the real data reference (depicted by the black dashed line) in the non-private setting. As privacy levels increase, we expect a moderate increase in the distance—moving from \ephundred~ to \epfive. This examination aims to substantiate the assertions made by prior study~\cite{park2018data} that this metric has the potential to quantify privacy. However, the results illustrated in ~\figureautorefname{~\ref{fig:closeness_approximation_combined}} present a counter-intuitive observation. All models, excluding the graphical-based models \pgm and \privsyn[], demonstrate distances below the real data reference. This holds true for both private and non-private scenarios. Notably, the \vae model stands out, exhibiting a low distance to the closest test record (i.e closest to the real data reference but still falls below the black dashed line). This shows a relative drop of 48\% when contrasted with the reference established by real data. Notably, the \pgm and \privsyn models, which yield unsatisfactory outcomes in the histogram intersection metric, also exhibit the most substantial distances to the real data reference. This persistent distance above the black dashed line further indicates that the reverse discretization process could lead to a loss of precision in these models. Additionally, for the \vae model, across the \epfive~ to \epfifty~ range, there's a pronounced variance in scores across different experimental random seeds. This variance might offer insights into the model's sensitivity behavior in the private case.

\subsubsection{Summary of Utility and Statistical Evaluation} 
The observed results of~\autoref{fig:utility_and_statistical_combined} underscore the necessity of assessing diverse metrics when evaluating synthetic data. Moreover, it brings to light an intriguing revelation: even if the synthetic data strays from both marginal and joint distributions, it still exhibits the capacity to maintain substantial downstream utility tasks. This observation reinforces the significance of a comprehensive evaluation approach that considers various aspects of data behavior and performance. 

\subsection{Biological Evaluation}
To evaluate the different models for biological soundness, we assessed their capabilities of maintaining two biological aspects in the generated synthetic data: (1)  
the preservation of differential expression by assessing the TPR and FPR of reconstructed DE-genes per model and across privacy parameters and (2) the preservation of co-expressions between genes, i.e., their Pearson Correlation Coefficients \emph{r} as well as the activation of co-expressed modules. The models were evaluated once without the constraint of DP and then with DP using $\varepsilon$ = 5, 10, 20, 50, 100.

\begin{figure*}
\centering
\newcommand{\figwidth}{0.49\textwidth}
\subfigure[$\varepsilon=5$]{
\begin{minipage}[b]{\figwidth} \includegraphics[,width=1.0\textwidth,trim={0 0 12cm 0},clip]{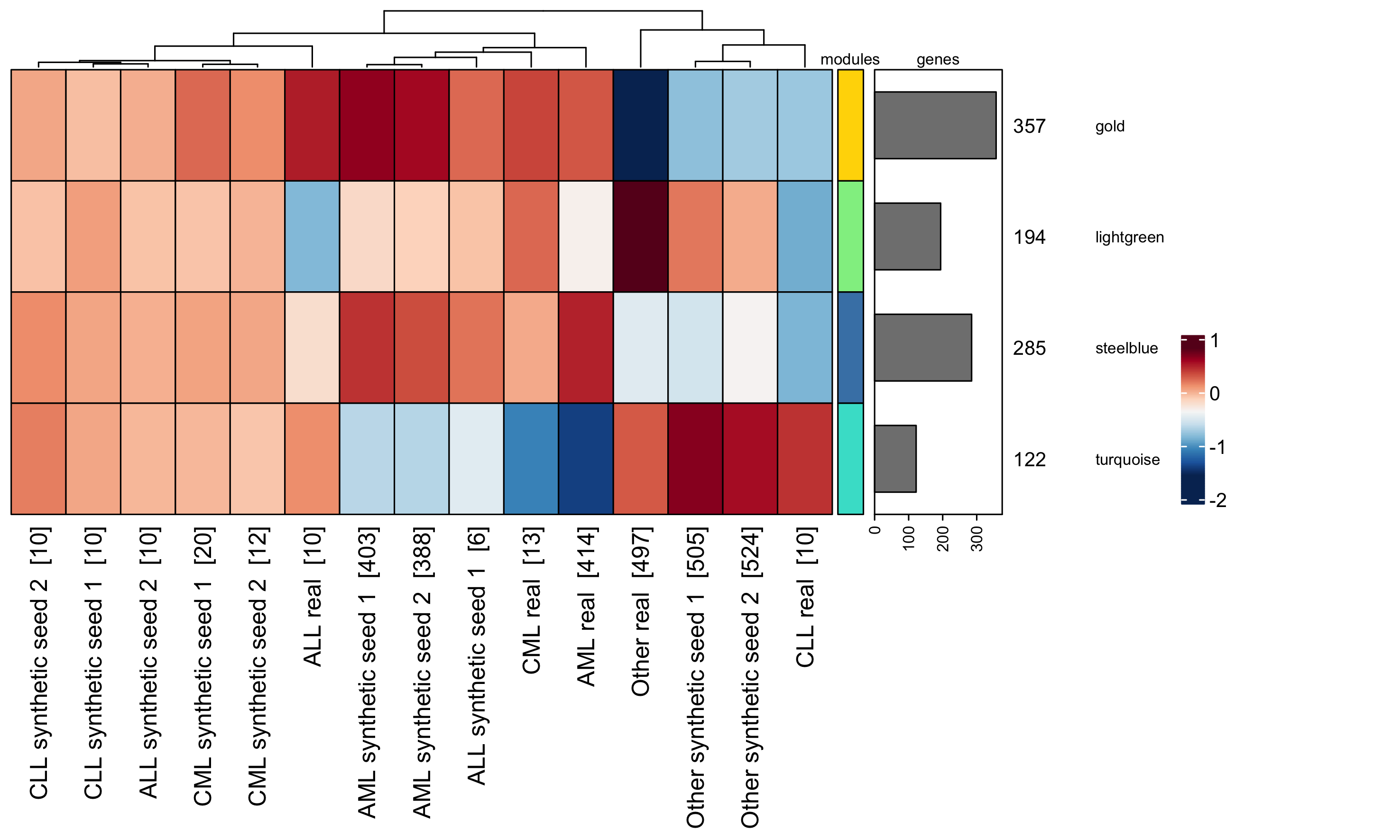}
\vspace{-16pt}
\end{minipage}}
\subfigure[$\varepsilon=10$]{
\begin{minipage}[b]{\figwidth} \includegraphics[,width=1.0\textwidth,trim={0 0 12cm 0},clip]{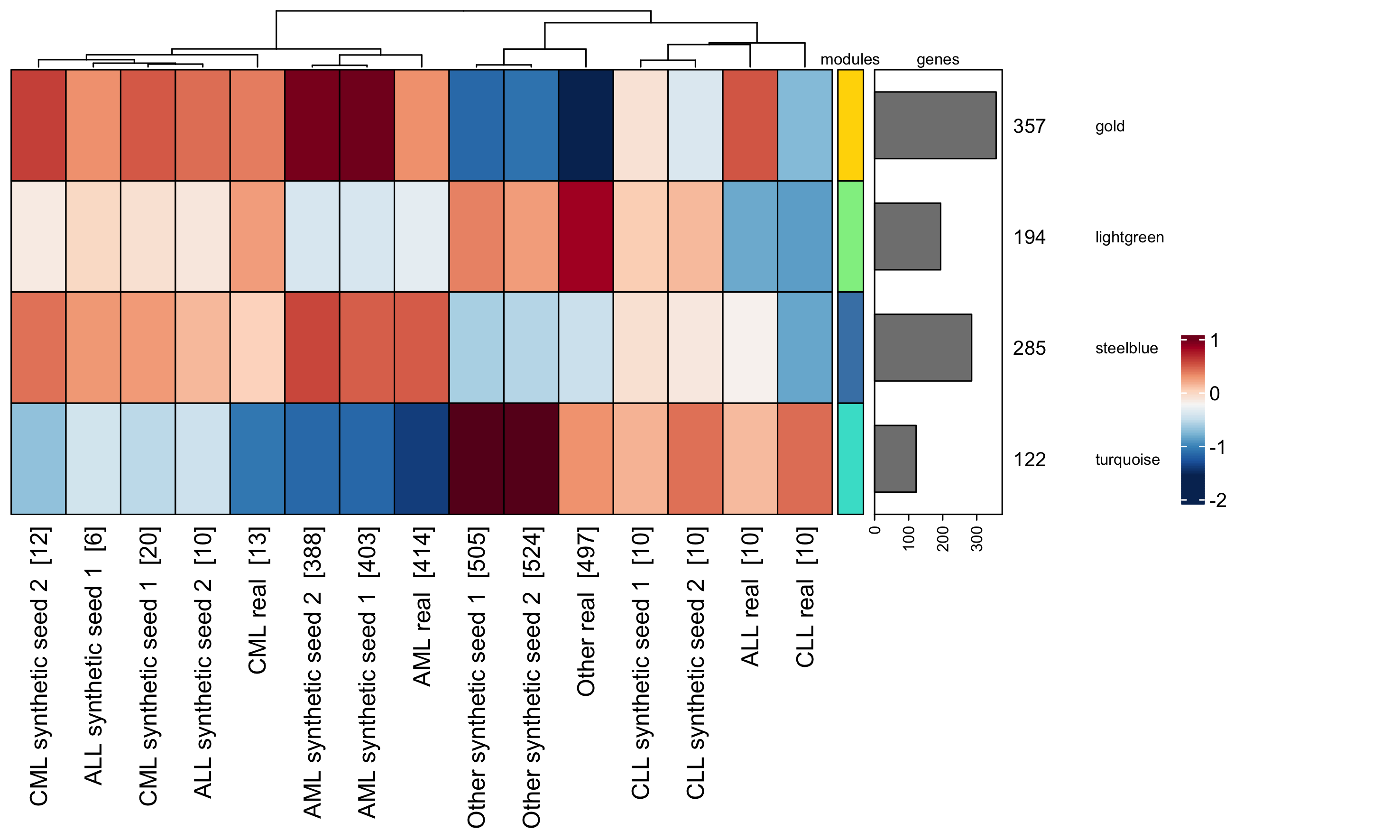}
\vspace{-16pt}
\end{minipage}}
\subfigure[$\varepsilon=20$]{
\begin{minipage}[b]{\figwidth} \includegraphics[,width=1.0\textwidth,trim={0 0 12cm 0},clip]{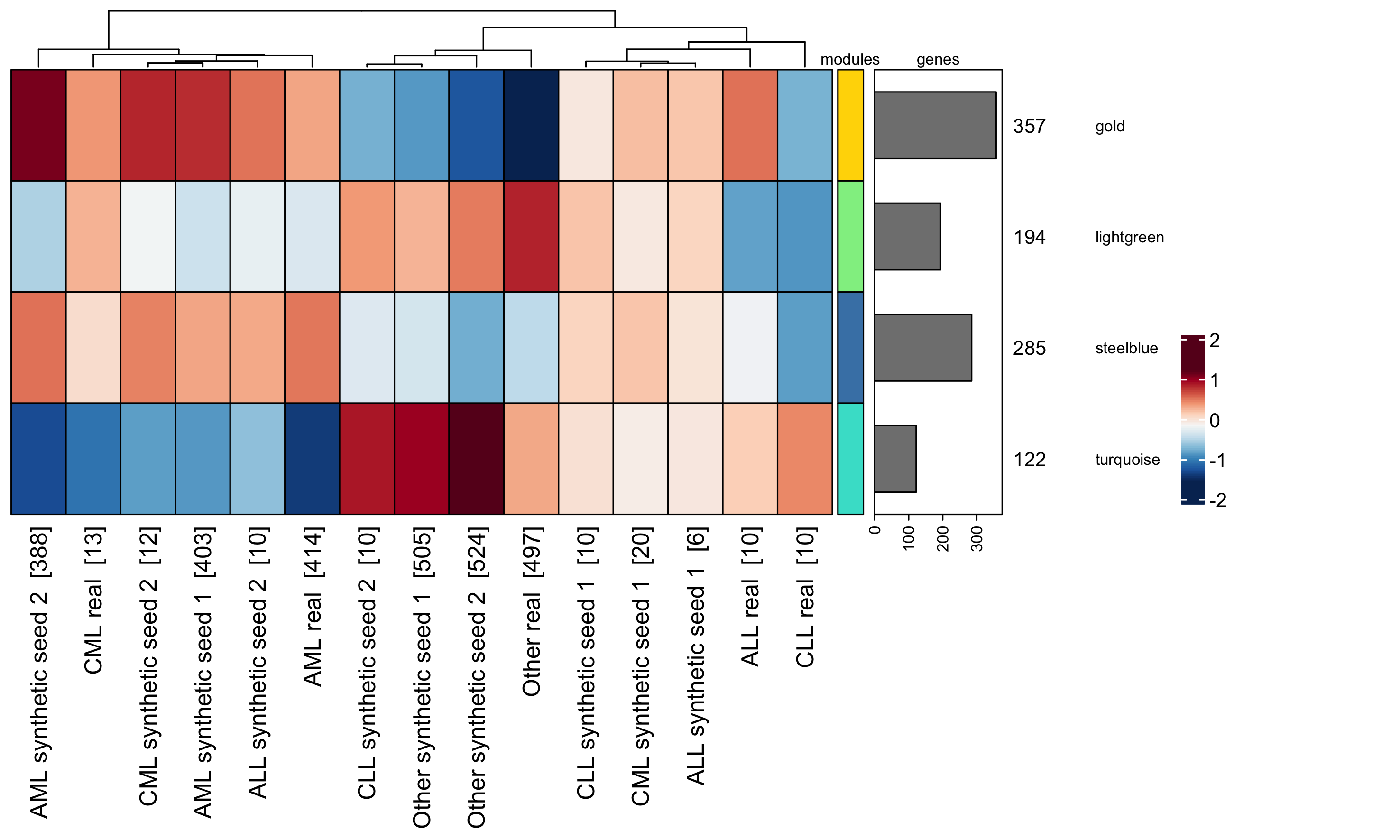}
\vspace{-16pt}
\end{minipage}}
\subfigure[$\varepsilon=50$]{
\begin{minipage}[b]{\figwidth} \includegraphics[,width=1.0\textwidth,trim={0 0 12cm 0},clip]{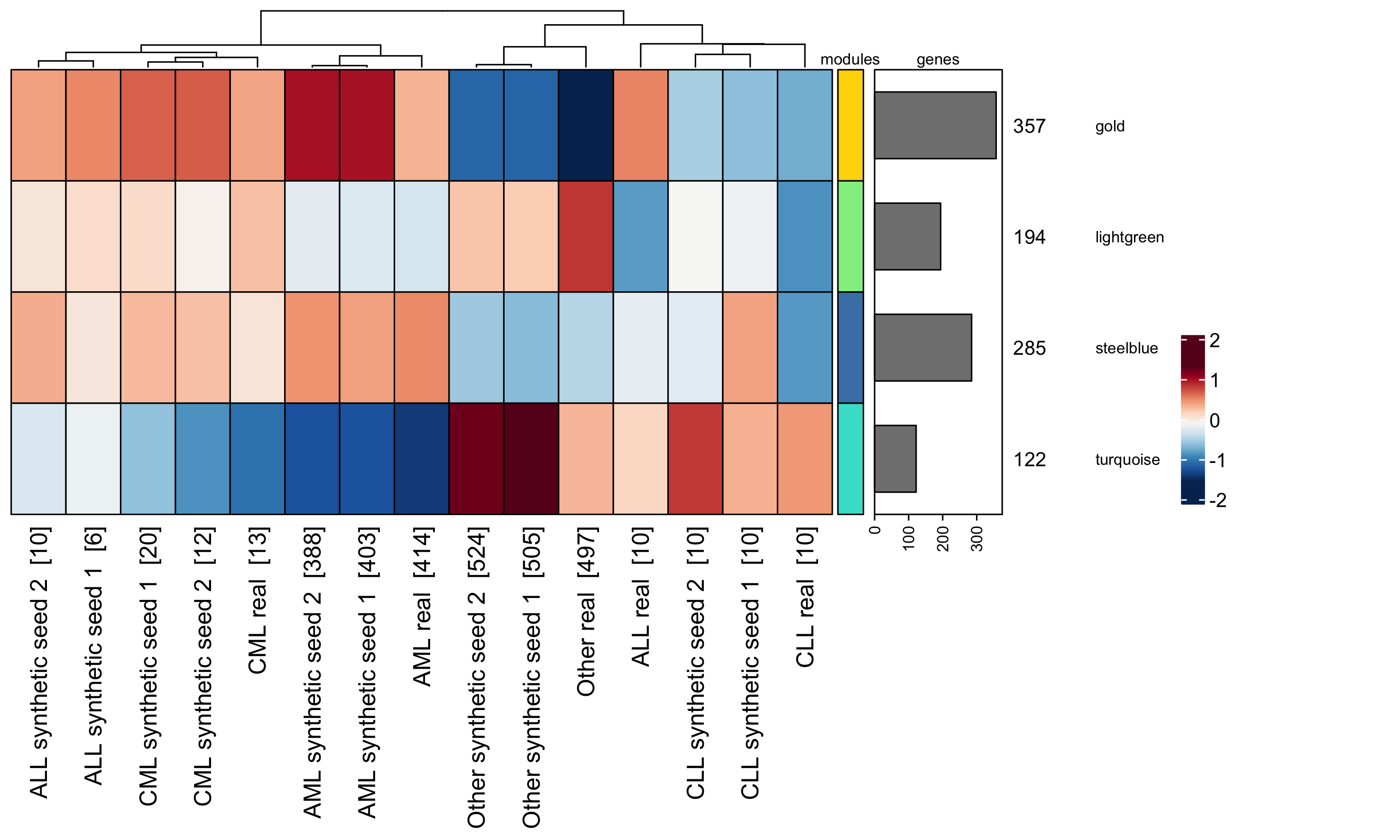}
\vspace{-16pt}
\end{minipage}}
\subfigure[$\varepsilon=100$]{
\begin{minipage}[b]{\figwidth} \includegraphics[,width=1.0\textwidth,trim={0 0 12cm 0},clip]{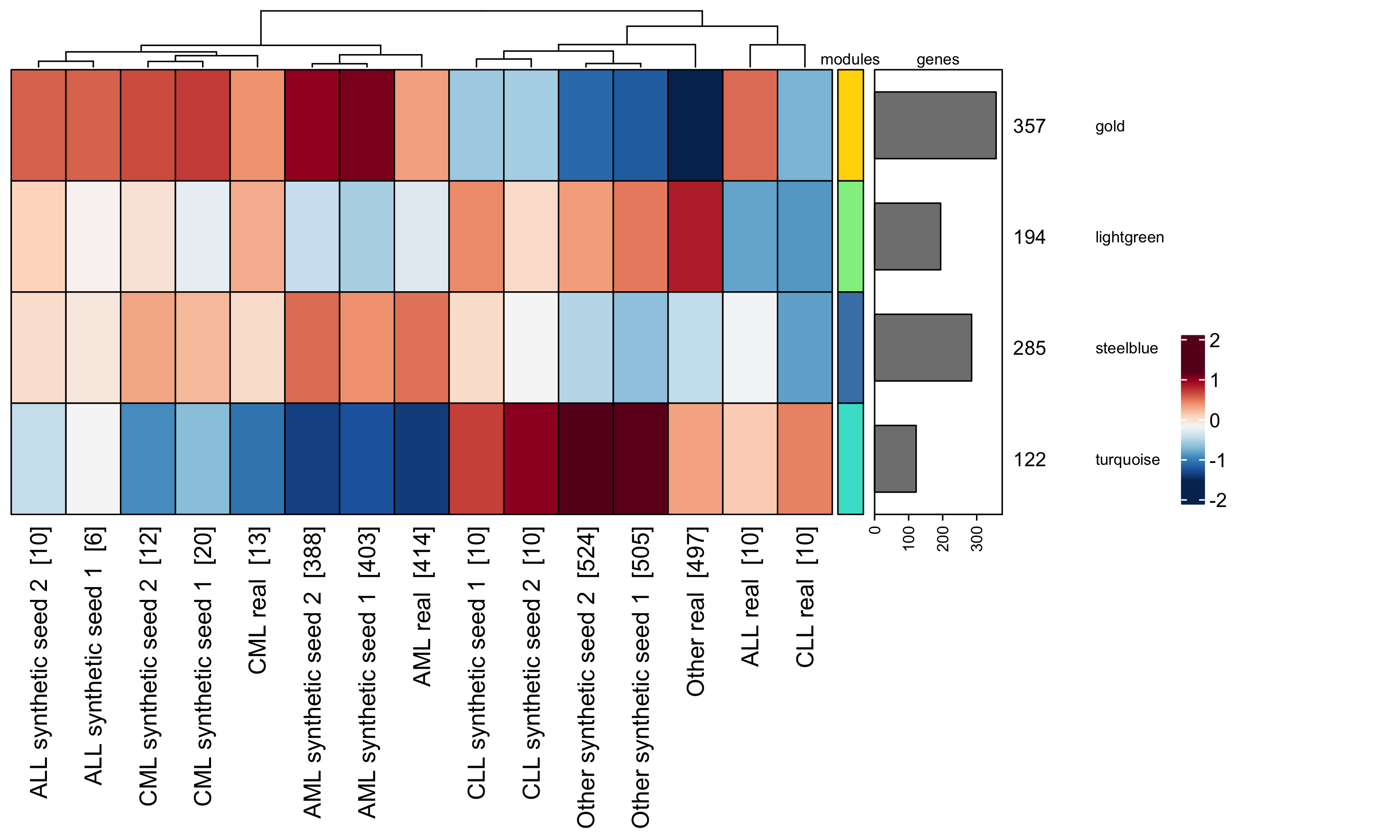}
\vspace{-16pt}
\end{minipage}}
\subfigure[Non-private]{
\begin{minipage}[b]{\figwidth} \includegraphics[trim={0 0 12cm 0},clip,width=1.0\textwidth]{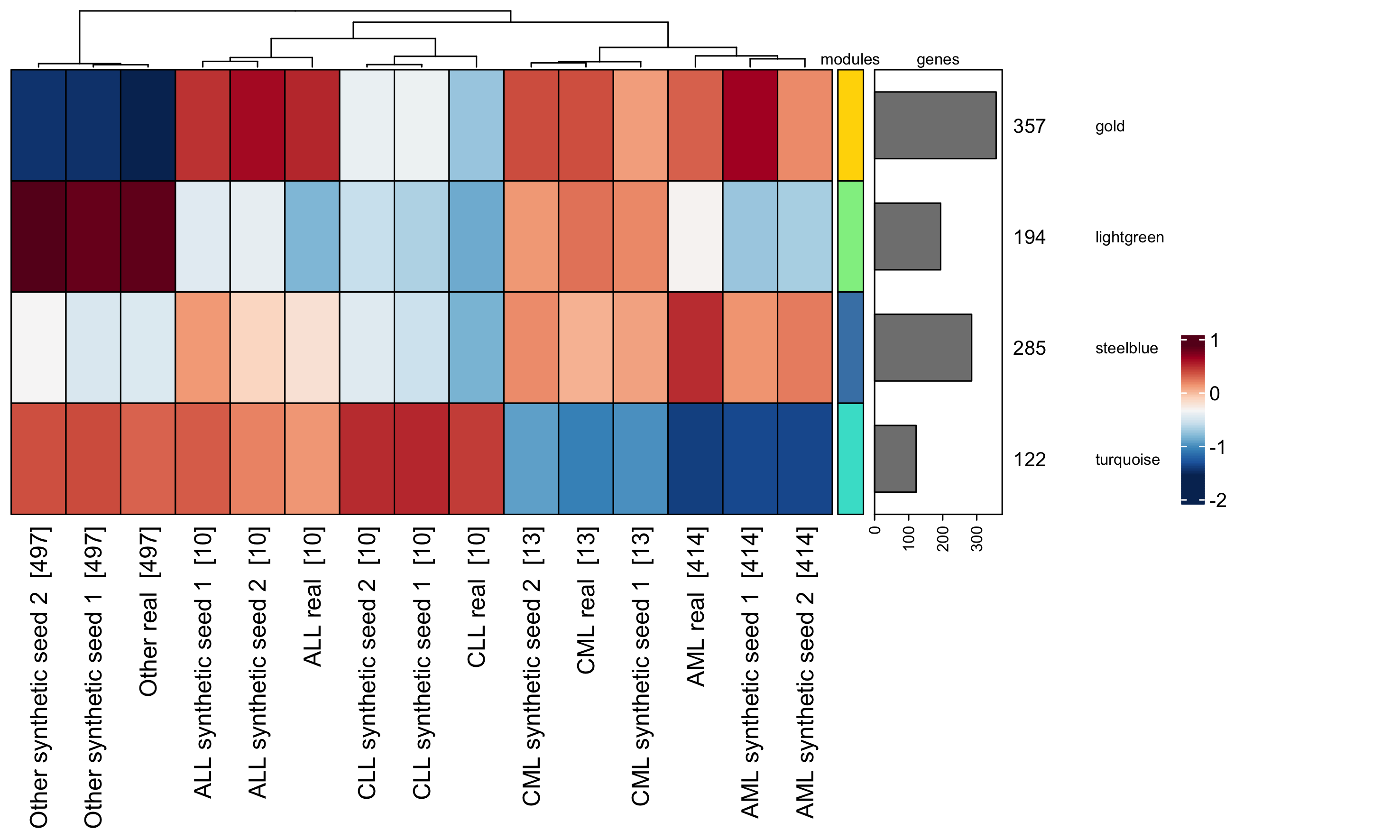}
\vspace{-16pt}
\end{minipage}}
\vspace{-14pt}
\caption{Activation patterns of co-expressed gene modules in \vae for $r$ > 0. \textmd{Shown are the Group Fold
Changes (GFCs) of gene modules (\textit{rows}) in the real and the synthetic data sampled with two different seeds. The dendrograms representing the hierarchical clustering of the sample groups  differentiated by label class and seed, with each \textit{column} corresponding to a distinct group. Optimally, samples with the same label classes should be adjacent, indicating that they are clustered together. Numbers on the right indicate the number of genes per module, numbers in square brackets on the bottom indicate the number of samples per condition and dataset. Darker shades of red imply activation of the gene module, while darker shades of blue indicate deactivation. }}
\label{figure:modules-VAE}
\end{figure*}

\subsubsection{Differential Expression}
We first compared the models’ ability to maintain DE-genes in a \textbf{non-private} case. As shown in \figureautorefname~\ref{figure:DE-Gene Preservation}, it can be observed that the TPR was high for \privsyn and \vae models, reaching more than 75\% on both data split seeds. \rongauss, \pgm and \gan showed subpar results, with the \gan model performing particularly poorly. Regarding the FPR, all models maintained rates below 25\%, with \privsyn and \pgm reaching FPRs close to zero.
\noindent For the \textbf{DP}-case, we observe from \figureautorefname~\ref{figure:DE-Gene Preservation} the following:
\begin{itemize}[leftmargin=*]
    \item \vae[]: At a privacy parameter $\varepsilon$ = 100, the TPR decreases noticeably in comparison to the non-DP setting from around  75\% on average to approximately 50\%. 
    As $\varepsilon$ is reduced further, the TPR continues to show a decreasing tendency, albeit at a less steep rate. Even at the lowest privacy budget of $\varepsilon=5$, the TPR of \vae remains higher than that of the \gan in the non-DP setting. Moreover, \vae shows better or equal TPR than \privsyn at low $\varepsilon$ values, and outperforms \rongauss across all $\varepsilon$ but underperforms \pgm once DP is introduced.
   
    The FPR increases slightly when introducing DP but remains largely stable for different values of $\varepsilon$.
        \item \gan[]:The TPR of DE-genes in the \gan model observed under non-DP conditions remains poor at the introduction of DP and decreasing $\varepsilon$, staying below 20\%, while the FPR remains stable (around 10\%) across all $\varepsilon$.
    \item \rongauss[]: The TPR of the \rongauss model drops when introducing DP. Intriguingly, and somewhat against expectations, it exhibits a slight improvement as the privacy loss $\varepsilon$ decreases, yet it still only attains low values (around 30\%). Concurrently,  
    the FPR steadily increases with decreasing $\varepsilon$, eventually approaching the TPR.
    
    \item \pgm[]: 
    The TPR of 
    \pgm exhibits a slight increase with decreasing $\varepsilon$, with \pgm outperforming all other models for $\varepsilon \le $ 50. Conversely, the FPR rate also increases drastically, reaching around 35\%.
    
    \item \privsyn[]: While \privsyn showed near perfect TPR in the non-DP setting, it is strongly impacted by the introduction of DP, falling below \vae and \pgm for $\varepsilon \leq $ 50. This performance loss is similarly reflected in the increasing FPR with decreasing $ \varepsilon $ values.
    
    \end{itemize}
In summary, while \privsyn and \vae demonstrate good preservation of DE genes in the non-DP case, their performances drop when introducing DP and they are  surpassed by the \pgm model. However, this boost in DE-gene preservation of the \texttt{Private}-\texttt{PGM} model is accompanied by an increasing false positive rate, possibly indicating that the \pgm model generally tends to generate more DE genes, however without biological correctness.  
Both the \gan and the \rongauss models perform poorly on this metric, especially in the DP case.

\subsubsection{Co-expression}
Here, we investigated both the general preservation of co-expressed genes as well as the activation and deactivation of strongly co-expressed gene sets, so-called \emph{modules}, detected in the real data. The preserved co-expressions as well as the activation patterns of co-expressed gene modules were assessed once for all positive correlations identified in the data (r > 0) (\autoref{figure:Co-Expression Preservation}, Appendix  \autoref{figure:S1}-\ref{figure:S5}) 
and once after filtering for only highly co-expressed genes (r > 0.7) (Appendix \autoref{figure:S6}-\ref{figure:S8}). The latter is motivated by the typical interest in strongly correlated genes during co-expression analyses. The detection of gene modules was performed on the real data for these respective filtering thresholds. In both cases, co-expressions were filtered for associated p-values < 0.05.

We first investigate the  \textbf{non-private} case. When considering all co-expressions with $r$ > 0, the \vae reconstructed most of the them while only introducing few false ones that did not exist in the real data (\figureautorefname~\ref{figure:Co-Expression Preservation}). It further maintained highly similar patterns of \textit{up}- and \textit{down-regulation} in the gene modules (Appendix \figureautorefname~\ref{figure:S1}), with samples clustering by class rather than dataset. The \gan model had less correctly and more incorrectly reconstructed co-expressions than the \vae (\figureautorefname~\ref{figure:Co-Expression Preservation}) and the patterns of activation in the gene modules do not match the real data (Appendix \figureautorefname~\ref{figure:S2}). The \pgm and \privsyn models had very similar performances, with more correctly than incorrectly reconstructed co-expressions, however only reconstructing half of the co-expressions found in the real data (\figureautorefname~\ref{figure:Co-Expression Preservation}). The activation of the gene modules was well reconstructed (Appendix \figureautorefname~\ref{figure:S3}, \ref{figure:S4}). The number of correctly and incorrectly reconstructed co-expressions was almost equal for the \rongauss model (\figureautorefname~\ref{figure:Co-Expression Preservation}) and activation patterns in the gene modules were almost entirely lost (Appendix \figureautorefname~\ref{figure:S5}).
Reducing the co-expressions to only those with $r$ > 0.7, only the \vae model and one of the sampling seeds for the \gan yielded any results. The \vae correctly reconstructed most co-expressions from the real set but additionally introduced an almost equal number of incorrect co-expressions (Appendix \figureautorefname~\ref{figure:S6}). Activation patterns in gene modules were well preserved (Appendix \figureautorefname~\ref{figure:S7}). In the data generated by the \gan[], the number of incorrectly introduced co-expressions was very high (Appendix \figureautorefname~\ref{figure:S6}) and activation patterns in the gene modules remained poor (Appendix \figureautorefname~\ref{figure:S8}).

\noindent For the \textbf{DP}-case, we list below our findings:
\begin{itemize}[leftmargin=*]
    \item \vae[]: For all co-expressions with $r$ > 0, the number of correct co-expressions reconstructed by the \vae reduced gradually when introducing DP with decreasing $\varepsilon$ (\figureautorefname~\ref{figure:Co-Expression Preservation}). Meanwhile, the number of incorrect co-expressions more than doubled. Activation patterns of gene modules were well maintained for the classes \textit{CML}, \textit{AML} and \textit{Other} at $\varepsilon$ = 100 and 50. For lower $\varepsilon$, the characteristic patterns of the modules were increasingly lost as illustrated in \figureautorefname~\ref{figure:modules-VAE}, 
    indicated by the increasing lack of distinctive colors. While the order of gene modules (rows) is fixed to improve comparability, the order of sample groups (columns) is dictated by their hierarchical clustering. This is intended, since it illustrates similarity between module expression of different conditions in the different datasets. In the case of biologically high-quality synthetic data, synthetic samples  are expected to co-locate with real samples of the same condition. Note that the results are only shown for one seed used for splitting the dataset for training. Detailed illustrations for all $\varepsilon$ and seeds can be found in Appendix \figureautorefname~\ref{figure:S1}. If the synthetic data successfully captured the co-expression modules, disease classes are expected to cluster together across synthetic and real data. Focusing only on highly co-expressed genes with $r$ > 0.7, a high number of co-expressions is introduced that do not occur in the real data (Appendix \figureautorefname~\ref{figure:S6}). Preservation of module activation is comparable to that observed when selecting co-expressions with $r$ > 0 (Appendix \figureautorefname~\ref{figure:S7}).
\item \gan[]: When considering all co-expressions with $r$ > 0, the number of correctly reconstructed co-expressions decreased and the number of incorrect ones increased when introducing DP with $\varepsilon=100$ and reducing this value did not impact the metric further (\figureautorefname~\ref{figure:Co-Expression Preservation}). The module activation patterns from the real data are almost entirely lost with the modules demonstrating homogeneous activation (Appendix \figureautorefname~\ref{figure:S2}). When filtering for $r$ > 0.7 there were no co-expressions left for any of the $\varepsilon$-values.

\item \pgm \& \privsyn[]: The \pgm and \privsyn models demonstrated similar behavior, with the number of reconstructed co-expressions barely being affected by introducing varying levels of privacy in comparison to the non-DP case (\figureautorefname~\ref{figure:Co-Expression Preservation}).
\pgm  maintained the module activation patters for very high $\varepsilon$-values (100 and 50) and for lower $\varepsilon$ (20, 10, 5) patterns of large classes such as \textit{AML} and \textit{Other} where maintained, but degraded for the smaller classes (Appendix \figureautorefname~\ref{figure:S3}). Similar results are observed for \privsyn[], with the exception that the degradation of activation patterns already starts at $\varepsilon=50$ (Appendix \figureautorefname~\ref{figure:S4}). Like the \gan[], both models did not generate any significant co-expressions exceeding $r$ > 0.7.

\item \rongauss[]: While in the non-DP case, the number of incorrect co-expressions was still slightly lower than that of correct ones, this changes in the DP-case (\figureautorefname~\ref{figure:Co-Expression Preservation}). Decreasing values of $\varepsilon$, however, not only gradually increased the number of incorrectly reconstructed ones, but also that of the correctly reconstructed co-expressions. The gene modules lose their distinctive patterns, showing uniform activation and thus the synthetic data is clustering distinctly away from the real data for all $\varepsilon$ (Appendix \figureautorefname~\ref{figure:S5}). As was the case for all models but the \vae[], no high co-expressions with $r$ > 0.7 were generated by the \rongauss model.
\end{itemize}

\noindent In summary, all models except the \vae struggled at correctly recreating strong co-expressions and even the \vae was prone to introducing a high number of incorrect co-expressions for $r$ > 0.7. Also for weaker co-expressions, introducing DP strongly impaired the utility of the data both in terms of general co-expressions as well as the activation and inactivation of highly co-expressed modules, with only high $\varepsilon$-values of 100 and 50 maintaining the co-expression structure in the data of some models but not offering any considerable privacy.

\section{Discussion and Future Directions}
\label{sec:discussion}
\myparagraph{Private vs. non-private synthetic data} The biological evaluation of the different models yielded that some model types are capable of generating synthetic data with high biological utility in the non-DP case. 
However, the incorporation of DP, though essential for maintaining privacy, significantly hampers their performance. In examining the generally top-performing \vae models through membership inference attacks (\sectionautorefname~\ref{sec:threat_model}), we found that non-DP training poses a considerable privacy risk, with AUC-ROC scores of 0.949 and 0.614 for white-box (implemented following \cite{hilprecht2019monte}) and black-box attacks (implemented following \cite{chen2020gan}), respectively. Notably, setting the privacy budget at a relative high level of $\varepsilon$=100 resulted in a rapid decline of AUC-ROC scores to around 0.52 in both scenarios. While such high privacy budgets $\varepsilon=$100/50 in some cases still allowed good reconstruction of biology properties as measured by our metrics, these budgets are generally too high to be considered strictly privacy-preserving.

\myparagraph{Challenges of low sample regime} As has become apparent in the analysis of activation patterns of co-expressed modules, classes with low sample counts were the first to lose their activation patterns with decreasing privacy budgets. However, such low sample sizes are highly common in gene expression datasets given the often low availability of sampling material. This is particularly the case for rare diseases or samples that can only be acquired with invasive and/or risky medical procedures. Another point that requires addressing is the feature space. The results presented here were achieved on a strongly reduced feature space of approximately 1000 genes, with gene expression datasets often comprising 20-times as many features. 
The observed limitations of differentially private data generation can thus be expected to increase further when attempting to generate full sets. 

\myparagraph{Comparing models} The biological evaluation indicated that some model architectures (\vae[], \privsyn[] and
sometimes \pgm[]) are better than others (\gan[], \rongauss[]) at learning and generating such highly complex, non-normally distributed data like gene expressions. In general, \vae[] stands out with the best overall performance, likely because of their substantial expressive capacity, which outperforms simpler probabilistic models like \rongauss[] and methods dependent on low-dimensional approximations, such as \privsyn[] and \pgm[]. Moreover, VAEs benefit from stable training processes, advantageous in scenarios with limited samples, unlike the less stable GANs.  However, incorporating privacy into this process presents challenges, while maintaining biological utility in a privacy-preserving manner requires further research and possibly more data.

\myparagraph{Dependent data} In certain scenarios where the dataset used contains dependent records—such as those associated with the same individual (e.g., single-cell data), a transition to a more advanced level of protection becomes imperative, wherein the goal shifts to preserving each group of dependent records (referred to as Group-level DP). However, this elevation in privacy protection comes with the trade-off of injecting more noise, potentially leading to a greater compromise in the quality of the synthetic data. Furthermore, the task of defining a set of dependent records is not always straightforward. For instance, while it is evident that individuals within the same family often share a common genomic heritage, the extent of relatedness to consider when forming such groups remains ambiguous. Determining whether to include only immediate family members like parents and siblings or to encompass more distant relatives poses an additional challenge. Due to these intricate aspects of privacy considerations, we opt to exclude single-cell datasets from our analysis, despite their potential size advantage for assessing non-DP generative models.

\myparagraph{General-purpose synthetic data vs. task-specific data} Providing general-purpose private synthetic data that is useful for all kinds of downstream tasks while preserving statistical and biological properties is still a highly challenging task. Having accurate generators would also imply a strong model and  insights for the respective domain, which is often not the case for many bio-medical applications. In addition, small sub-population might not be represented and suffer from mode collapse issues of the generator. It has also been recently questions to what extend such an ultimate solution can be achieved at all ~\cite{groundhog,Stadler}. While it is difficult to predict how these trade-off develop in the future, the increased available of such medical data will have a positive effect. In addition, task-specific data generation (e.g. \cite{chen22neurips}) in a data distillation approach can relax the objectives, but is also departing from the goal of preserving statistic and biological properties by mostly focusing on downstream utility.

\section{Conclusions}
We provide the first systematic analysis of non-private and differentially private generation of gene expression data that covers five diverse modeling approaches ranging from simple density estimation over graphical models to deep generative models. Our analysis encompasses a diverse set of metrics that shed light on the quality of the generated data in terms of statistical and biological properties as well as down-stream utility. A key message of our work is that such a broad evaluation is necessary in order to understand the limitations of current generators. Overall, simple estimators fall behind in performance but equally very complex models like GAN are suffering from the low sample regime as typically encountered in bio-medical applications. While downstream utility can be strong, the synthetic data itself might not retain statistical nor biological properties. Adding privacy preserving estimation and learning of the generators amplifies these problems. A general model recommendation is difficult to provide, as these trade-offs will shift as more data is going to become available in the future. However, we see a tendency that the evaluated graphical models have retained better the differential expression and the variational autoencoder retained better the co-expression - in particular when privacy is added. We release our setup and evaluation framework in order to further drive progress in this domain\footnote{https://github.com/MarieOestreich/PRO-GENE-GEN}.

\begin{acks}
This work is supported by the Helmholtz Association within the project Protecting Genetic Data with Synthetic Cohorts from Deep Generative Models (PRO-GENE-GEN), grant No. ZT-I-PF-5-23 and Bundesministeriums fur Bildung und Forschung (PriSyn), grant No. 16KISAO29K. Additionally, this work is partially funded by ELSA – European Lighthouse on Secure and Safe AI funded by the European Union under grant agreement No. 101070617. Views and opinions expressed are however those of the authors only and do not necessarily reflect those of the European Union or European Commission. Neither the European Union nor the European Commission can be held responsible for them. Moreover, Dingfan Chen was partially supported by Qualcomm Innovation Fellowship Europe.
\end{acks}

\bibliographystyle{ACM-Reference-Format}
\bibliography{main}


\begin{thebibliography}{46}


\ifx \showCODEN    \undefined \def \showCODEN     #1{\unskip}     \fi
\ifx \showDOI      \undefined \def \showDOI       #1{#1}\fi
\ifx \showISBNx    \undefined \def \showISBNx     #1{\unskip}     \fi
\ifx \showISBNxiii \undefined \def \showISBNxiii  #1{\unskip}     \fi
\ifx \showISSN     \undefined \def \showISSN      #1{\unskip}     \fi
\ifx \showLCCN     \undefined \def \showLCCN      #1{\unskip}     \fi
\ifx \shownote     \undefined \def \shownote      #1{#1}          \fi
\ifx \showarticletitle \undefined \def \showarticletitle #1{#1}   \fi
\ifx \showURL      \undefined \def \showURL       {\relax}        \fi
\providecommand\bibfield[2]{#2}
\providecommand\bibinfo[2]{#2}
\providecommand\natexlab[1]{#1}
\providecommand\showeprint[2][]{arXiv:#2}

\bibitem[Abadi et~al\mbox{.}(2016)]%
        {abadi2016deep}
\bibfield{author}{\bibinfo{person}{Martin Abadi}, \bibinfo{person}{Andy Chu},
  \bibinfo{person}{Ian Goodfellow}, \bibinfo{person}{H~Brendan McMahan},
  \bibinfo{person}{Ilya Mironov}, \bibinfo{person}{Kunal Talwar}, {and}
  \bibinfo{person}{Li Zhang}.} \bibinfo{year}{2016}\natexlab{}.
\newblock \showarticletitle{Deep learning with differential privacy}. In
  \bibinfo{booktitle}{\emph{Proceedings of the 2016 ACM SIGSAC conference on
  computer and communications security}}. \bibinfo{pages}{308--318}.
\newblock


\bibitem[Afonja et~al\mbox{.}(2023)]%
        {afonja2023margctgan}
\bibfield{author}{\bibinfo{person}{Tejumade Afonja}, \bibinfo{person}{Dingfan
  Chen}, {and} \bibinfo{person}{Mario Fritz}.} \bibinfo{year}{2023}\natexlab{}.
\newblock \showarticletitle{MargCTGAN: A "Marginally" Better CTGAN for the Low
  Sample Regime}.
\newblock \bibinfo{journal}{\emph{arXiv preprint arXiv:2307.07997}}
  (\bibinfo{year}{2023}).
\newblock


\bibitem[Alzantot and Srivastava(2019)]%
        {uclanesl_dp_wgan}
\bibfield{author}{\bibinfo{person}{Moustafa Alzantot} {and}
  \bibinfo{person}{Mani Srivastava}.} \bibinfo{year}{2019}\natexlab{}.
\newblock \bibinfo{title}{{Differential Privacy Synthetic Data Generation using
  WGANs}}.
\newblock
\newblock
\urldef\tempurl%
\url{https://github.com/nesl/nist_differential_privacy_synthetic_data_challenge/}
\showURL{%
\tempurl}


\bibitem[Anders and Huber(2010)]%
        {Anders_Huber_2010}
\bibfield{author}{\bibinfo{person}{Simon Anders} {and}
  \bibinfo{person}{Wolfgang Huber}.} \bibinfo{year}{2010}\natexlab{}.
\newblock \showarticletitle{Differential expression analysis for sequence count
  data.}
\newblock \bibinfo{journal}{\emph{Genome Biology}} \bibinfo{volume}{11},
  \bibinfo{number}{10} (\bibinfo{date}{Oct} \bibinfo{year}{2010}),
  \bibinfo{pages}{R106}.
\newblock
\urldef\tempurl%
\url{https://doi.org/10.1186/gb-2010-11-10-r106}
\showDOI{\tempurl}


\bibitem[Arjovsky et~al\mbox{.}(2017)]%
        {arjovsky2017wasserstein}
\bibfield{author}{\bibinfo{person}{Martin Arjovsky}, \bibinfo{person}{Soumith
  Chintala}, {and} \bibinfo{person}{L{\'e}on Bottou}.}
  \bibinfo{year}{2017}\natexlab{}.
\newblock \showarticletitle{Wasserstein generative adversarial networks}. In
  \bibinfo{booktitle}{\emph{International conference on machine learning}}.
  PMLR, \bibinfo{pages}{214--223}.
\newblock


\bibitem[Assefa et~al\mbox{.}(2020)]%
        {Assefa_Vandesompele_Thas_2020}
\bibfield{author}{\bibinfo{person}{Alemu~Takele Assefa}, \bibinfo{person}{Jo
  Vandesompele}, {and} \bibinfo{person}{Olivier Thas}.}
  \bibinfo{year}{2020}\natexlab{}.
\newblock \showarticletitle{SPsimSeq: semi-parametric simulation of bulk and
  single-cell RNA-sequencing data.}
\newblock \bibinfo{journal}{\emph{Bioinformatics}} \bibinfo{volume}{36},
  \bibinfo{number}{10} (\bibinfo{date}{May} \bibinfo{year}{2020}),
  \bibinfo{pages}{3276–3278}.
\newblock
\urldef\tempurl%
\url{https://doi.org/10.1093/bioinformatics/btaa105}
\showDOI{\tempurl}


\bibitem[Chanyaswad et~al\mbox{.}(2019)]%
        {chanyaswad2019ron}
\bibfield{author}{\bibinfo{person}{Thee Chanyaswad},
  \bibinfo{person}{Changchang Liu}, {and} \bibinfo{person}{Prateek Mittal}.}
  \bibinfo{year}{2019}\natexlab{}.
\newblock \showarticletitle{RON-Gauss: Enhancing Utility in Non-Interactive
  Private Data Release}.
\newblock \bibinfo{journal}{\emph{Proceedings on Privacy Enhancing
  Technologies}}  \bibinfo{volume}{1} (\bibinfo{year}{2019}),
  \bibinfo{pages}{26--46}.
\newblock


\bibitem[Chen et~al\mbox{.}(2022)]%
        {chen22neurips}
\bibfield{author}{\bibinfo{person}{Dingfan Chen}, \bibinfo{person}{Raouf
  Kerkouche}, {and} \bibinfo{person}{Mario Fritz}.}
  \bibinfo{year}{2022}\natexlab{}.
\newblock \showarticletitle{Private Set Generation with Discriminative
  Information}. In \bibinfo{booktitle}{\emph{Neural Information Processing
  Systems (NeurIPS)}} (2022-12-01).
\newblock
\urldef\tempurl%
\url{https://arxiv.org/abs/2211.04446 https://arxiv.org/pdf/2211.04446.pdf}
\showURL{%
\tempurl}


\bibitem[Chen et~al\mbox{.}(2020)]%
        {chen2020gan}
\bibfield{author}{\bibinfo{person}{Dingfan Chen}, \bibinfo{person}{Ning Yu},
  \bibinfo{person}{Yang Zhang}, {and} \bibinfo{person}{Mario Fritz}.}
  \bibinfo{year}{2020}\natexlab{}.
\newblock \showarticletitle{Gan-leaks: A taxonomy of membership inference
  attacks against generative models}. In \bibinfo{booktitle}{\emph{Proceedings
  of the 2020 ACM SIGSAC conference on computer and communications security
  (CCS)}}. \bibinfo{pages}{343--362}.
\newblock


\bibitem[Conesa et~al\mbox{.}(2016)]%
        {Conesa}
\bibfield{author}{\bibinfo{person}{Ana Conesa}, \bibinfo{person}{Pedro
  Madrigal}, \bibinfo{person}{Sonia Tarazona}, \bibinfo{person}{David
  Gomez-Cabrero}, \bibinfo{person}{Alejandra Cervera}, \bibinfo{person}{Andrew
  McPherson}, \bibinfo{person}{Michał~Wojciech Szcześniak},
  \bibinfo{person}{Daniel~J Gaffney}, \bibinfo{person}{Laura~L Elo},
  \bibinfo{person}{Xuegong Zhang}, {and} \bibinfo{person}{et al.}}
  \bibinfo{year}{2016}\natexlab{}.
\newblock \showarticletitle{A survey of best practices for RNA-seq data
  analysis.}
\newblock \bibinfo{journal}{\emph{Genome Biology}} \bibinfo{volume}{17},
  \bibinfo{number}{1} (\bibinfo{date}{Jan} \bibinfo{year}{2016}),
  \bibinfo{pages}{13}.
\newblock
\urldef\tempurl%
\url{https://doi.org/10.1186/s13059-016-0881-8}
\showDOI{\tempurl}


\bibitem[Costa-Silva et~al\mbox{.}(2017)]%
        {Costa-Silva}
\bibfield{author}{\bibinfo{person}{Juliana Costa-Silva},
  \bibinfo{person}{Douglas Domingues}, {and} \bibinfo{person}{Fabricio~Martins
  Lopes}.} \bibinfo{year}{2017}\natexlab{}.
\newblock \showarticletitle{RNA-Seq differential expression analysis: An
  extended review and a software tool.}
\newblock \bibinfo{journal}{\emph{Plos One}} \bibinfo{volume}{12},
  \bibinfo{number}{12} (\bibinfo{date}{Dec} \bibinfo{year}{2017}),
  \bibinfo{pages}{e0190152}.
\newblock
\urldef\tempurl%
\url{https://doi.org/10.1371/journal.pone.0190152}
\showDOI{\tempurl}


\bibitem[de~Torrenté et~al\mbox{.}(2020)]%
        {de_Torrenté_Zimmerman_Suzuki_Christopeit_Greally_Mar_2020}
\bibfield{author}{\bibinfo{person}{Laurence de Torrenté},
  \bibinfo{person}{Samuel Zimmerman}, \bibinfo{person}{Masako Suzuki},
  \bibinfo{person}{Maximilian Christopeit}, \bibinfo{person}{John~M Greally},
  {and} \bibinfo{person}{Jessica~C Mar}.} \bibinfo{year}{2020}\natexlab{}.
\newblock \showarticletitle{The shape of gene expression distributions matter:
  how incorporating distribution shape improves the interpretation of cancer
  transcriptomic data.}
\newblock \bibinfo{journal}{\emph{BMC Bioinformatics}} \bibinfo{volume}{21},
  \bibinfo{number}{Suppl 21} (\bibinfo{date}{Dec} \bibinfo{year}{2020}),
  \bibinfo{pages}{562}.
\newblock
\urldef\tempurl%
\url{https://doi.org/10.1186/s12859-020-03892-w}
\showDOI{\tempurl}


\bibitem[Dwork et~al\mbox{.}(2014)]%
        {dwork2014algorithmic}
\bibfield{author}{\bibinfo{person}{Cynthia Dwork}, \bibinfo{person}{Aaron
  Roth}, {et~al\mbox{.}}} \bibinfo{year}{2014}\natexlab{}.
\newblock \showarticletitle{The algorithmic foundations of differential
  privacy}.
\newblock \bibinfo{journal}{\emph{Foundations and Trends{\textregistered} in
  Theoretical Computer Science}} \bibinfo{volume}{9}, \bibinfo{number}{3--4}
  (\bibinfo{year}{2014}).
\newblock


\bibitem[Gillenwater et~al\mbox{.}(2021)]%
        {gillenwater21a}
\bibfield{author}{\bibinfo{person}{Jennifer Gillenwater},
  \bibinfo{person}{Matthew Joseph}, {and} \bibinfo{person}{Alex Kulesza}.}
  \bibinfo{year}{2021}\natexlab{}.
\newblock \showarticletitle{Differentially Private Quantiles}. In
  \bibinfo{booktitle}{\emph{Proceedings of the 38th International Conference on
  Machine Learning (ICML)}} \emph{(\bibinfo{series}{Proceedings of Machine
  Learning Research}, Vol.~\bibinfo{volume}{139})},
  \bibfield{editor}{\bibinfo{person}{Marina Meila} {and} \bibinfo{person}{Tong
  Zhang}} (Eds.). \bibinfo{publisher}{PMLR}, \bibinfo{pages}{3713--3722}.
\newblock


\bibitem[Goodfellow et~al\mbox{.}(2014)]%
        {goodfellow2014generative}
\bibfield{author}{\bibinfo{person}{Ian Goodfellow}, \bibinfo{person}{Jean
  Pouget-Abadie}, \bibinfo{person}{Mehdi Mirza}, \bibinfo{person}{Bing Xu},
  \bibinfo{person}{David Warde-Farley}, \bibinfo{person}{Sherjil Ozair},
  \bibinfo{person}{Aaron Courville}, {and} \bibinfo{person}{Yoshua Bengio}.}
  \bibinfo{year}{2014}\natexlab{}.
\newblock \showarticletitle{Generative adversarial nets}.
\newblock \bibinfo{journal}{\emph{Advances in neural information processing
  systems}}  \bibinfo{volume}{27} (\bibinfo{year}{2014}).
\newblock


\bibitem[Hilprecht et~al\mbox{.}(2019)]%
        {hilprecht2019monte}
\bibfield{author}{\bibinfo{person}{Benjamin Hilprecht}, \bibinfo{person}{Martin
  H{\"a}rterich}, {and} \bibinfo{person}{Daniel Bernau}.}
  \bibinfo{year}{2019}\natexlab{}.
\newblock \showarticletitle{Monte Carlo and Reconstruction Membership Inference
  Attacks against Generative Models.}
\newblock \bibinfo{journal}{\emph{Proc. Priv. Enhancing Technol.}}
  \bibinfo{volume}{2019}, \bibinfo{number}{4} (\bibinfo{year}{2019}),
  \bibinfo{pages}{232--249}.
\newblock


\bibitem[Kantorovich(1960)]%
        {kantorovich1960mathematical}
\bibfield{author}{\bibinfo{person}{Leonid~V Kantorovich}.}
  \bibinfo{year}{1960}\natexlab{}.
\newblock \showarticletitle{Mathematical methods of organizing and planning
  production}.
\newblock \bibinfo{journal}{\emph{Management science}} \bibinfo{volume}{6},
  \bibinfo{number}{4} (\bibinfo{year}{1960}), \bibinfo{pages}{366--422}.
\newblock


\bibitem[Kingma and Welling(2014)]%
        {Kingma2014}
\bibfield{author}{\bibinfo{person}{Diederik~P. Kingma} {and}
  \bibinfo{person}{Max Welling}.} \bibinfo{year}{2014}\natexlab{}.
\newblock \showarticletitle{{Auto-Encoding Variational Bayes}}. In
  \bibinfo{booktitle}{\emph{2nd International Conference on Learning
  Representations (ICLR)}}.
\newblock


\bibitem[Lall et~al\mbox{.}(2022)]%
        {Lall_Ray_Bandyopadhyay_2022}
\bibfield{author}{\bibinfo{person}{Snehalika Lall}, \bibinfo{person}{Sumanta
  Ray}, {and} \bibinfo{person}{Sanghamitra Bandyopadhyay}.}
  \bibinfo{year}{2022}\natexlab{}.
\newblock \showarticletitle{LSH-GAN enables in-silico generation of cells for
  small sample high dimensional scRNA-seq data.}
\newblock \bibinfo{journal}{\emph{Communications Biology}} \bibinfo{volume}{5},
  \bibinfo{number}{1} (\bibinfo{date}{Jun} \bibinfo{year}{2022}),
  \bibinfo{pages}{577}.
\newblock
\urldef\tempurl%
\url{https://doi.org/10.1038/s42003-022-03473-y}
\showDOI{\tempurl}


\bibitem[Langfelder and Horvath(2008)]%
        {Langfelder_Horvath_2008}
\bibfield{author}{\bibinfo{person}{Peter Langfelder} {and}
  \bibinfo{person}{Steve Horvath}.} \bibinfo{year}{2008}\natexlab{}.
\newblock \showarticletitle{WGCNA: an R package for weighted correlation
  network analysis.}
\newblock \bibinfo{journal}{\emph{BMC Bioinformatics}}  \bibinfo{volume}{9}
  (\bibinfo{date}{Dec} \bibinfo{year}{2008}), \bibinfo{pages}{559}.
\newblock
\urldef\tempurl%
\url{https://doi.org/10.1186/1471-2105-9-559}
\showDOI{\tempurl}


\bibitem[Li and Wang(2021)]%
        {Li_Wang_2021}
\bibfield{author}{\bibinfo{person}{Xinmin Li} {and} \bibinfo{person}{Cun-Yu
  Wang}.} \bibinfo{year}{2021}\natexlab{}.
\newblock \showarticletitle{From bulk, single-cell to spatial RNA sequencing.}
\newblock \bibinfo{journal}{\emph{International journal of oral science}}
  \bibinfo{volume}{13}, \bibinfo{number}{1} (\bibinfo{date}{Nov}
  \bibinfo{year}{2021}), \bibinfo{pages}{36}.
\newblock
\urldef\tempurl%
\url{https://doi.org/10.1038/s41368-021-00146-0}
\showDOI{\tempurl}


\bibitem[Lin(1991)]%
        {Lin1991DivergenceMB}
\bibfield{author}{\bibinfo{person}{Jianhua Lin}.}
  \bibinfo{year}{1991}\natexlab{}.
\newblock \showarticletitle{Divergence measures based on the Shannon entropy}.
\newblock \bibinfo{journal}{\emph{IEEE Trans. Inf. Theory}}
  \bibinfo{volume}{37} (\bibinfo{year}{1991}), \bibinfo{pages}{145--151}.
\newblock
\urldef\tempurl%
\url{https://api.semanticscholar.org/CorpusID:12121632}
\showURL{%
\tempurl}


\bibitem[Love et~al\mbox{.}(2014)]%
        {Love_Huber_Anders_2014}
\bibfield{author}{\bibinfo{person}{Michael~I Love}, \bibinfo{person}{Wolfgang
  Huber}, {and} \bibinfo{person}{Simon Anders}.}
  \bibinfo{year}{2014}\natexlab{}.
\newblock \showarticletitle{Moderated estimation of fold change and dispersion
  for RNA-seq data with DESeq2.}
\newblock \bibinfo{journal}{\emph{Genome Biology}} \bibinfo{volume}{15},
  \bibinfo{number}{12} (\bibinfo{year}{2014}), \bibinfo{pages}{550}.
\newblock
\urldef\tempurl%
\url{https://doi.org/10.1186/s13059-014-0550-8}
\showDOI{\tempurl}


\bibitem[Marouf et~al\mbox{.}(2020)]%
        {Marouf_Machart_Bansal_Kilian_Magruder_Krebs_Bonn_2020}
\bibfield{author}{\bibinfo{person}{Mohamed Marouf}, \bibinfo{person}{Pierre
  Machart}, \bibinfo{person}{Vikas Bansal}, \bibinfo{person}{Christoph Kilian},
  \bibinfo{person}{Daniel~S Magruder}, \bibinfo{person}{Christian~F Krebs},
  {and} \bibinfo{person}{Stefan Bonn}.} \bibinfo{year}{2020}\natexlab{}.
\newblock \showarticletitle{Realistic in silico generation and augmentation of
  single-cell RNA-seq data using generative adversarial networks.}
\newblock   \bibinfo{volume}{11} (\bibinfo{date}{Jan} \bibinfo{year}{2020}),
  \bibinfo{pages}{166}.
\newblock
\urldef\tempurl%
\url{https://doi.org/10.1038/s41467-019-14018-z}
\showDOI{\tempurl}


\bibitem[McKenna et~al\mbox{.}(2019)]%
        {mckenna2019graphical}
\bibfield{author}{\bibinfo{person}{Ryan McKenna}, \bibinfo{person}{Daniel
  Sheldon}, {and} \bibinfo{person}{Gerome Miklau}.}
  \bibinfo{year}{2019}\natexlab{}.
\newblock \showarticletitle{Graphical-model based estimation and inference for
  differential privacy}. In \bibinfo{booktitle}{\emph{International Conference
  on Machine Learning}}. PMLR, \bibinfo{pages}{4435--4444}.
\newblock


\bibitem[Oestreich et~al\mbox{.}(2021)]%
        {Oestreich_Chen_Schultze_Fritz_Becker_2021}
\bibfield{author}{\bibinfo{person}{Marie Oestreich}, \bibinfo{person}{Dingfan
  Chen}, \bibinfo{person}{Joachim~L Schultze}, \bibinfo{person}{Mario Fritz},
  {and} \bibinfo{person}{Matthias Becker}.} \bibinfo{year}{2021}\natexlab{}.
\newblock \showarticletitle{Privacy considerations for sharing genomics data.}
\newblock \bibinfo{journal}{\emph{EXCLI journal}}  \bibinfo{volume}{20}
  (\bibinfo{date}{Jul} \bibinfo{year}{2021}), \bibinfo{pages}{1243–1260}.
\newblock
\urldef\tempurl%
\url{https://doi.org/10.17179/excli2021-4002}
\showDOI{\tempurl}


\bibitem[Oestreich et~al\mbox{.}(2022)]%
        {Oestreich_Holsten_Agrawal_Dahm_Koch_Jin_Becker_Ulas_2022}
\bibfield{author}{\bibinfo{person}{Marie Oestreich}, \bibinfo{person}{Lisa
  Holsten}, \bibinfo{person}{Shobhit Agrawal}, \bibinfo{person}{Kilian Dahm},
  \bibinfo{person}{Philipp Koch}, \bibinfo{person}{Han Jin},
  \bibinfo{person}{Matthias Becker}, {and} \bibinfo{person}{Thomas Ulas}.}
  \bibinfo{year}{2022}\natexlab{}.
\newblock \showarticletitle{hCoCena: horizontal integration and analysis of
  transcriptomics datasets.}
\newblock   \bibinfo{volume}{38} (\bibinfo{date}{Oct} \bibinfo{year}{2022}),
  \bibinfo{pages}{4727–4734}.
\newblock
\urldef\tempurl%
\url{https://doi.org/10.1093/bioinformatics/btac589}
\showDOI{\tempurl}


\bibitem[Pandey and Onkara(2023)]%
        {Pandey_Onkara_2023}
\bibfield{author}{\bibinfo{person}{Diksha Pandey} {and}
  \bibinfo{person}{Perumal~P Onkara}.} \bibinfo{year}{2023}\natexlab{}.
\newblock \showarticletitle{Improved downstream functional analysis of
  single-cell RNA-sequence data using DGAN.}
\newblock \bibinfo{journal}{\emph{Scientific Reports}} \bibinfo{volume}{13},
  \bibinfo{number}{1} (\bibinfo{date}{Jan} \bibinfo{year}{2023}),
  \bibinfo{pages}{1618}.
\newblock
\urldef\tempurl%
\url{https://doi.org/10.1038/s41598-023-28952-y}
\showDOI{\tempurl}


\bibitem[Papernot and Steinke(2021)]%
        {papernot2021hyperparameter}
\bibfield{author}{\bibinfo{person}{Nicolas Papernot} {and}
  \bibinfo{person}{Thomas Steinke}.} \bibinfo{year}{2021}\natexlab{}.
\newblock \showarticletitle{Hyperparameter Tuning with Renyi Differential
  Privacy}. In \bibinfo{booktitle}{\emph{International Conference on Learning
  Representations (ICLR)}}.
\newblock


\bibitem[Park et~al\mbox{.}(2018)]%
        {park2018data}
\bibfield{author}{\bibinfo{person}{Noseong Park}, \bibinfo{person}{Mahmoud
  Mohammadi}, \bibinfo{person}{Kshitij Gorde}, \bibinfo{person}{Sushil
  Jajodia}, \bibinfo{person}{Hongkyu Park}, {and} \bibinfo{person}{Youngmin
  Kim}.} \bibinfo{year}{2018}\natexlab{}.
\newblock \showarticletitle{Data synthesis based on generative adversarial
  networks}.
\newblock \bibinfo{journal}{\emph{arXiv preprint arXiv:1806.03384}}
  (\bibinfo{year}{2018}).
\newblock


\bibitem[Ritchie et~al\mbox{.}(2015)]%
        {Ritchie_2015}
\bibfield{author}{\bibinfo{person}{Matthew~E Ritchie}, \bibinfo{person}{Belinda
  Phipson}, \bibinfo{person}{Di Wu}, \bibinfo{person}{Yifang Hu},
  \bibinfo{person}{Charity~W Law}, \bibinfo{person}{Wei Shi}, {and}
  \bibinfo{person}{Gordon~K Smyth}.} \bibinfo{year}{2015}\natexlab{}.
\newblock \showarticletitle{limma powers differential expression analyses for
  RNA-sequencing and microarray studies.}
\newblock   \bibinfo{volume}{43} (\bibinfo{date}{Apr} \bibinfo{year}{2015}),
  \bibinfo{pages}{e47}.
\newblock
\urldef\tempurl%
\url{https://doi.org/10.1093/nar/gkv007}
\showDOI{\tempurl}


\bibitem[Robinson et~al\mbox{.}(2010)]%
        {Robinson_McCarthy_Smyth_2010}
\bibfield{author}{\bibinfo{person}{Mark~D Robinson}, \bibinfo{person}{Davis~J
  McCarthy}, {and} \bibinfo{person}{Gordon~K Smyth}.}
  \bibinfo{year}{2010}\natexlab{}.
\newblock \showarticletitle{edgeR: a Bioconductor package for differential
  expression analysis of digital gene expression data.}
\newblock \bibinfo{journal}{\emph{Bioinformatics}} \bibinfo{volume}{26},
  \bibinfo{number}{1} (\bibinfo{date}{Jan} \bibinfo{year}{2010}),
  \bibinfo{pages}{139–140}.
\newblock
\urldef\tempurl%
\url{https://doi.org/10.1093/bioinformatics/btp616}
\showDOI{\tempurl}


\bibitem[Sohn et~al\mbox{.}(2015)]%
        {sohn2015learning}
\bibfield{author}{\bibinfo{person}{Kihyuk Sohn}, \bibinfo{person}{Honglak Lee},
  {and} \bibinfo{person}{Xinchen Yan}.} \bibinfo{year}{2015}\natexlab{}.
\newblock \showarticletitle{Learning structured output representation using
  deep conditional generative models}.
\newblock \bibinfo{journal}{\emph{Advances in neural information processing
  systems}}  \bibinfo{volume}{28} (\bibinfo{year}{2015}).
\newblock


\bibitem[Stadler et~al\mbox{.}(2022)]%
        {groundhog}
\bibfield{author}{\bibinfo{person}{Theresa Stadler}, \bibinfo{person}{Bristena
  Oprisanu}, {and} \bibinfo{person}{Carmela Troncoso}.}
  \bibinfo{year}{2022}\natexlab{}.
\newblock \showarticletitle{Synthetic Data -- Anonymisation Groundhog Day}. In
  \bibinfo{booktitle}{\emph{USENIX Security Symposium}}.
\newblock


\bibitem[Stadler and Troncoso(2022)]%
        {Stadler}
\bibfield{author}{\bibinfo{person}{Theresa Stadler} {and}
  \bibinfo{person}{Carmela Troncoso}.} \bibinfo{year}{2022}\natexlab{}.
\newblock \showarticletitle{Why the search for a privacy-preserving data
  sharing mechanism is failing}.
\newblock \bibinfo{journal}{\emph{Nature Computational Science}}
  \bibinfo{volume}{2}, \bibinfo{number}{4} (\bibinfo{year}{2022}),
  \bibinfo{pages}{208--210}.
\newblock


\bibitem[Subramanian et~al\mbox{.}(2017)]%
        {Subramanian}
\bibfield{author}{\bibinfo{person}{Aravind Subramanian}, \bibinfo{person}{Rajiv
  Narayan}, \bibinfo{person}{Steven~M Corsello}, \bibinfo{person}{David~D
  Peck}, \bibinfo{person}{Ted~E Natoli}, \bibinfo{person}{Xiaodong Lu},
  \bibinfo{person}{Joshua Gould}, \bibinfo{person}{John~F Davis},
  \bibinfo{person}{Andrew~A Tubelli}, \bibinfo{person}{Jacob~K Asiedu}, {and}
  \bibinfo{person}{et al.}} \bibinfo{year}{2017}\natexlab{}.
\newblock \showarticletitle{A next generation connectivity map: L1000 platform
  and the first 1,000,000 profiles.}
\newblock \bibinfo{journal}{\emph{Cell}} \bibinfo{volume}{171},
  \bibinfo{number}{6} (\bibinfo{date}{Nov} \bibinfo{year}{2017}),
  \bibinfo{pages}{1437--1452.e17}.
\newblock
\urldef\tempurl%
\url{https://doi.org/10.1016/j.cell.2017.10.049}
\showDOI{\tempurl}


\bibitem[Tang et~al\mbox{.}(2009)]%
        {Tang_Barbacioru_Wang_Nordman_Lee_Xu_Wang_Bodeau_Tuch_Siddiqui_et_al}
\bibfield{author}{\bibinfo{person}{Fuchou Tang}, \bibinfo{person}{Catalin
  Barbacioru}, \bibinfo{person}{Yangzhou Wang}, \bibinfo{person}{Ellen
  Nordman}, \bibinfo{person}{Clarence Lee}, \bibinfo{person}{Nanlan Xu},
  \bibinfo{person}{Xiaohui Wang}, \bibinfo{person}{John Bodeau},
  \bibinfo{person}{Brian~B Tuch}, \bibinfo{person}{Asim Siddiqui}, {and}
  \bibinfo{person}{et al.}} \bibinfo{year}{2009}\natexlab{}.
\newblock \showarticletitle{mRNA-Seq whole-transcriptome analysis of a single
  cell.}
\newblock \bibinfo{journal}{\emph{Nature Methods}} \bibinfo{volume}{6},
  \bibinfo{number}{5} (\bibinfo{date}{May} \bibinfo{year}{2009}),
  \bibinfo{pages}{377–382}.
\newblock
\urldef\tempurl%
\url{https://doi.org/10.1038/nmeth.1315}
\showDOI{\tempurl}


\bibitem[Torfi et~al\mbox{.}(2022)]%
        {torfi2022differentially}
\bibfield{author}{\bibinfo{person}{Amirsina Torfi}, \bibinfo{person}{Edward~A
  Fox}, {and} \bibinfo{person}{Chandan~K Reddy}.}
  \bibinfo{year}{2022}\natexlab{}.
\newblock \showarticletitle{Differentially private synthetic medical data
  generation using convolutional GANs}.
\newblock \bibinfo{journal}{\emph{Information Sciences}}  \bibinfo{volume}{586}
  (\bibinfo{year}{2022}), \bibinfo{pages}{485--500}.
\newblock


\bibitem[Tramer and Boneh(2020)]%
        {tramer2020differentially}
\bibfield{author}{\bibinfo{person}{Florian Tramer} {and} \bibinfo{person}{Dan
  Boneh}.} \bibinfo{year}{2020}\natexlab{}.
\newblock \showarticletitle{Differentially Private Learning Needs Better
  Features (or Much More Data)}. In \bibinfo{booktitle}{\emph{International
  Conference on Learning Representations (ICLR)}}.
\newblock


\bibitem[Treppner et~al\mbox{.}(2021)]%
        {Treppner_Salas-Bastos_Hess_Lenz_Vogel_Binder_2021}
\bibfield{author}{\bibinfo{person}{Martin Treppner}, \bibinfo{person}{Adrián
  Salas-Bastos}, \bibinfo{person}{Moritz Hess}, \bibinfo{person}{Stefan Lenz},
  \bibinfo{person}{Tanja Vogel}, {and} \bibinfo{person}{Harald Binder}.}
  \bibinfo{year}{2021}\natexlab{}.
\newblock \showarticletitle{Synthetic single cell RNA sequencing data from
  small pilot studies using deep generative models.}
\newblock \bibinfo{journal}{\emph{Scientific Reports}} \bibinfo{volume}{11},
  \bibinfo{number}{1} (\bibinfo{date}{Apr} \bibinfo{year}{2021}),
  \bibinfo{pages}{9403}.
\newblock
\urldef\tempurl%
\url{https://doi.org/10.1038/s41598-021-88875-4}
\showDOI{\tempurl}


\bibitem[Viñas et~al\mbox{.}(2022)]%
        {Viñas_Andrés-Terré_Liò_Bryson_2022}
\bibfield{author}{\bibinfo{person}{Ramon Viñas}, \bibinfo{person}{Helena
  Andrés-Terré}, \bibinfo{person}{Pietro Liò}, {and} \bibinfo{person}{Kevin
  Bryson}.} \bibinfo{year}{2022}\natexlab{}.
\newblock \showarticletitle{Adversarial generation of gene expression data.}
\newblock \bibinfo{journal}{\emph{Bioinformatics}} \bibinfo{volume}{38},
  \bibinfo{number}{3} (\bibinfo{date}{Jan} \bibinfo{year}{2022}),
  \bibinfo{pages}{730–737}.
\newblock
\urldef\tempurl%
\url{https://doi.org/10.1093/bioinformatics/btab035}
\showDOI{\tempurl}


\bibitem[Wang et~al\mbox{.}(2009)]%
        {Wang_Gerstein_Snyder_2009}
\bibfield{author}{\bibinfo{person}{Zhong Wang}, \bibinfo{person}{Mark
  Gerstein}, {and} \bibinfo{person}{Michael Snyder}.}
  \bibinfo{year}{2009}\natexlab{}.
\newblock \showarticletitle{RNA-Seq: a revolutionary tool for transcriptomics.}
\newblock \bibinfo{journal}{\emph{Nature Reviews. Genetics}}
  \bibinfo{volume}{10}, \bibinfo{number}{1} (\bibinfo{date}{Jan}
  \bibinfo{year}{2009}), \bibinfo{pages}{57–63}.
\newblock
\urldef\tempurl%
\url{https://doi.org/10.1038/nrg2484}
\showDOI{\tempurl}


\bibitem[Warnat-Herresthal et~al\mbox{.}(2020)]%
        {Warnat-Herresthal}
\bibfield{author}{\bibinfo{person}{Stefanie Warnat-Herresthal},
  \bibinfo{person}{Konstantinos Perrakis}, \bibinfo{person}{Bernd Taschler},
  \bibinfo{person}{Matthias Becker}, \bibinfo{person}{Kevin Baßler},
  \bibinfo{person}{Marc Beyer}, \bibinfo{person}{Patrick Günther},
  \bibinfo{person}{Jonas Schulte-Schrepping}, \bibinfo{person}{Lea Seep},
  \bibinfo{person}{Kathrin Klee}, {and} \bibinfo{person}{et al.}}
  \bibinfo{year}{2020}\natexlab{}.
\newblock \showarticletitle{Scalable Prediction of Acute Myeloid Leukemia Using
  High-Dimensional Machine Learning and Blood Transcriptomics.}
\newblock \bibinfo{journal}{\emph{iScience}} \bibinfo{volume}{23},
  \bibinfo{number}{1} (\bibinfo{date}{Jan} \bibinfo{year}{2020}),
  \bibinfo{pages}{100780}.
\newblock
\urldef\tempurl%
\url{https://doi.org/10.1016/j.isci.2019.100780}
\showDOI{\tempurl}


\bibitem[Wilcoxon(1945)]%
        {Wilcoxon_1945}
\bibfield{author}{\bibinfo{person}{Frank Wilcoxon}.}
  \bibinfo{year}{1945}\natexlab{}.
\newblock \showarticletitle{Individual comparisons by ranking methods}.
\newblock \bibinfo{journal}{\emph{Biometrics Bulletin}} \bibinfo{volume}{1},
  \bibinfo{number}{6} (\bibinfo{date}{Dec} \bibinfo{year}{1945}),
  \bibinfo{pages}{80}.
\newblock
\urldef\tempurl%
\url{https://doi.org/10.2307/3001968}
\showDOI{\tempurl}


\bibitem[Zappia et~al\mbox{.}(2017)]%
        {Zappia_Phipson_Oshlack_2017}
\bibfield{author}{\bibinfo{person}{Luke Zappia}, \bibinfo{person}{Belinda
  Phipson}, {and} \bibinfo{person}{Alicia Oshlack}.}
  \bibinfo{year}{2017}\natexlab{}.
\newblock \showarticletitle{Splatter: simulation of single-cell RNA sequencing
  data.}
\newblock \bibinfo{journal}{\emph{Genome Biology}} \bibinfo{volume}{18},
  \bibinfo{number}{1} (\bibinfo{date}{Sep} \bibinfo{year}{2017}),
  \bibinfo{pages}{174}.
\newblock
\urldef\tempurl%
\url{https://doi.org/10.1186/s13059-017-1305-0}
\showDOI{\tempurl}


\bibitem[Zhang et~al\mbox{.}(2021)]%
        {zhang2021privsyn}
\bibfield{author}{\bibinfo{person}{Zhikun Zhang}, \bibinfo{person}{Tianhao
  Wang}, \bibinfo{person}{Ninghui Li}, \bibinfo{person}{Jean Honorio},
  \bibinfo{person}{Michael Backes}, \bibinfo{person}{Shibo He},
  \bibinfo{person}{Jiming Chen}, {and} \bibinfo{person}{Yang Zhang}.}
  \bibinfo{year}{2021}\natexlab{}.
\newblock \showarticletitle{$\{$PrivSyn$\}$: Differentially Private Data
  Synthesis}. In \bibinfo{booktitle}{\emph{30th USENIX Security Symposium
  (USENIX Security 21)}}. \bibinfo{pages}{929--946}.
\newblock


\end{thebibliography}

\appendix
\onecolumn

\section{Utility and Statistical Evaluation}
\subsection{Plot of Evaluation Metrics Across Individual Models without Fixing y-axis limit}
\begin{figure*}[!htbp]
\centering
\newcommand{\figwidth}{0.18\textwidth}
\subfigure[\vae]{
\begin{minipage}[b]{\figwidth} \includegraphics[width=1.0\textwidth]{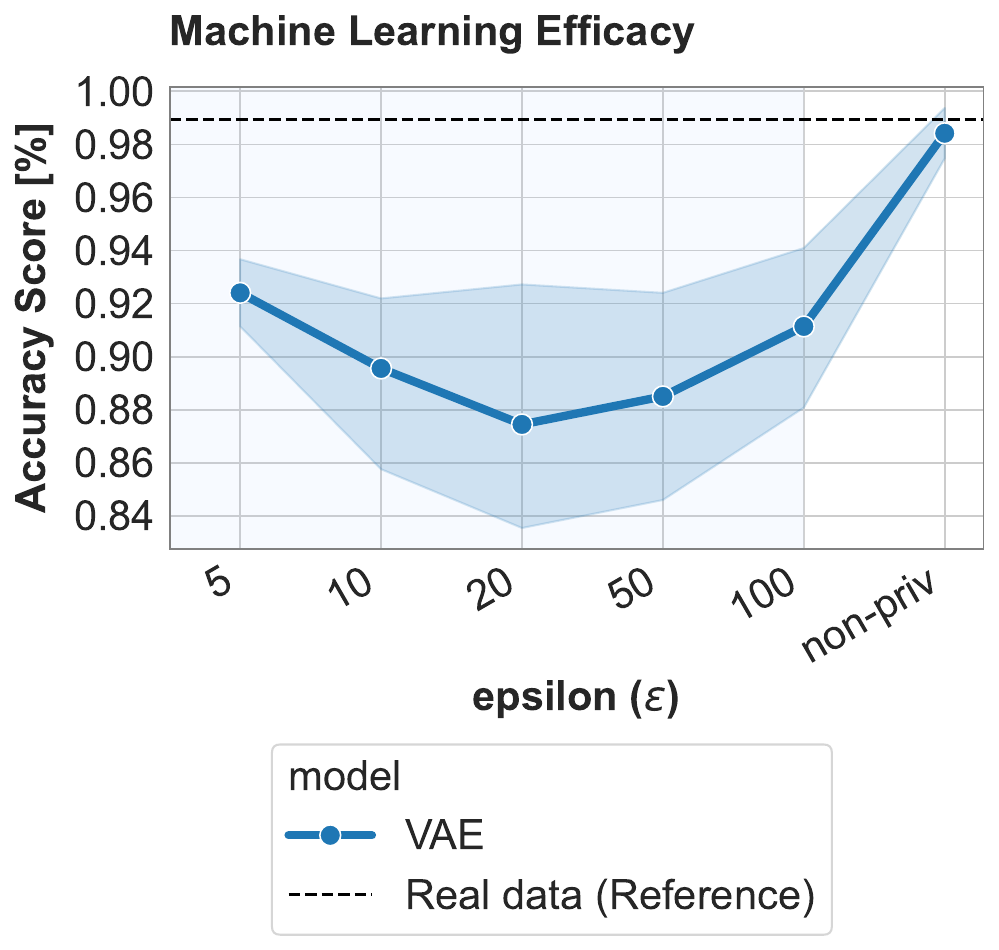}
\end{minipage}
}
\subfigure[\gan]{
\begin{minipage}[b]{\figwidth} \includegraphics[width=1.0\textwidth]{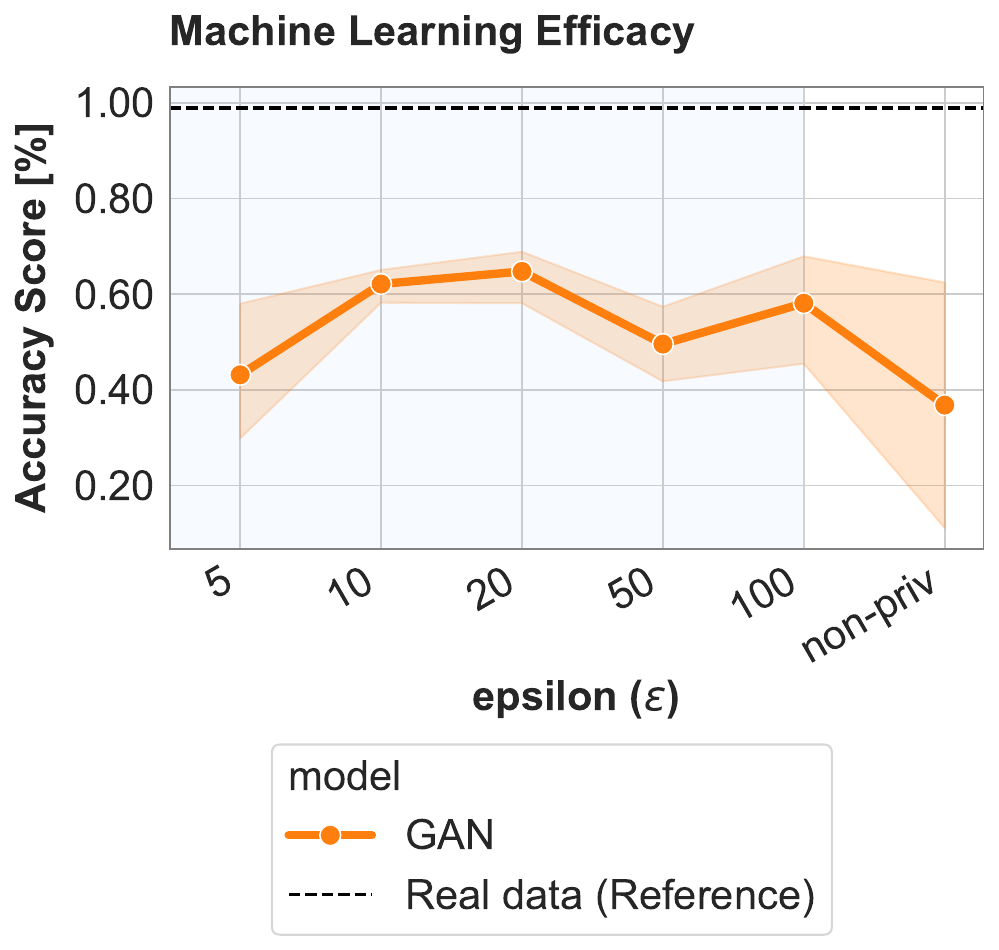}
\end{minipage}
}
\subfigure[\pgm]{
\begin{minipage}[b]{\figwidth} \includegraphics[width=1.0\textwidth]{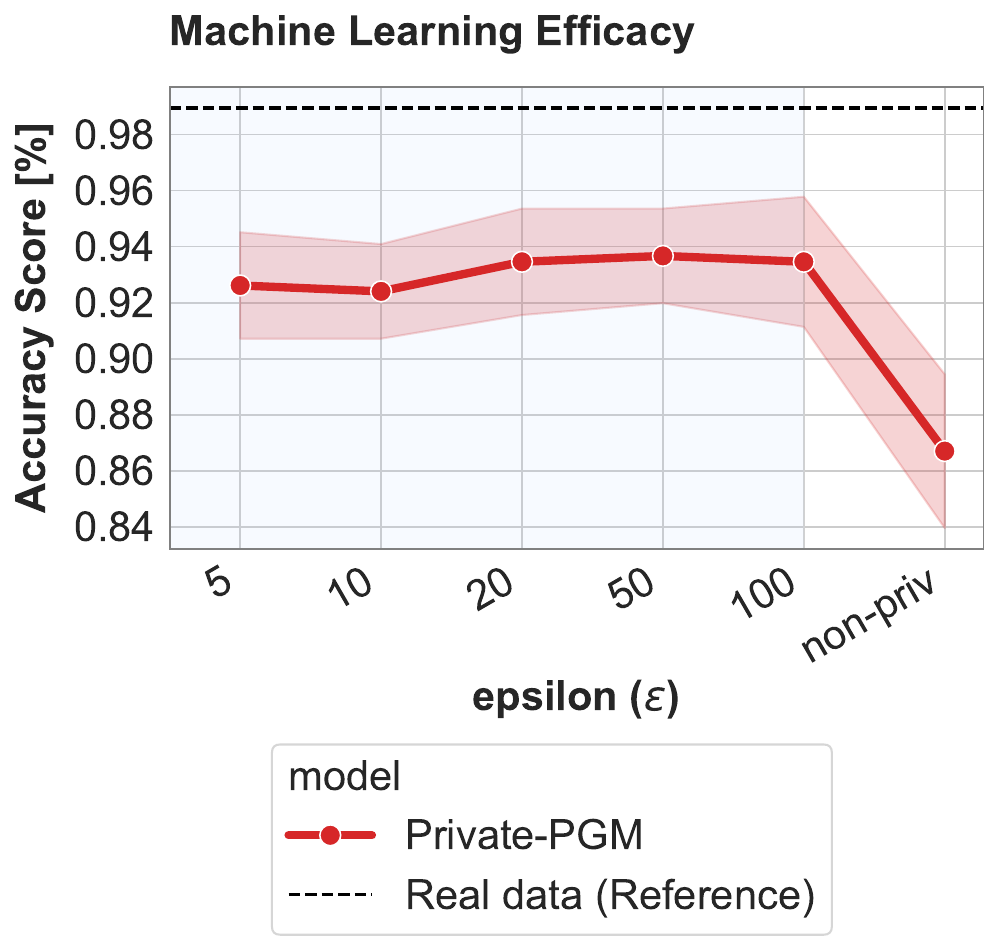}
\end{minipage}
}
\subfigure[\privsyn]{
\begin{minipage}[b]{\figwidth} \includegraphics[width=1.0\textwidth]{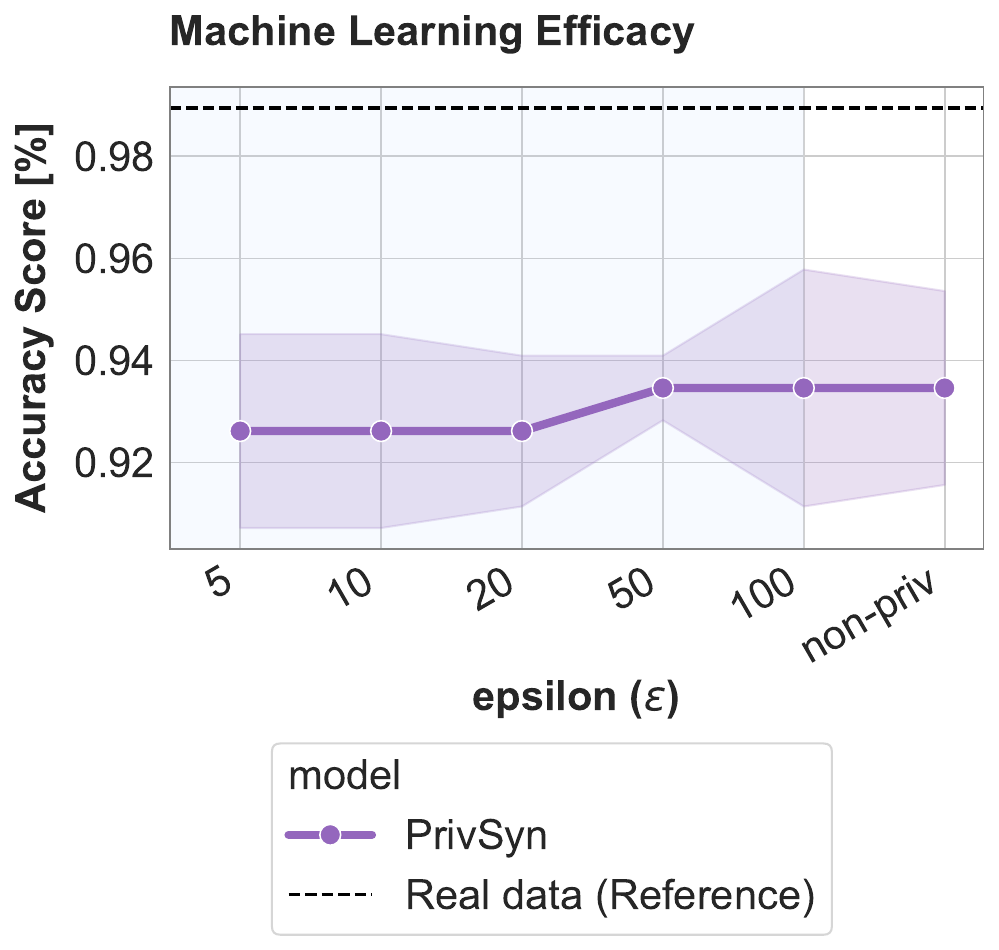}
\end{minipage}
}
\subfigure[\rongauss]{
\begin{minipage}[b]{\figwidth} \includegraphics[width=1.0\textwidth]{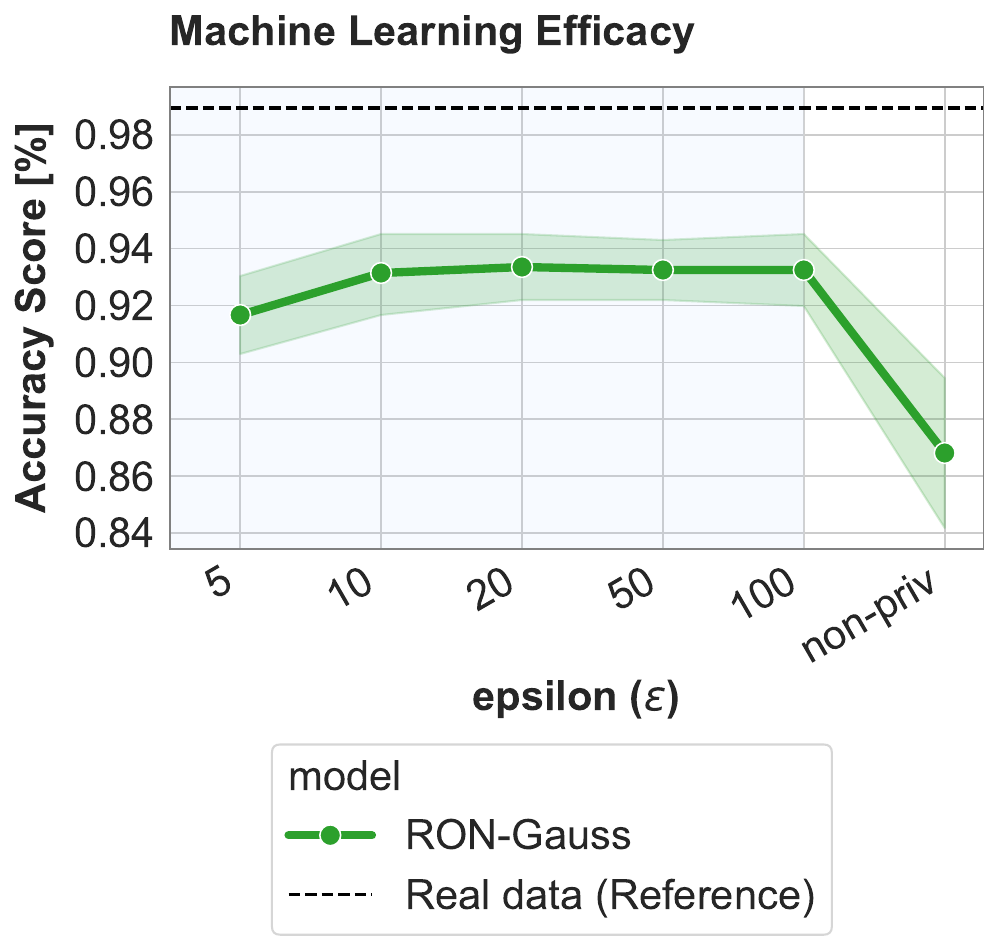}
\end{minipage}
}
\caption{Utility Evaluation by Machine Learning Efficacy. }
\vspace{5pt}
\end{figure*}
\begin{figure*}[!htbp]
\centering
\newcommand{\figwidth}{0.18\textwidth}
\subfigure[\vae]{
\begin{minipage}[b]{\figwidth} \includegraphics[width=1.0\textwidth]{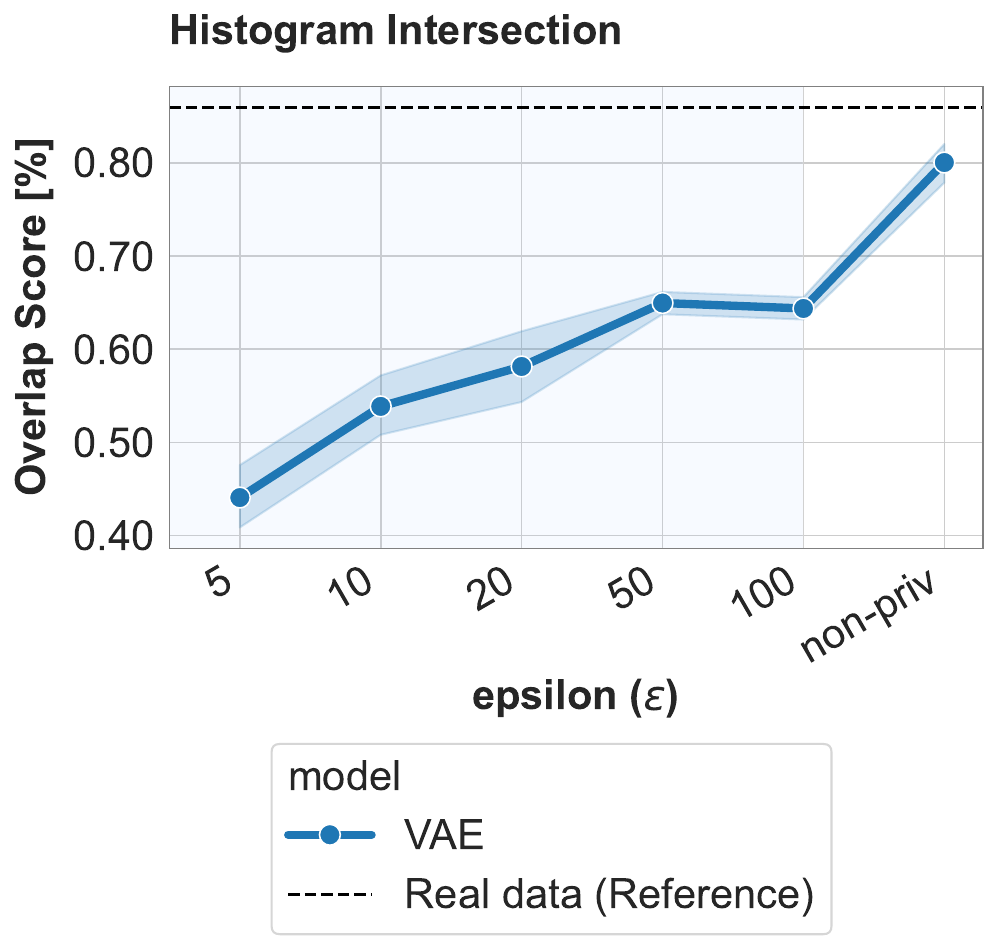}
\end{minipage}
}
\subfigure[\gan]{
\begin{minipage}[b]{\figwidth} \includegraphics[width=1.0\textwidth]{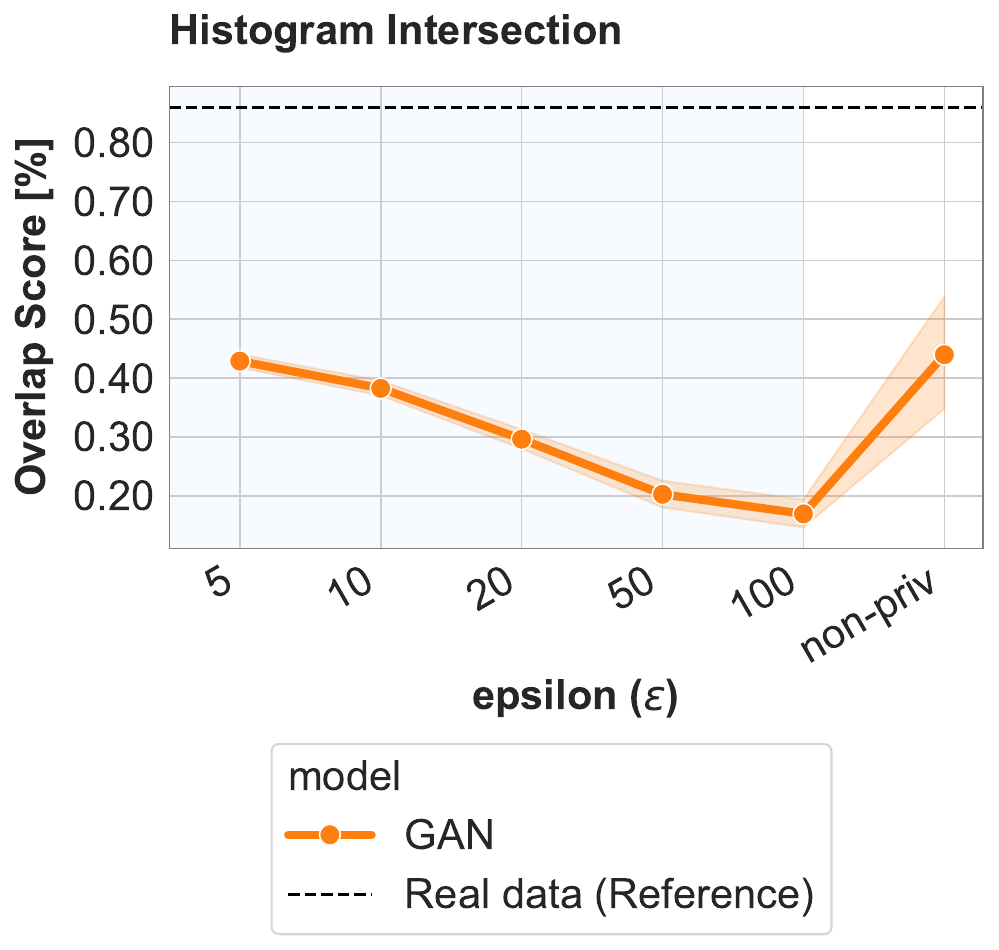}
\end{minipage}
}
\subfigure[\pgm]{
\begin{minipage}[b]{\figwidth} \includegraphics[width=1.0\textwidth]{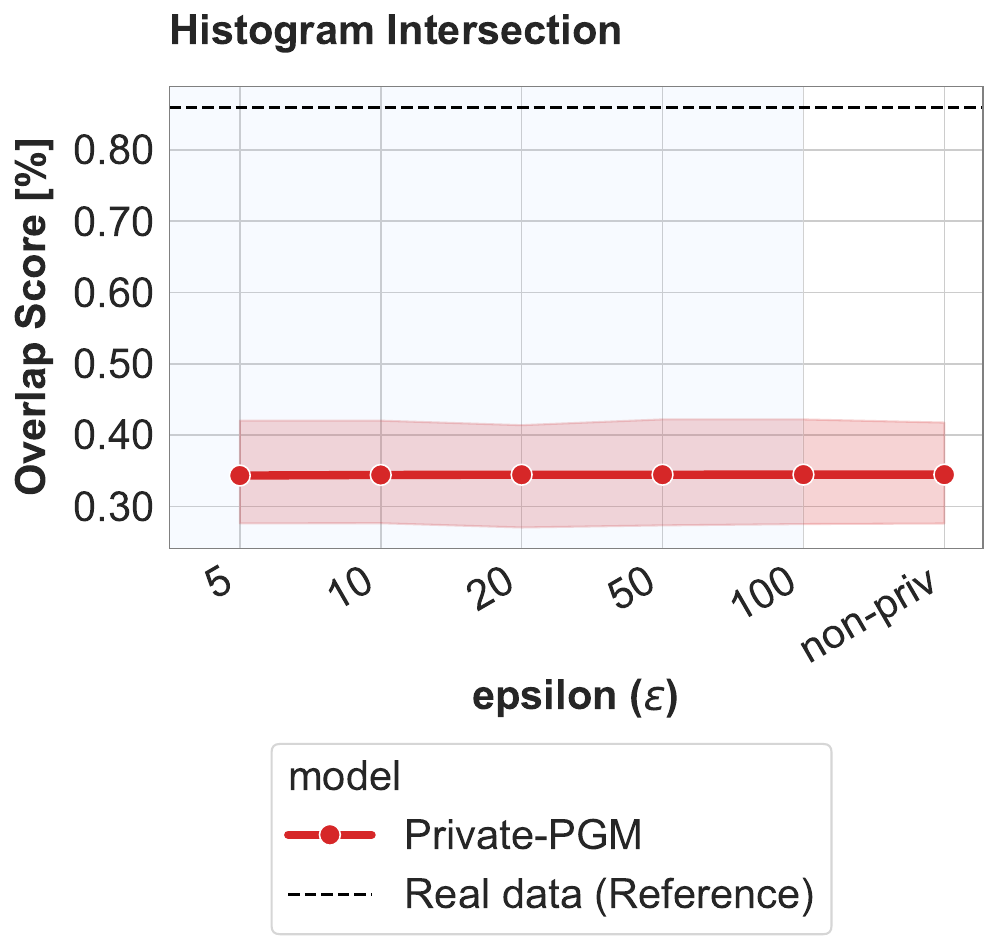}
\end{minipage}
}
\subfigure[\privsyn]{
\begin{minipage}[b]{\figwidth} \includegraphics[width=1.0\textwidth]{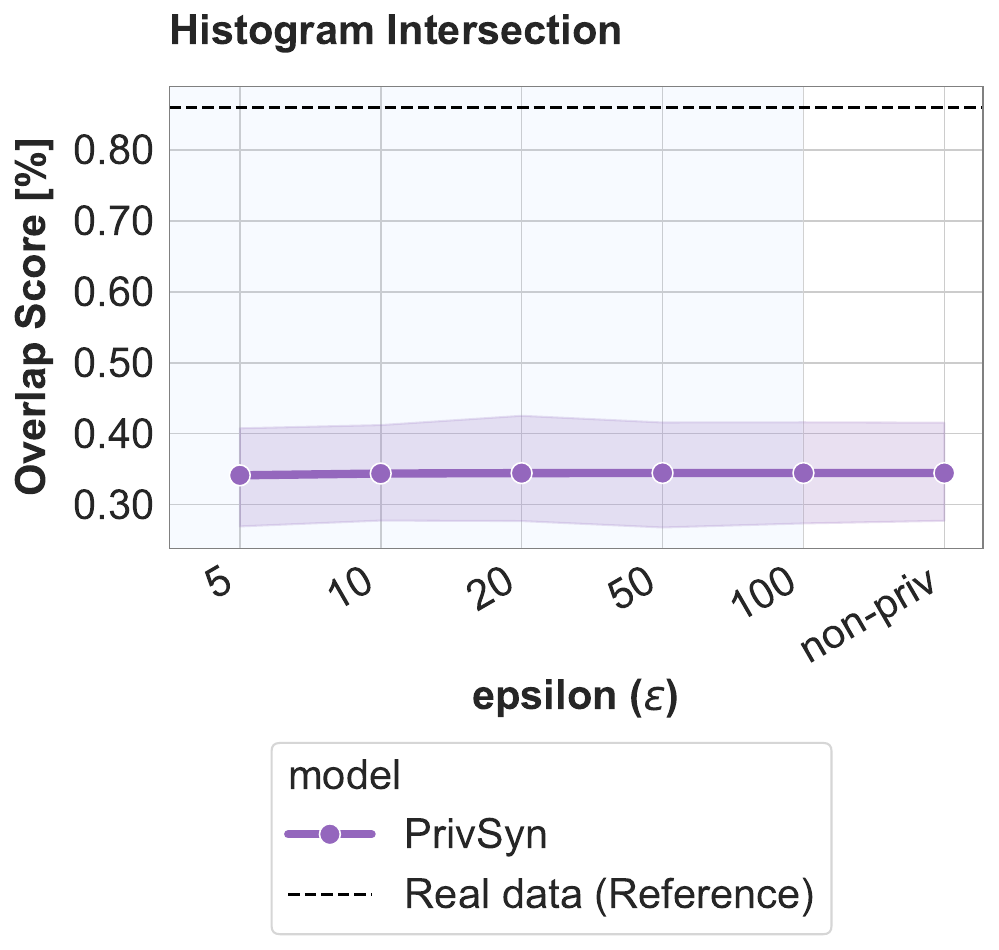}
\end{minipage}
}
\subfigure[\rongauss]{
\begin{minipage}[b]{\figwidth} \includegraphics[width=1.0\textwidth]{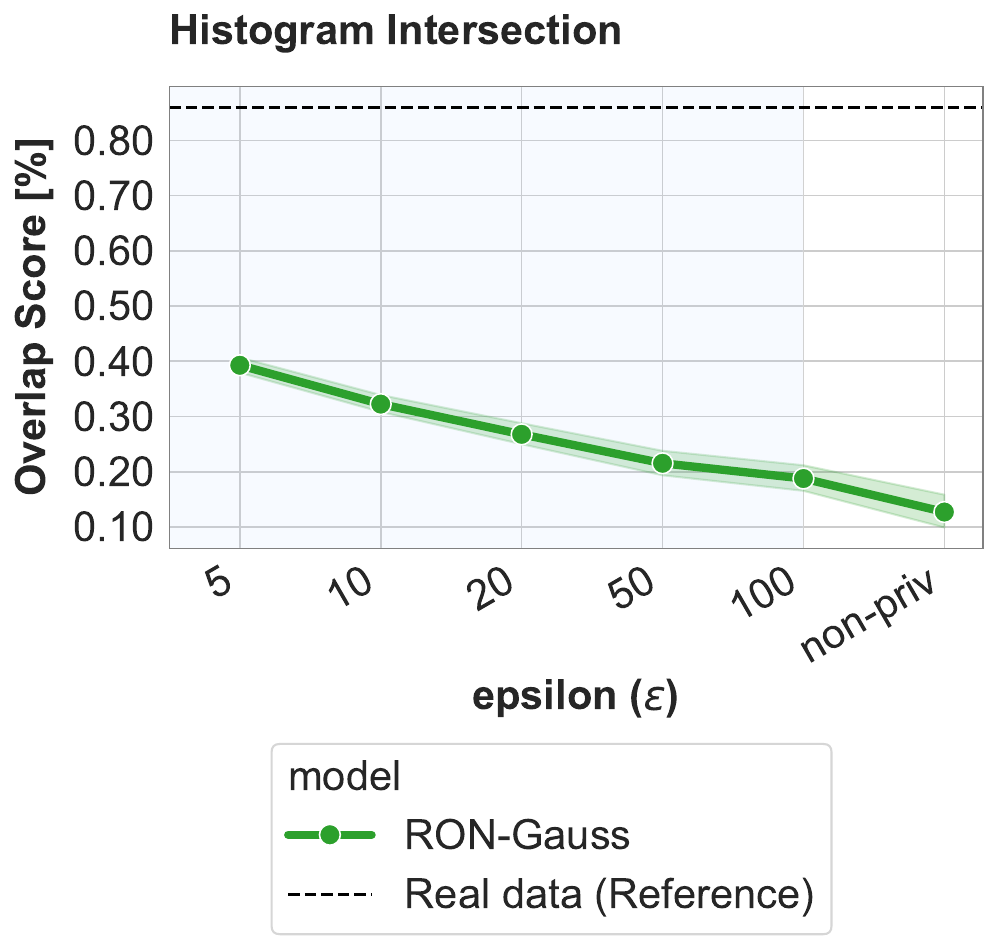}
\end{minipage}
}
\caption{Statistical Evaluation by Histogram Intersection. }
\vspace{5pt}
\end{figure*}

\begin{figure*}[!htbp]
\centering
\newcommand{\figwidth}{0.18\textwidth}
\subfigure[\vae]{
\begin{minipage}[b]{\figwidth} \includegraphics[width=1.0\textwidth]{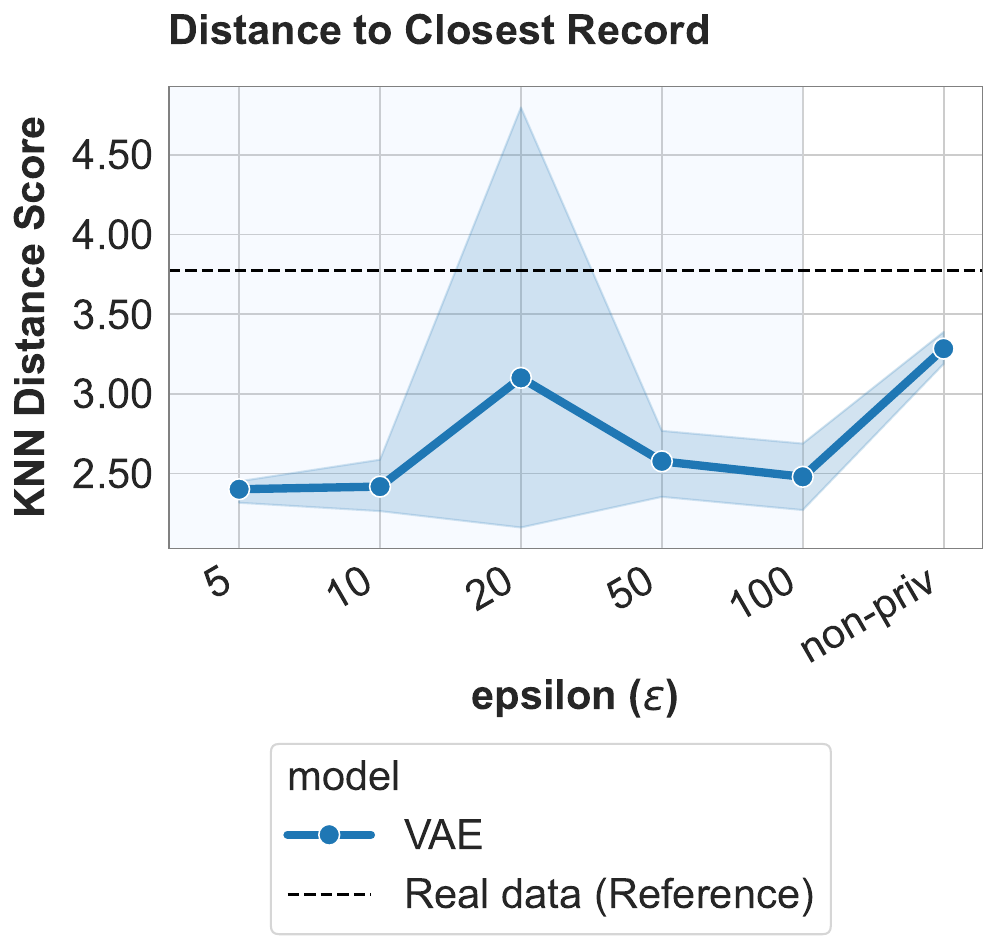}
\end{minipage}
}
\subfigure[\gan]{
\begin{minipage}[b]{\figwidth} \includegraphics[width=1.0\textwidth]{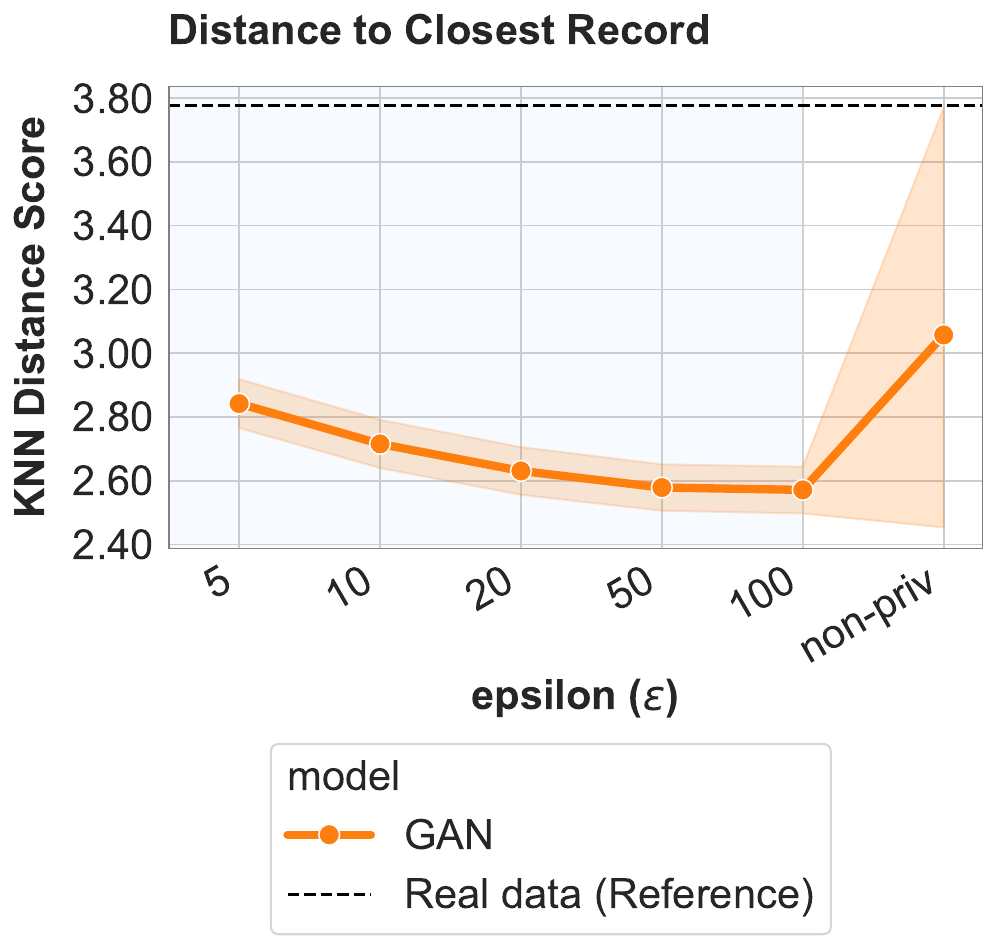}
\end{minipage}
}
\subfigure[\pgm]{
\begin{minipage}[b]{\figwidth} \includegraphics[width=1.0\textwidth]{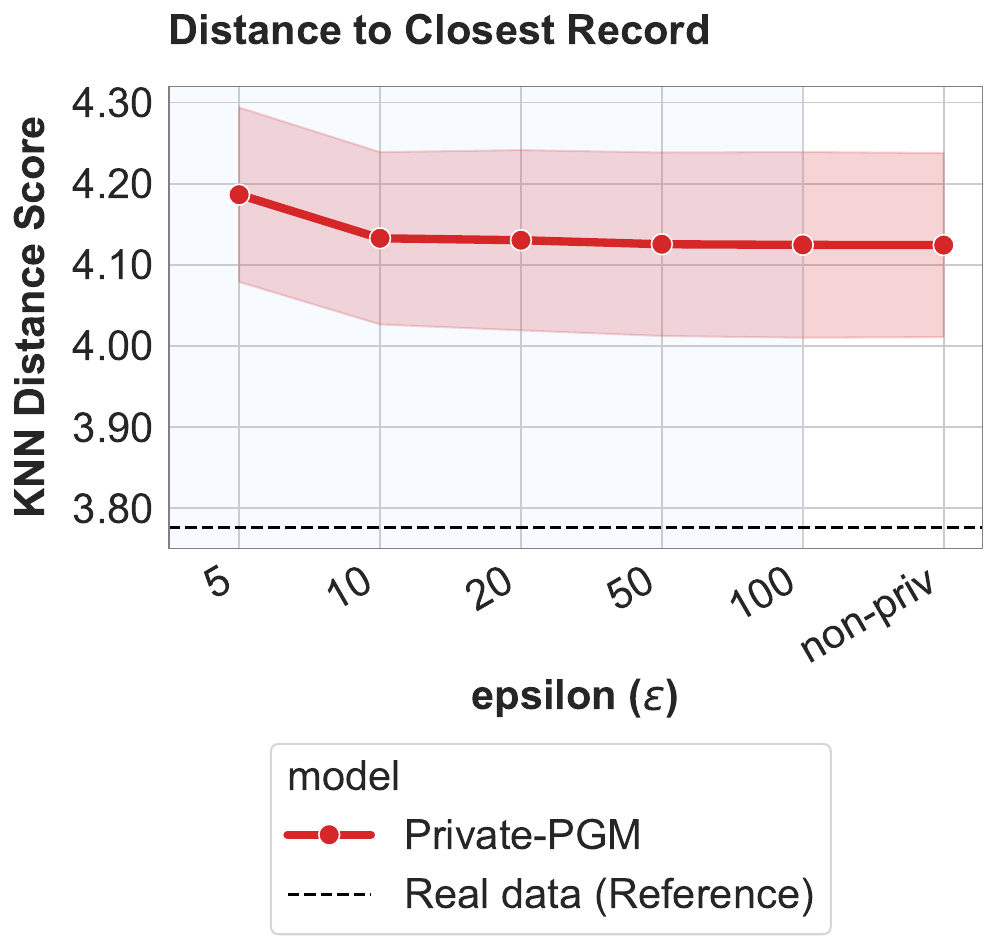}
\end{minipage}
}
\subfigure[\privsyn]{
\begin{minipage}[b]{\figwidth} \includegraphics[width=1.0\textwidth]{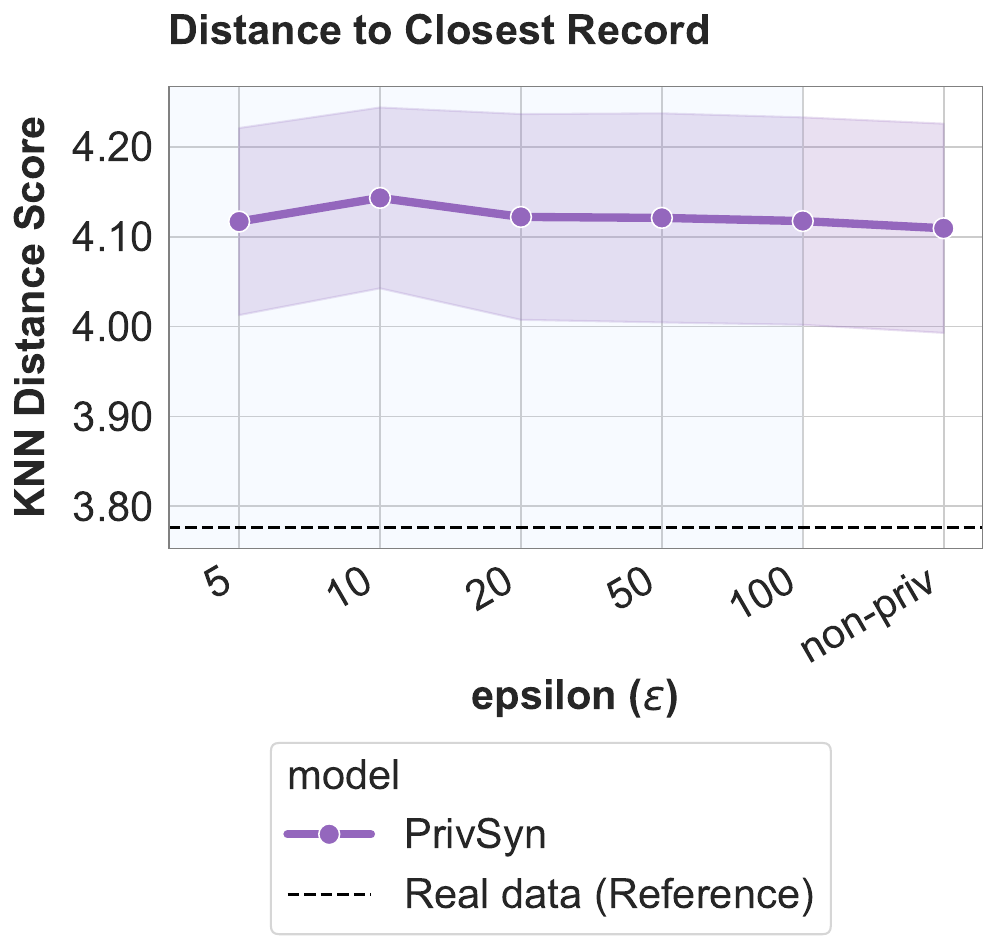}
\end{minipage}
}
\subfigure[\rongauss]{
\begin{minipage}[b]{\figwidth} \includegraphics[width=1.0\textwidth]{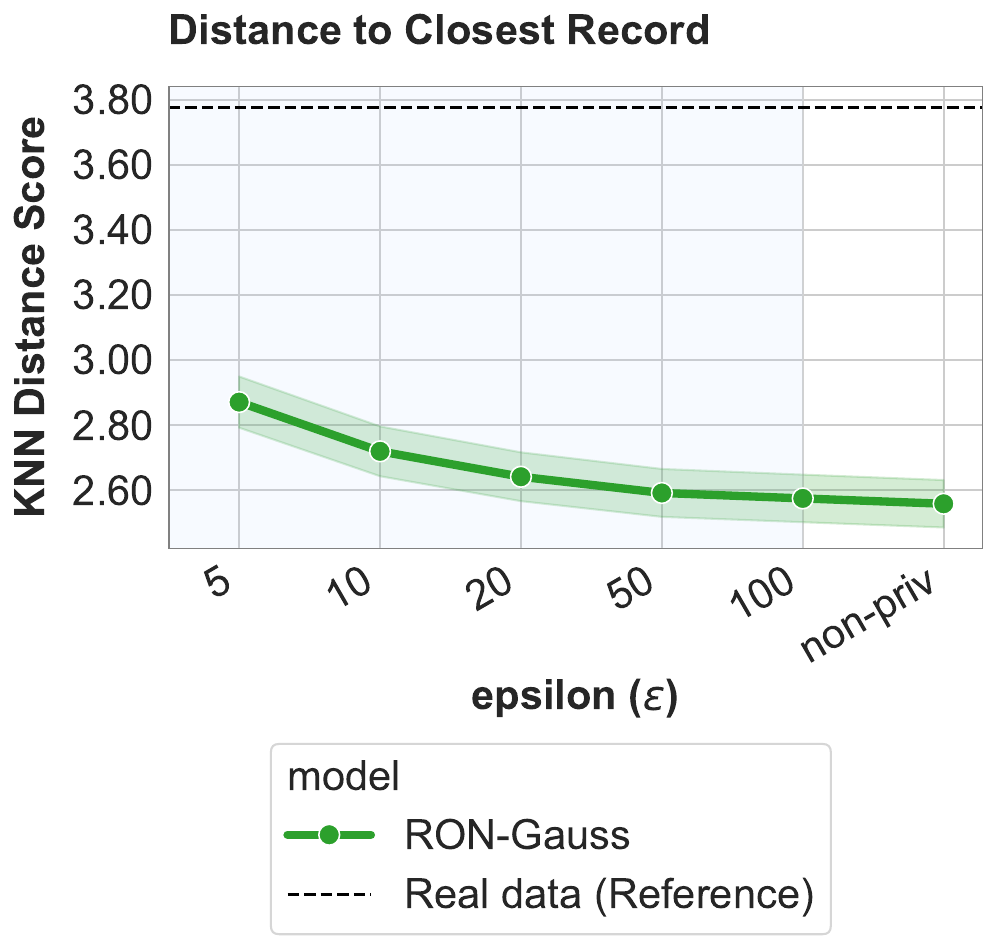}
\end{minipage}
}
\caption{Distance to Closest Record. }
\vspace{5pt}
\end{figure*}

\begin{figure*}[!htbp]
\centering
\newcommand{\figwidth}{0.18\textwidth}
\subfigure[\vae]{
\begin{minipage}[b]{\figwidth} \includegraphics[width=1.0\textwidth]{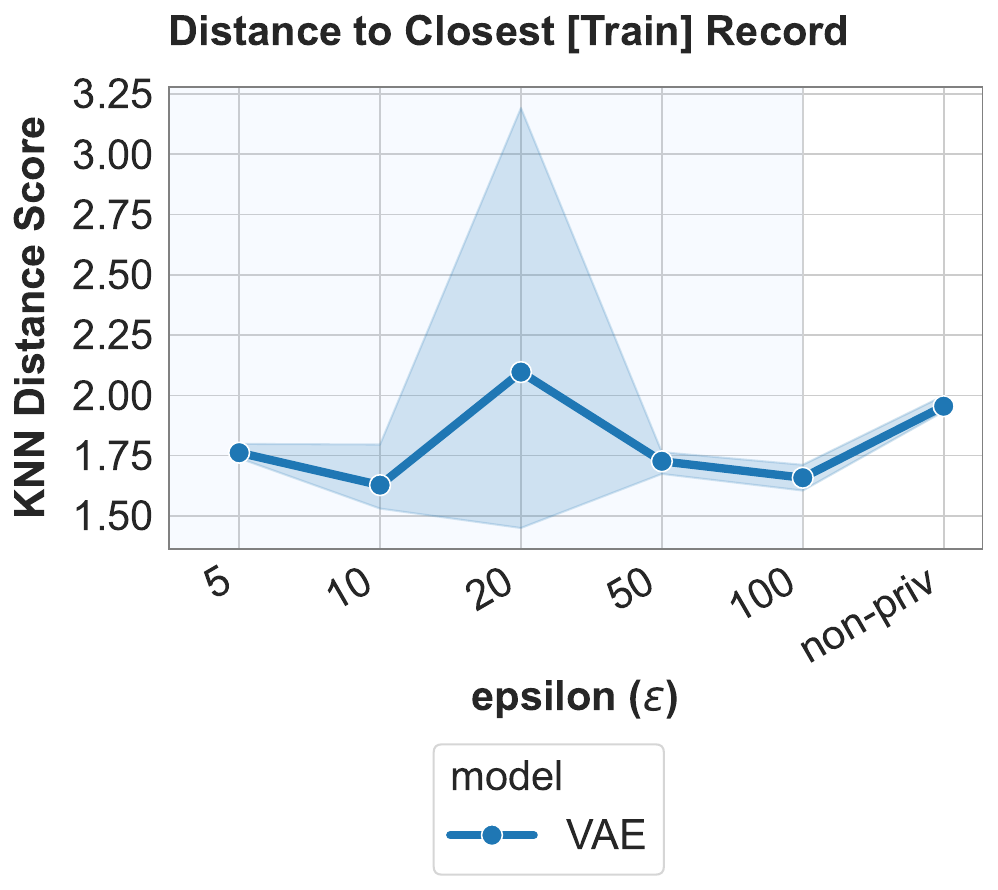}
\end{minipage}
}
\subfigure[\gan]{
\begin{minipage}[b]{\figwidth} \includegraphics[width=1.0\textwidth]{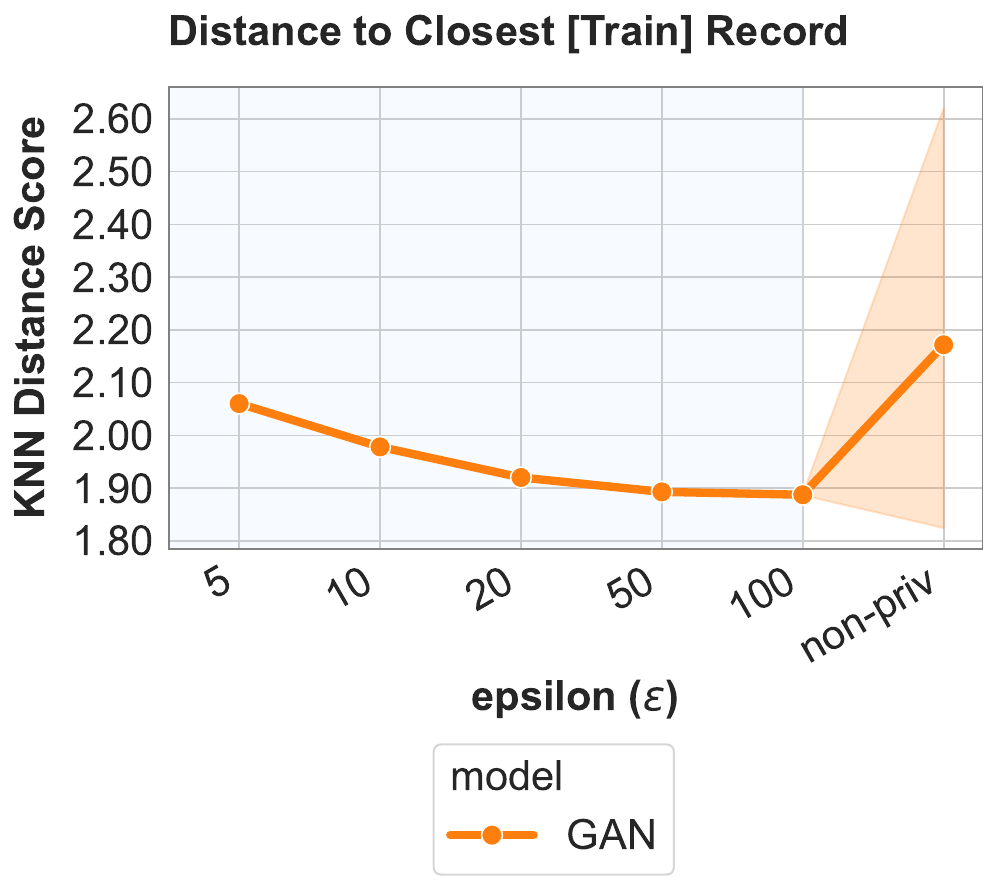}
\end{minipage}
}
\subfigure[\pgm]{
\begin{minipage}[b]{\figwidth} \includegraphics[width=1.0\textwidth]{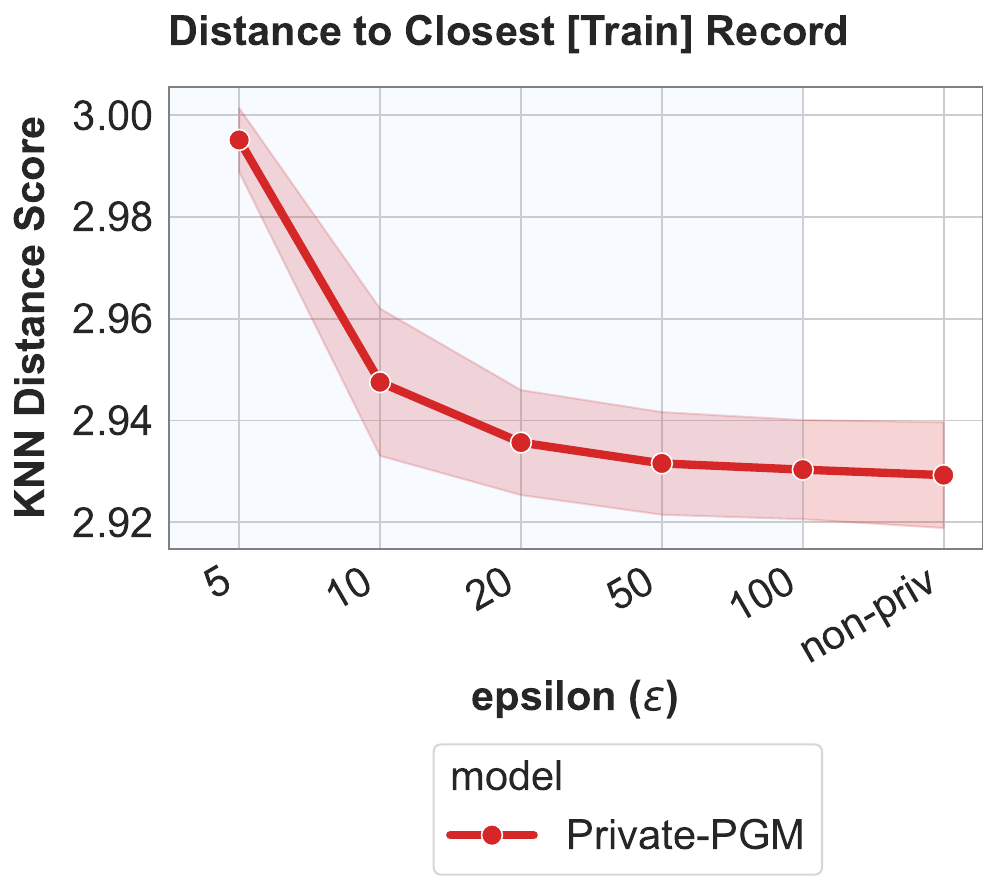}
\end{minipage}
}
\subfigure[\privsyn]{
\begin{minipage}[b]{\figwidth} \includegraphics[width=1.0\textwidth]{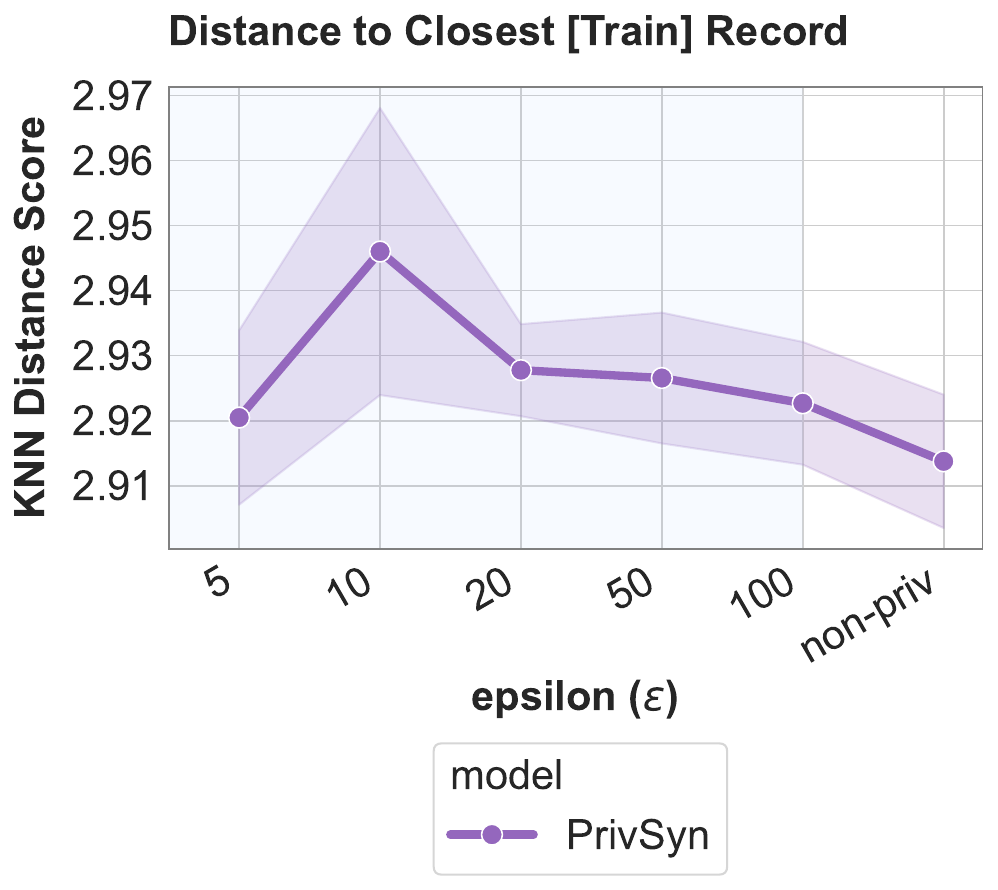}
\end{minipage}
}
\subfigure[\rongauss]{
\begin{minipage}[b]{\figwidth} \includegraphics[width=1.0\textwidth]{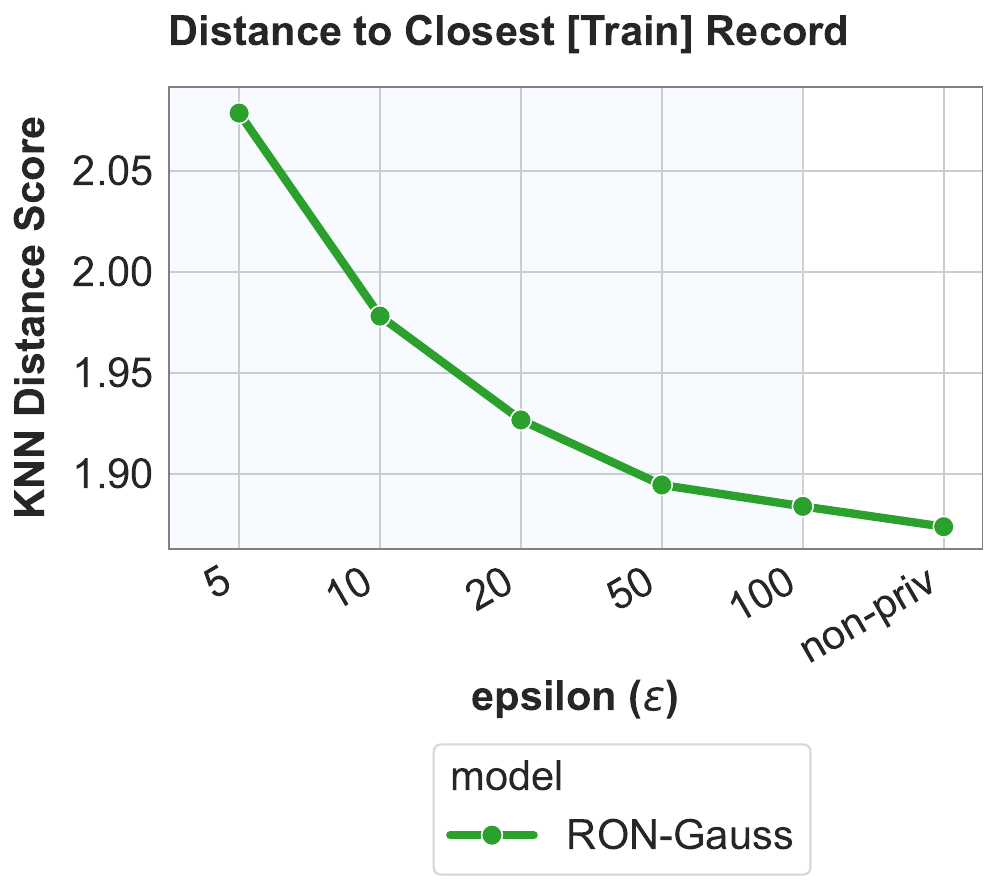}
\end{minipage}
}
\caption{Distance to Closest (Train) Record. }
\vspace{5pt}
\end{figure*}

\clearpage
\subsection{Distance to Closest Train Record}
\begin{figure*}[!htbp]
\centering
\includegraphics[width=0.5\textwidth]{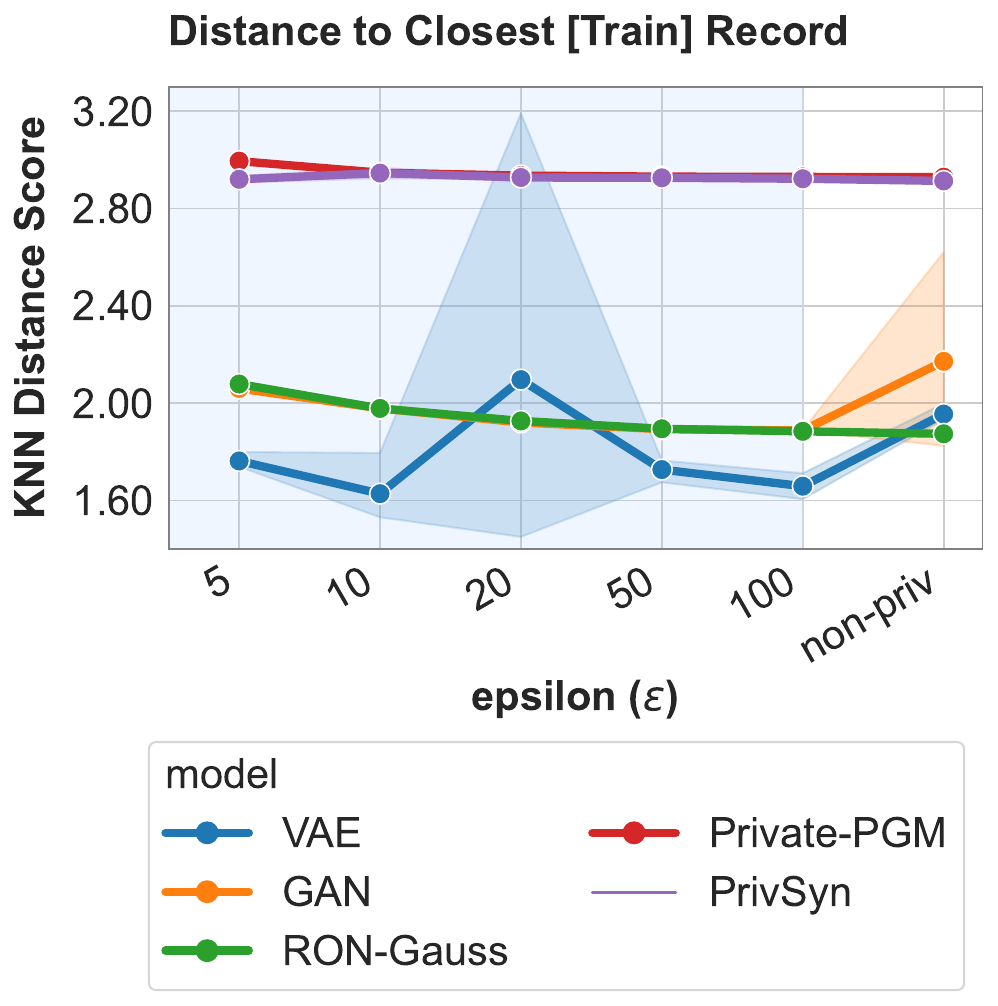}
\vspace{5pt}
\caption{Distance to the Closest Train Record. \textmd{There is no real data (Reference) since the reference is to the train data. A Score of 0 would signify synthetic data that exactly replicates the training data.}}
\end{figure*}

\subsection{Correlation between Marginal Metrics}
\begin{figure*}[!htbp]
\centering
\includegraphics[width=0.6\textwidth]{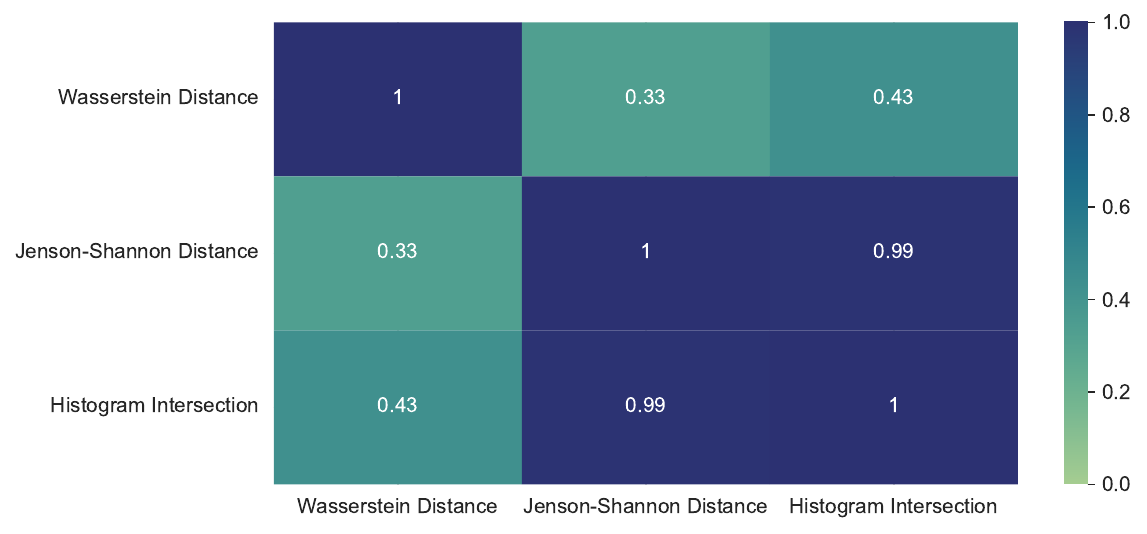}
\vspace{5pt}
\caption[Correlation Coefficients between Marginal Metrics]{Correlation Coefficients between Marginal Metrics in Absolute Value. \textmd{A higher score  closer to 1 indicates stronger correlation.}
}
\label{fig:marginal_corr}
\end{figure*}

\clearpage
\section{Additional Plots for Gene Co-expression Evaluation}
\subsection{$r>0$ (Default Setting)}
\vspace{-15pt}
\begin{figure*}[!htbp]
\centering
\newcommand{\figwidth}{0.32\textwidth}
\subfigure[$\varepsilon=5$, seed 1]{
\begin{minipage}[b]{\figwidth} \includegraphics[width=1.0\textwidth,trim={0 0 12cm 0},clip]{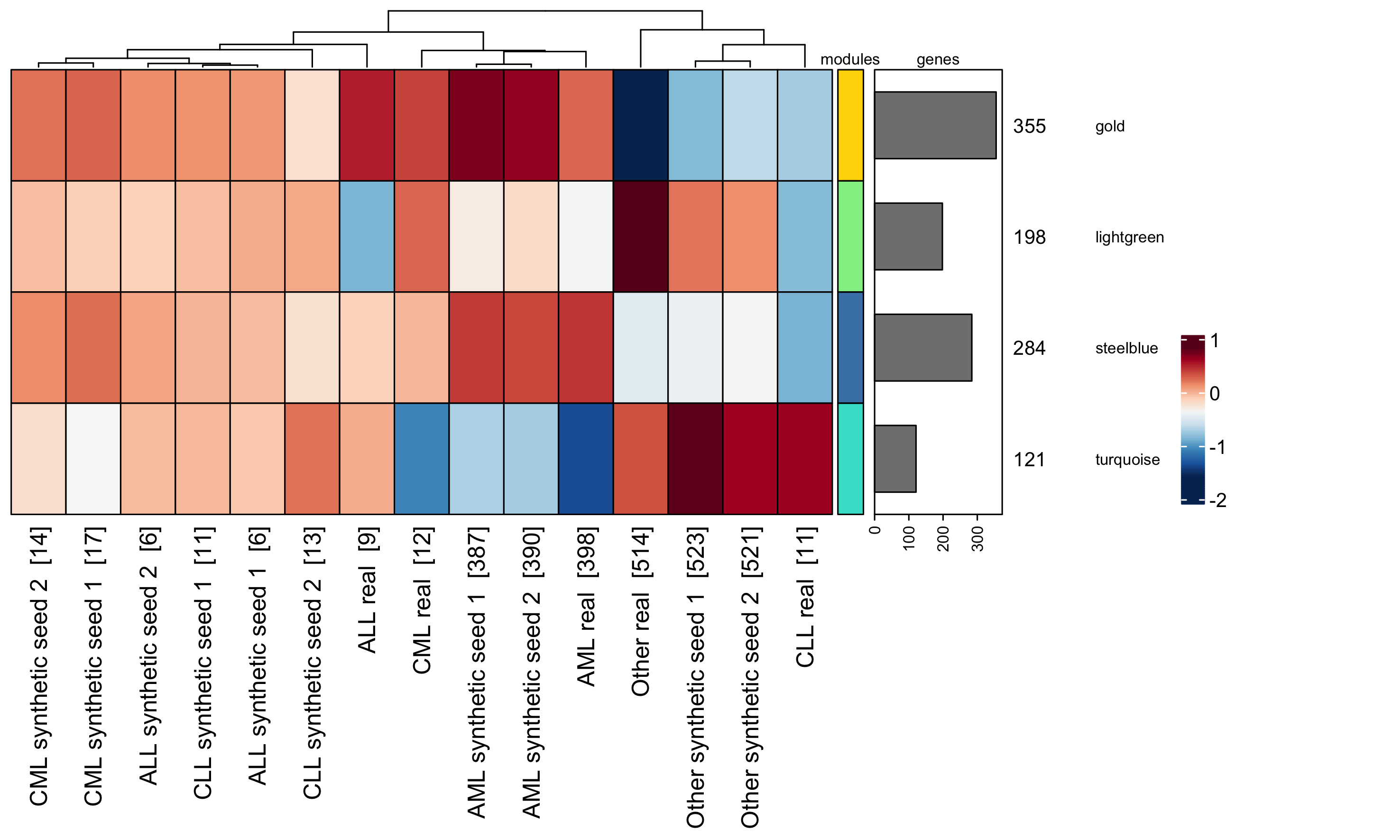}
\end{minipage}
}
\subfigure[$\varepsilon=10$, seed 1]{
\begin{minipage}[b]{\figwidth} \includegraphics[width=1.0\textwidth,trim={0 0 12cm 0},clip]{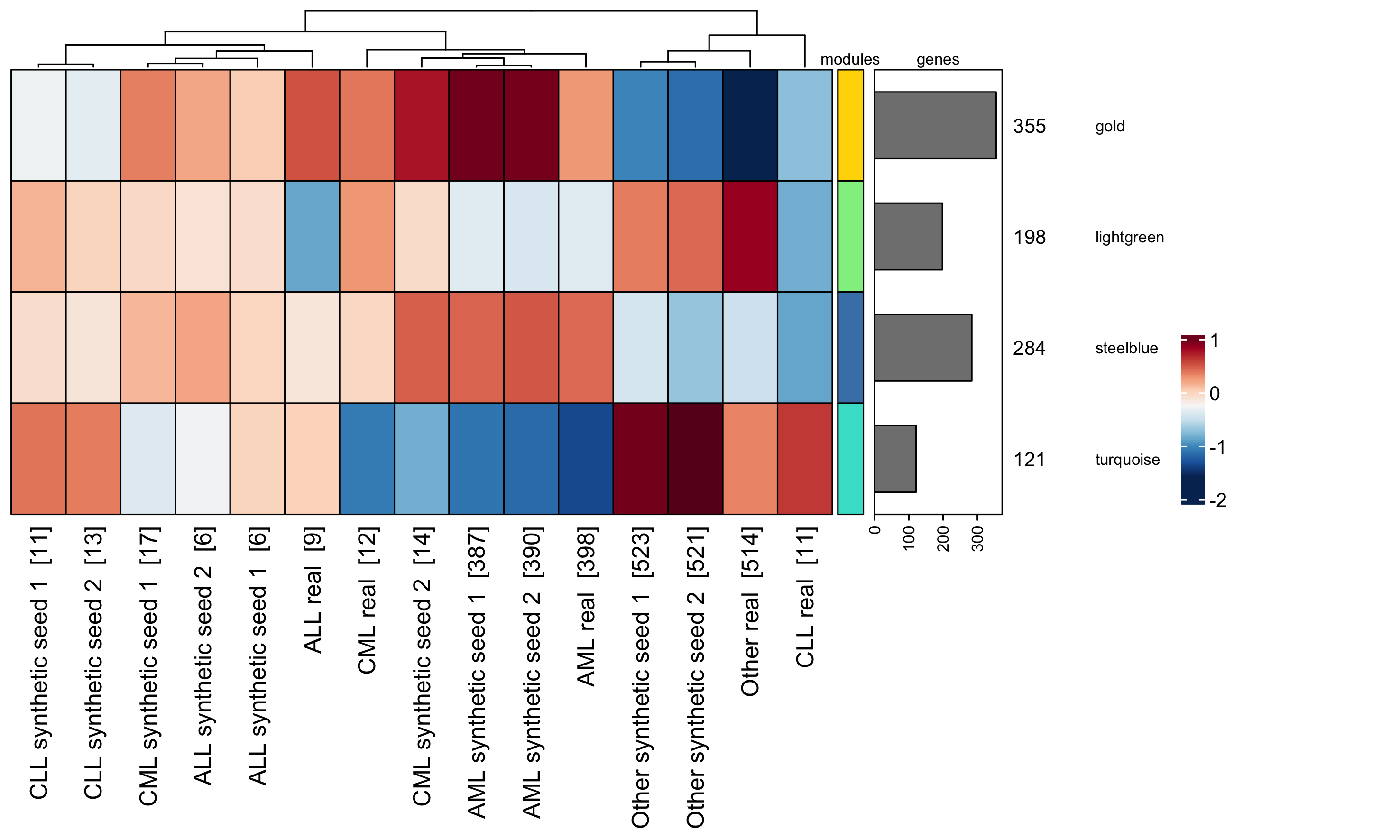}
\end{minipage}
}
\subfigure[$\varepsilon=20$, seed 1]{
\begin{minipage}[b]{\figwidth} \includegraphics[width=1.0\textwidth,trim={0 0 12cm 0},clip]{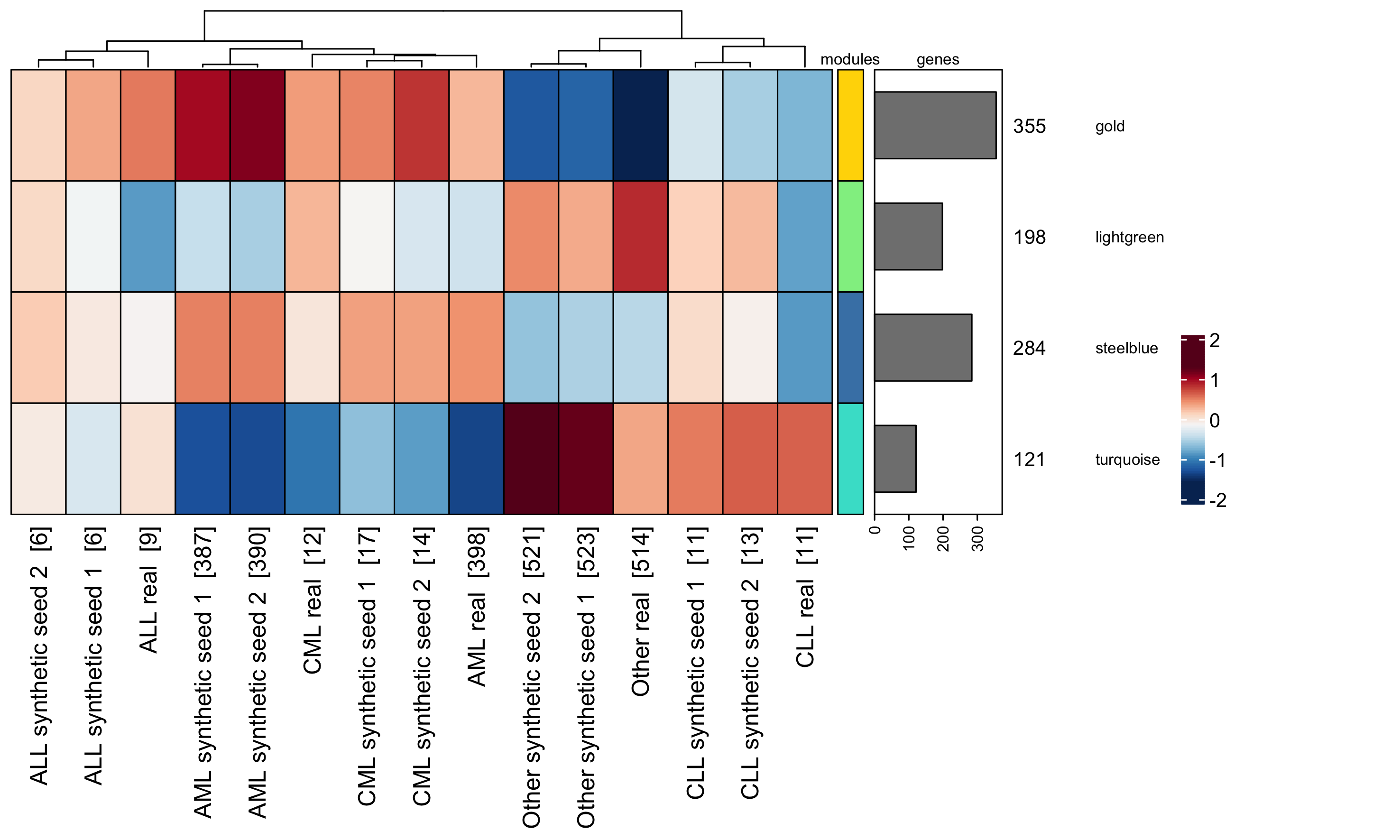}
\end{minipage}
}
\subfigure[$\varepsilon=50$, seed 1]{
\begin{minipage}[b]{\figwidth} \includegraphics[width=1.0\textwidth,trim={0 0 12cm 0},clip]{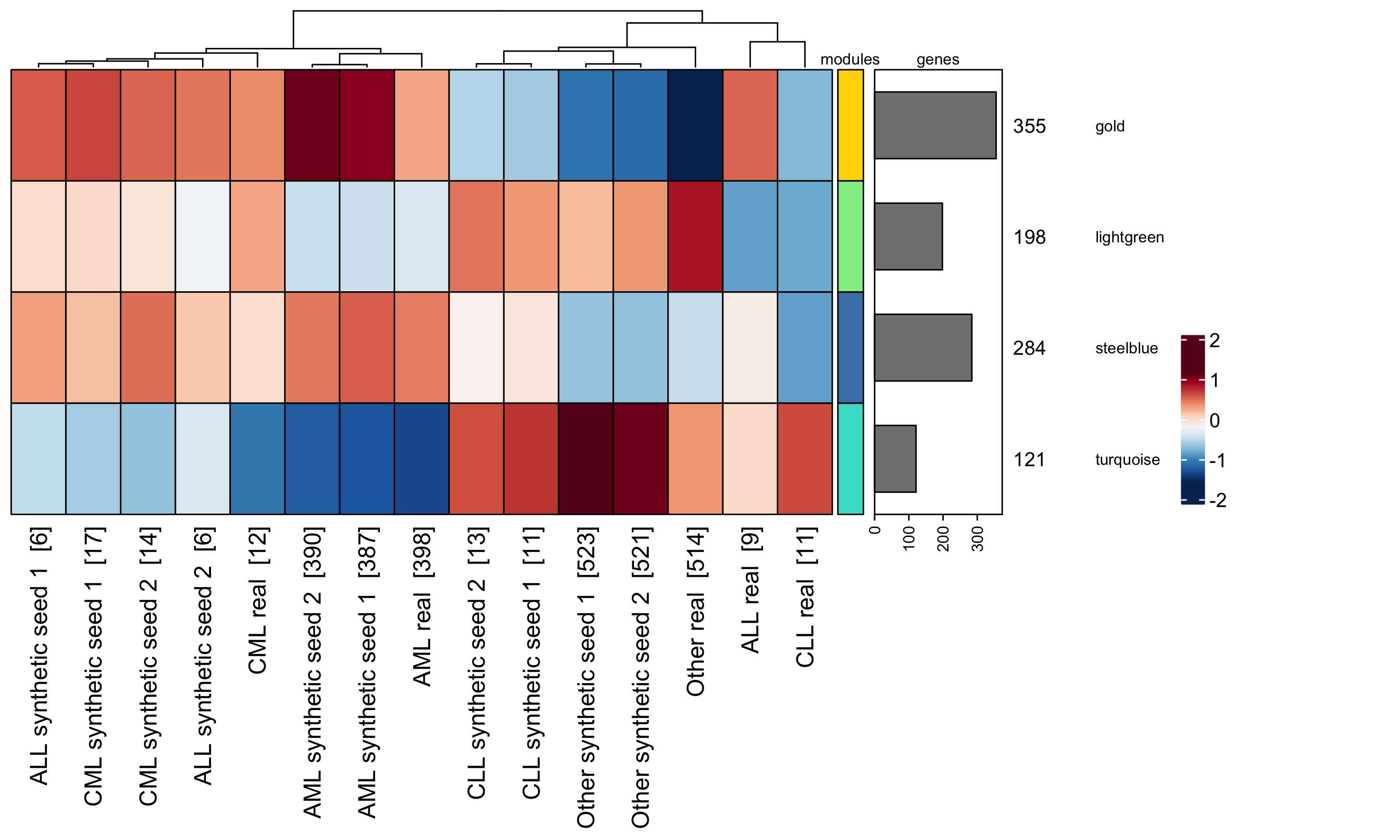}
\end{minipage}
}
\subfigure[$\varepsilon=100$, seed 1]{
\begin{minipage}[b]{\figwidth} \includegraphics[width=1.0\textwidth,trim={0 0 12cm 0},clip]{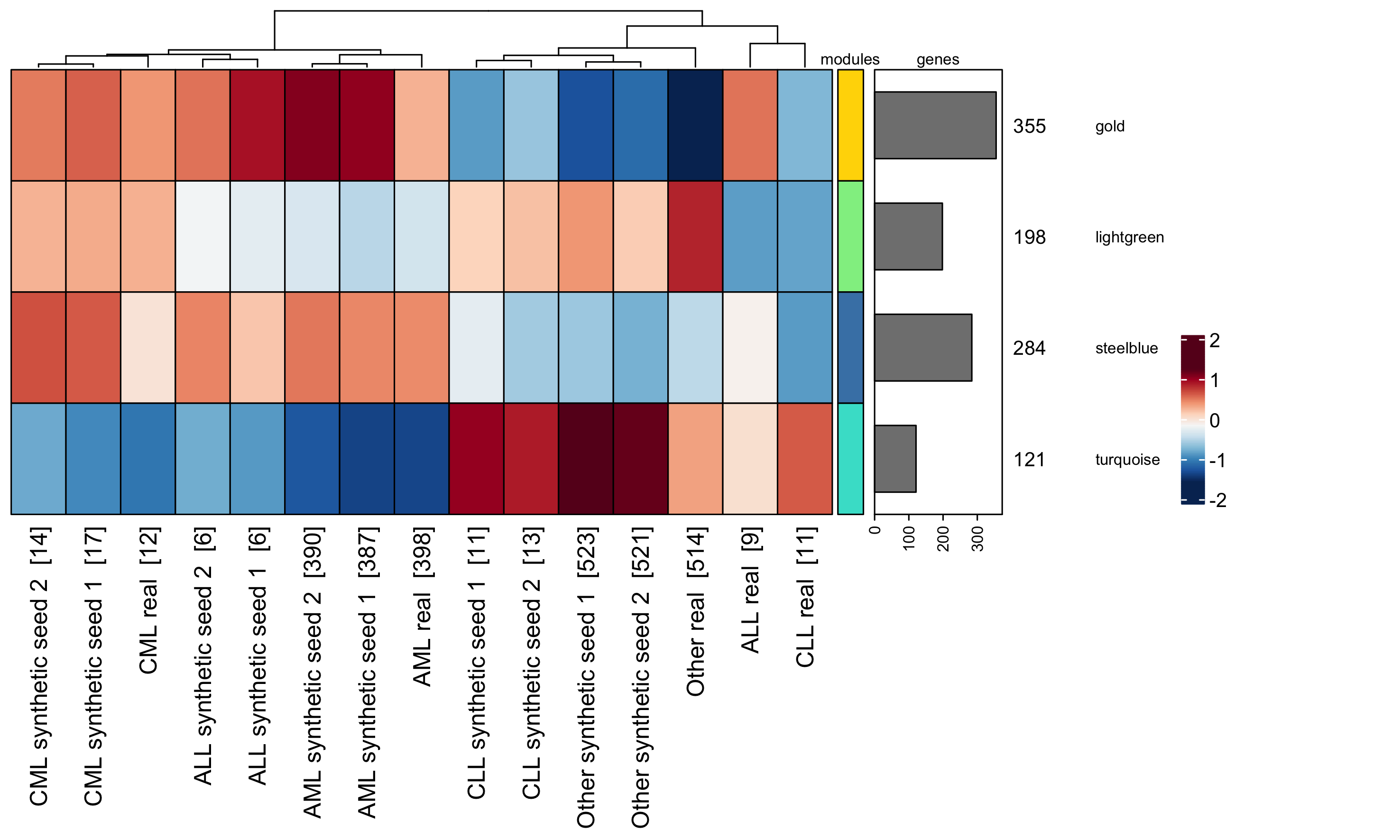}
\end{minipage}
}
\subfigure[non-priv, seed 1]{
\begin{minipage}[b]{\figwidth} \includegraphics[width=1.0\textwidth,trim={0 0 12cm 0},clip]{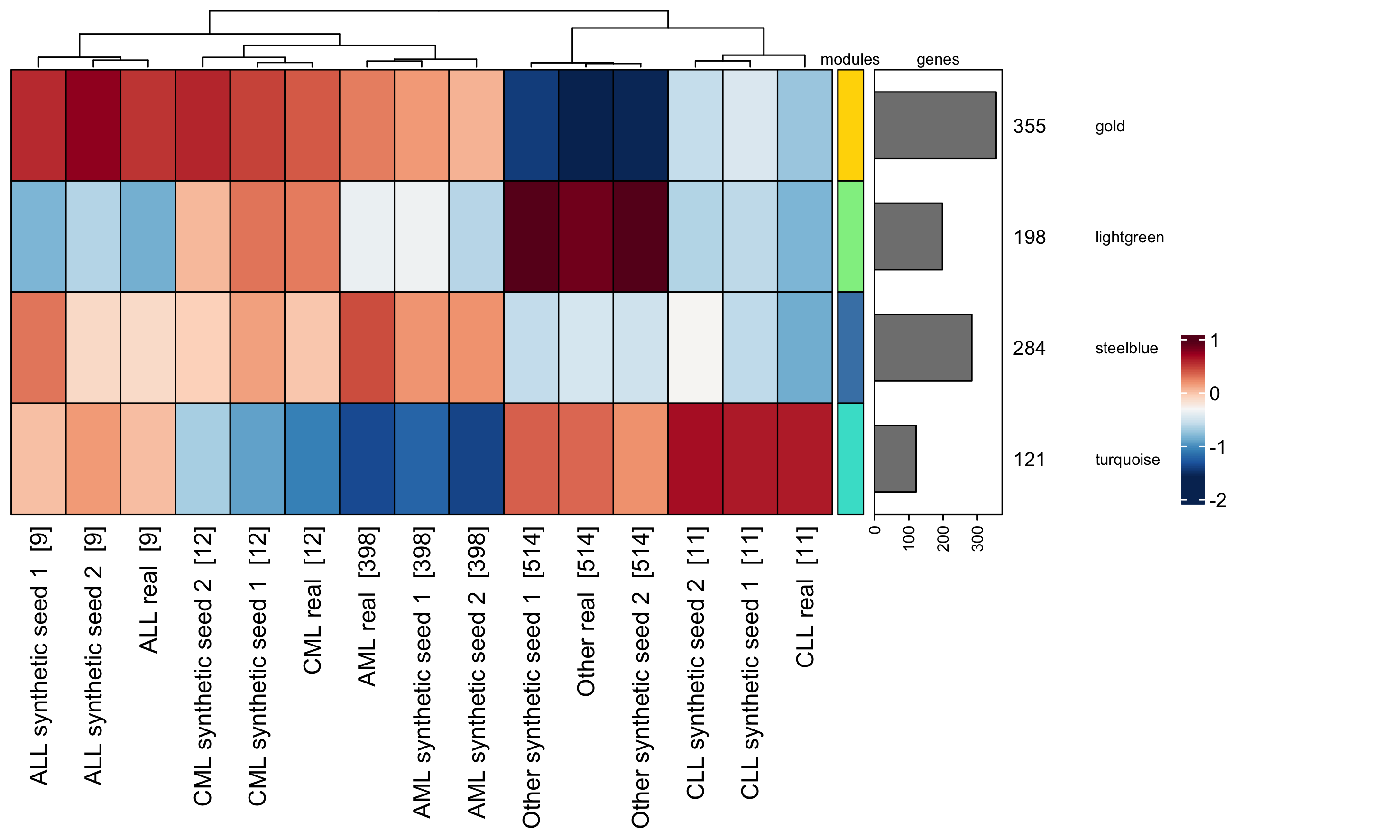}
\end{minipage}
}
\subfigure[$\varepsilon=5$, seed 2]{
\begin{minipage}[b]{\figwidth} \includegraphics[width=1.0\textwidth,trim={0 0 12cm 0},clip]{figures/module_heatmap_VAE_ep5_k1000.png}
\end{minipage}
}
\subfigure[$\varepsilon=10$, seed 2]{
\begin{minipage}[b]{\figwidth} \includegraphics[width=1.0\textwidth,trim={0 0 12cm 0},clip]{figures/module_heatmap_VAE_ep10_k1000.png}
\end{minipage}
}
\subfigure[$\varepsilon=20$, seed 2]{
\begin{minipage}[b]{\figwidth} \includegraphics[width=1.0\textwidth,trim={0 0 12cm 0},clip]{figures/module_heatmap_VAE_ep20_k1000.png}
\end{minipage}
}
\subfigure[$\varepsilon=50$, seed 2]{
\begin{minipage}[b]{\figwidth} \includegraphics[width=1.0\textwidth,trim={0 0 12cm 0},clip]{figures/module_heatmap_VAE_ep50_k1000.png}
\end{minipage}
}
\subfigure[$\varepsilon=100$, seed 2]{
\begin{minipage}[b]{\figwidth} \includegraphics[width=1.0\textwidth,trim={0 0 12cm 0},clip]{figures/module_heatmap_VAE_ep100_k1000.png}
\end{minipage}
}
\subfigure[non-priv, seed 2]{
\begin{minipage}[b]{\figwidth} \includegraphics[width=1.0\textwidth,trim={0 0 12cm 0},clip]{figures/module_heatmap_VAE_nonpriv_k1000.png}
\end{minipage}
}
\caption{Activation patterns of co-expressed gene modules in VAE after filtering co-expressions for $r$ > 0. \textmd{Shown are the Group Fold Changes (GFCs) of gene modules (rows) in the real and the synthetic data sampled with two different seeds. 
Darker shades of red imply activation of the gene module, while darker shades of blue indicate deactivation. 
A heatmap is shown for each $\varepsilon$ twice, once for each seed used to \emph{split} the training data. 
While the synthetic data maintains the expression patterns of modules in the non-private setting, they gradually decay with decreasing $\varepsilon$.}}
\label{figure:S1}
\end{figure*}

\begin{figure*}[!htbp]
\centering
\newcommand{\figwidth}{0.32\textwidth}
\subfigure[$\varepsilon=5$, seed 1]{
\begin{minipage}[b]{\figwidth} \includegraphics[width=1.0\textwidth,trim={0 0 12cm 0},clip]{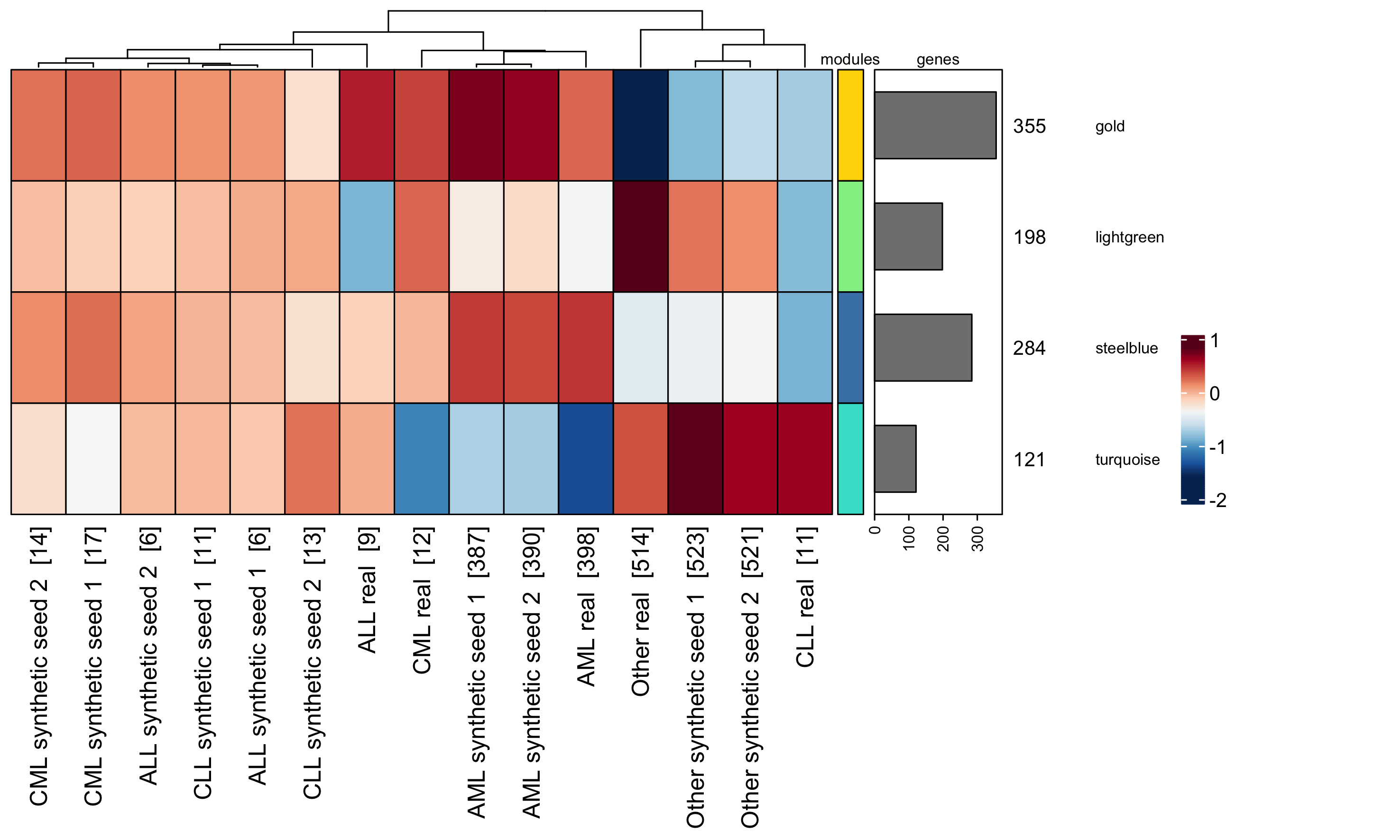}
\end{minipage}
}
\subfigure[$\varepsilon=10$, seed 1]{
\begin{minipage}[b]{\figwidth} \includegraphics[width=1.0\textwidth,trim={0 0 12cm 0},clip]{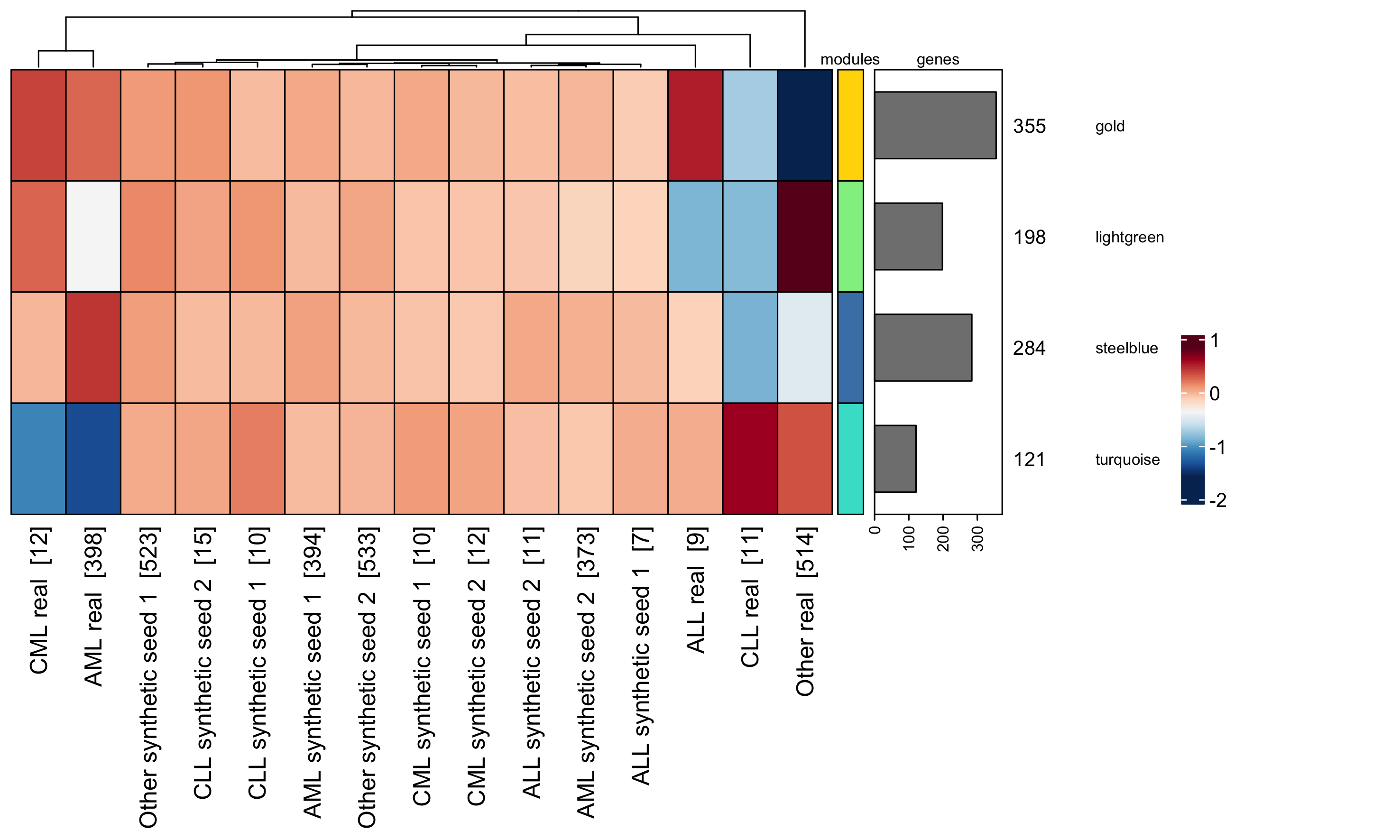}
\end{minipage}
}
\subfigure[$\varepsilon=20$, seed 1]{
\begin{minipage}[b]{\figwidth} \includegraphics[width=1.0\textwidth,trim={0 0 12cm 0},clip]{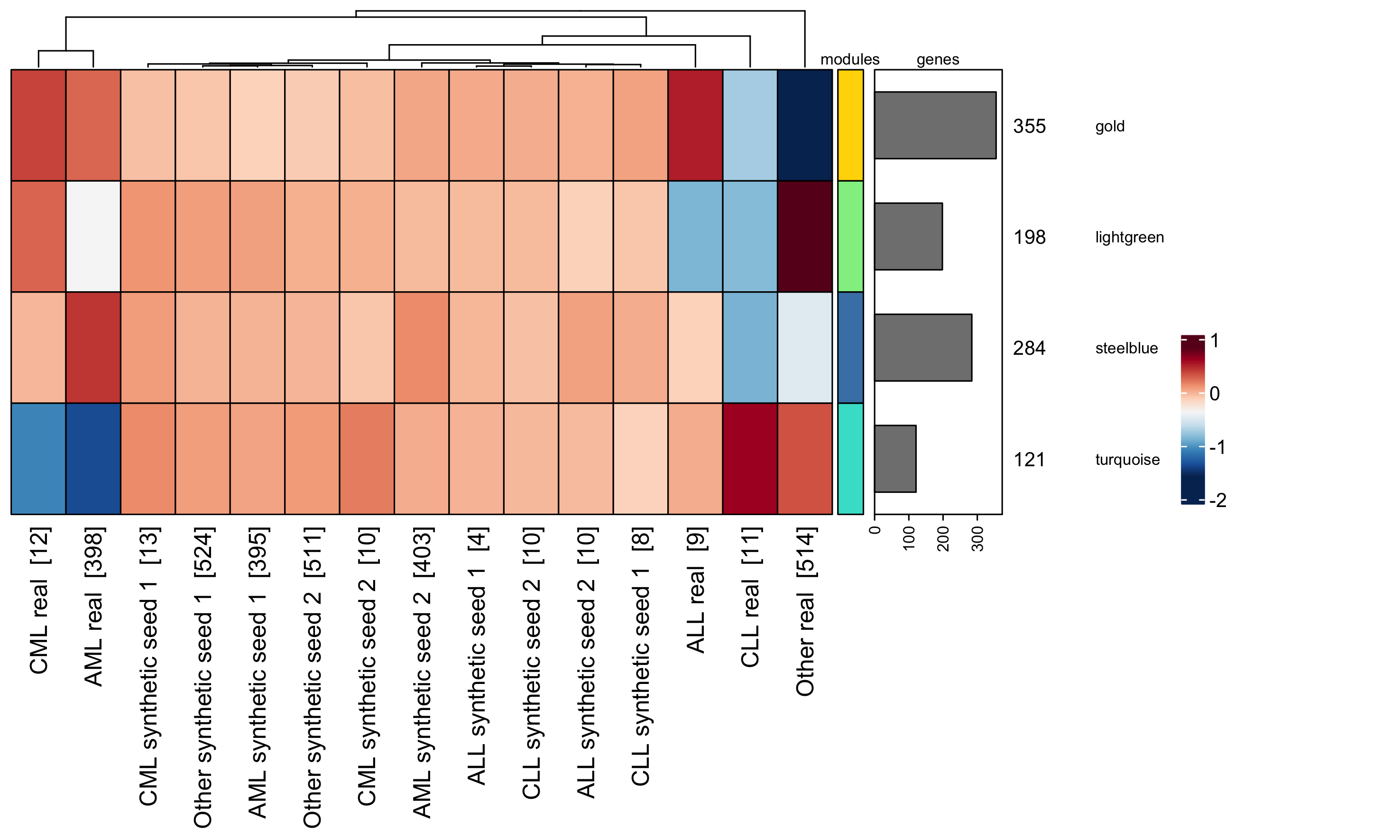}
\end{minipage}
}
\subfigure[$\varepsilon=50$, seed 1]{
\begin{minipage}[b]{\figwidth} \includegraphics[width=1.0\textwidth,trim={0 0 12cm 0},clip]{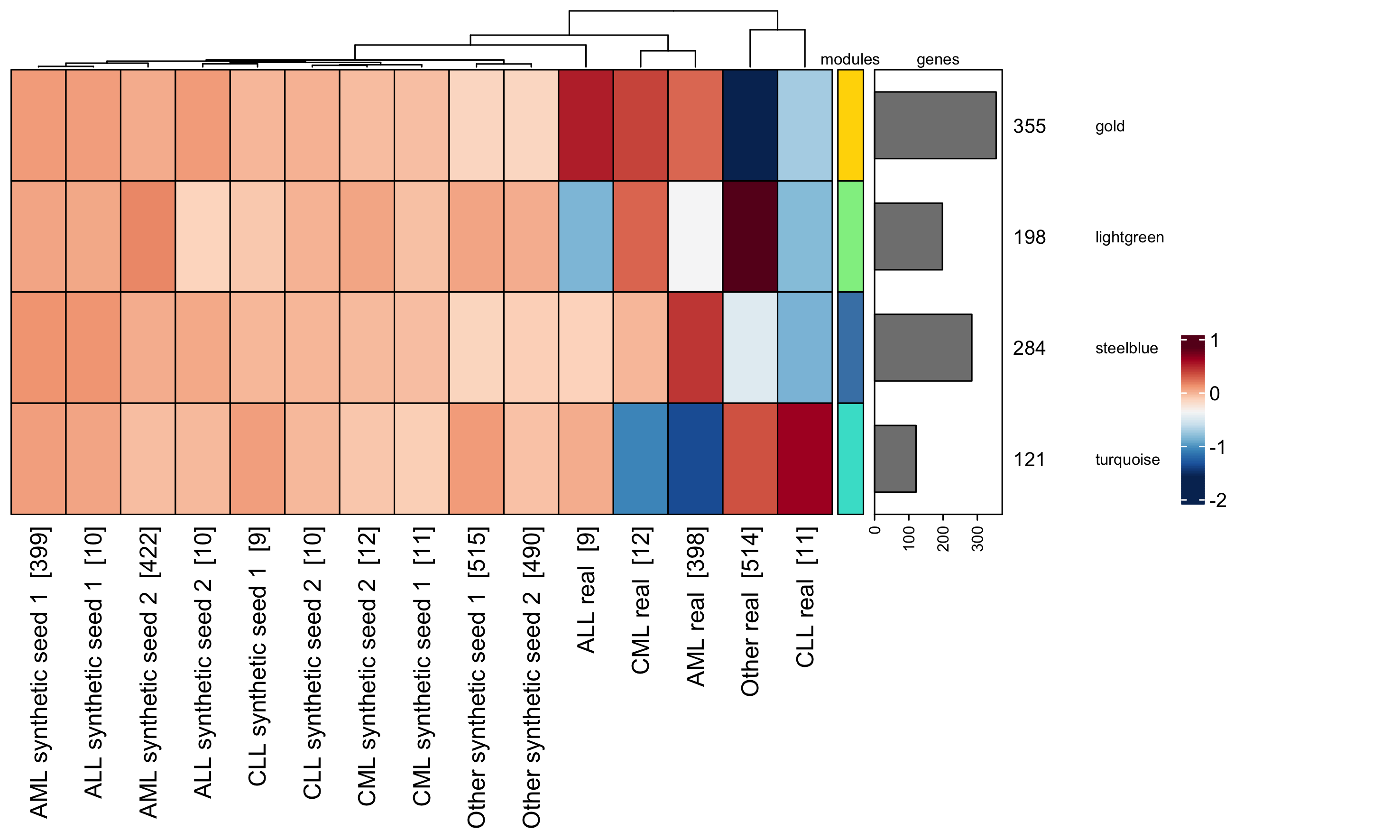}
\end{minipage}
}
\subfigure[$\varepsilon=100$, seed 1]{
\begin{minipage}[b]{\figwidth} \includegraphics[width=1.0\textwidth,trim={0 0 12cm 0},clip]{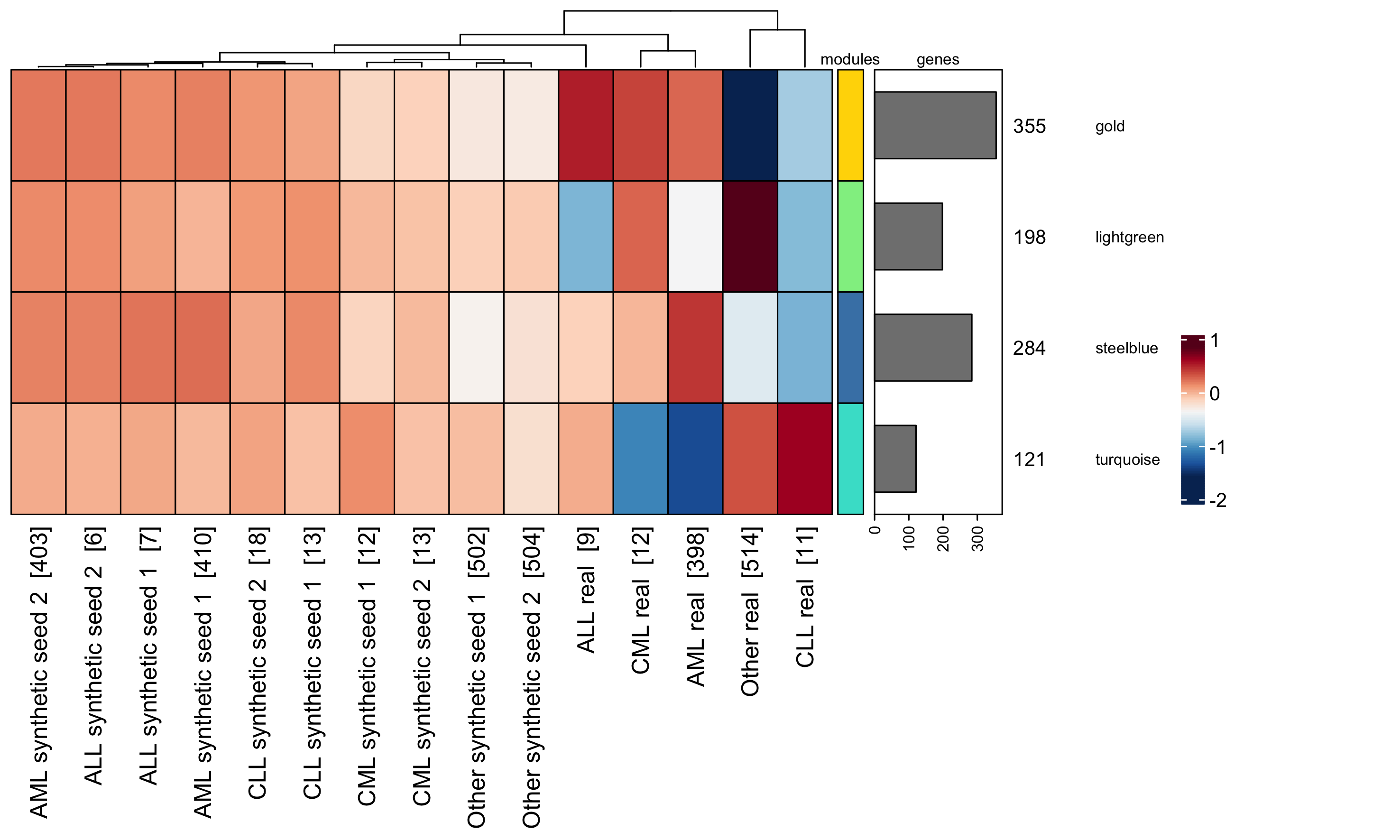}
\end{minipage}
}
\subfigure[non-priv, seed 1]{
\begin{minipage}[b]{\figwidth} \includegraphics[width=1.0\textwidth,trim={0 0 12cm 0},clip]{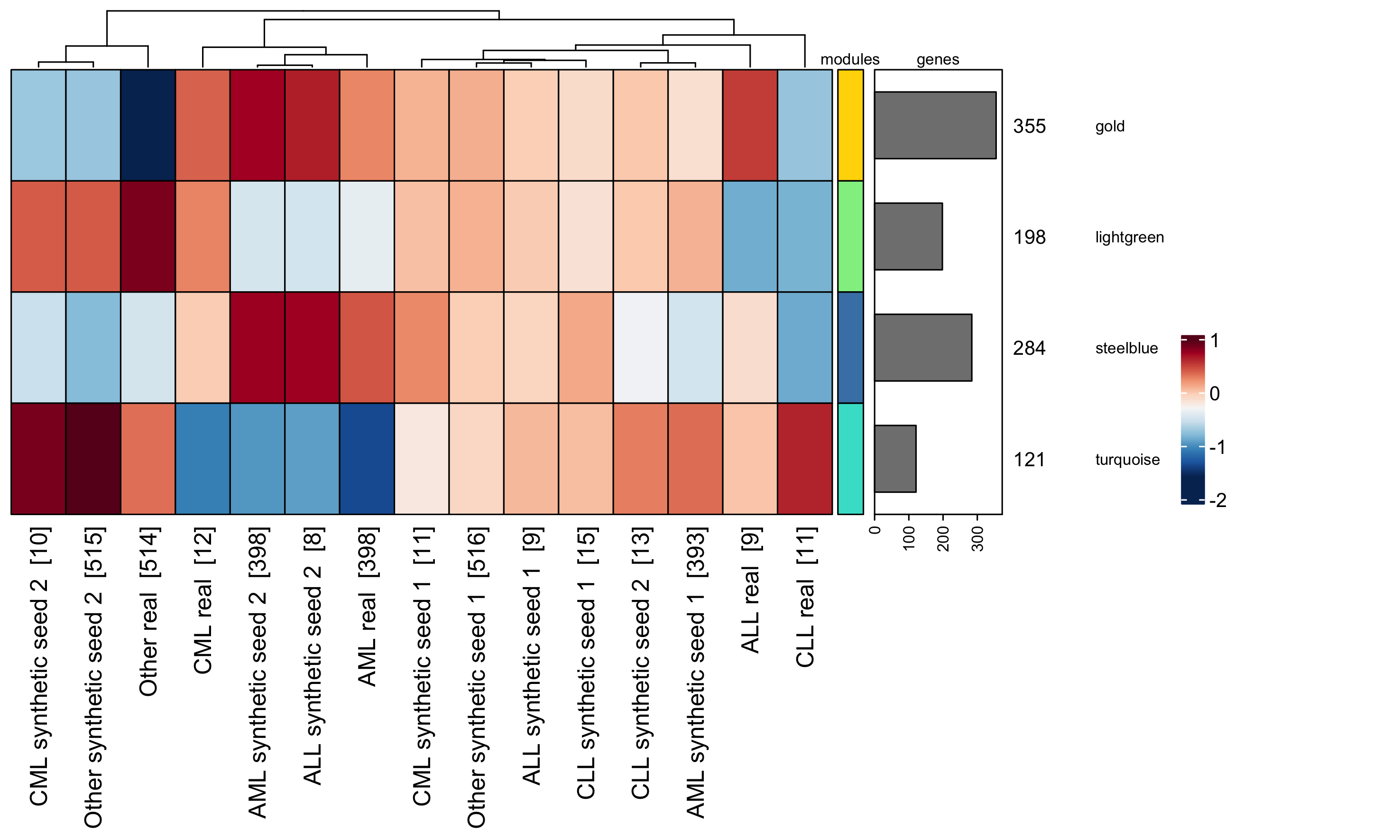}
\end{minipage}
}
\subfigure[$\varepsilon=5$, seed 2]{
\begin{minipage}[b]{\figwidth} \includegraphics[width=1.0\textwidth,trim={0 0 12cm 0},clip]{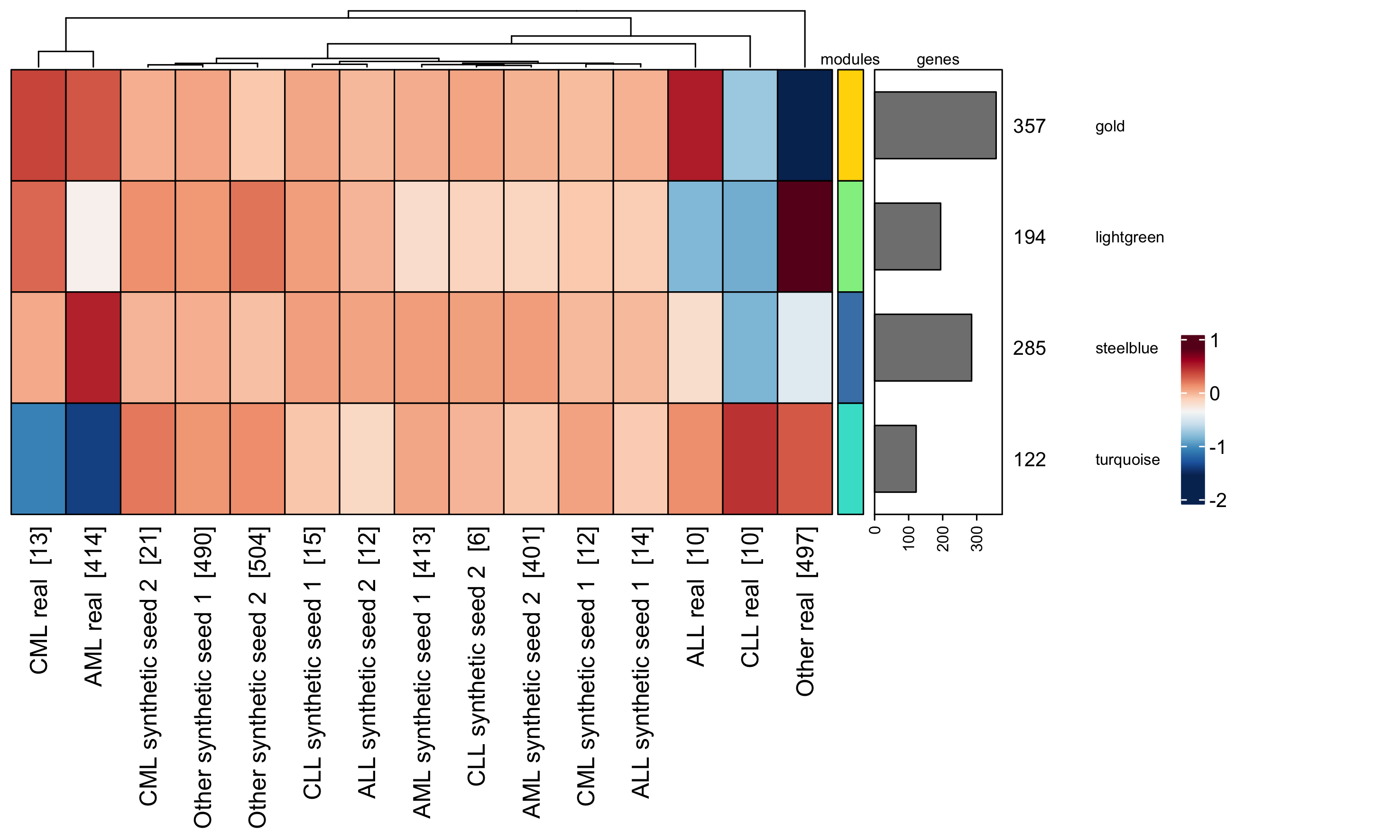}
\end{minipage}
}
\subfigure[$\varepsilon=10$, seed 2]{
\begin{minipage}[b]{\figwidth} \includegraphics[width=1.0\textwidth,trim={0 0 12cm 0},clip]{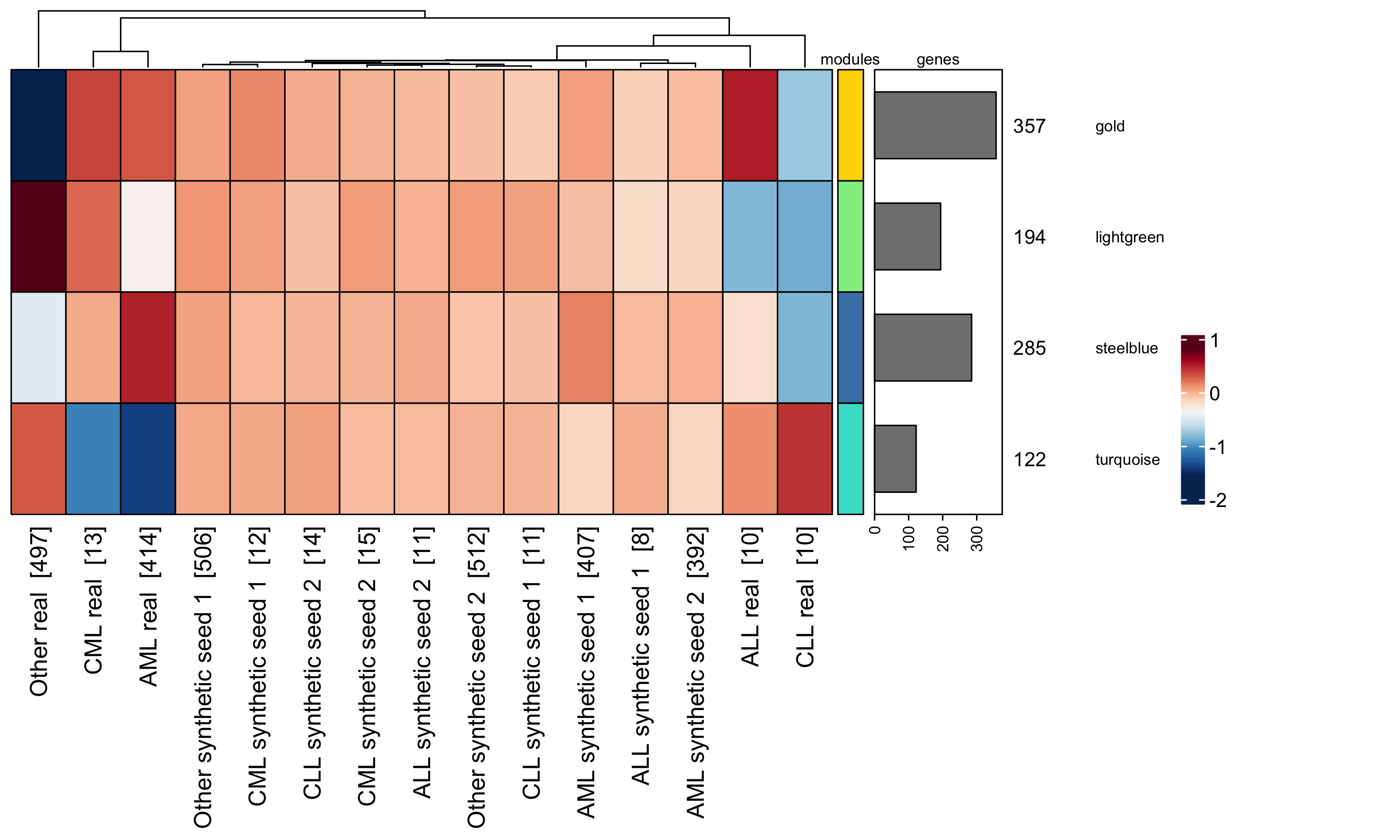}
\end{minipage}
}
\subfigure[$\varepsilon=20$, seed 2]{
\begin{minipage}[b]{\figwidth} \includegraphics[width=1.0\textwidth,trim={0 0 12cm 0},clip]{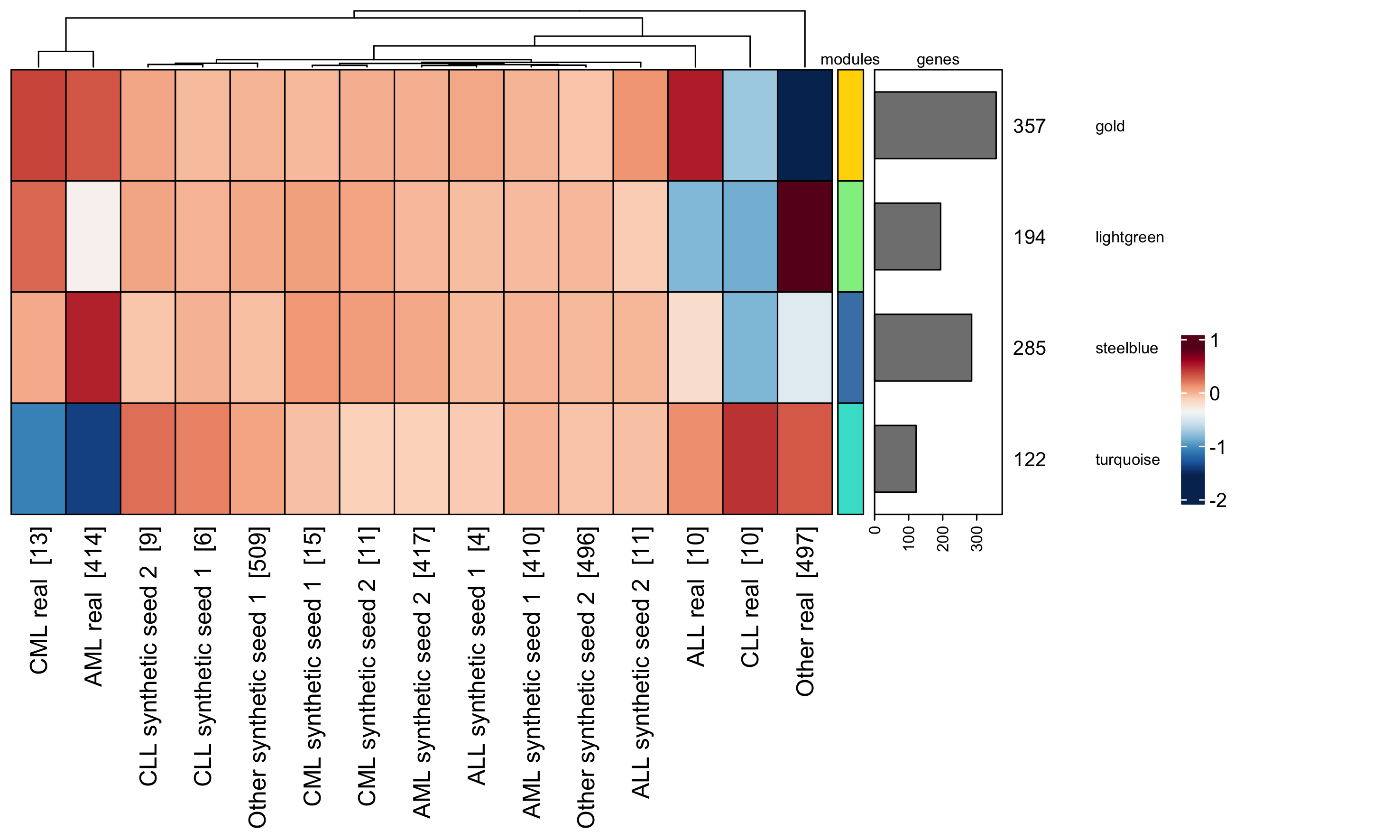}
\end{minipage}
}
\subfigure[$\varepsilon=50$, seed 2]{
\begin{minipage}[b]{\figwidth} \includegraphics[width=1.0\textwidth,trim={0 0 12cm 0},clip]{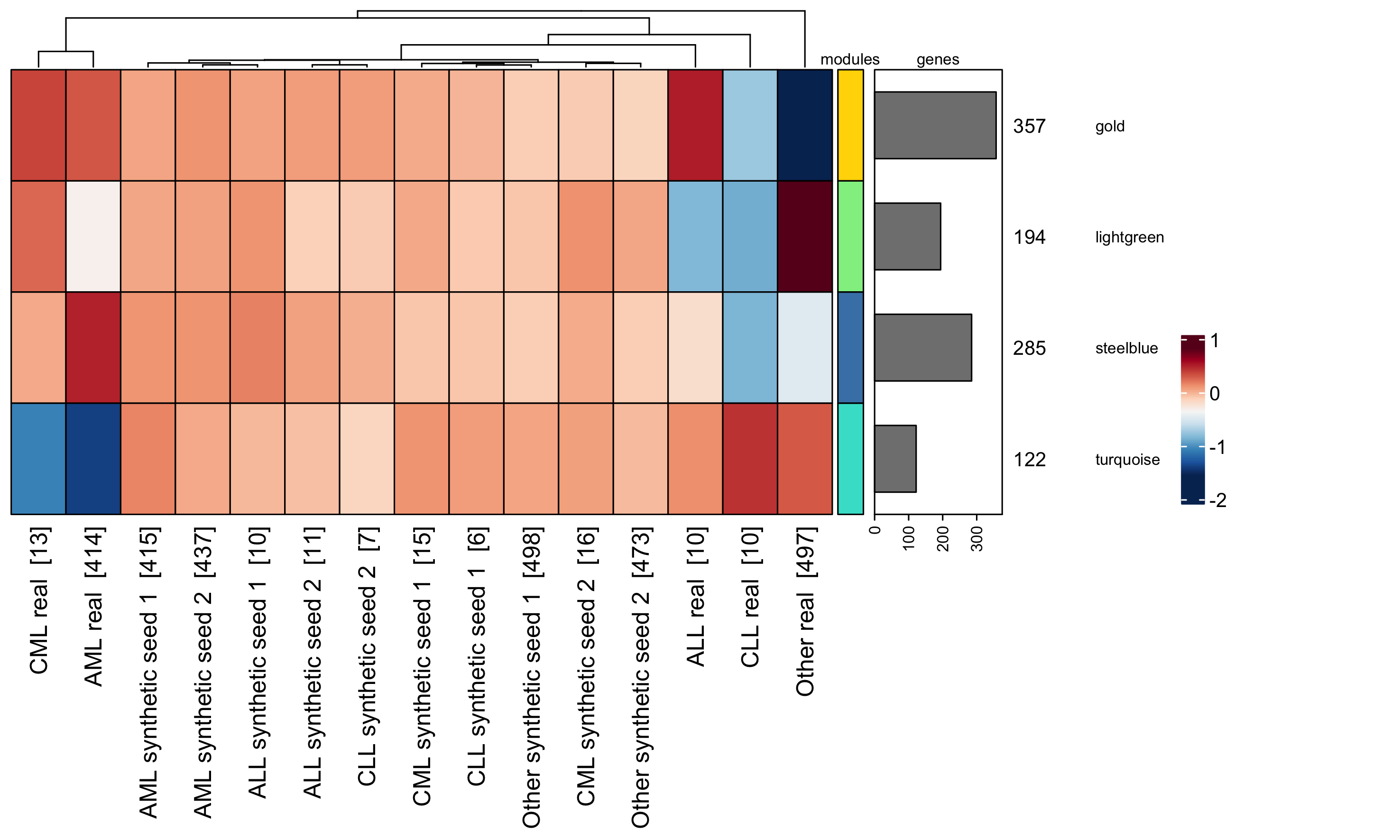}
\end{minipage}
}
\subfigure[$\varepsilon=100$, seed 2]{
\begin{minipage}[b]{\figwidth} \includegraphics[width=1.0\textwidth,trim={0 0 12cm 0},clip]{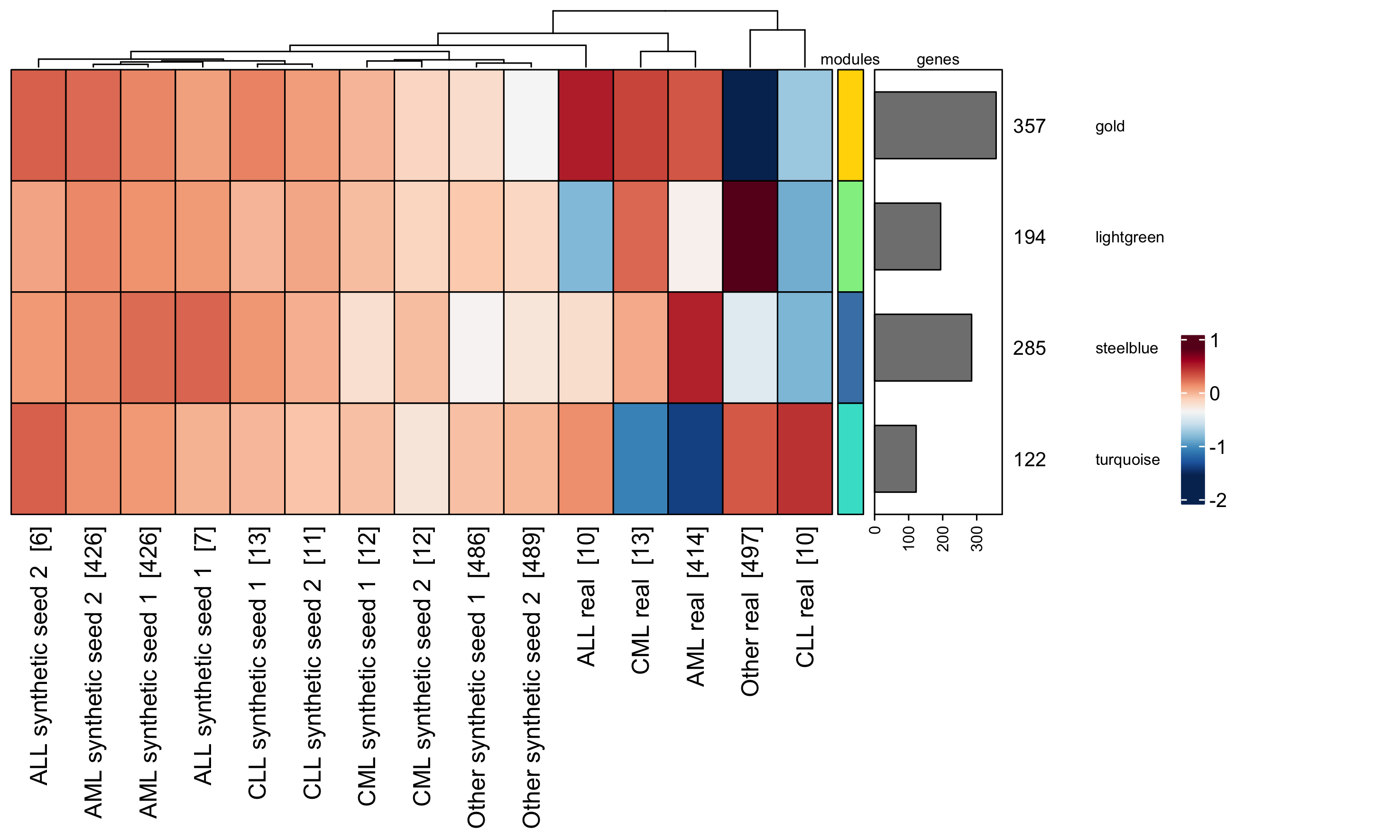}
\end{minipage}
}
\subfigure[non-priv, seed 2]{
\begin{minipage}[b]{\figwidth} \includegraphics[width=1.0\textwidth,trim={0 0 12cm 0},clip]{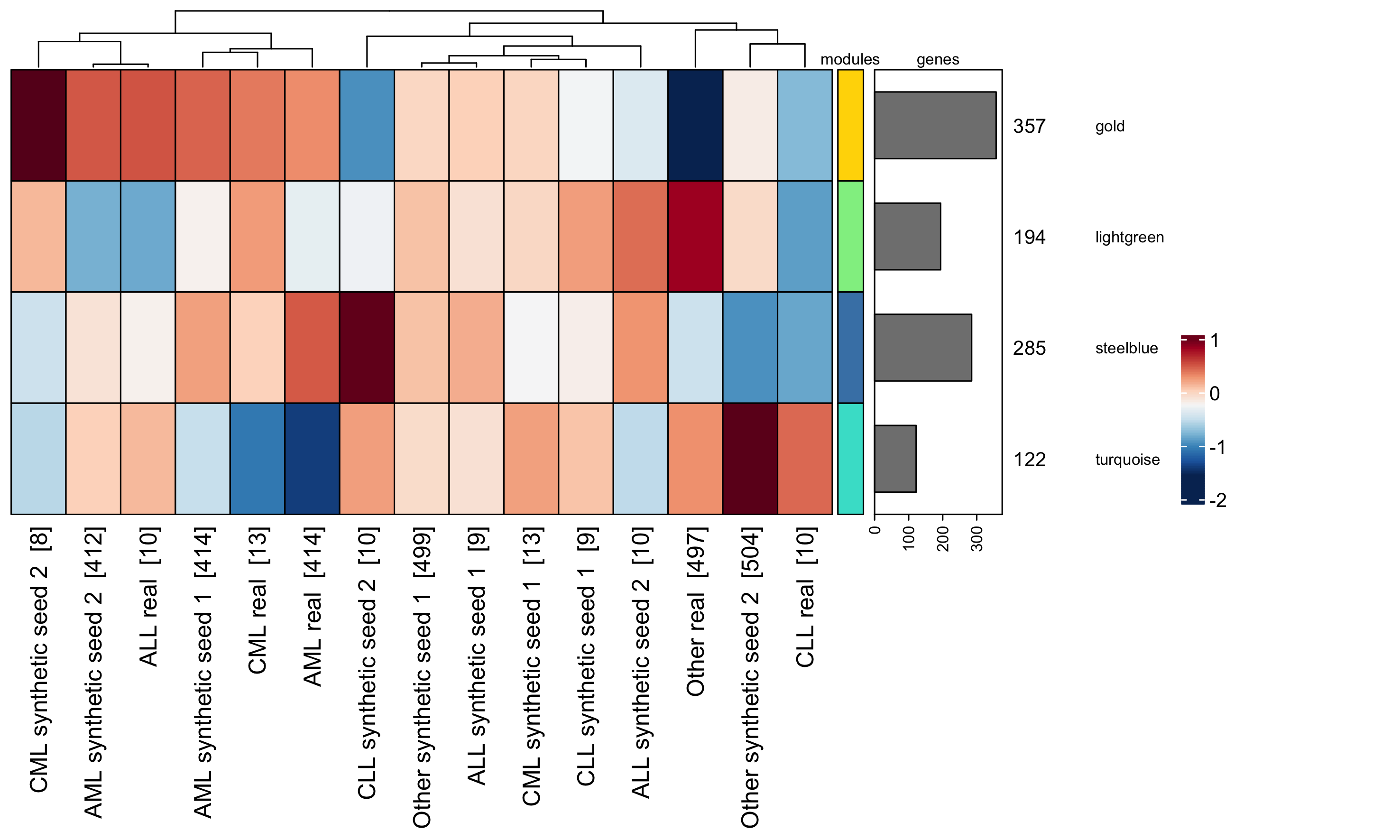}
\end{minipage}
}
\caption{Activation patterns of co-expressed gene modules in GAN after filtering co-expressions for $r$ > 0. \textmd{Shown are the Group Fold Changes (GFCs) of gene modules (rows) in the real and the synthetic data sampled with two different seeds. Numbers on the right indicate the number of genes per module. Darker shades of red imply activation of the gene module, while darker shades of blue indicate deactivation. The dendrograms show the hierarchical clustering of the classes in the different data sets. A heatmap is shown for each $\varepsilon$ twice, once for each seed used to \emph{split} the training data. Each heatmap further features, in addition to the real data, data from two synthetic sets, one for each seed used to \textit{generate} the data. The loss of structure in the module activation patterns of the synthetic data is striking, even for high privacy budgets and the non-private setting.}}
\label{figure:S2}
\end{figure*}

\begin{figure*}[!htbp]
\centering
\newcommand{\figwidth}{0.32\textwidth}
\subfigure[$\varepsilon=5$, seed 1]{
\begin{minipage}[b]{\figwidth} \includegraphics[width=1.0\textwidth,trim={0 0 12cm 0},clip]{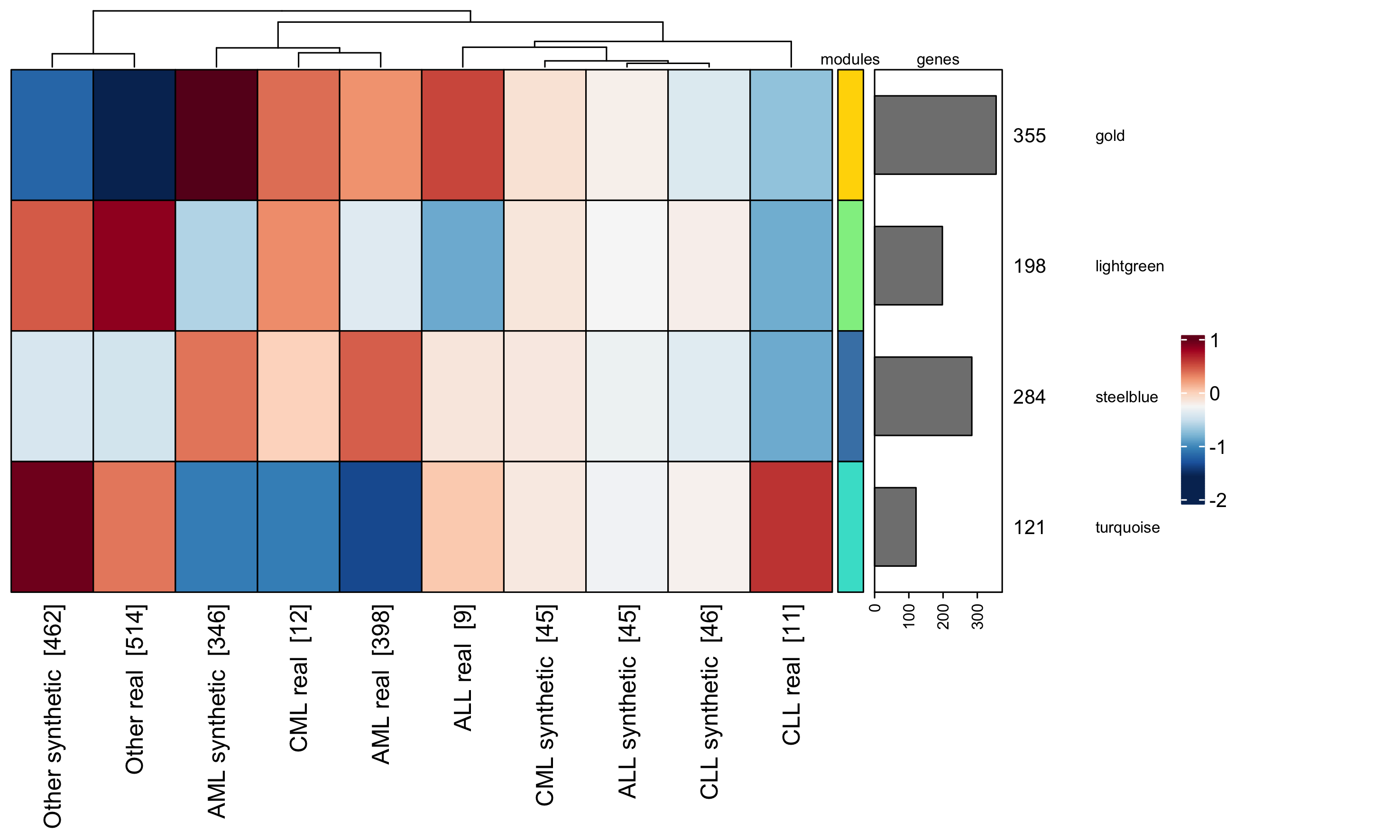}
\end{minipage}
}
\subfigure[$\varepsilon=10$, seed 1]{
\begin{minipage}[b]{\figwidth} \includegraphics[width=1.0\textwidth,trim={0 0 12cm 0},clip]{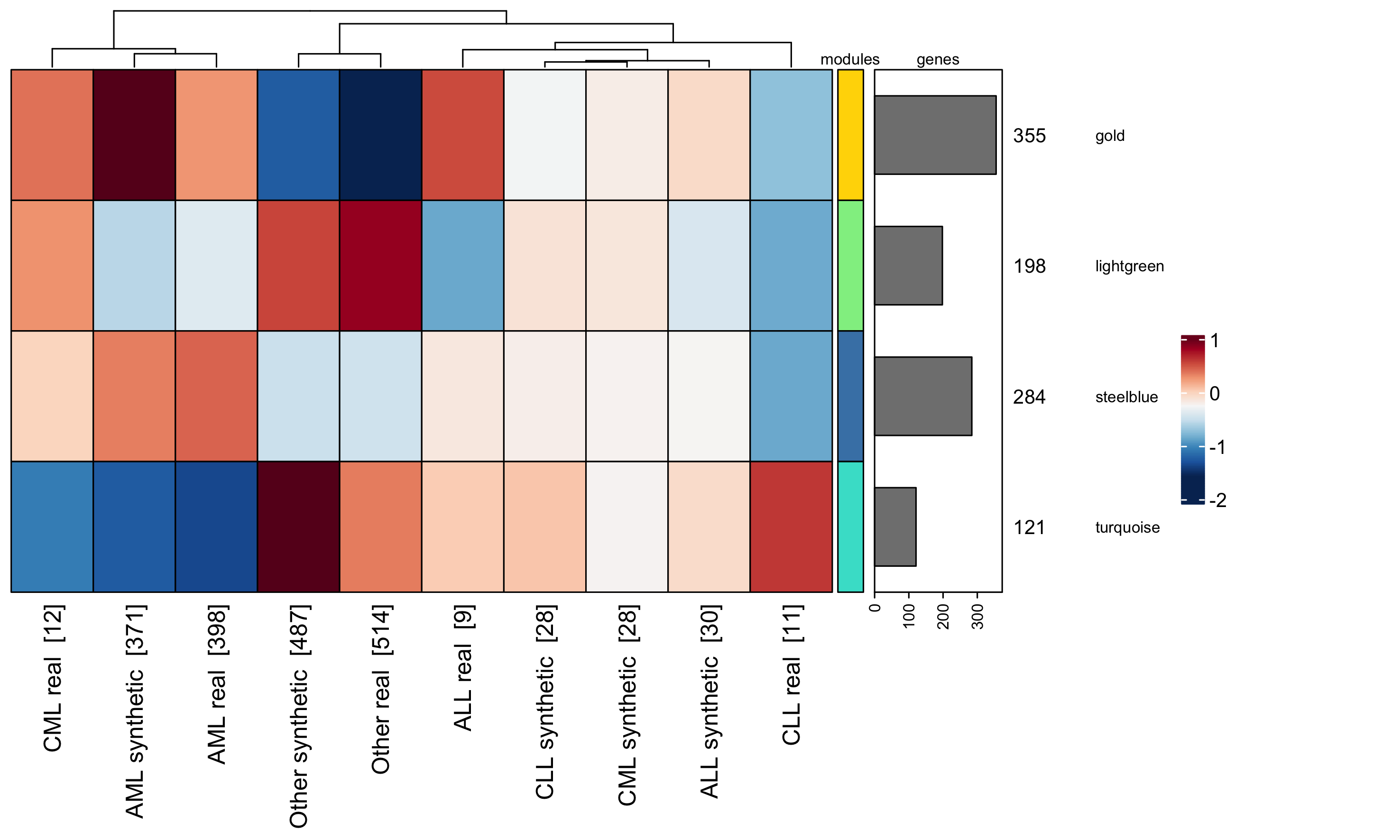}
\end{minipage}
}
\subfigure[$\varepsilon=20$, seed 1]{
\begin{minipage}[b]{\figwidth} \includegraphics[width=1.0\textwidth,trim={0 0 12cm 0},clip]{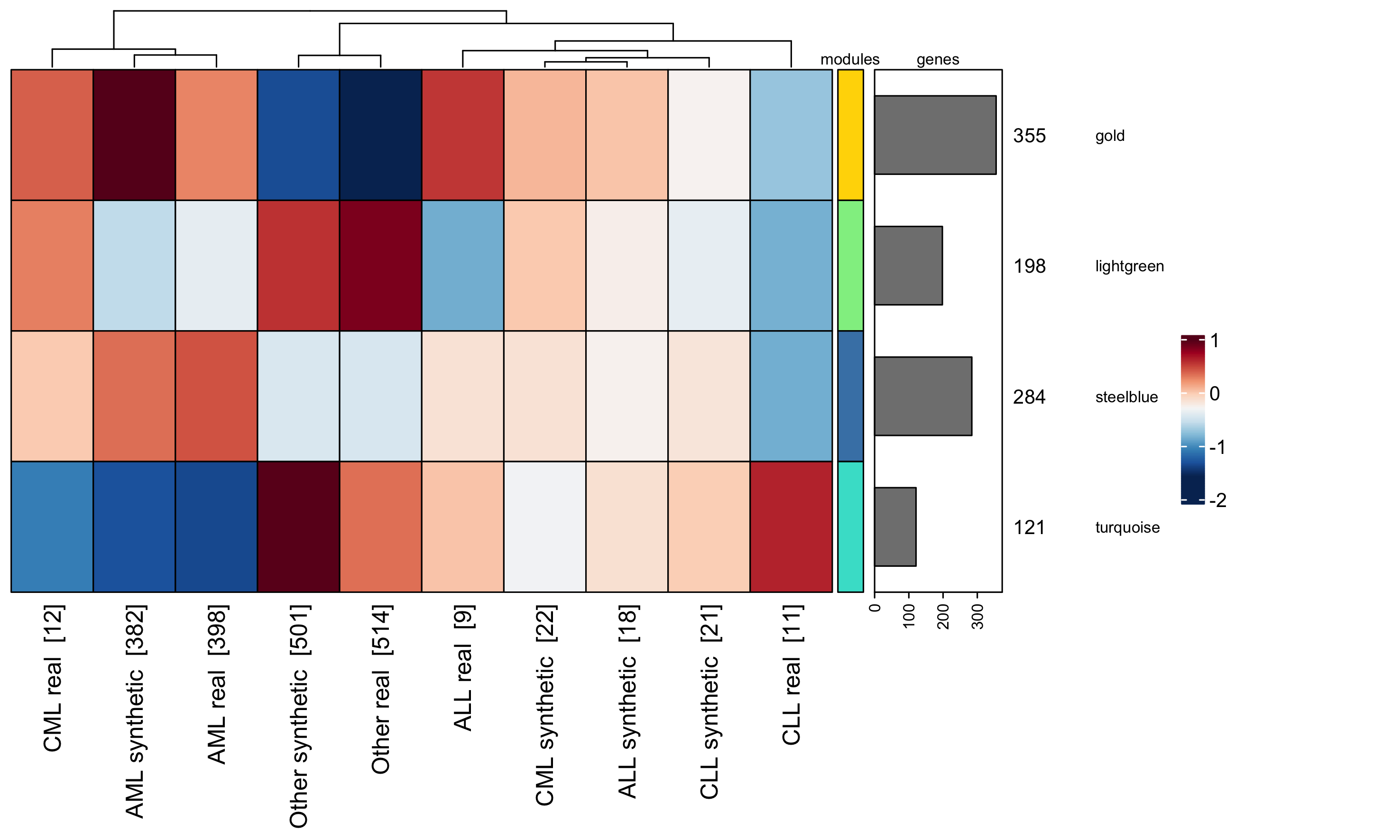}
\end{minipage}
}
\subfigure[$\varepsilon=50$, seed 1]{
\begin{minipage}[b]{\figwidth} \includegraphics[width=1.0\textwidth,trim={0 0 12cm 0},clip]{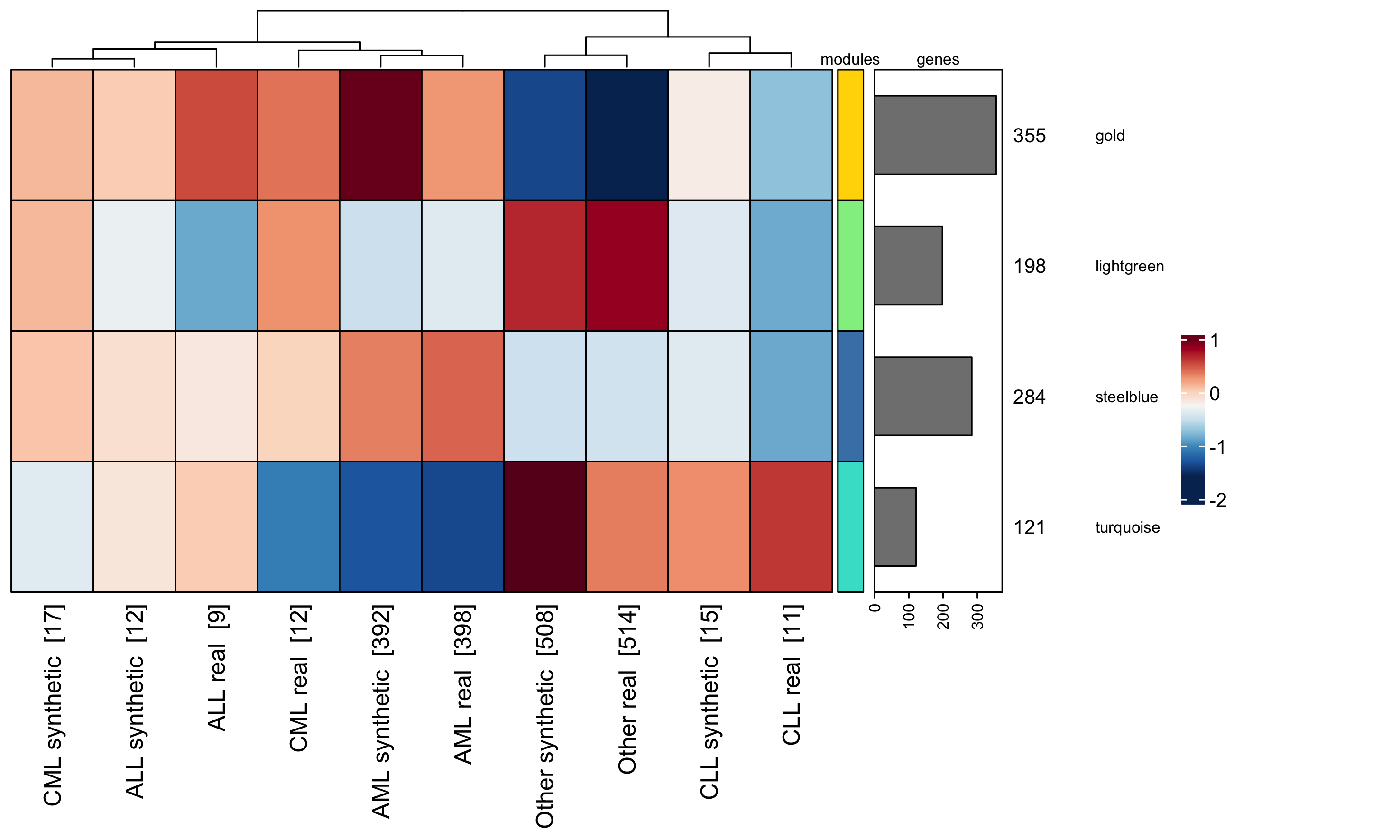}
\end{minipage}
}
\subfigure[$\varepsilon=100$, seed 1]{
\begin{minipage}[b]{\figwidth} \includegraphics[width=1.0\textwidth,trim={0 0 12cm 0},clip]{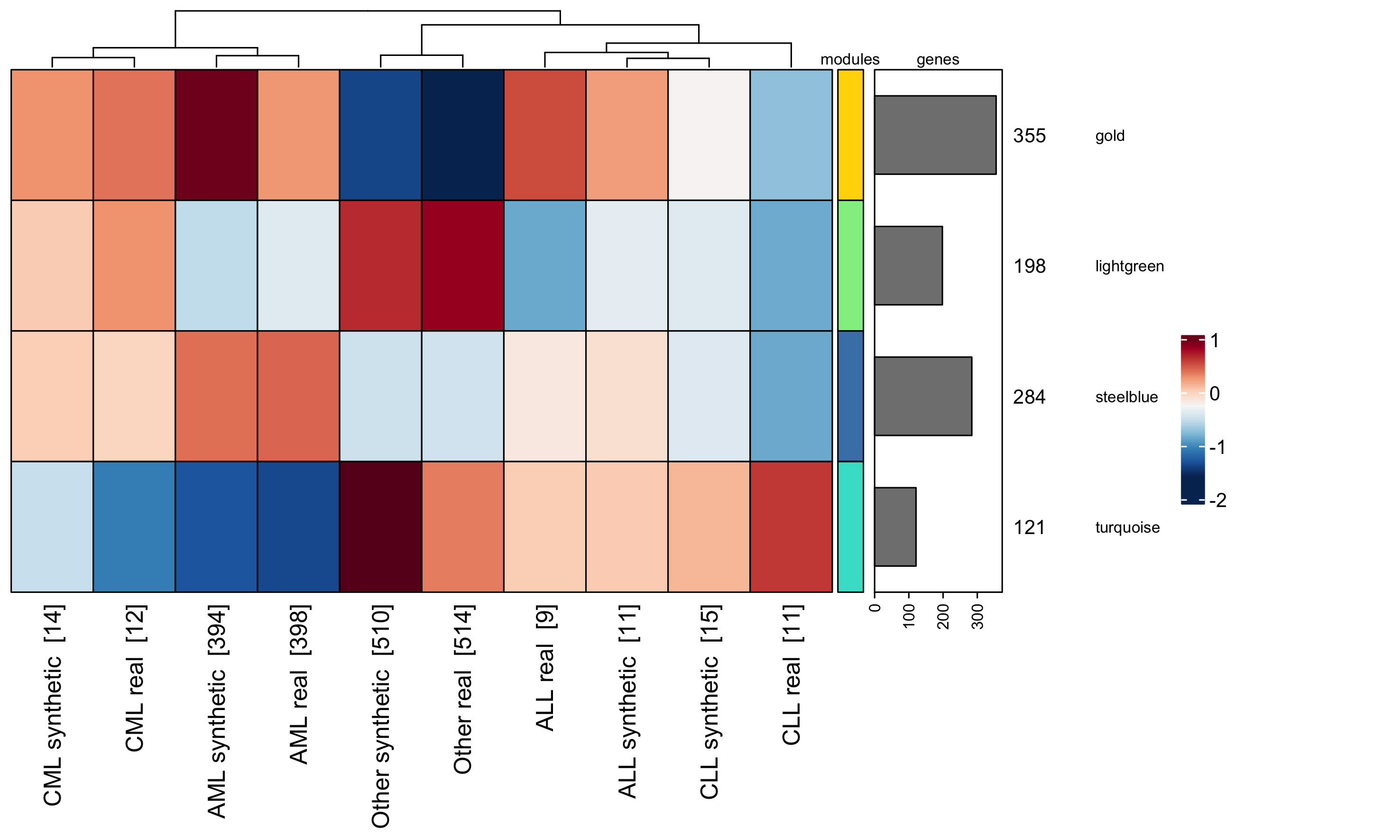}
\end{minipage}
}
\subfigure[non-priv, seed 1]{
\begin{minipage}[b]{\figwidth} \includegraphics[width=1.0\textwidth,trim={0 0 12cm 0},clip]{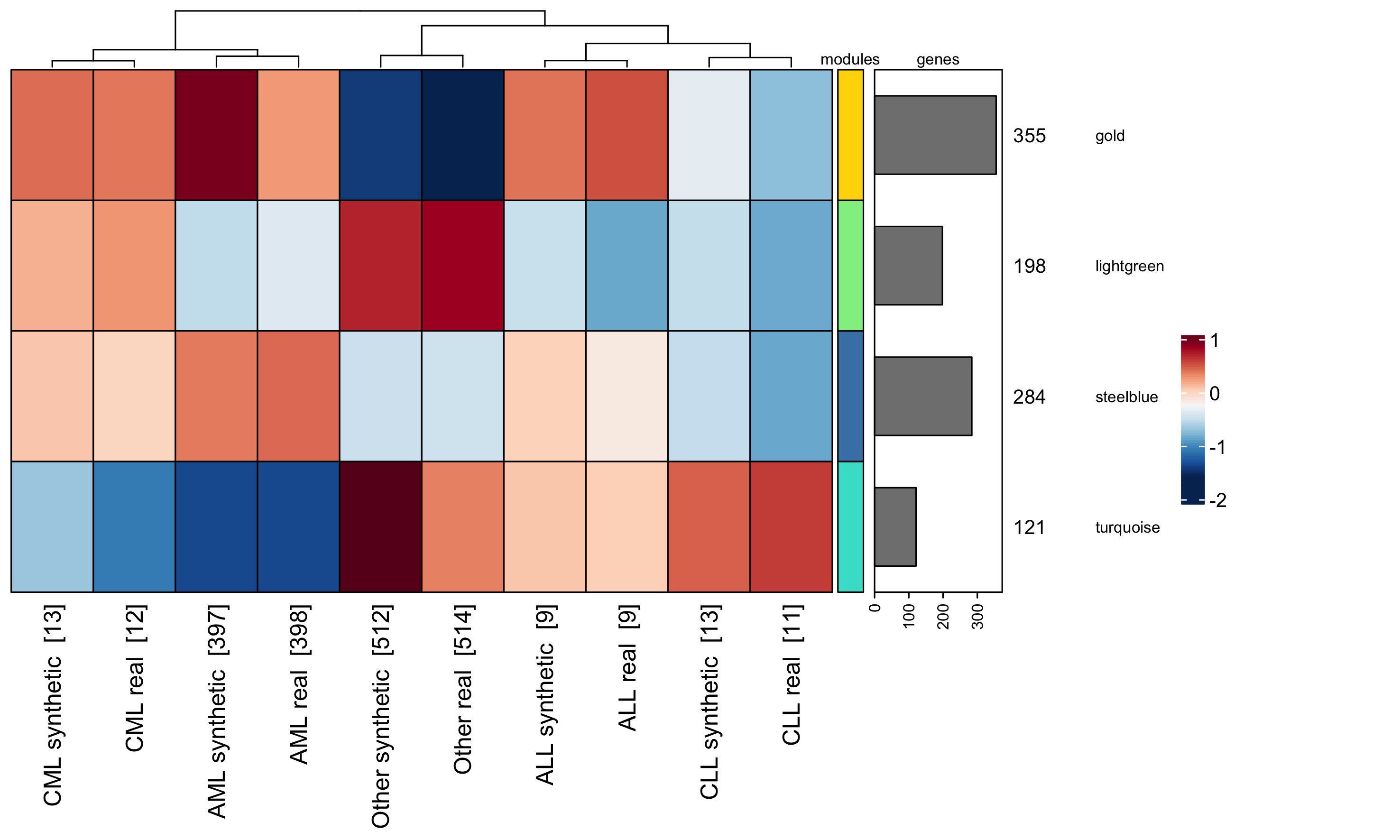}
\end{minipage}
}
\subfigure[$\varepsilon=5$, seed 2]{
\begin{minipage}[b]{\figwidth} \includegraphics[width=1.0\textwidth,trim={0 0 12cm 0},clip]{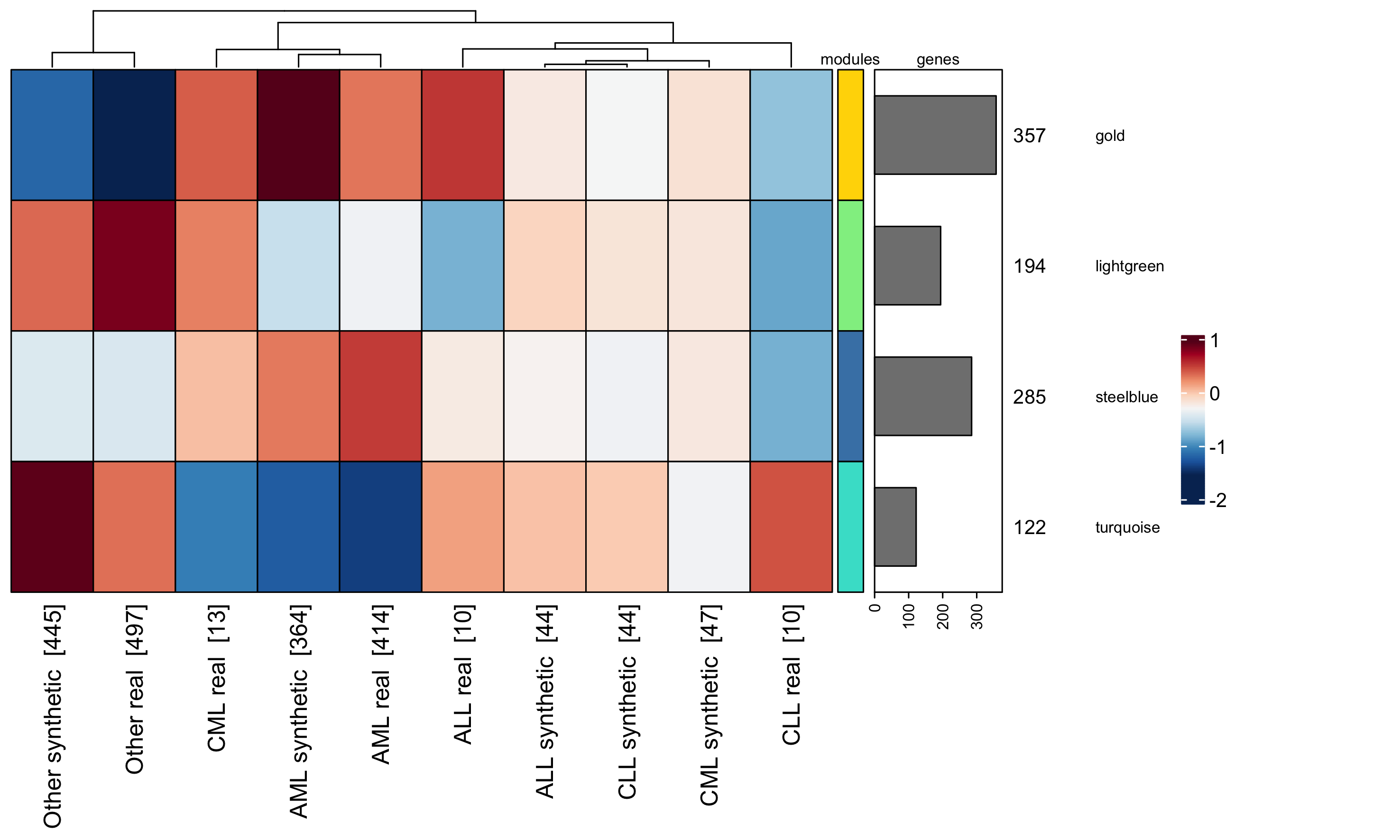}
\end{minipage}
}
\subfigure[$\varepsilon=10$, seed 2]{
\begin{minipage}[b]{\figwidth} \includegraphics[width=1.0\textwidth,trim={0 0 12cm 0},clip]{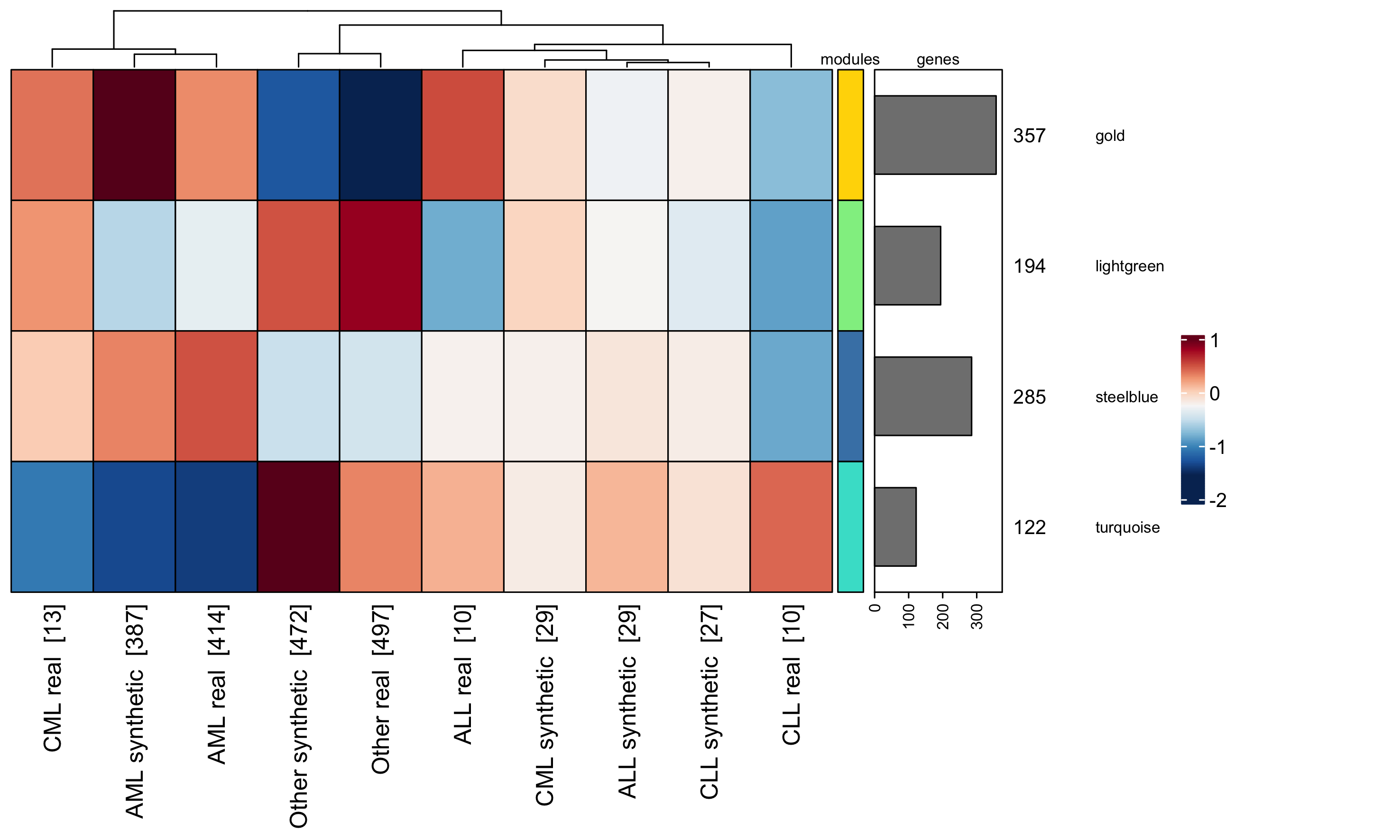}
\end{minipage}
}
\subfigure[$\varepsilon=20$, seed 2]{
\begin{minipage}[b]{\figwidth} \includegraphics[width=1.0\textwidth,trim={0 0 12cm 0},clip]{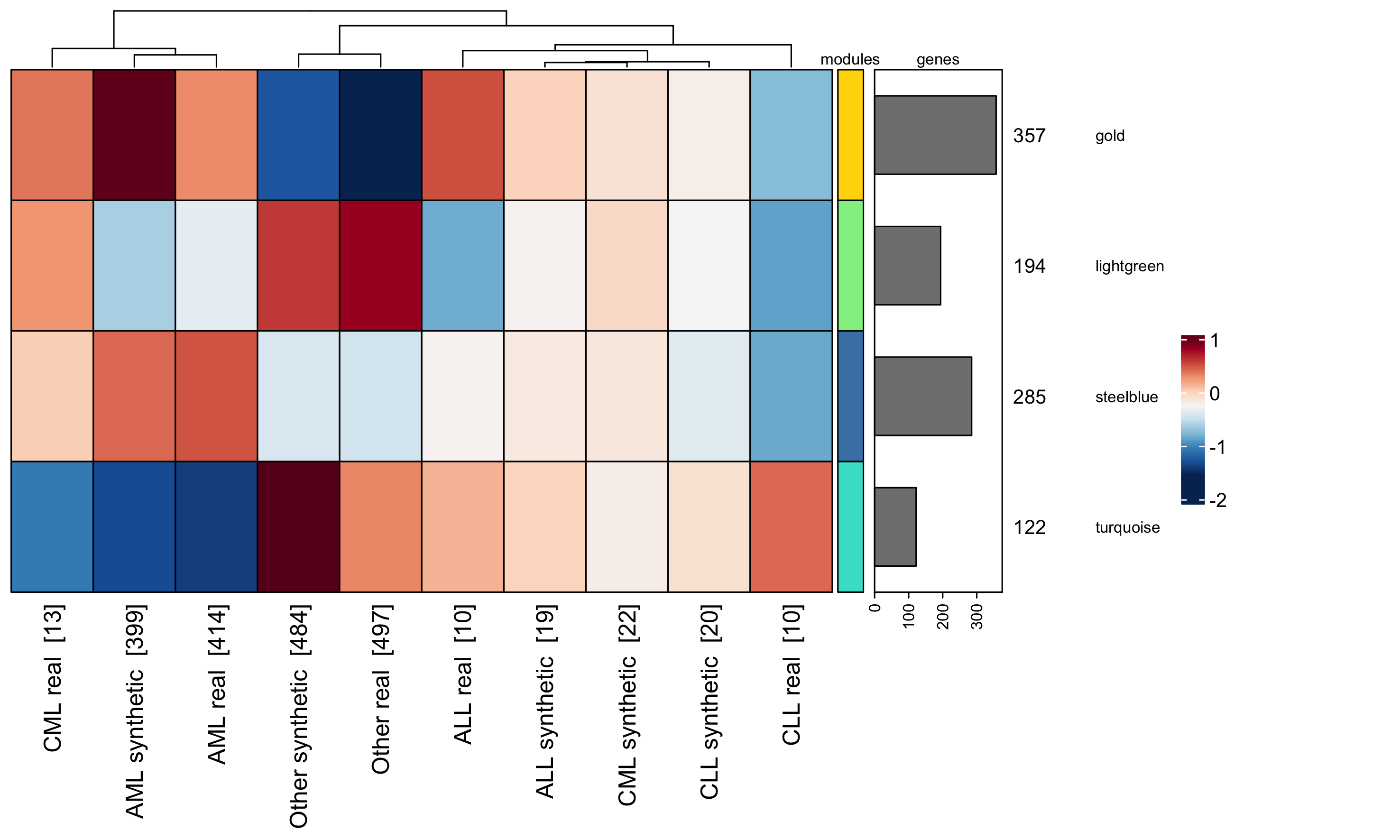}
\end{minipage}
}
\subfigure[$\varepsilon=50$, seed 2]{
\begin{minipage}[b]{\figwidth} \includegraphics[width=1.0\textwidth,trim={0 0 12cm 0},clip]{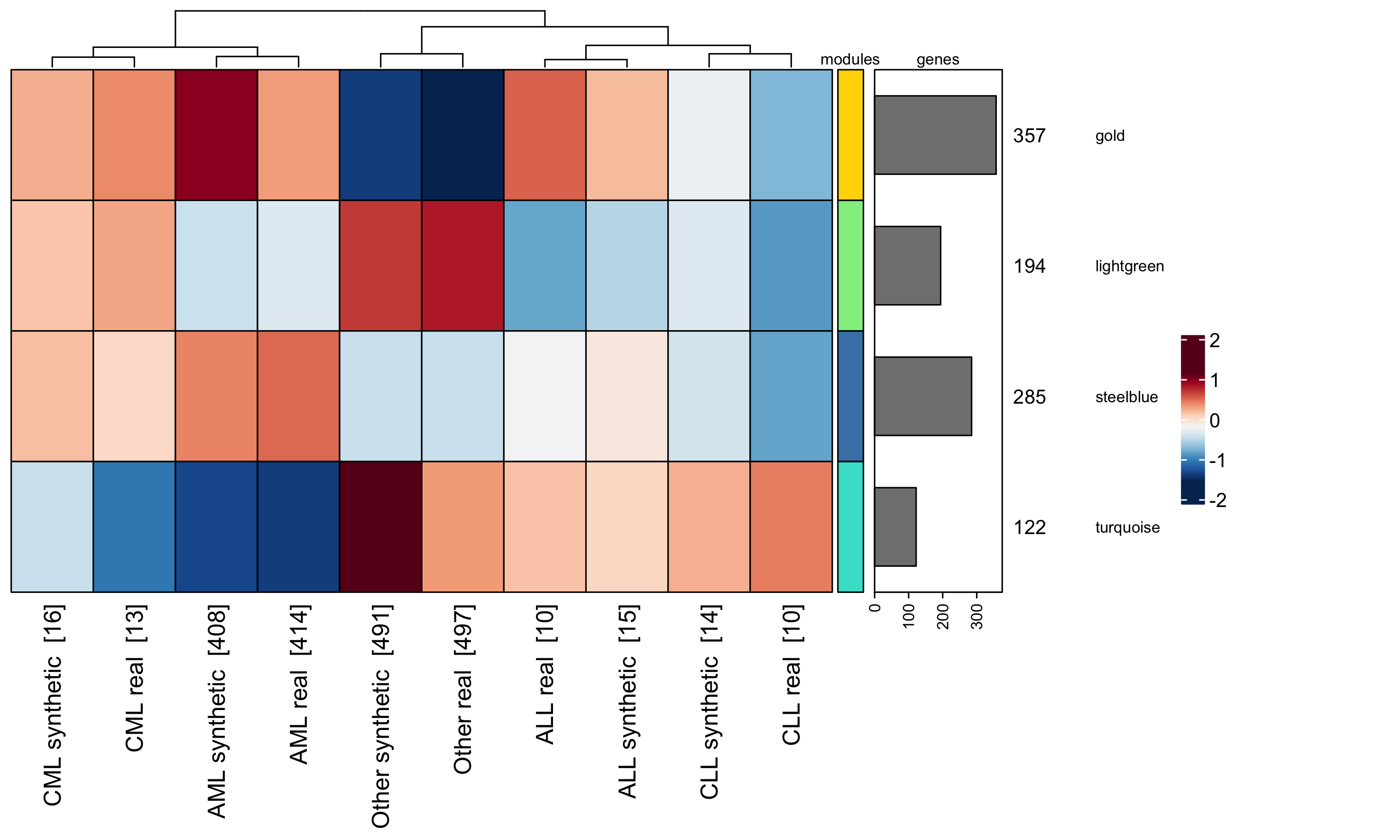}
\end{minipage}
}
\subfigure[$\varepsilon=100$, seed 2]{
\begin{minipage}[b]{\figwidth} \includegraphics[width=1.0\textwidth,trim={0 0 12cm 0},clip]{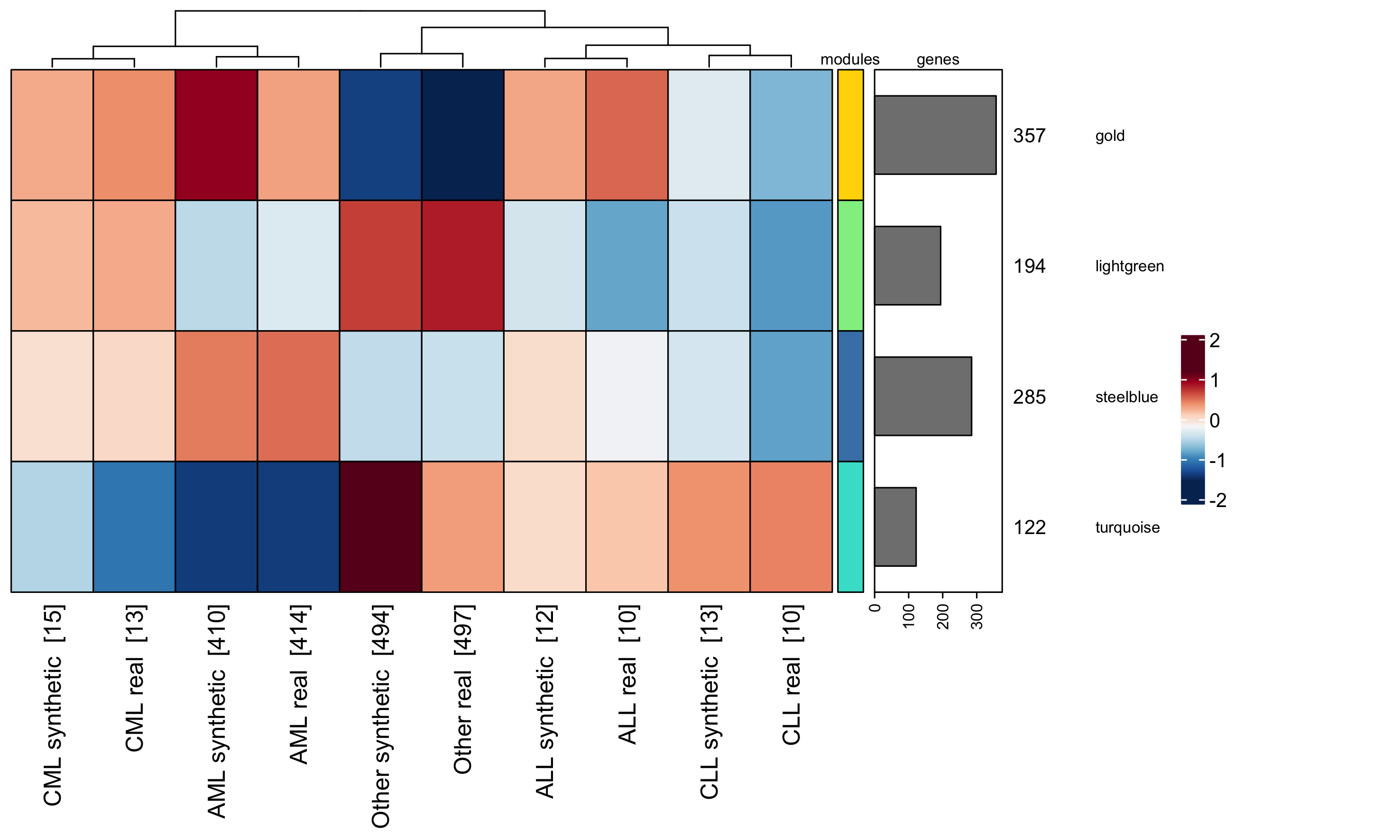}
\end{minipage}
}
\subfigure[non-priv, seed 2]{
\begin{minipage}[b]{\figwidth} \includegraphics[width=1.0\textwidth,trim={0 0 12cm 0},clip]{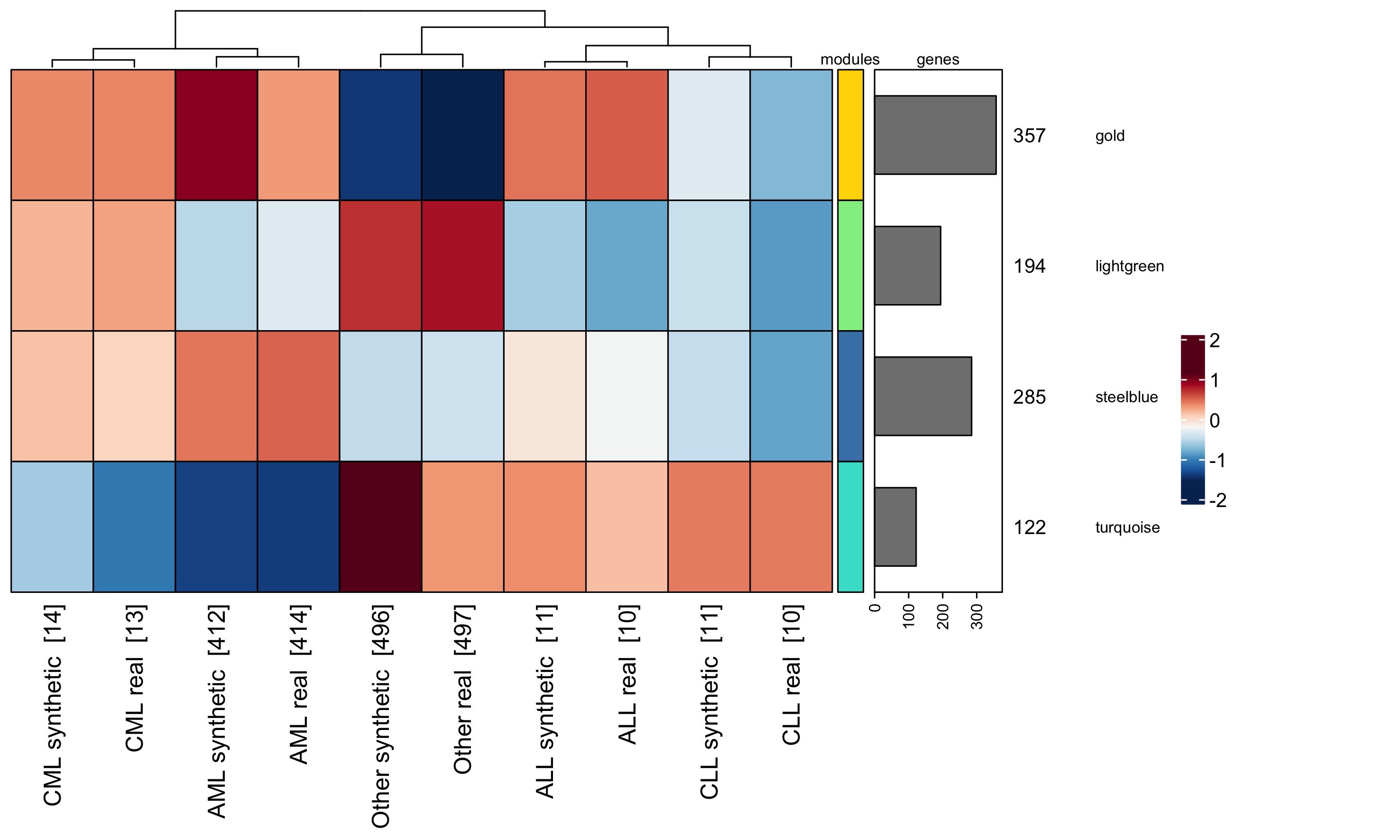}
\end{minipage}
}
\caption{Activation patterns of co-expressed gene modules in PGM after filtering co-expressions for $r$ > 0. \textmd{Shown are the Group Fold Changes (GFCs) of gene modules (rows) in the real and the synthetic data sampled with two different seeds. Numbers on the right indicate the number of genes per module. Darker shades of red imply activation of the gene module, while darker shades of blue indicate deactivation. The dendrograms show the hierarchical clustering of the classes in the different data sets. A heatmap is shown for each $\varepsilon$ twice, once for each seed used to \emph{split} the training data. Each heatmap further features, in addition to the real data, data from two synthetic sets, one for each seed used to \textit{generate} the data. The synthetic data exhibits a visible loss of module activation patterns for $\varepsilon \leq 20$.}}
\label{figure:S3}
\end{figure*}

\begin{figure*}[!htbp]
\centering
\newcommand{\figwidth}{0.32\textwidth}
\subfigure[$\varepsilon=5$, seed 1]{
\begin{minipage}[b]{\figwidth} \includegraphics[width=1.0\textwidth,trim={0 0 12cm 0},clip]{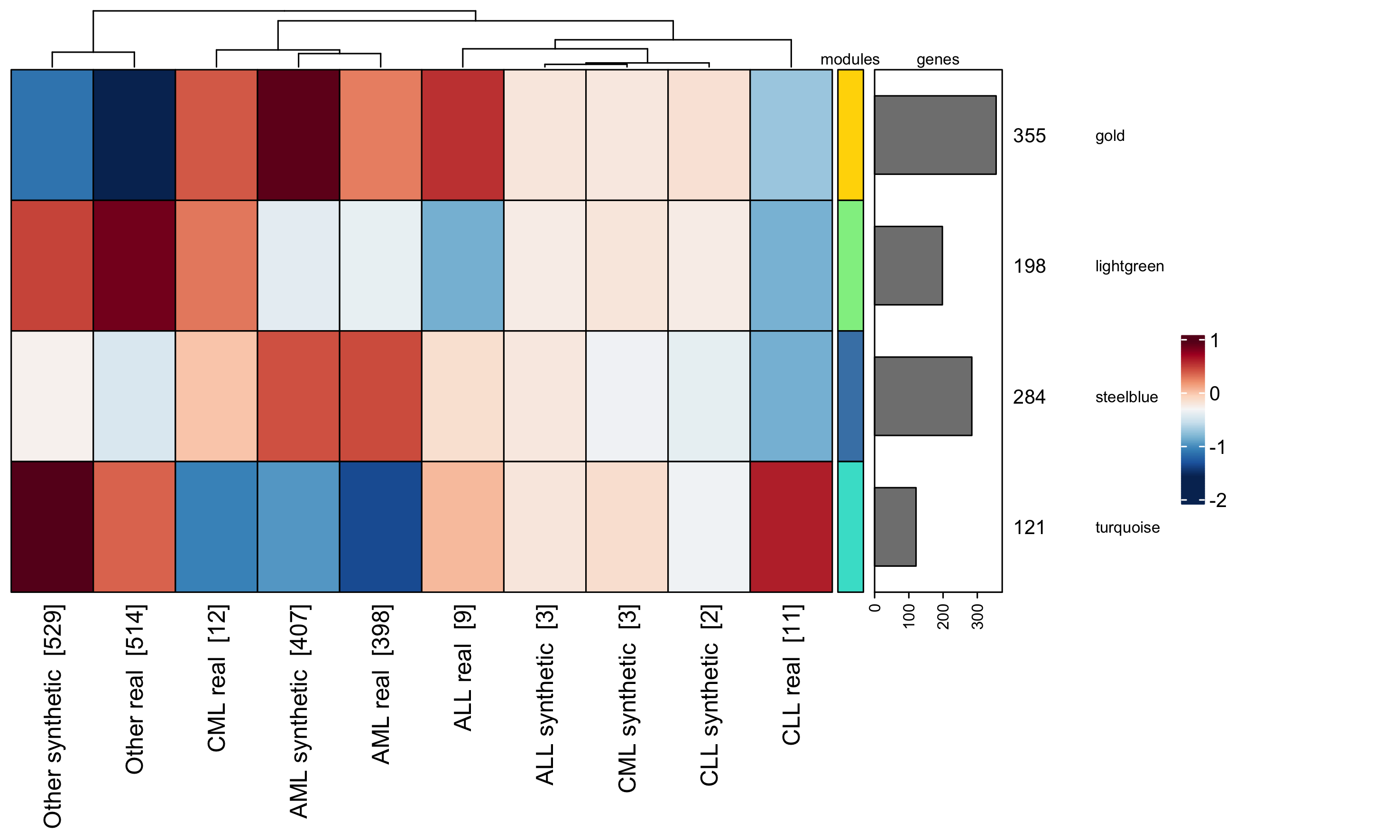}
\end{minipage}
}
\subfigure[$\varepsilon=10$, seed 1]{
\begin{minipage}[b]{\figwidth} \includegraphics[width=1.0\textwidth,trim={0 0 12cm 0},clip]{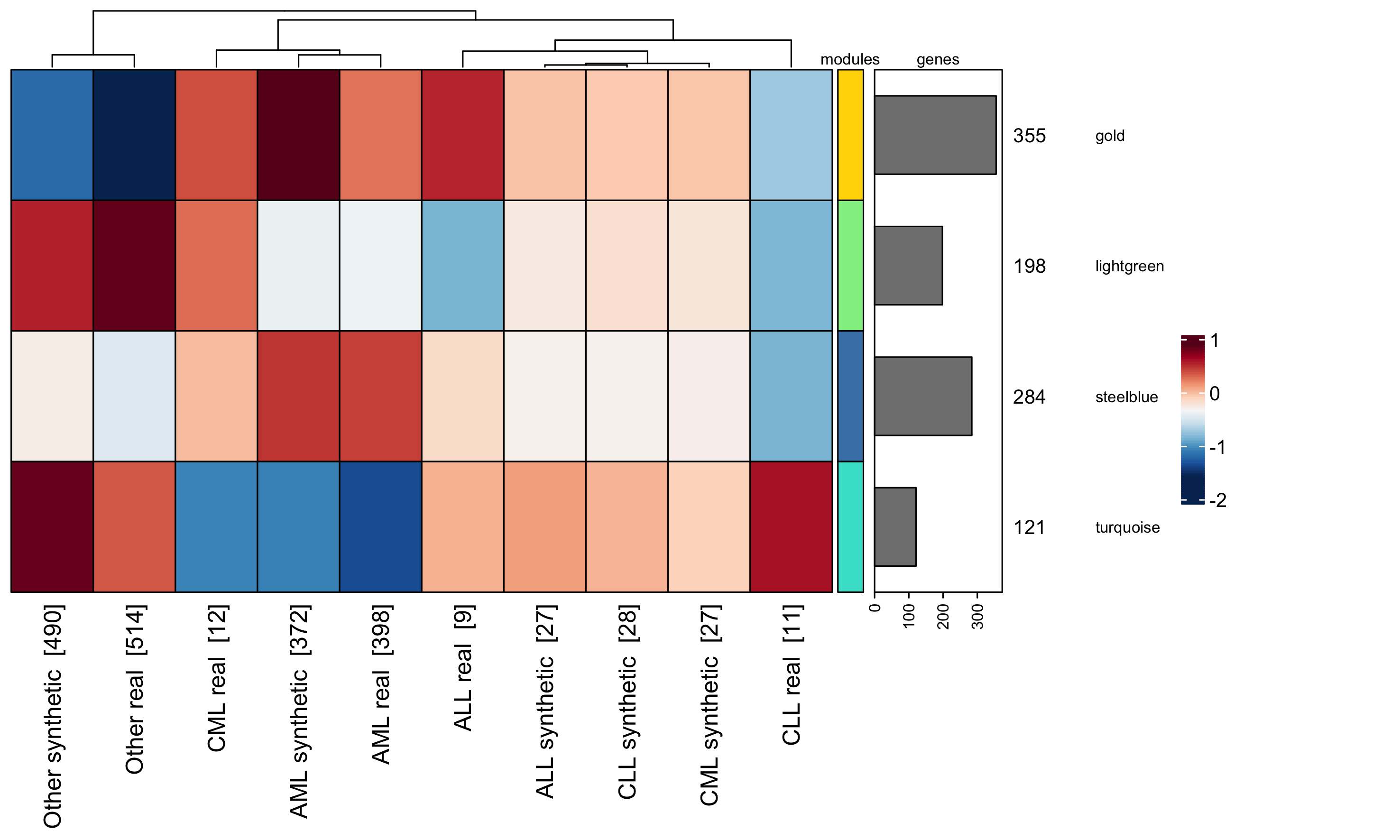}
\end{minipage}
}
\subfigure[$\varepsilon=20$, seed 1]{
\begin{minipage}[b]{\figwidth} \includegraphics[width=1.0\textwidth,trim={0 0 12cm 0},clip]{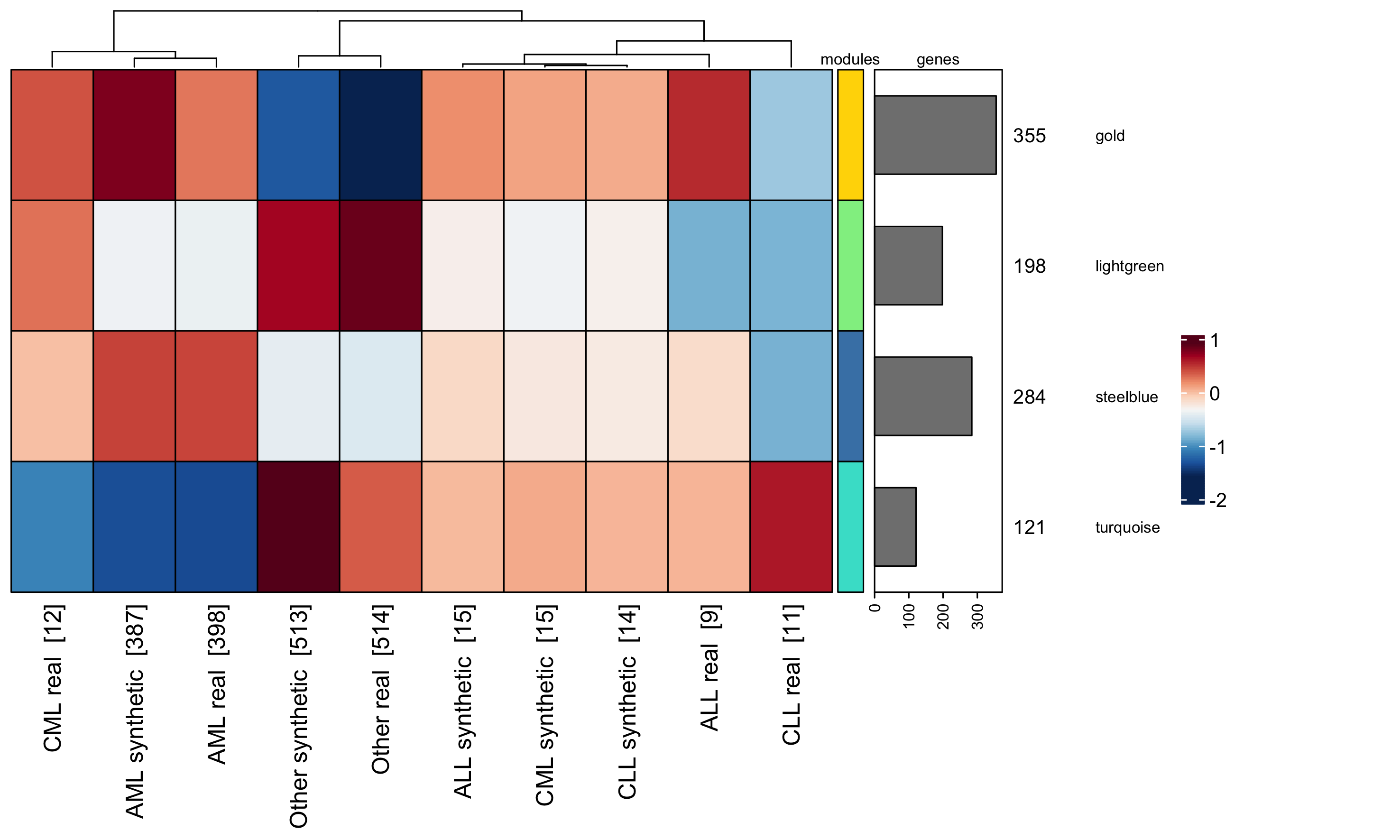}
\end{minipage}
}
\subfigure[$\varepsilon=50$, seed 1]{
\begin{minipage}[b]{\figwidth} \includegraphics[width=1.0\textwidth,trim={0 0 12cm 0},clip]{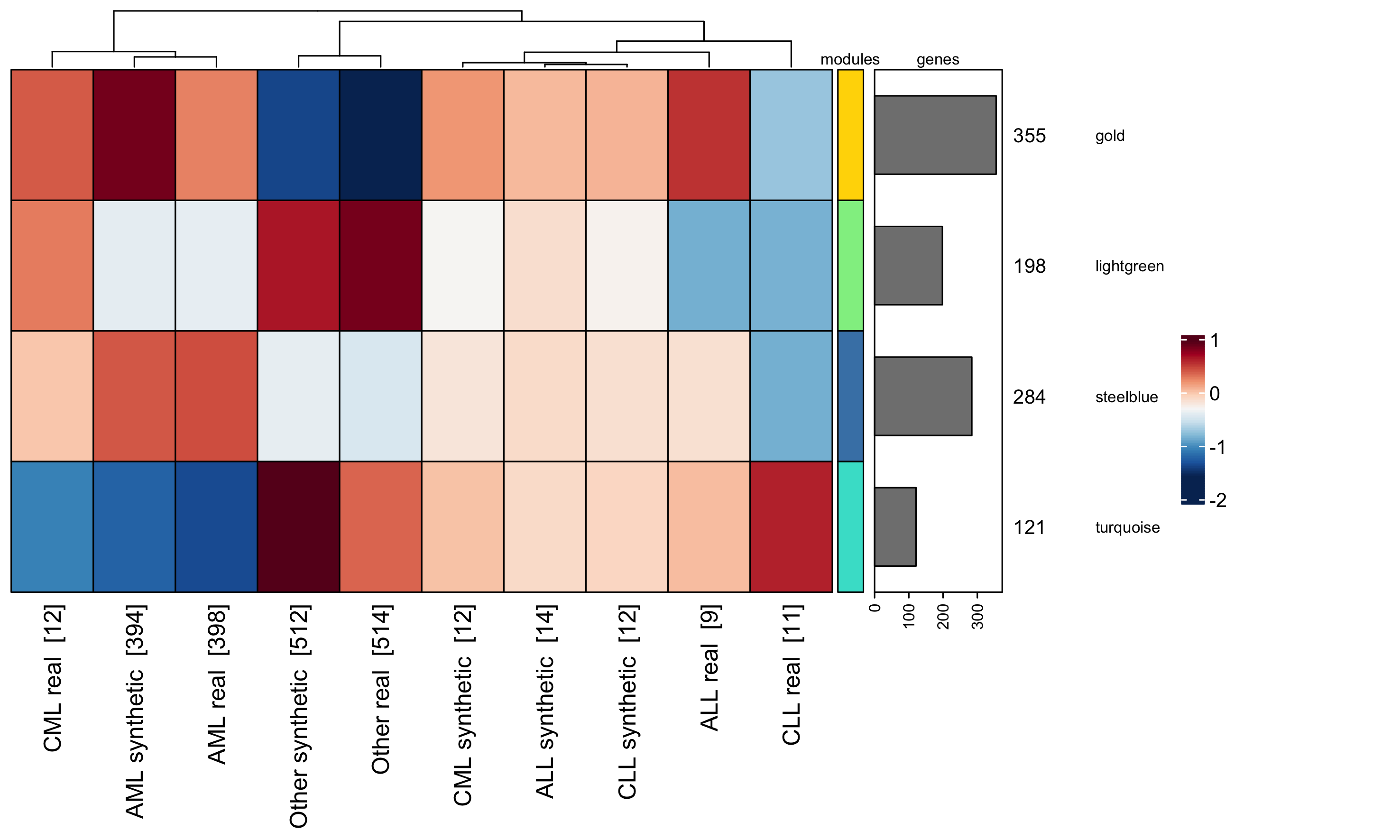}
\end{minipage}
}
\subfigure[$\varepsilon=100$, seed 1]{
\begin{minipage}[b]{\figwidth} \includegraphics[width=1.0\textwidth,trim={0 0 12cm 0},clip]{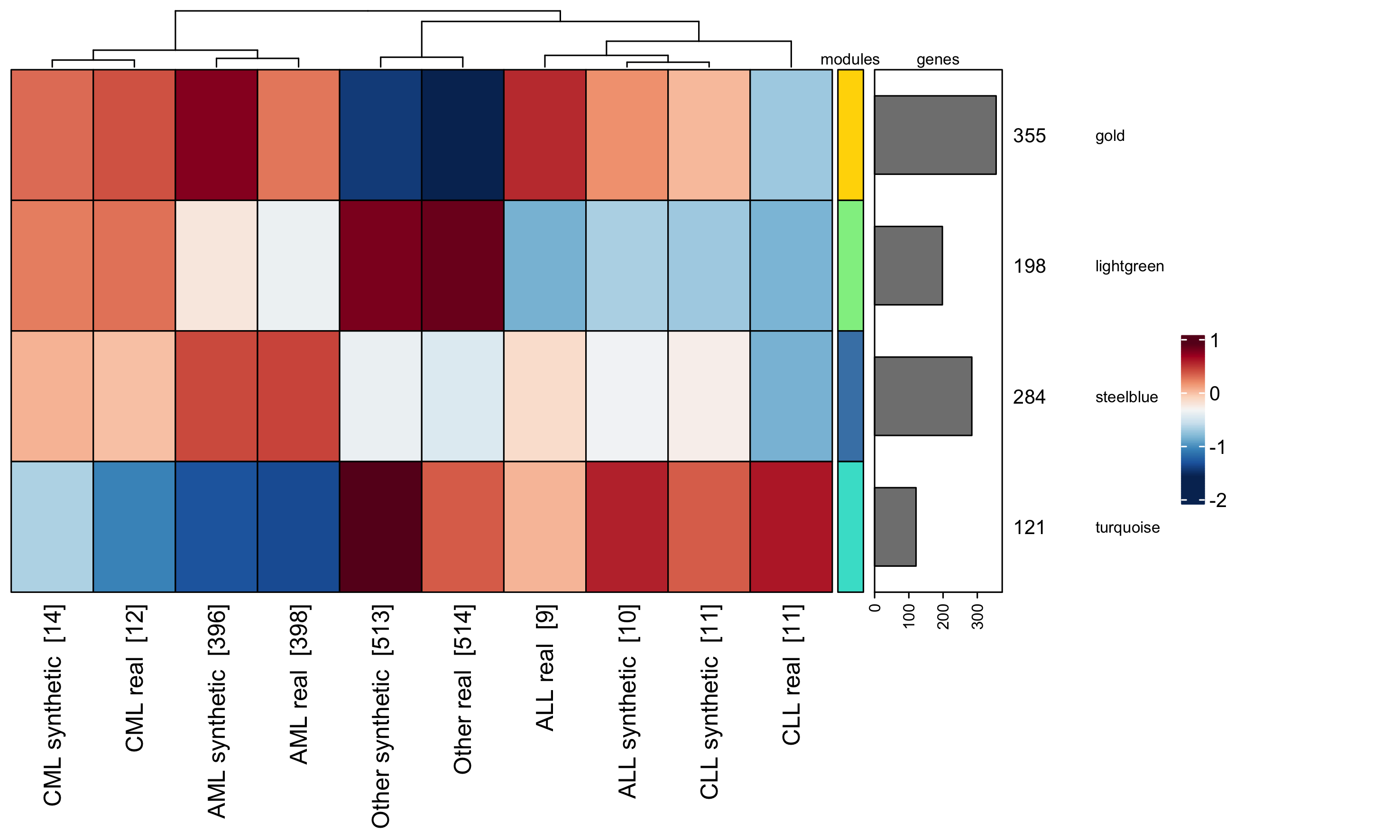}
\end{minipage}
}
\subfigure[non-priv, seed 1]{
\begin{minipage}[b]{\figwidth} \includegraphics[width=1.0\textwidth,trim={0 0 12cm 0},clip]{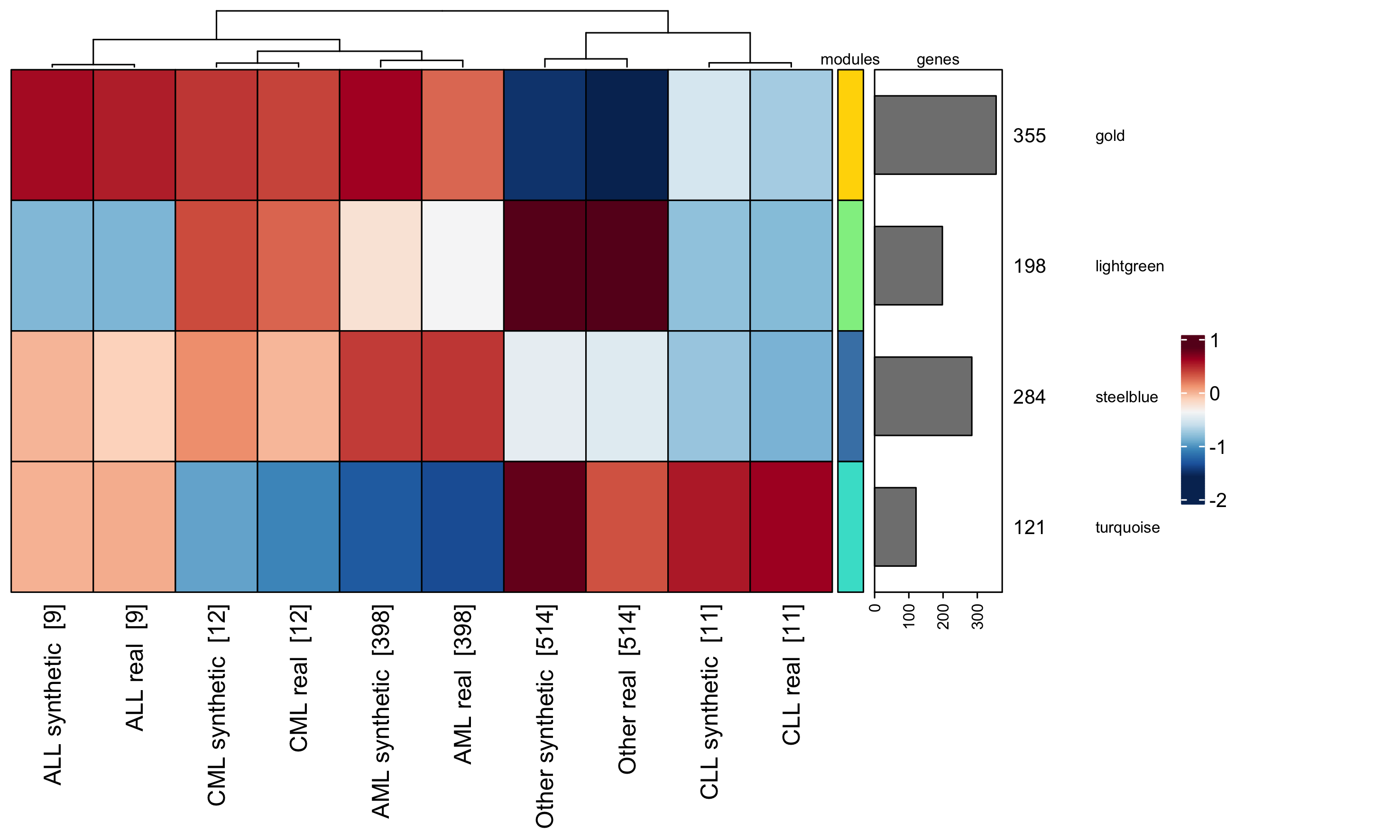}
\end{minipage}
}
\subfigure[$\varepsilon=5$, seed 2]{
\begin{minipage}[b]{\figwidth} \includegraphics[width=1.0\textwidth,trim={0 0 12cm 0},clip]{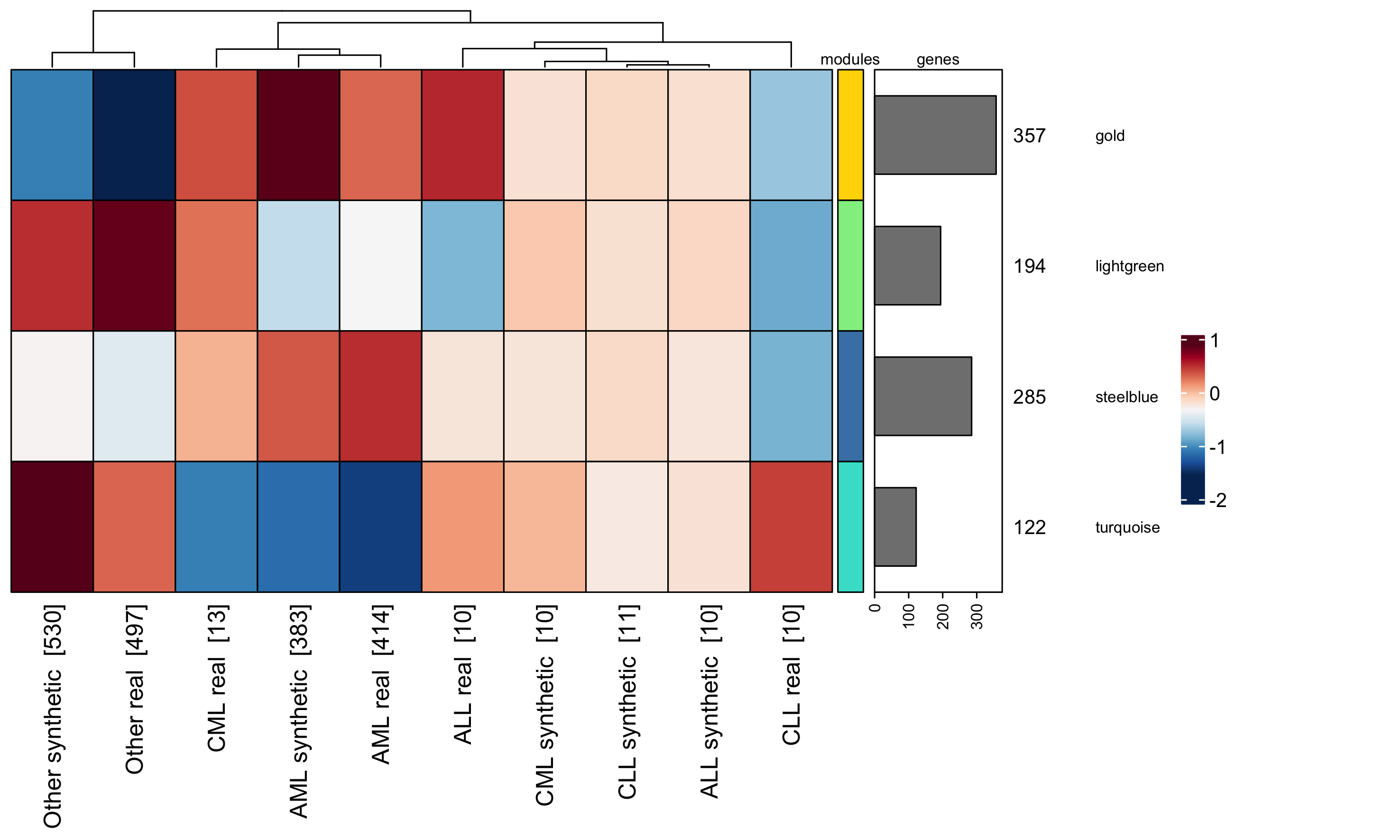}
\end{minipage}
}
\subfigure[$\varepsilon=10$, seed 2]{
\begin{minipage}[b]{\figwidth} \includegraphics[width=1.0\textwidth,trim={0 0 12cm 0},clip]{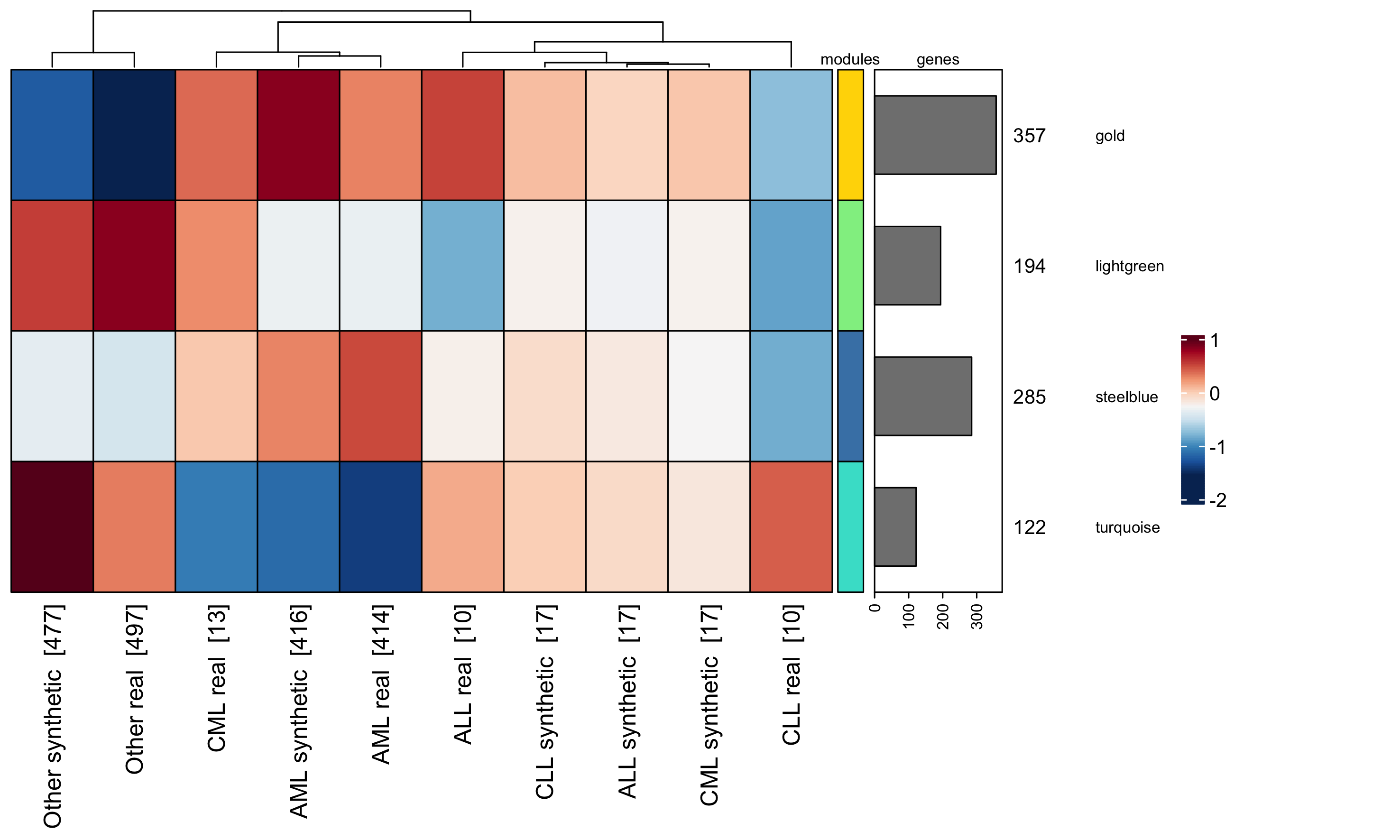}
\end{minipage}
}
\subfigure[$\varepsilon=20$, seed 2]{
\begin{minipage}[b]{\figwidth} \includegraphics[width=1.0\textwidth,trim={0 0 12cm 0},clip]{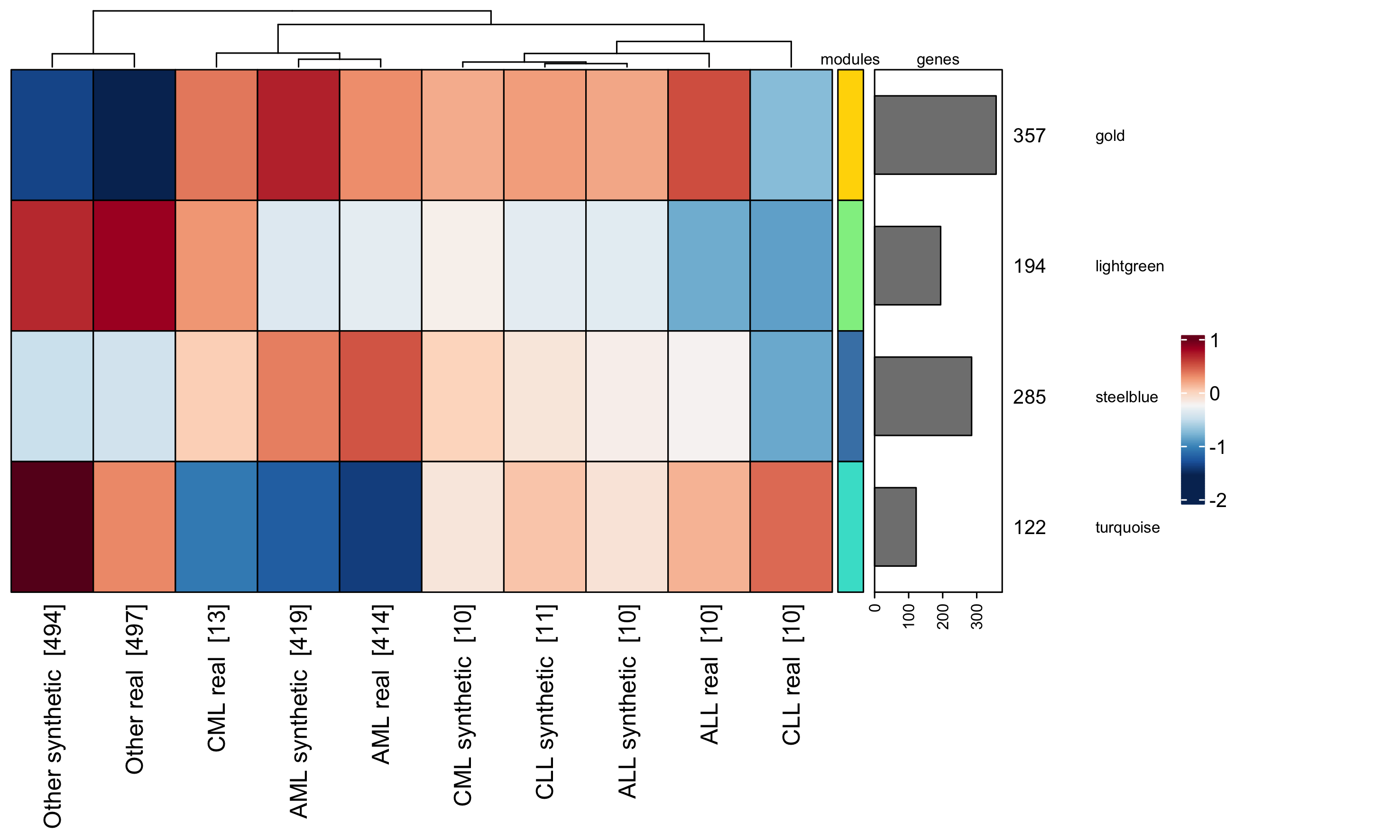}
\end{minipage}
}
\subfigure[$\varepsilon=50$, seed 2]{
\begin{minipage}[b]{\figwidth} \includegraphics[width=1.0\textwidth,trim={0 0 12cm 0},clip]{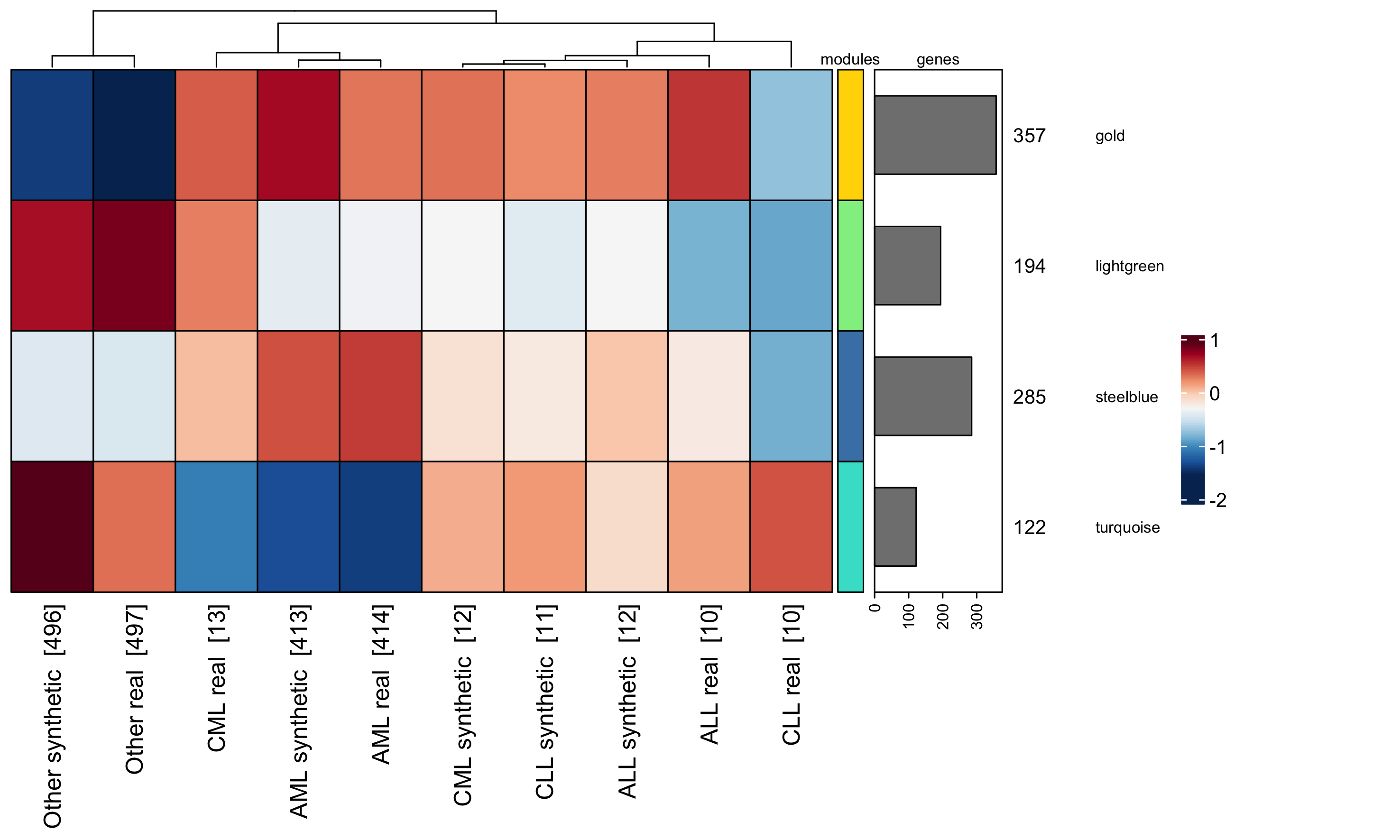}
\end{minipage}
}
\subfigure[$\varepsilon=100$, seed 2]{
\begin{minipage}[b]{\figwidth} \includegraphics[width=1.0\textwidth,trim={0 0 12cm 0},clip]{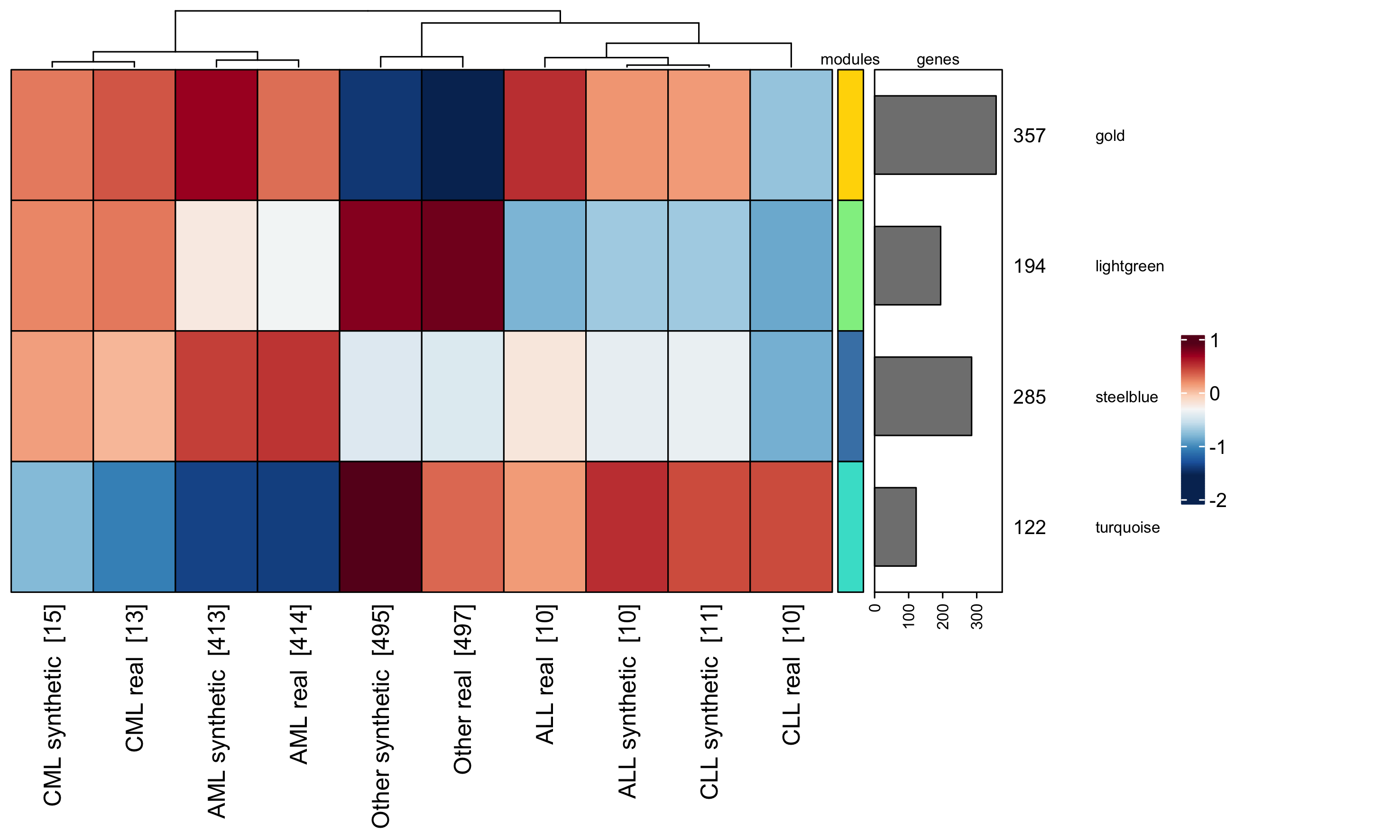}
\end{minipage}
}
\subfigure[non-priv, seed 2]{
\begin{minipage}[b]{\figwidth} \includegraphics[width=1.0\textwidth,trim={0 0 12cm 0},clip]{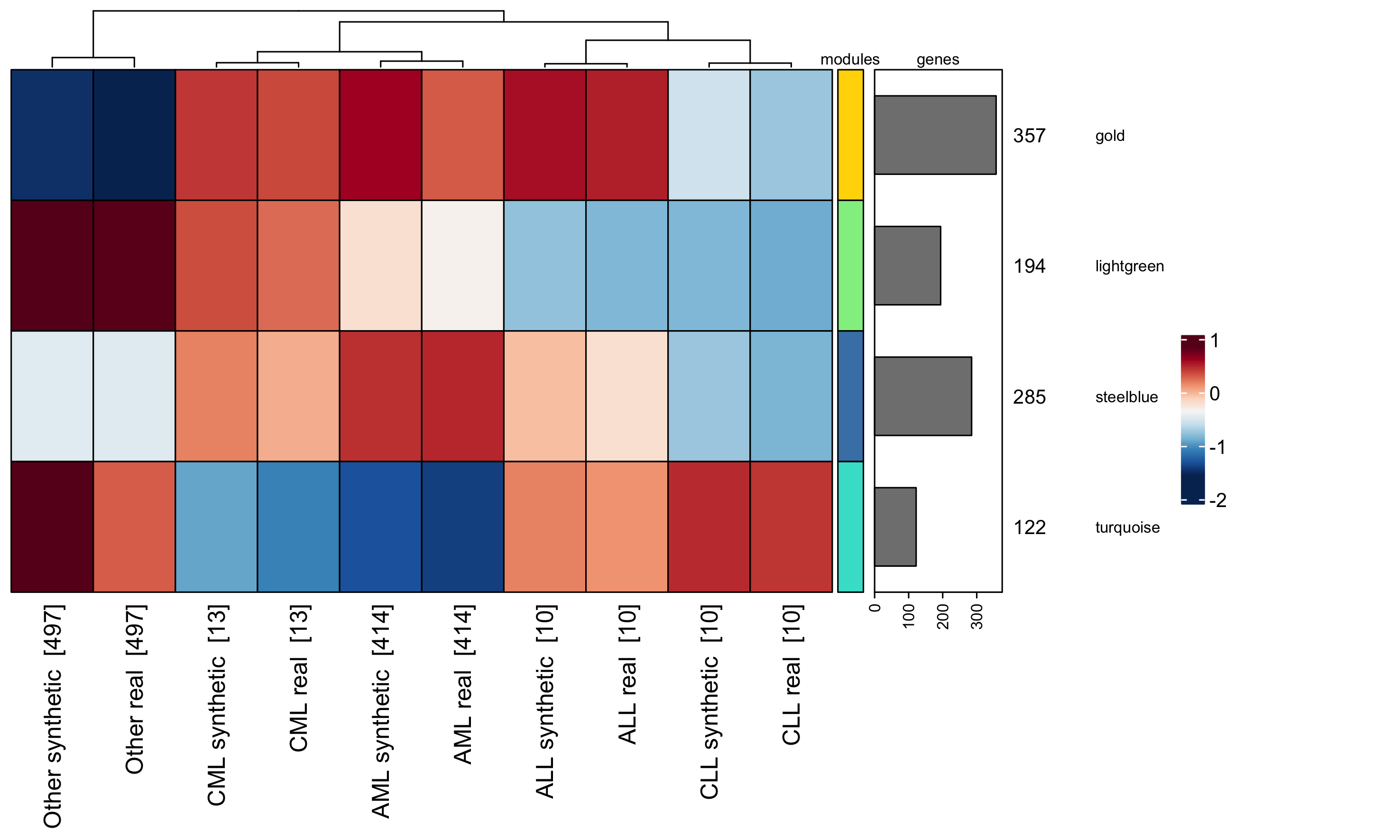}
\end{minipage}
}
\caption{Activation patterns of co-expressed gene modules in PrivSyn after filtering co-expressions for $r$ > 0. \textmd{Shown are the Group Fold Changes (GFCs) of gene modules (rows) in the real and the synthetic data sampled with two different seeds. Numbers on the right indicate the number of genes per module. Darker shades of red imply activation of the gene module, while darker shades of blue indicate deactivation. The dendrograms show the hierarchical clustering of the classes in the different data sets. A heatmap is shown for each $\varepsilon$ twice, once for each seed used to \emph{split} the training data. Each heatmap further features, in addition to the real data, data from two synthetic sets, one for each seed used to \textit{generate} the data. A loss of module activation patterns can be observed for all shown privacy budgets, becoming increasingly prominent with decreasing $\varepsilon$.}}
\label{figure:S4}
\end{figure*}

\begin{figure*}[!htbp]
\centering
\newcommand{\figwidth}{0.32\textwidth}
\subfigure[$\varepsilon=5$, seed 1]{
\begin{minipage}[b]{\figwidth} \includegraphics[width=1.0\textwidth,trim={0 0 12cm 0},clip]{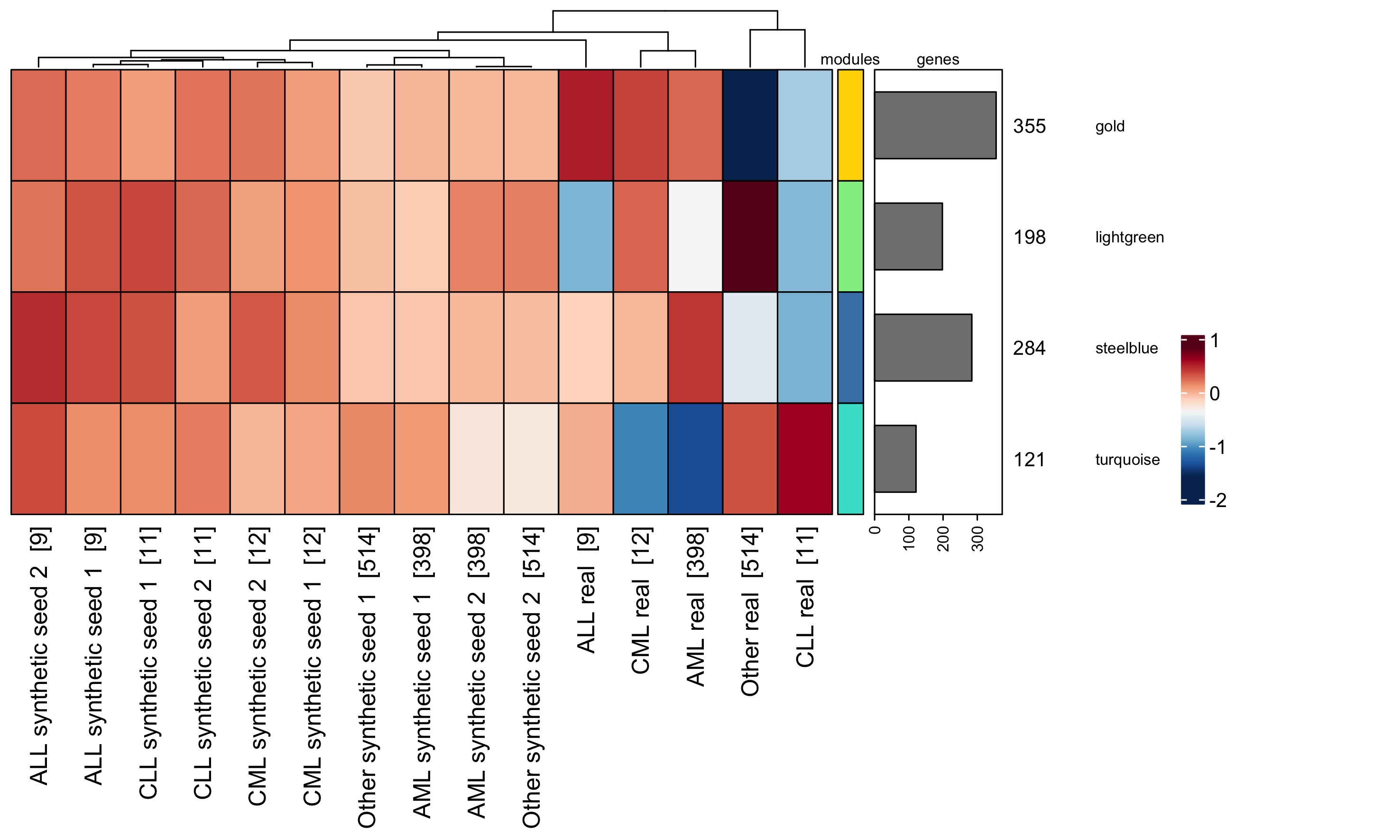}
\end{minipage}
}
\subfigure[$\varepsilon=10$, seed 1]{
\begin{minipage}[b]{\figwidth} \includegraphics[width=1.0\textwidth,trim={0 0 12cm 0},clip]{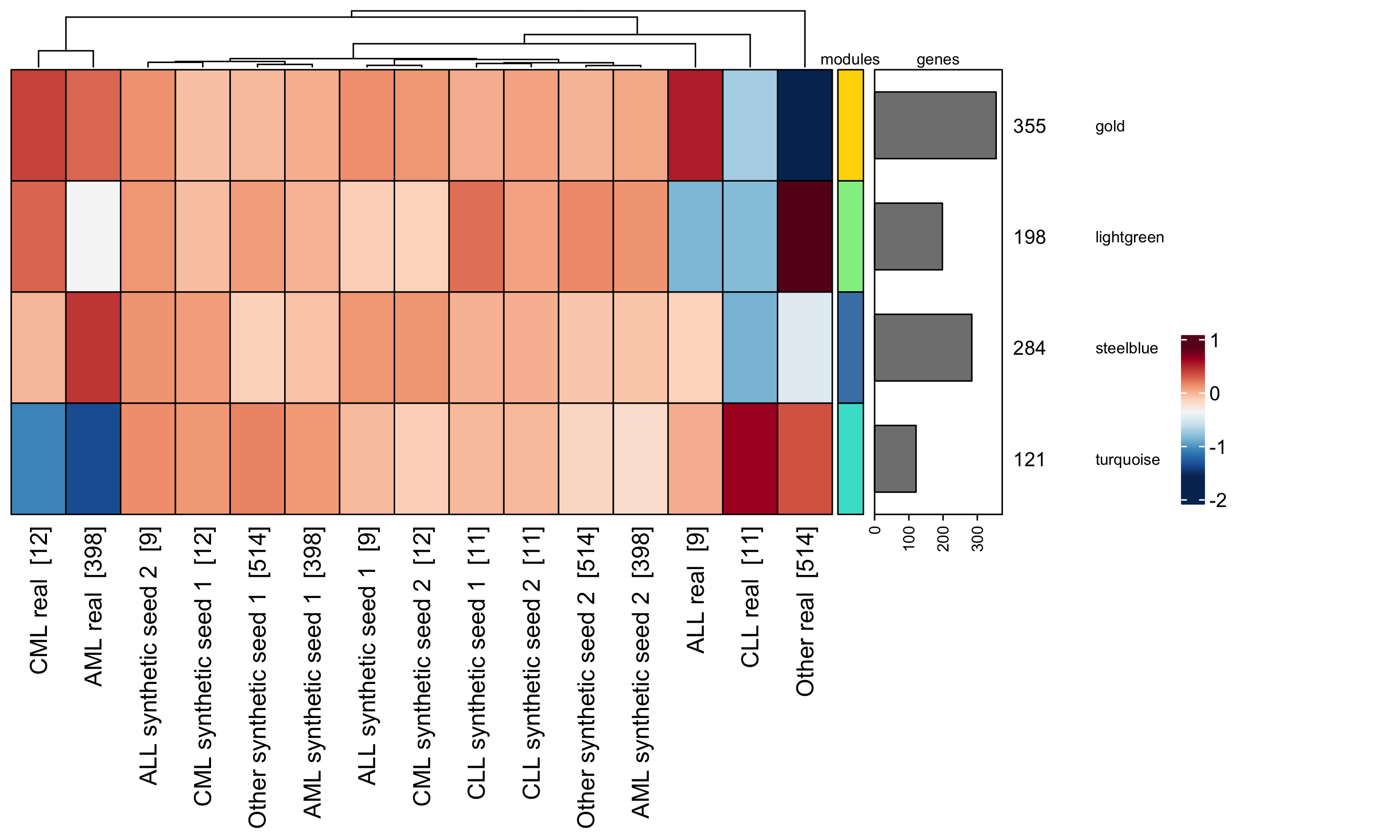}
\end{minipage}
}
\subfigure[$\varepsilon=20$, seed 1]{
\begin{minipage}[b]{\figwidth} \includegraphics[width=1.0\textwidth,trim={0 0 12cm 0},clip]{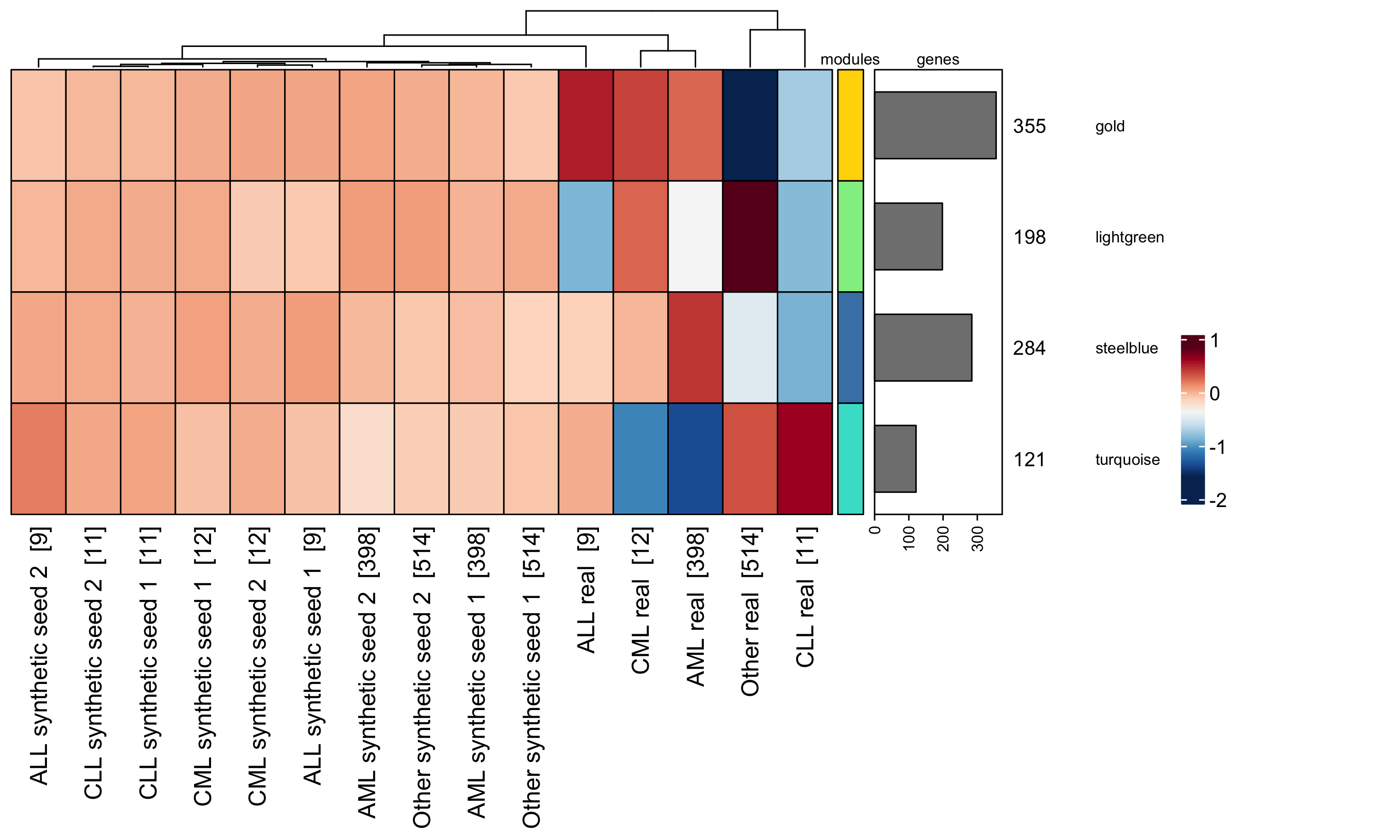}
\end{minipage}
}
\subfigure[$\varepsilon=50$, seed 1]{
\begin{minipage}[b]{\figwidth} \includegraphics[width=1.0\textwidth,trim={0 0 12cm 0},clip]{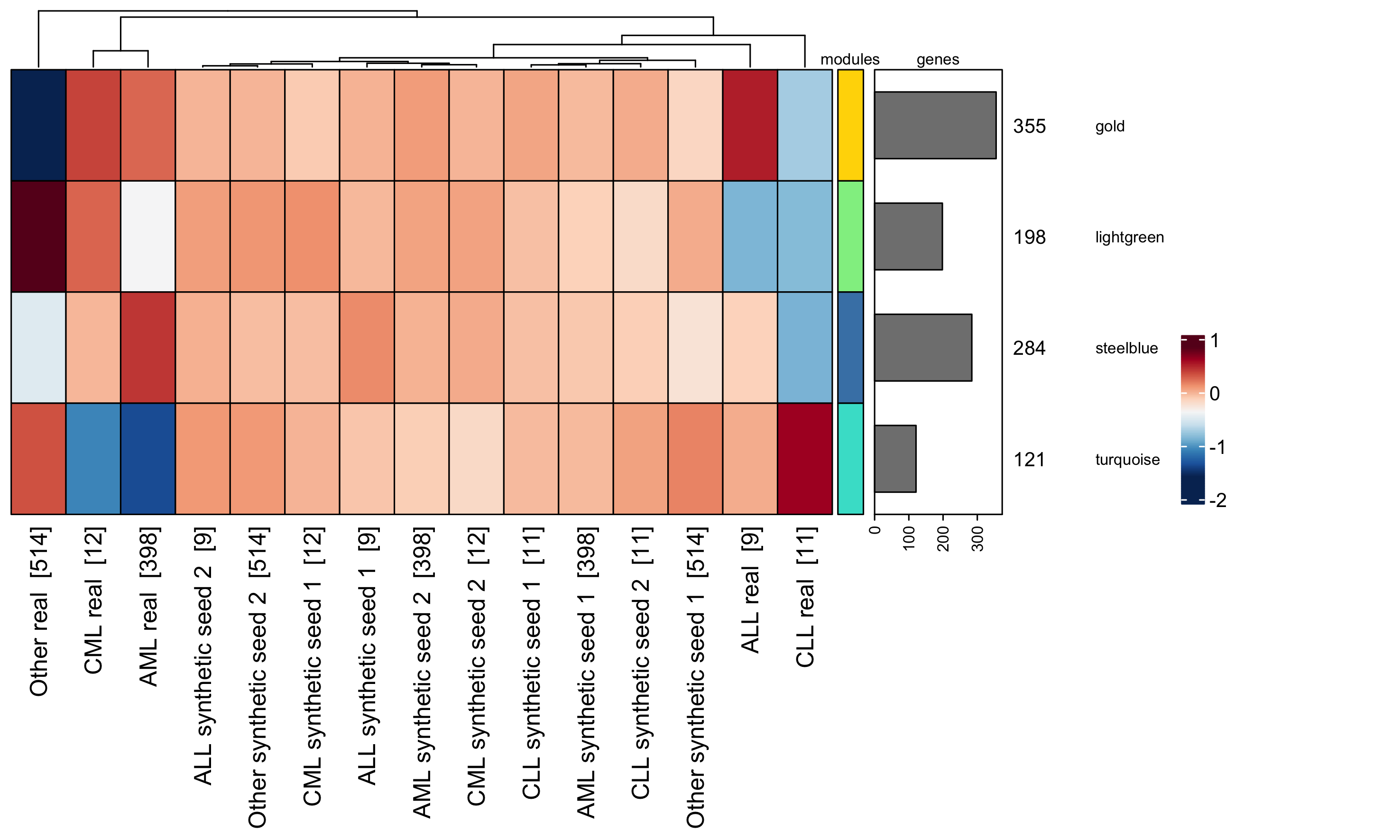}
\end{minipage}
}
\subfigure[$\varepsilon=100$, seed 1]{
\begin{minipage}[b]{\figwidth} \includegraphics[width=1.0\textwidth,trim={0 0 12cm 0},clip]{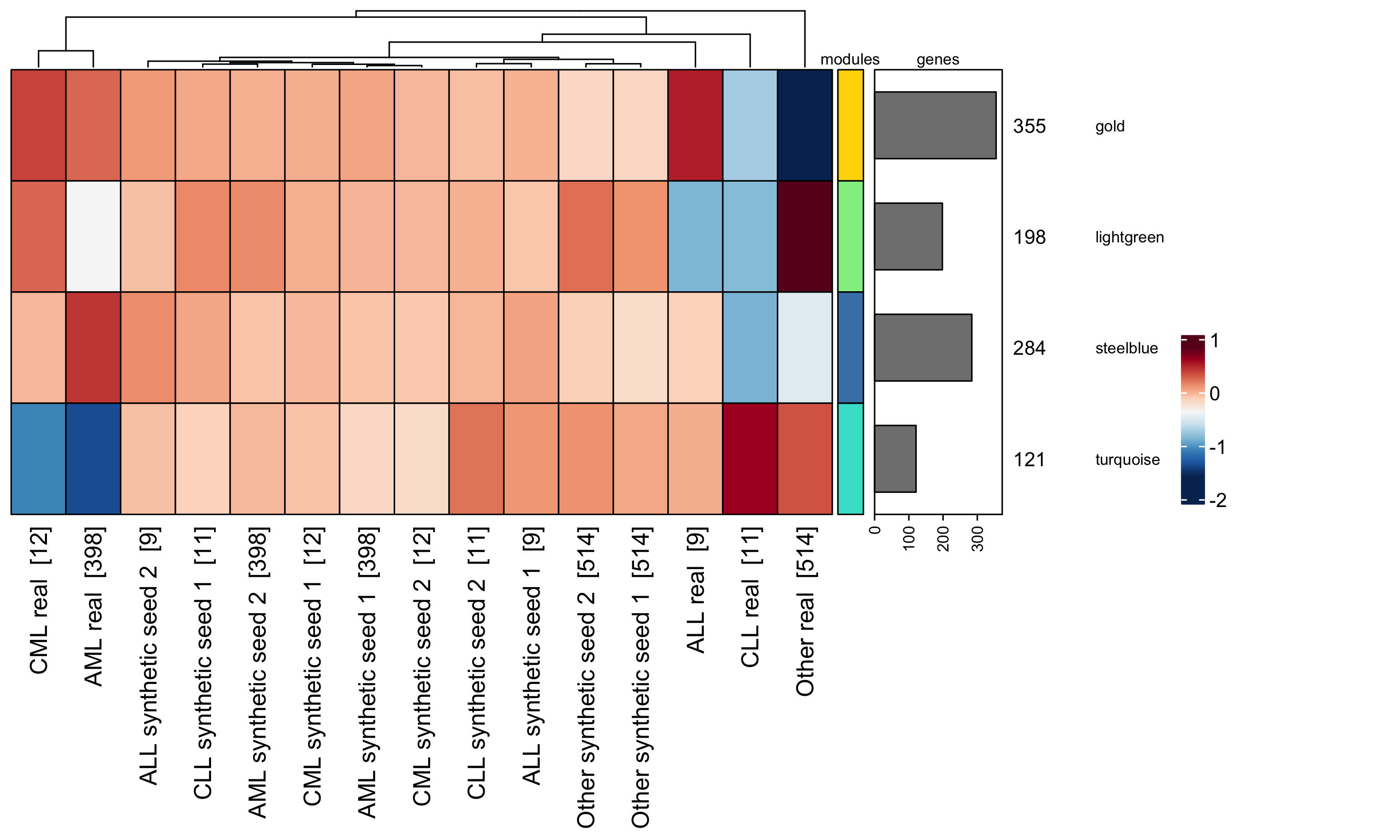}
\end{minipage}
}
\subfigure[non-priv, seed 1]{
\begin{minipage}[b]{\figwidth} \includegraphics[width=1.0\textwidth,trim={0 0 12cm 0},clip]{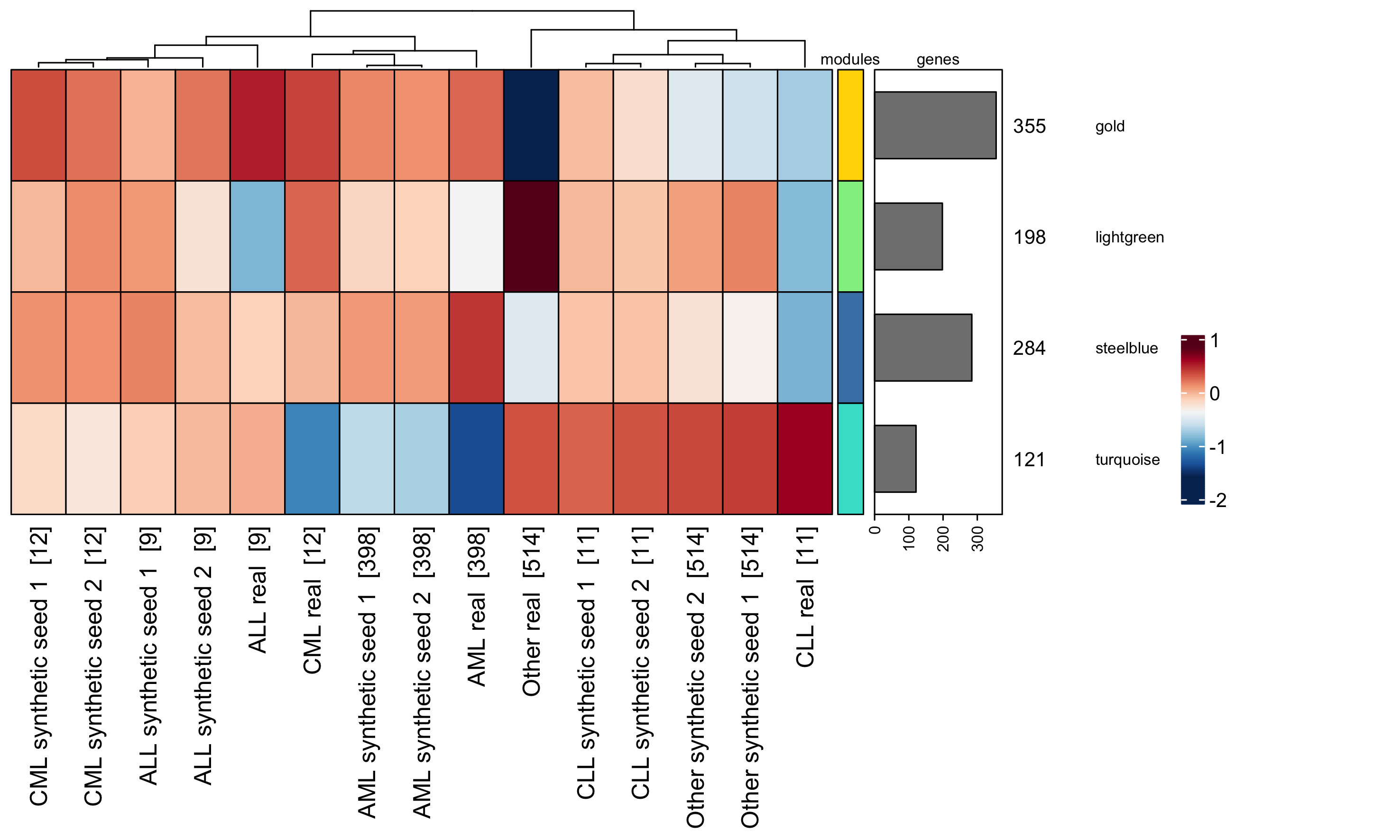}
\end{minipage}
}
\subfigure[$\varepsilon=5$, seed 2]{
\begin{minipage}[b]{\figwidth} \includegraphics[width=1.0\textwidth,trim={0 0 12cm 0},clip]{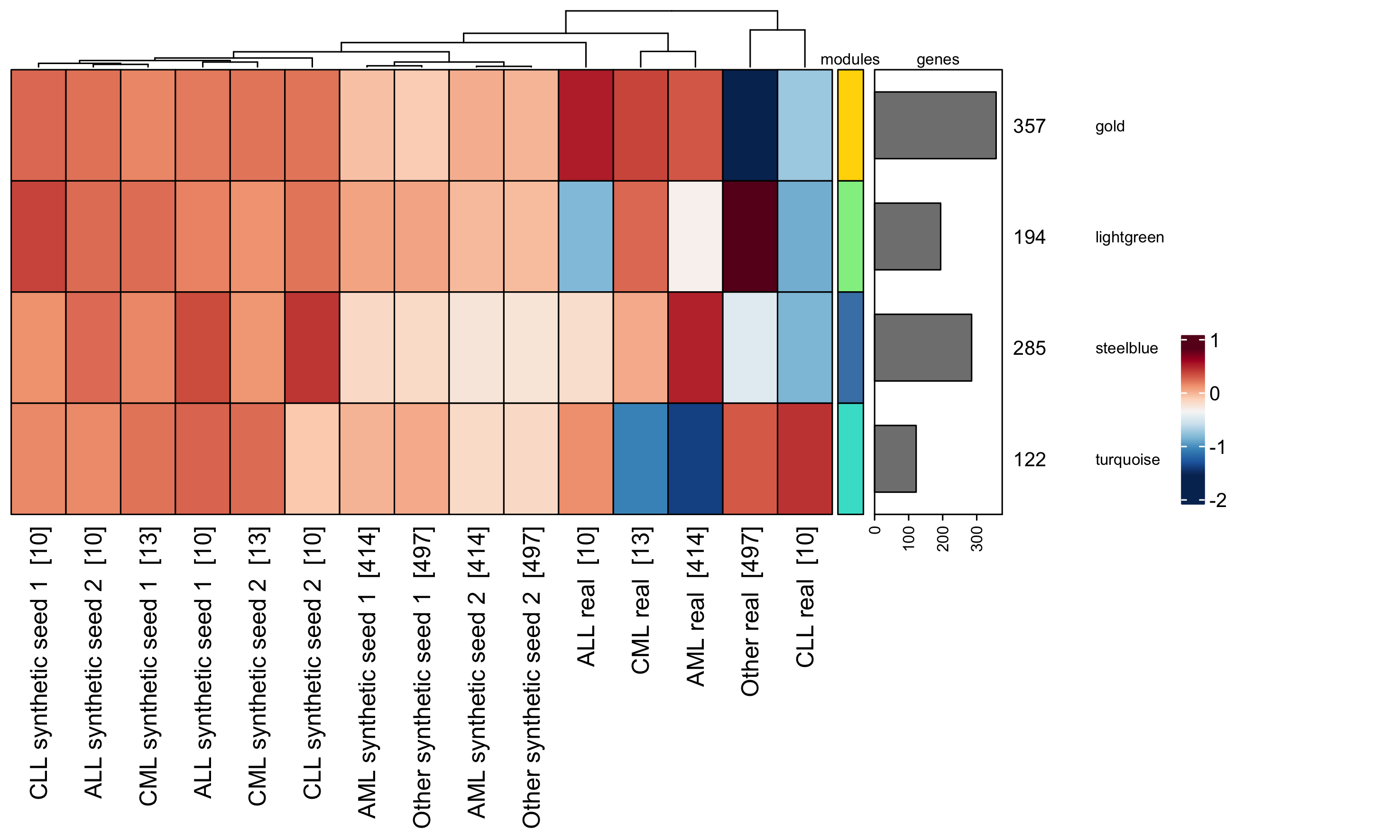}
\end{minipage}
}
\subfigure[$\varepsilon=10$, seed 2]{
\begin{minipage}[b]{\figwidth} \includegraphics[width=1.0\textwidth,trim={0 0 12cm 0},clip]{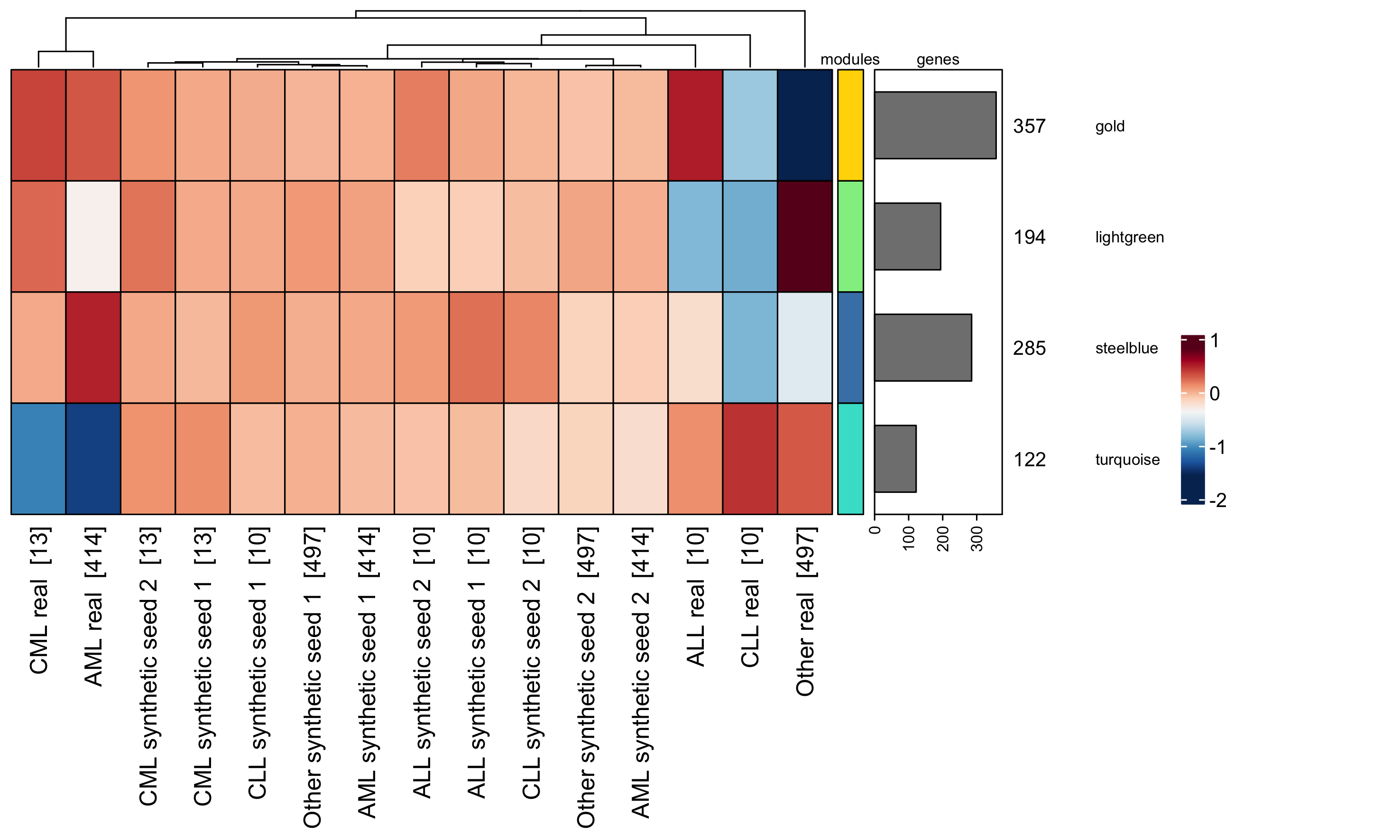}
\end{minipage}
}
\subfigure[$\varepsilon=20$, seed 2]{
\begin{minipage}[b]{\figwidth} \includegraphics[width=1.0\textwidth,trim={0 0 12cm 0},clip]{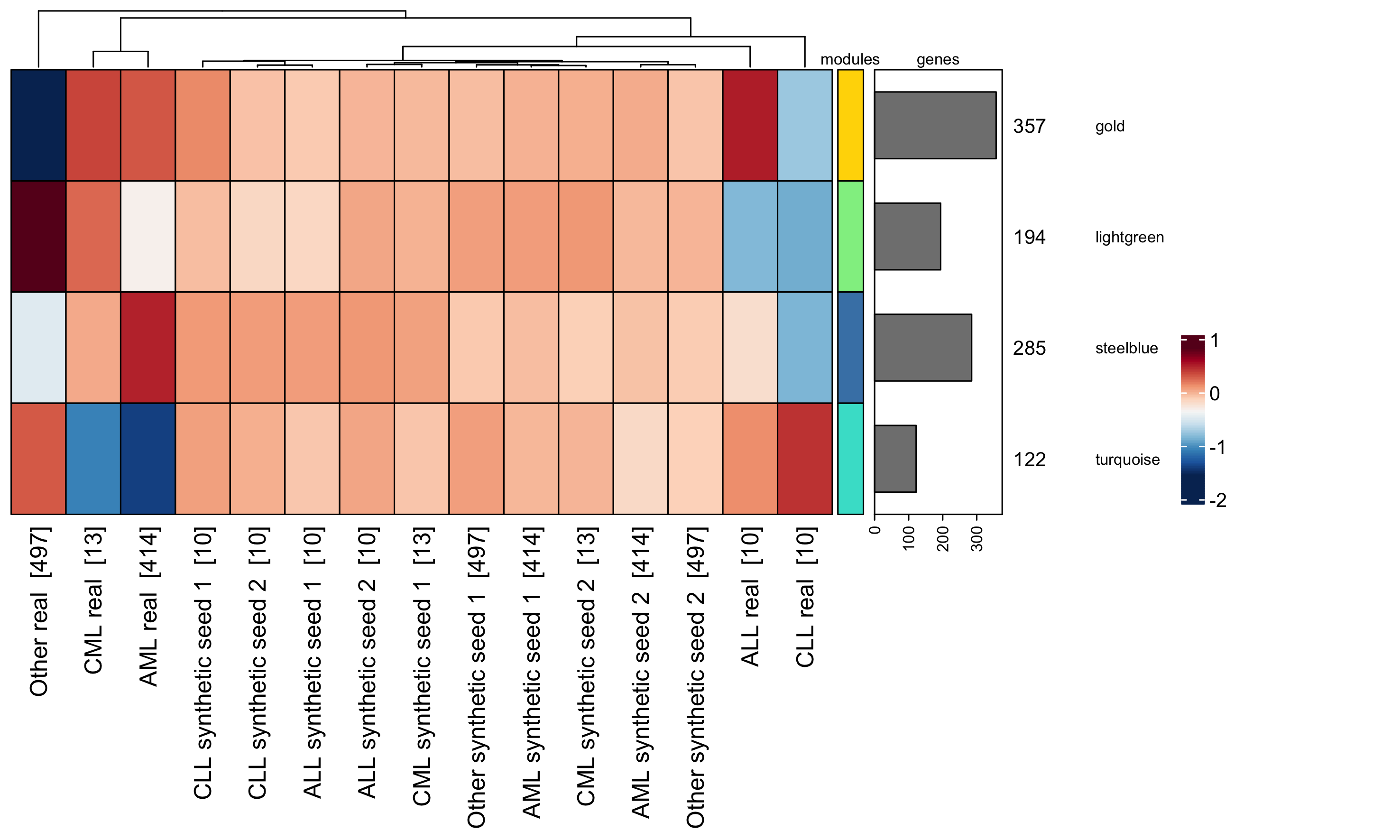}
\end{minipage}
}
\subfigure[$\varepsilon=50$, seed 2]{
\begin{minipage}[b]{\figwidth} \includegraphics[width=1.0\textwidth,trim={0 0 12cm 0},clip]{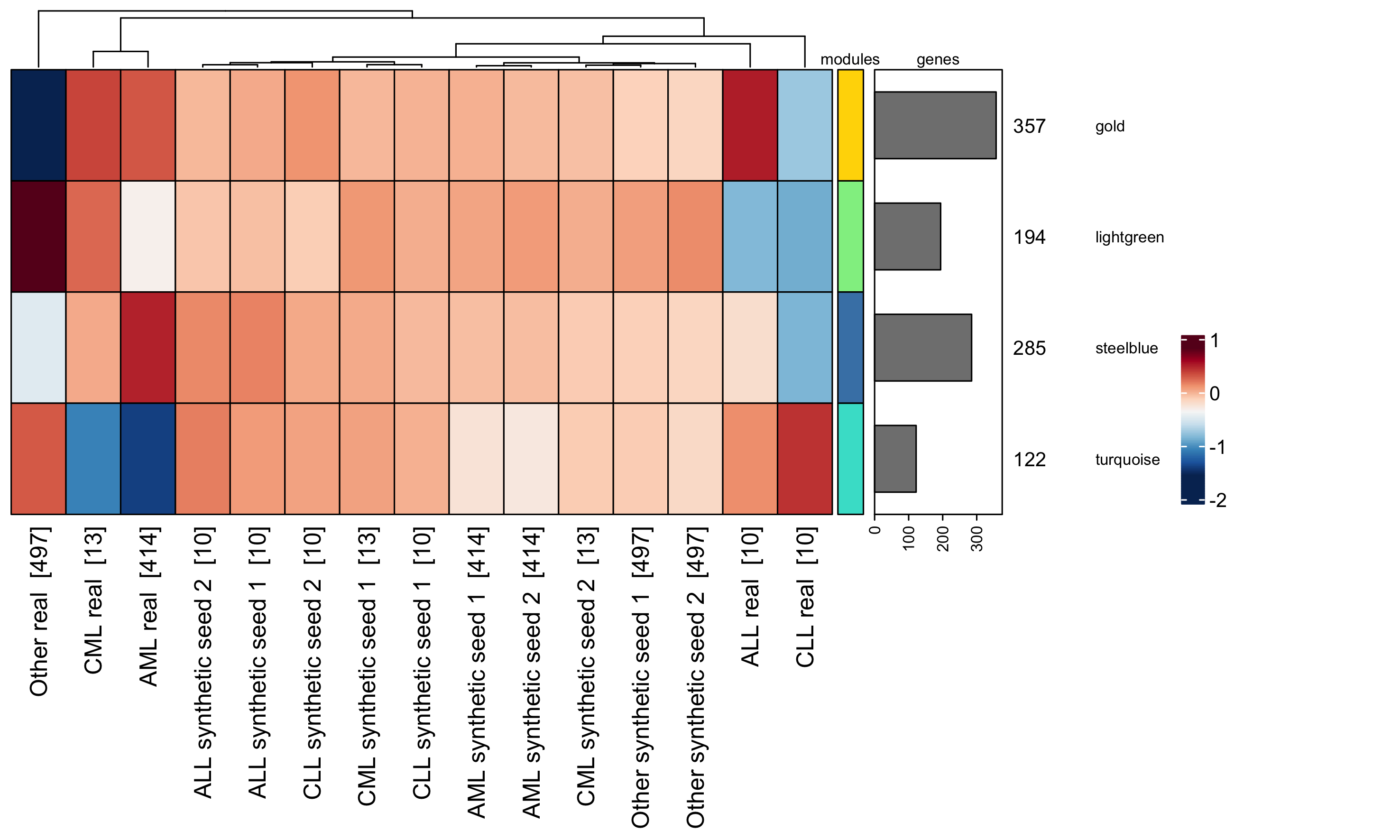}
\end{minipage}
}
\subfigure[$\varepsilon=100$, seed 2]{
\begin{minipage}[b]{\figwidth} \includegraphics[width=1.0\textwidth,trim={0 0 12cm 0},clip]{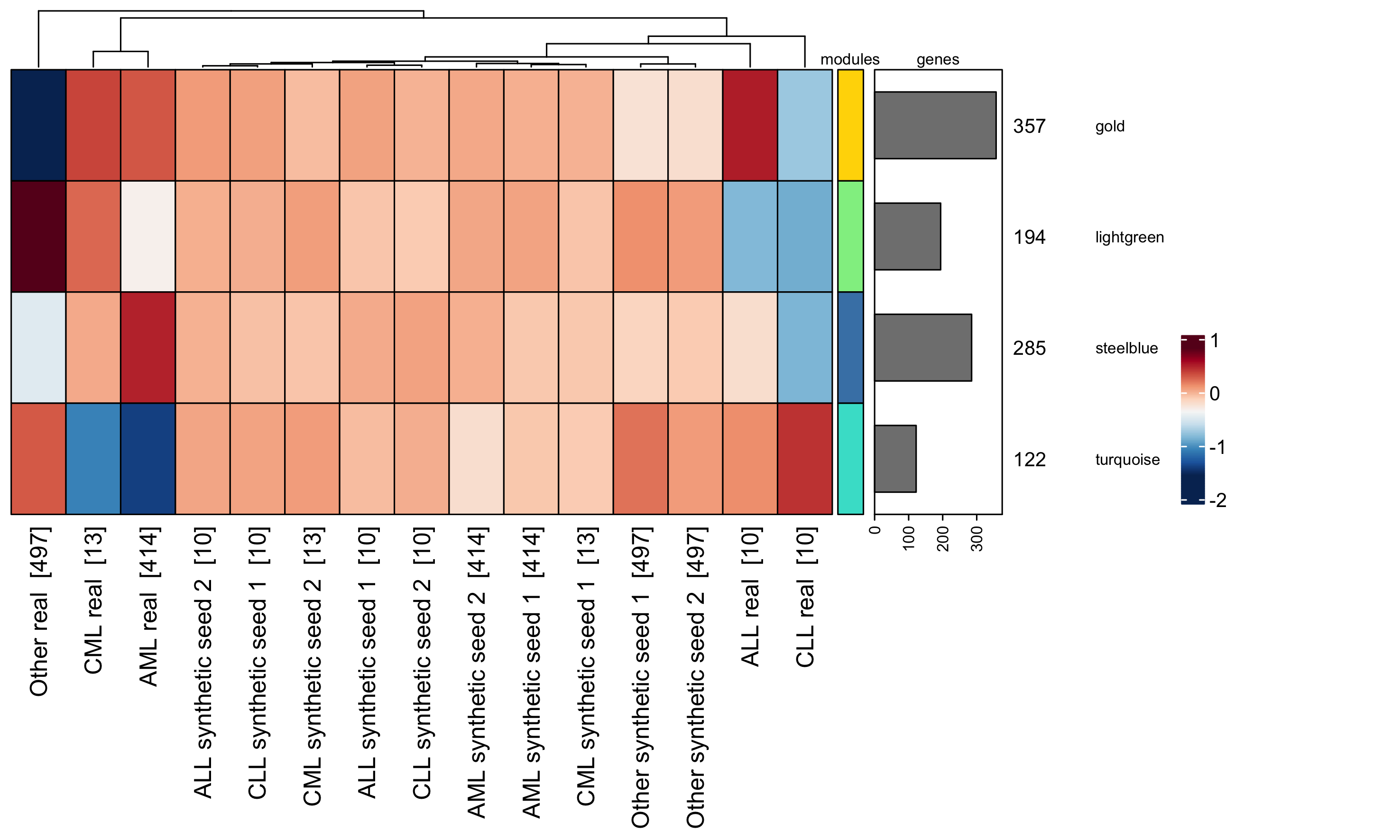}
\end{minipage}
}
\subfigure[non-priv, seed 2]{
\begin{minipage}[b]{\figwidth} \includegraphics[width=1.0\textwidth,trim={0 0 12cm 0},clip]{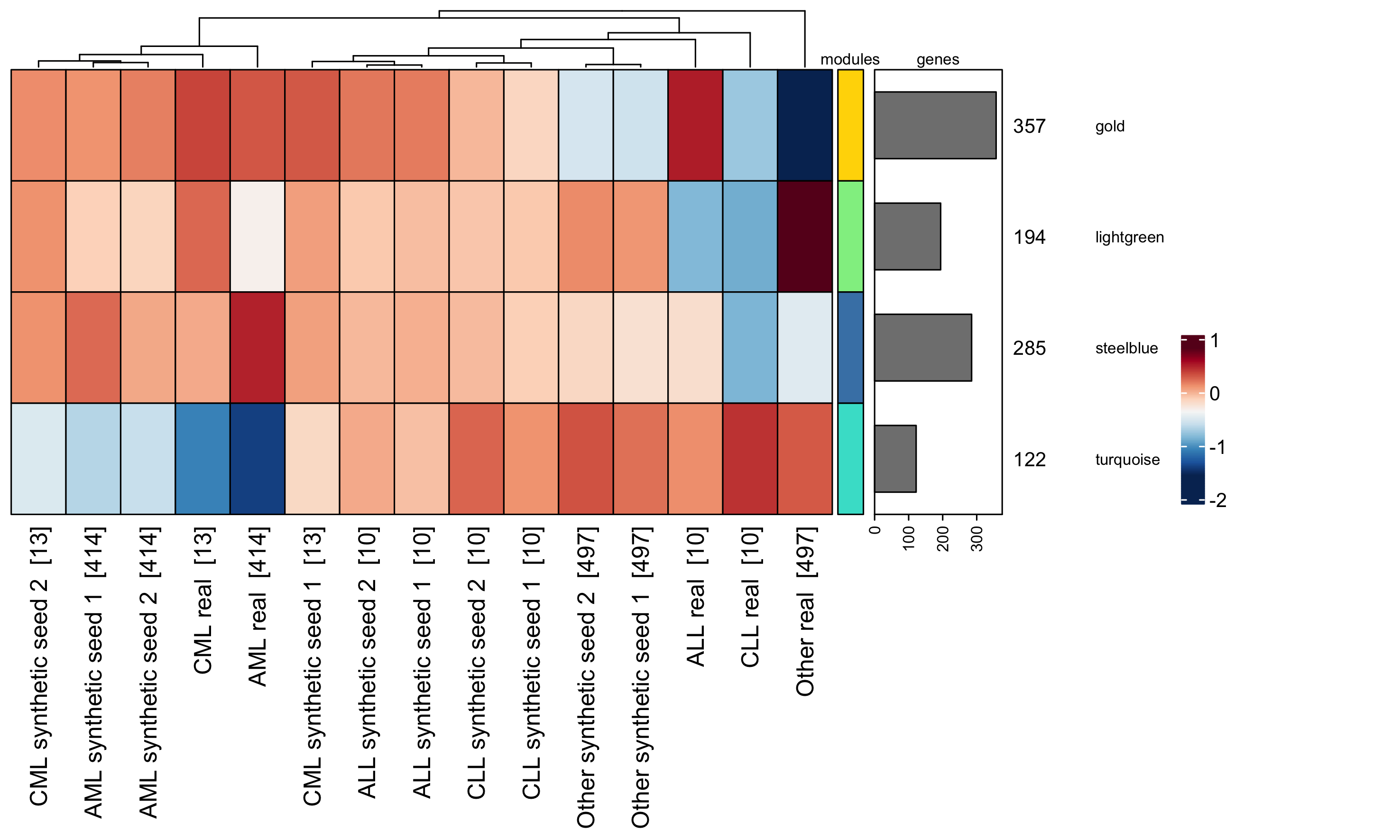}
\end{minipage}
}
\caption{Activation patterns of co-expressed gene modules in \rongauss after filtering co-expressions for $r$ > 0. \textmd{Shown are the Group Fold Changes (GFCs) of gene modules (rows) in the real and the synthetic data sampled with two different seeds. Numbers on the right indicate the number of genes per module. Darker shades of red imply activation of the gene module, while darker shades of blue indicate deactivation. The dendrograms show the hierarchical clustering of the classes in the different data sets. A heatmap is shown for each $\varepsilon$ twice, once for each seed used to \emph{split} the training data. Each heatmap further features, in addition to the real data, data from two synthetic sets, one for each seed used to \textit{generate} the data. Already in the non-private setting, the synthetic data exhibits mostly homogeneous activation of the different gene modules, maintaining almost none of the structure present in the real data.}}
\label{figure:S5}
\end{figure*}

\clearpage
\subsection{$r>0.7$}
\begin{figure*}[!htbp]
\centering
\newcommand{\figwidth}{0.9\textwidth}

\subfigure[Co-Expression Preservation]{
\begin{minipage}[b]{\figwidth} \includegraphics[width=1.0\textwidth]{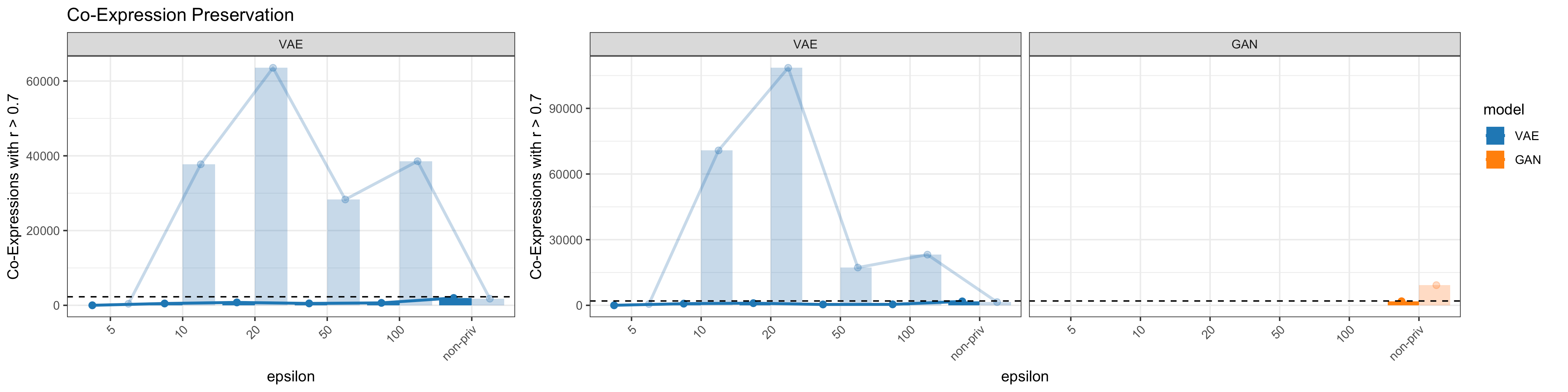}
\end{minipage}
}

\caption{Biological Evaluation by Co-Expression Preservation for $r$ > 0.7. \textmd{Shown is the co-expression preservation across the tested models for different values of $\varepsilon$ as well as the non-private case for two different seeds used for creating the training split (left and right plot). Note that in the first split (left) only the VAE model was capable of generating significant co-expressions with $r$ > 0.7, while in the second seed (right) also the GAN trained without DP yielded some co-expressions. Non-transparent bars give the number of correctly reconstructed co-expressions with Pearson Correlation Coefficient $r$ > 0.7 and an associated p-value < 0.05, while semi-transparent bars give the number of co-expressions introduced by the model that did not exist in the real data. The dashed black line indicates the number of co-expressions in the real data. All values shown are means across two different seeds set for generating the data. In the VAE, co-expressions that were falsely introduced in the synthetic data strongly outweigh the correctly reconstructed ones, while the GAN struggles to produce any co-expressions above 0.7, regardless of correct or incorrect.}}
\label{figure:S6}
\end{figure*}

\vspace{20pt}
\begin{figure*}[!htbp]
\centering
\newcommand{\figwidth}{0.5\textwidth}
\subfigure[non-priv, seed 2]{
\begin{minipage}[b]{\figwidth} \includegraphics[width=1.0\textwidth]{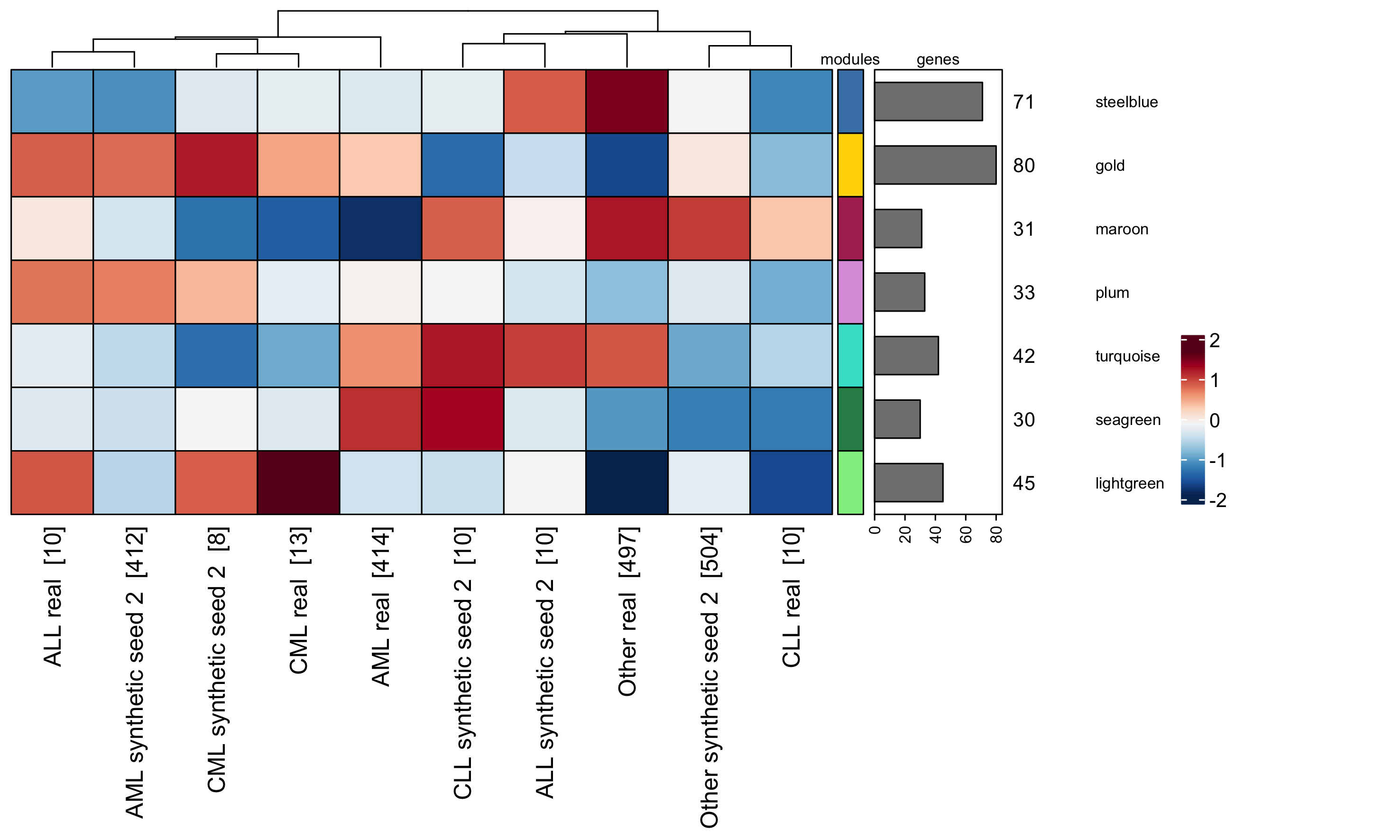}
\end{minipage}
}
\caption{Activation patterns of co-expressed gene modules in GAN after filtering co-expressions for $r$ > 0.7. \textmd{Shown are the Group Fold Changes (GFCs) of gene modules (rows) in the real and the synthetic data sampled with two different seeds. Numbers on the right indicate the number of genes per module. Darker shades of red imply activation of the gene module, while darker shades of blue indicate deactivation. The dendrograms show the hierarchical clustering of the classes in the different data sets. A heatmap is shown for each $\varepsilon$ twice, once for each seed used to \emph{split} the training data. Each heatmap further features, in addition to the real data, data from two synthetic sets, one for each seed used to \textit{generate} the data. The activation patterns of co-expression modules are poorly maintained in the synthetic data, indicated by the incorrect clustering of disease classes in the real and synthetic data. }}
\label{figure:S7}
\end{figure*}

\begin{figure*}[!htbp]
\centering
\newcommand{\figwidth}{0.32\textwidth}
\subfigure[$\varepsilon=5$, seed 1]{
\begin{minipage}[b]{\figwidth} \includegraphics[width=1.0\textwidth,trim={0 0 12cm 0},clip]{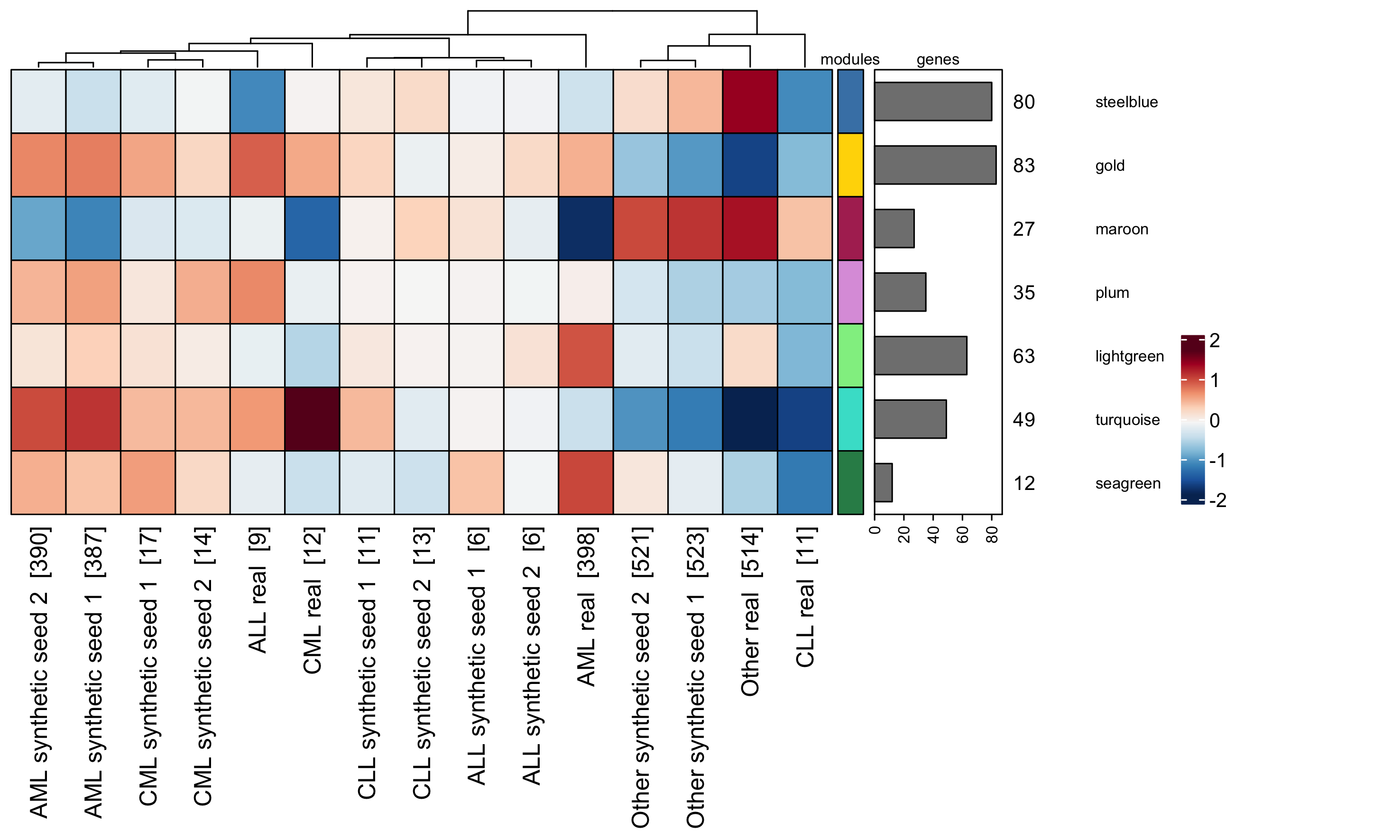}
\end{minipage}
}
\subfigure[$\varepsilon=10$, seed 1]{
\begin{minipage}[b]{\figwidth} \includegraphics[width=1.0\textwidth,trim={0 0 12cm 0},clip]{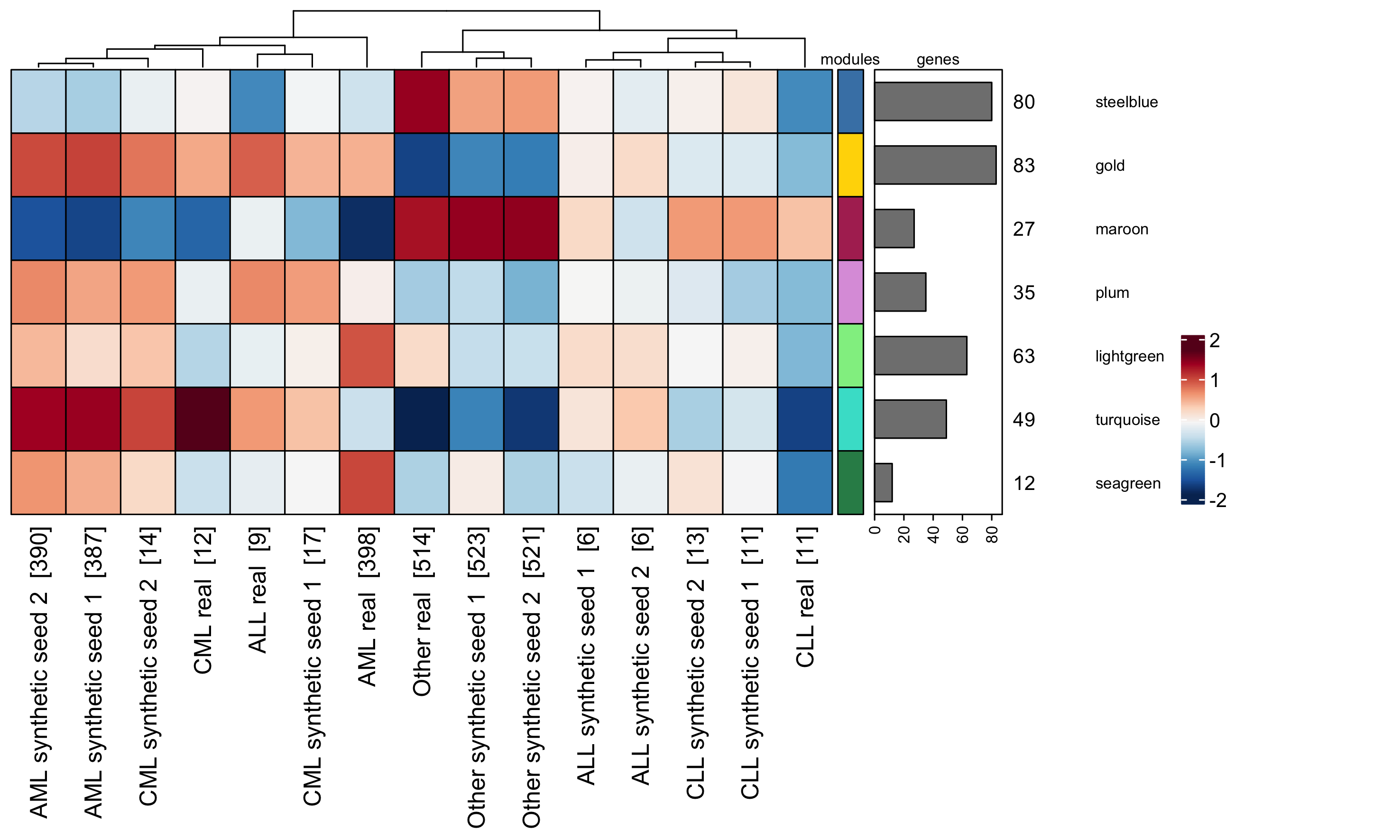}
\end{minipage}
}
\subfigure[$\varepsilon=20$, seed 1]{
\begin{minipage}[b]{\figwidth} \includegraphics[width=1.0\textwidth,trim={0 0 12cm 0},clip]{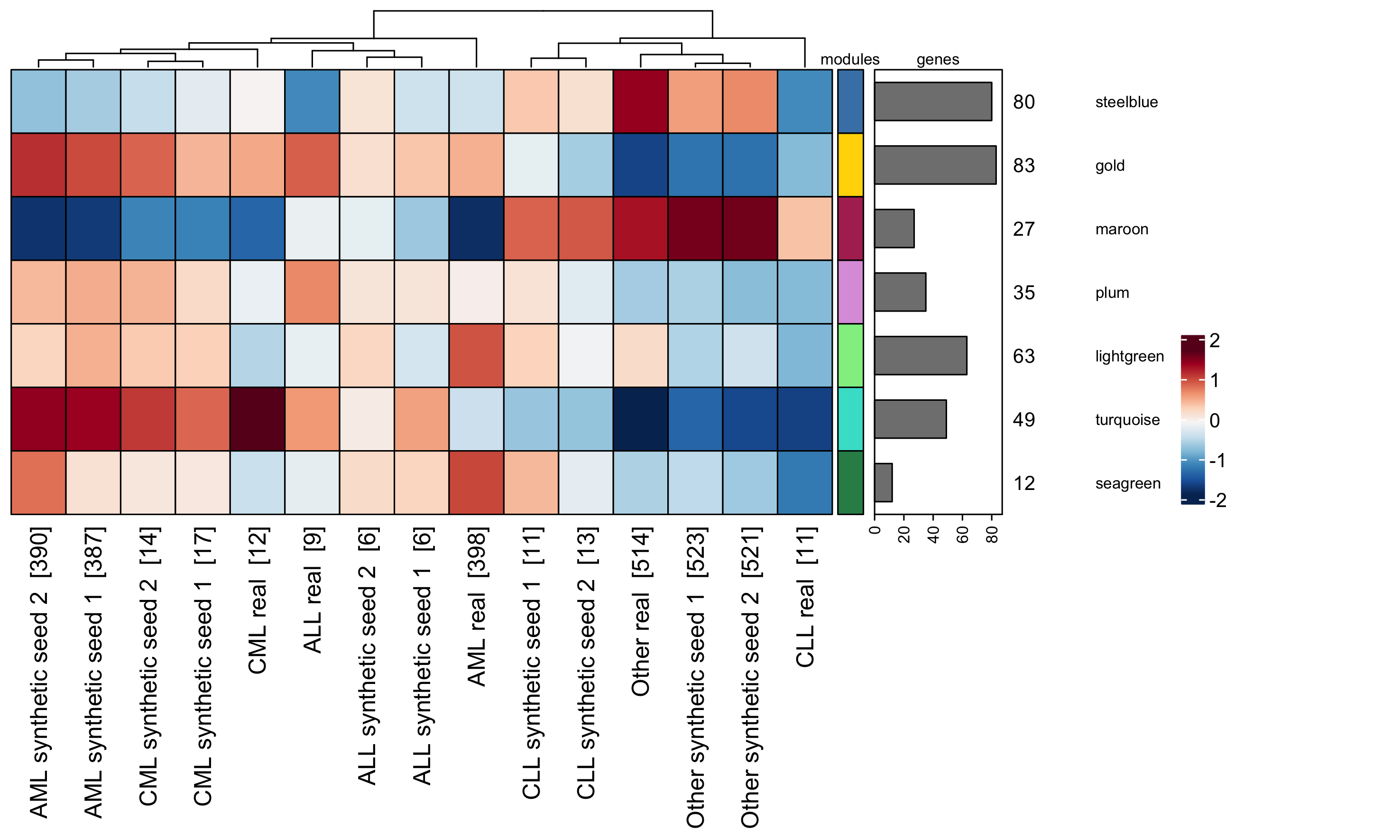}
\end{minipage}
}
\subfigure[$\varepsilon=50$, seed 1]{
\begin{minipage}[b]{\figwidth} \includegraphics[width=1.0\textwidth,trim={0 0 12cm 0},clip]{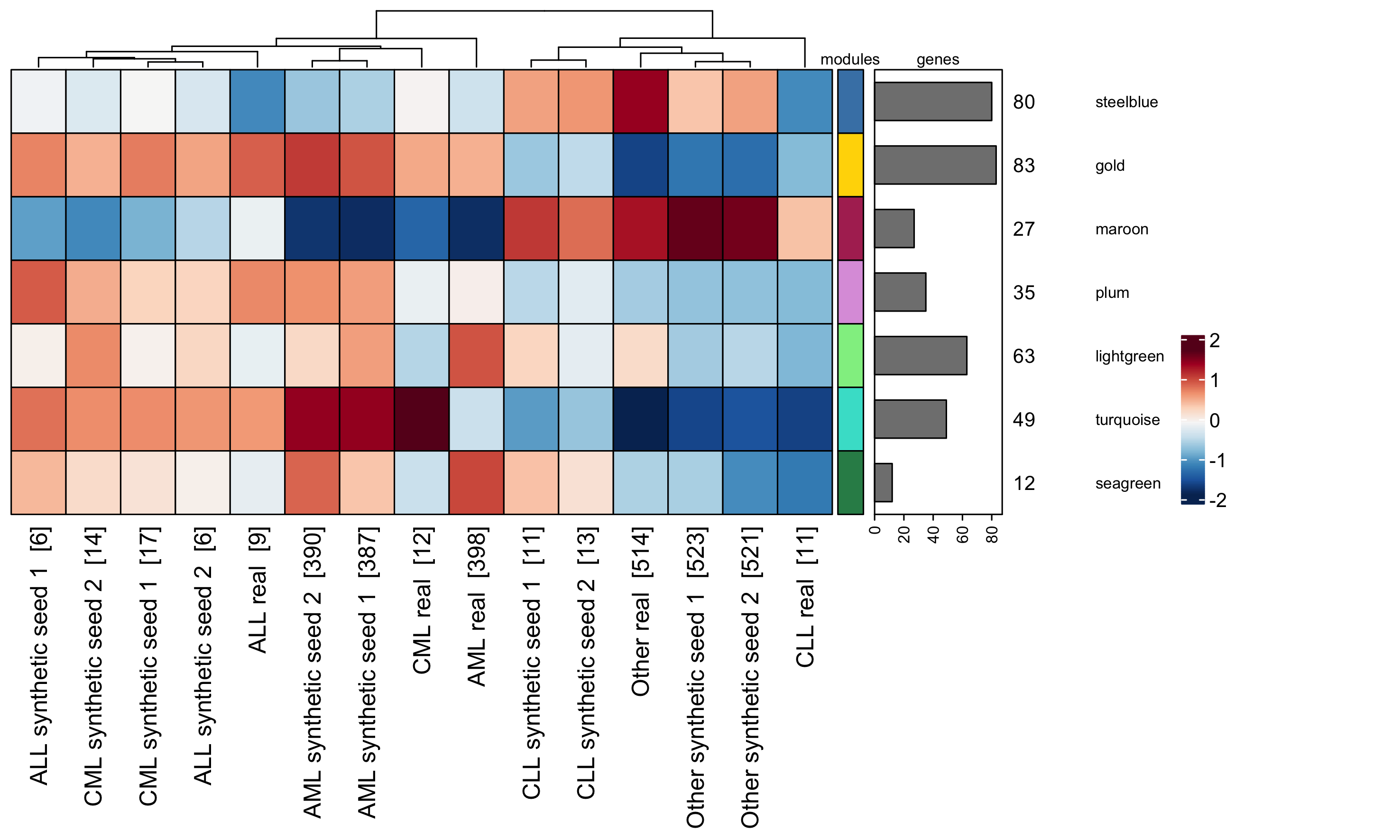}
\end{minipage}
}
\subfigure[$\varepsilon=100$, seed 1]{
\begin{minipage}[b]{\figwidth} \includegraphics[width=1.0\textwidth,trim={0 0 12cm 0},clip]{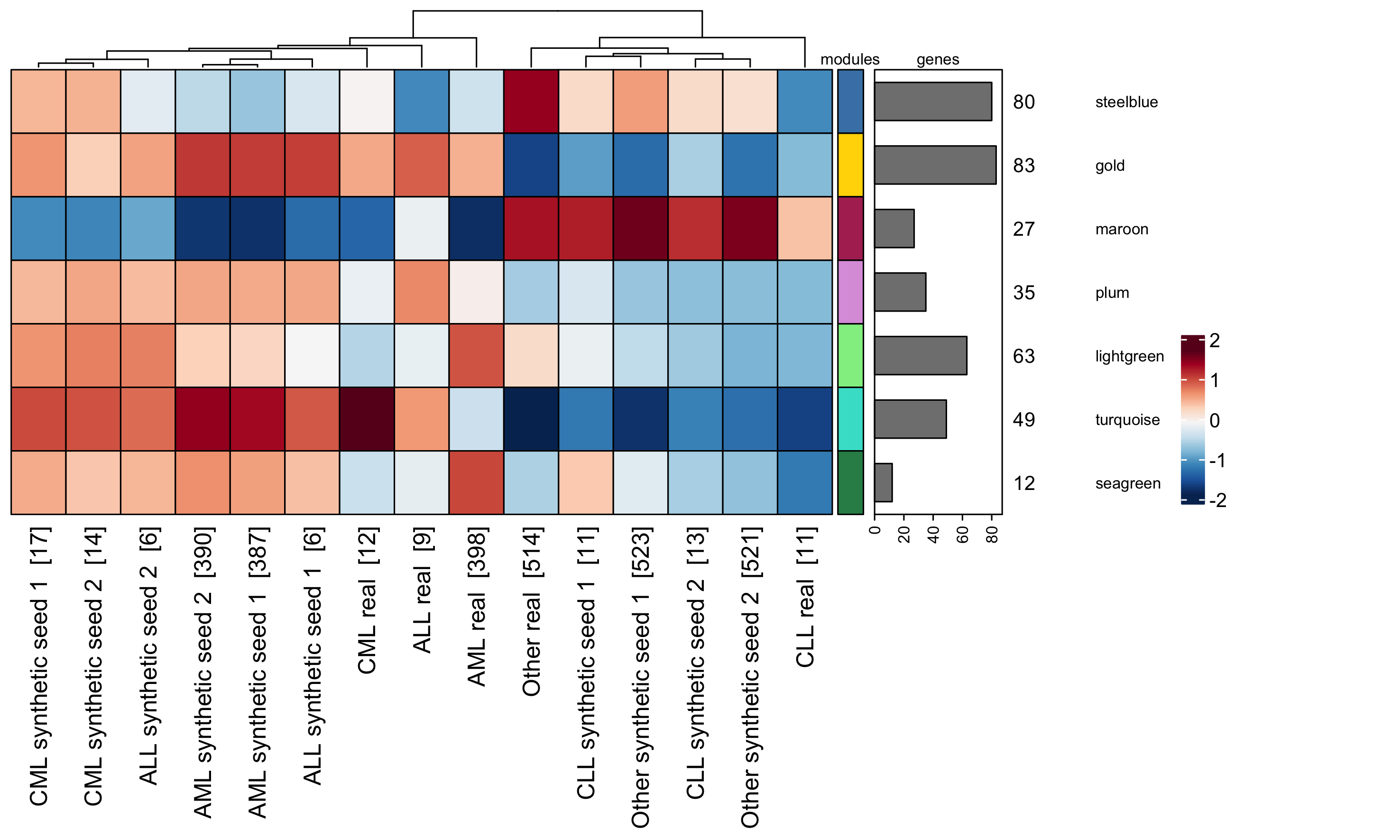}
\end{minipage}
}
\subfigure[non-priv, seed 1]{
\begin{minipage}[b]{\figwidth} \includegraphics[width=1.0\textwidth,trim={0 0 12cm 0},clip]{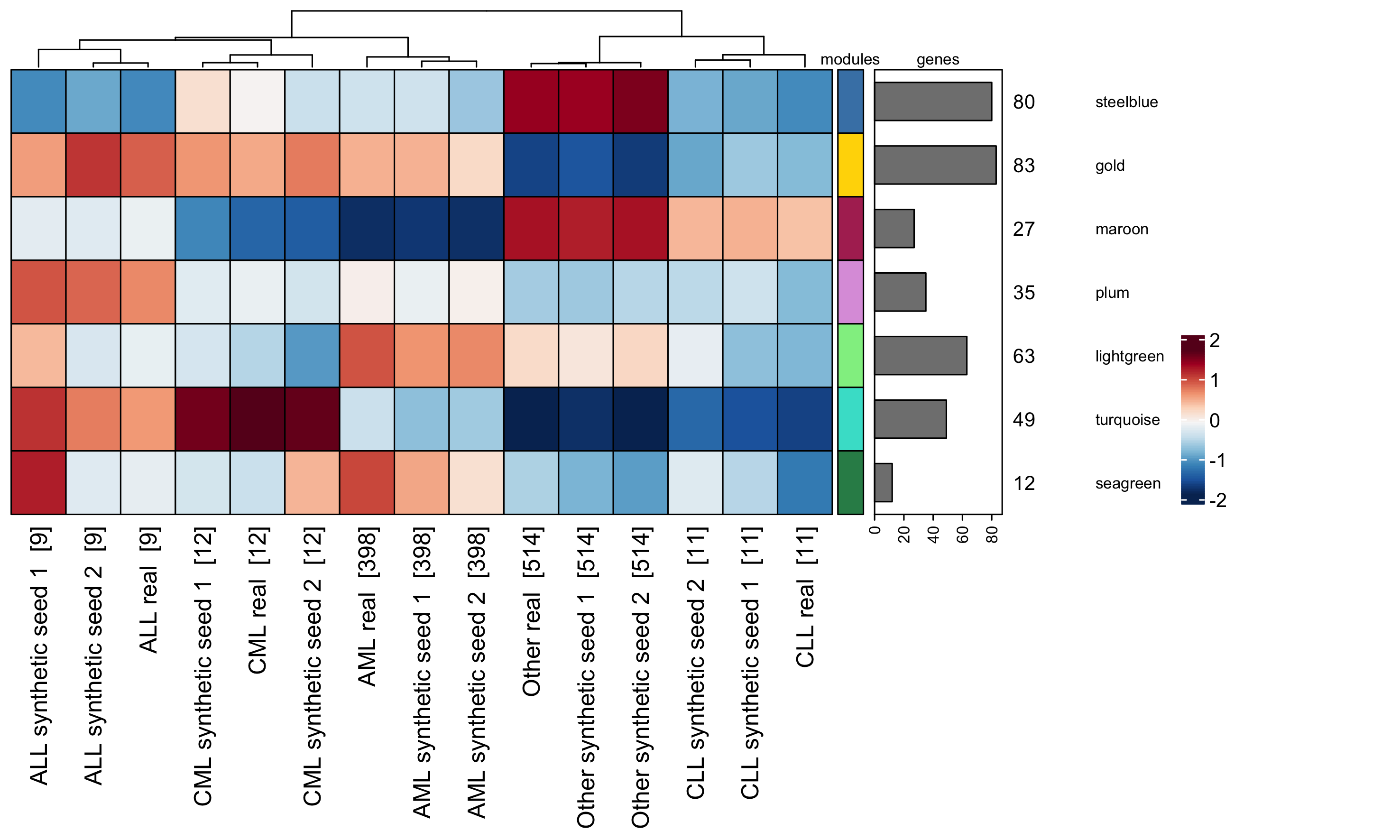}
\end{minipage}
}
\subfigure[$\varepsilon=5$, seed 2]{
\begin{minipage}[b]{\figwidth} \includegraphics[width=1.0\textwidth,trim={0 0 12cm 0},clip]{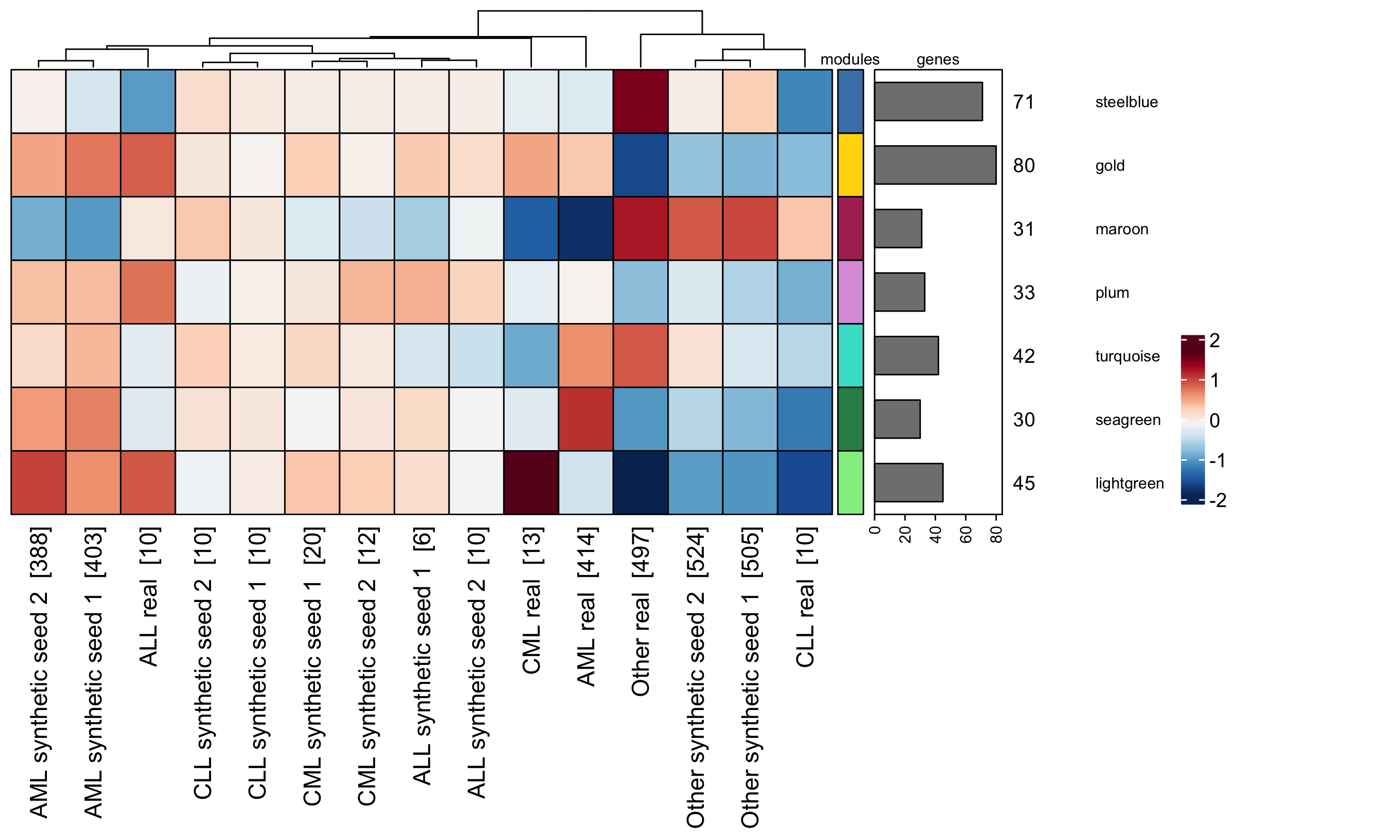}
\end{minipage}
}
\subfigure[$\varepsilon=10$, seed 2]{
\begin{minipage}[b]{\figwidth} \includegraphics[width=1.0\textwidth,trim={0 0 12cm 0},clip]{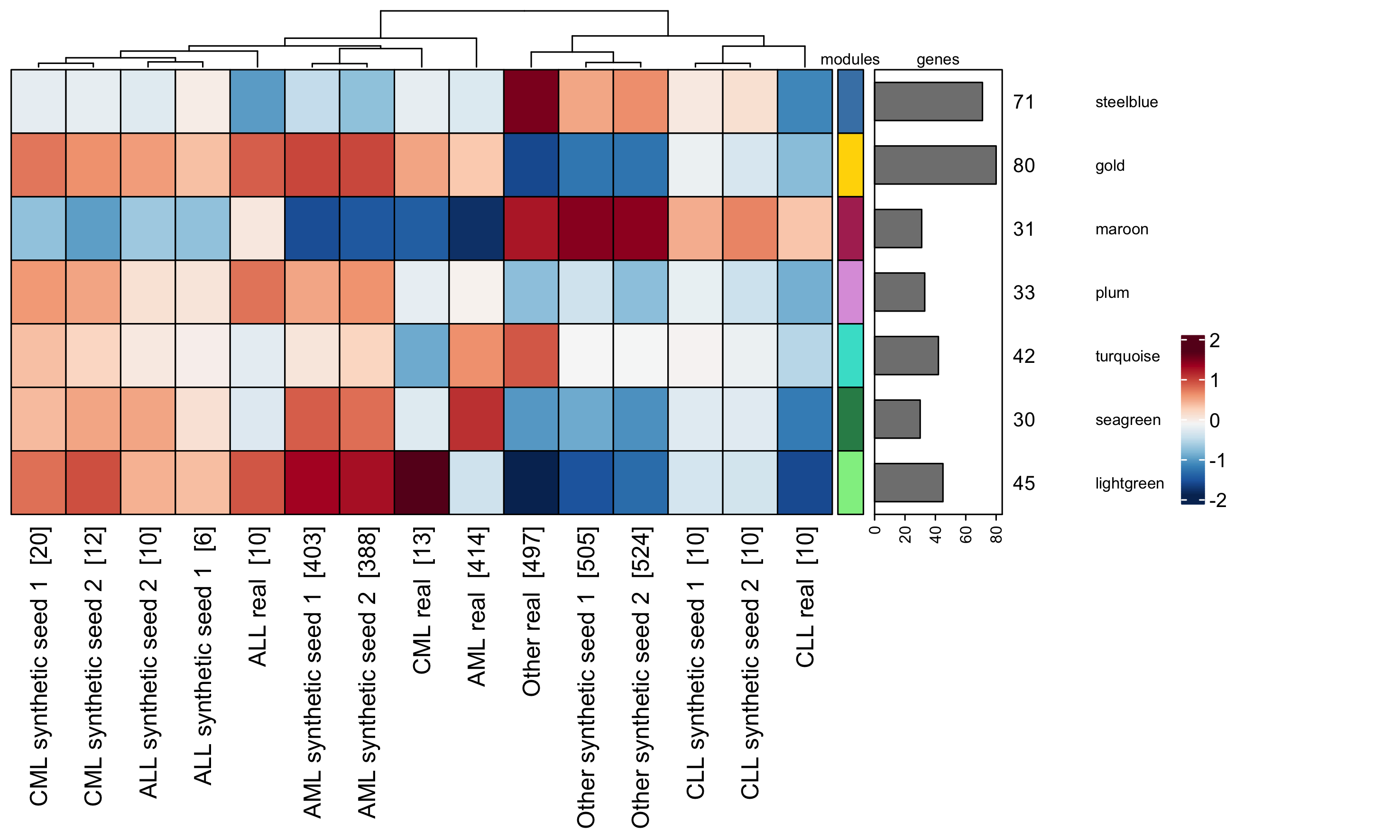}
\end{minipage}
}
\subfigure[$\varepsilon=20$, seed 2]{
\begin{minipage}[b]{\figwidth} \includegraphics[width=1.0\textwidth,trim={0 0 12cm 0},clip]{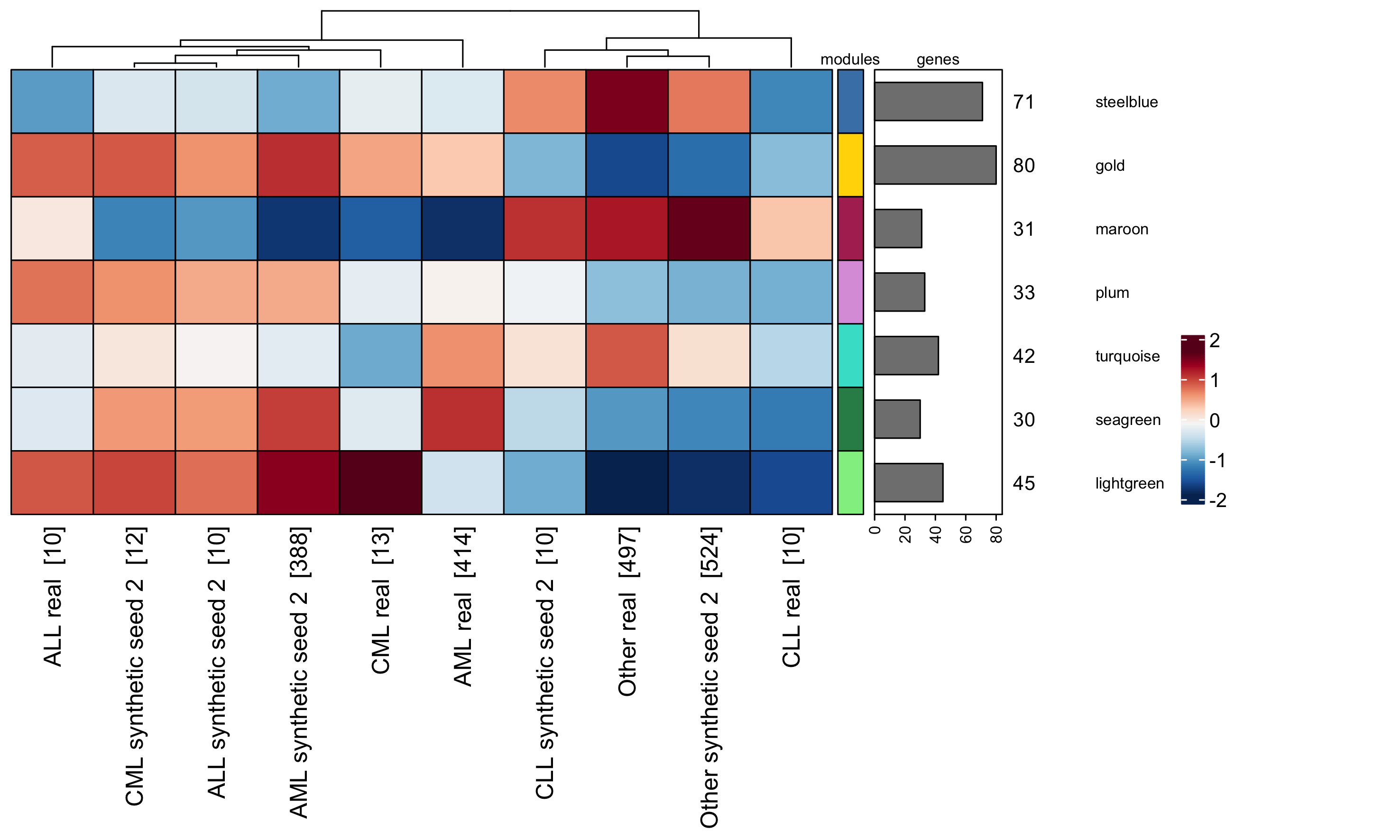}
\end{minipage}
}
\subfigure[$\varepsilon=50$, seed 2]{
\begin{minipage}[b]{\figwidth} \includegraphics[width=1.0\textwidth,trim={0 0 12cm 0},clip]{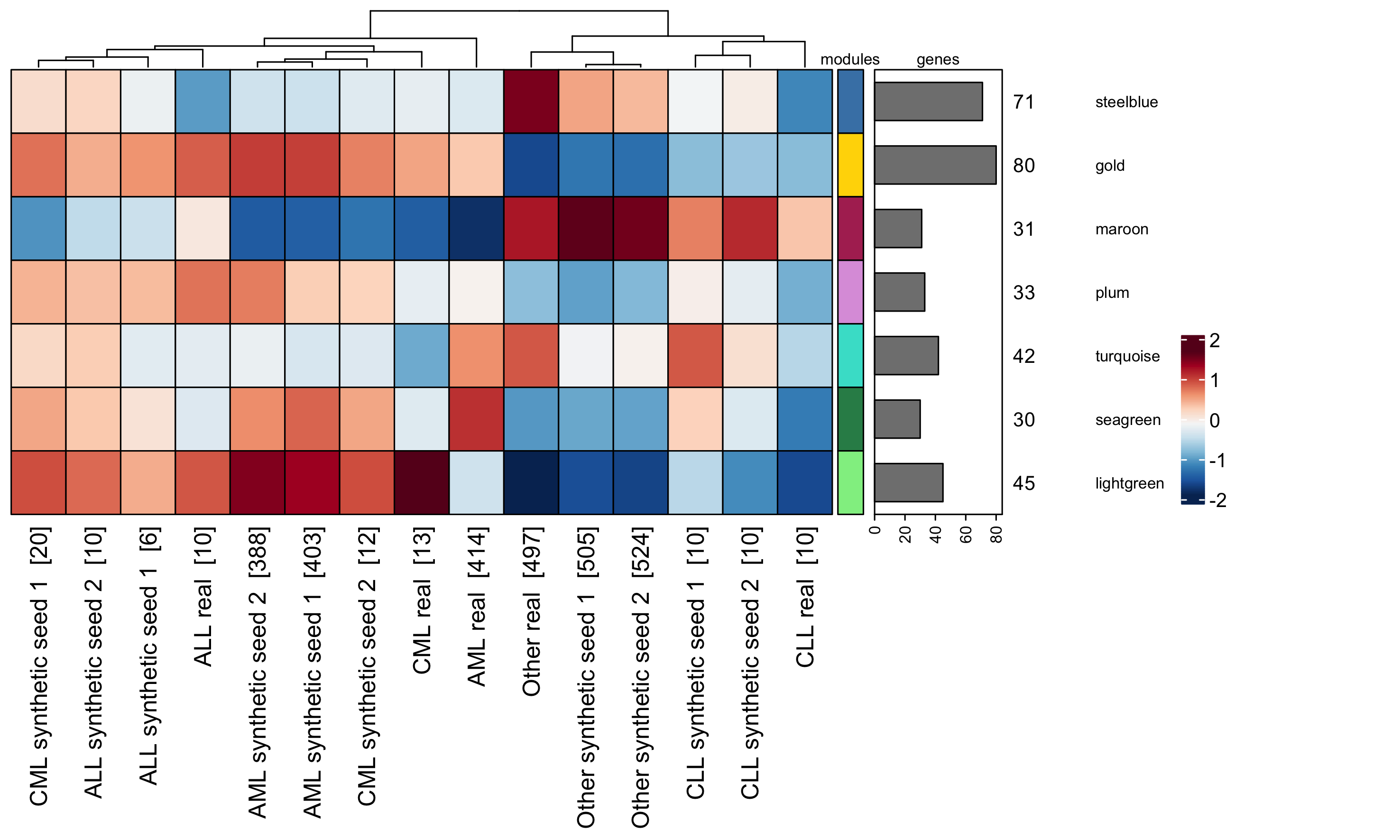}
\end{minipage}
}
\subfigure[$\varepsilon=100$, seed 2]{
\begin{minipage}[b]{\figwidth} \includegraphics[width=1.0\textwidth,trim={0 0 12cm 0},clip]{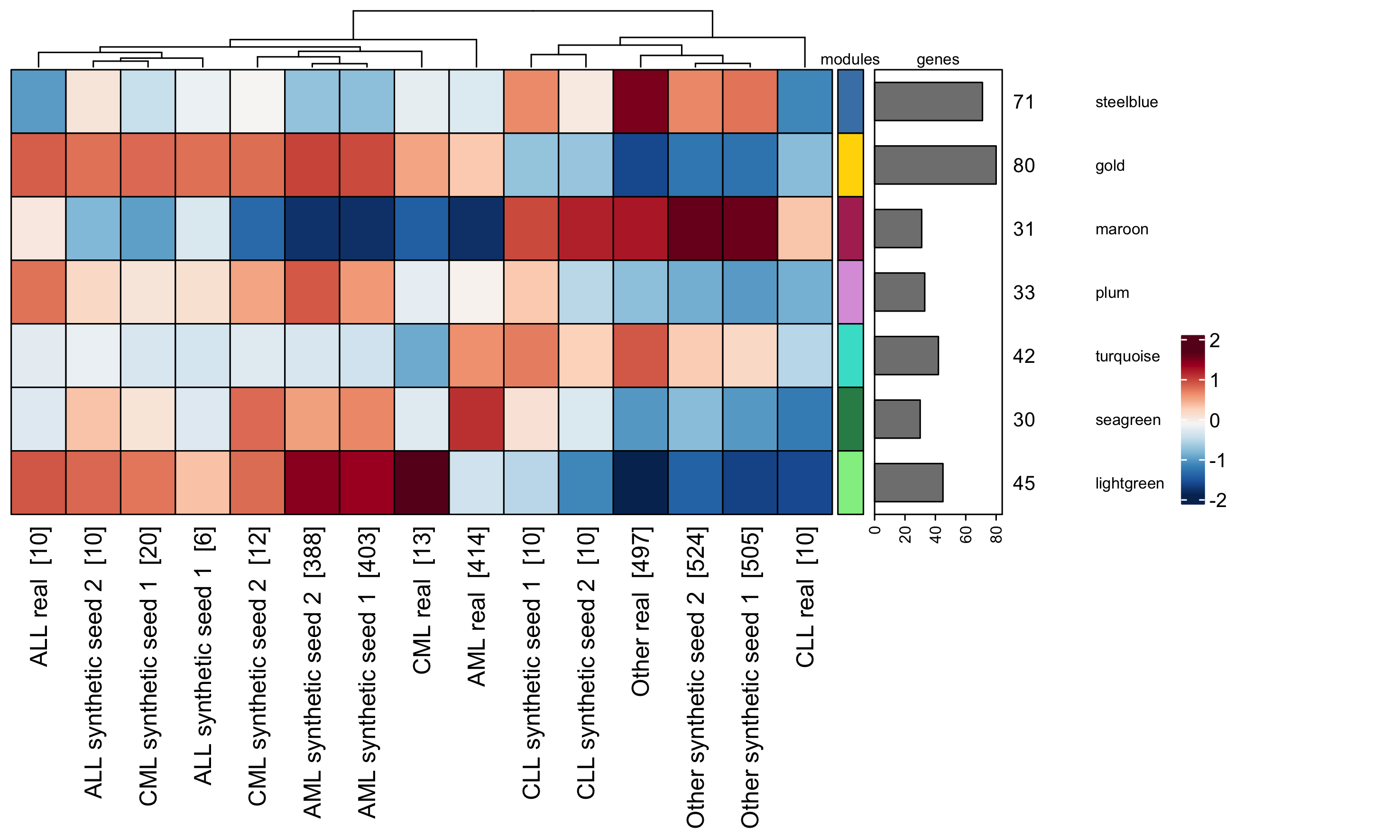}
\end{minipage}
}
\subfigure[non-priv, seed 2]{
\begin{minipage}[b]{\figwidth} \includegraphics[width=1.0\textwidth,trim={0 0 12cm 0},clip]{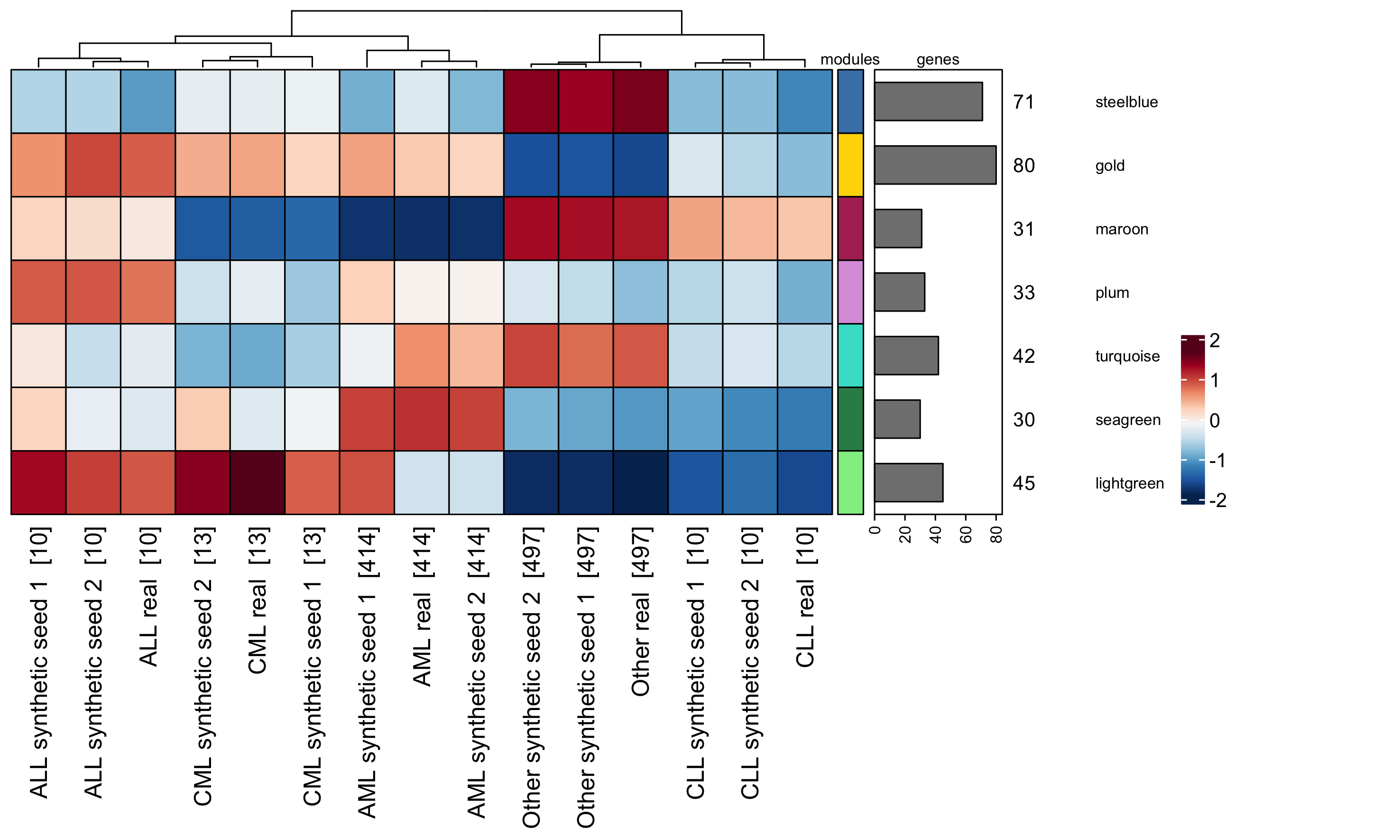}
\end{minipage}
}
\caption{Activation patterns of co-expressed gene modules in VAE after filtering co-expressions for $r$ > 0.7. \textmd{Shown are the Group Fold Changes (GFCs) of gene modules (rows) in the real and the synthetic data sampled with two different seeds. Numbers on the right indicate the number of genes per module. Darker shades of red imply activation of the gene module, while darker shades of blue indicate deactivation. The dendrograms show the hierarchical clustering of the classes in the different data sets. A heatmap is shown for each $\varepsilon$ twice, once for each seed used to \emph{split} the training data. Each heatmap further features, in addition to the real data, data from two synthetic sets, one for each seed used to \textit{generate} the data. Note that only one data generation seed yielded any co-expressions above > 0.7 in case of $\varepsilon$ = 20, data split seed 2. A general fading of module activation can be observed for decreasing privacy budgets, indicating poor reconstruction of module activation patterns in the synthetic data.}}
\label{figure:S8}
\end{figure*}

\clearpage

\end{document}